\documentclass[preprintnumbers,amsmath,amssymb,prd,superscriptaddress,notitlepage] {revtex4-1}
\pdfoutput=1

\usepackage{graphicx}
\usepackage{dcolumn}
\usepackage{bm}
\usepackage{subfigure}
\usepackage{longtable}

\newcommand{\fNL}{f_{\rm NL}}

\newcommand{\hMpc}{\ h^{-1}\text{Mpc}}

\newcommand{\ihMpc}{\ h\text{Mpc}^{-1}}
\newcommand{\iMpc}{\ \text{Mpc}^{-1}}

\newcommand{\be}{\begin{equation}}
\newcommand{\ee}{\end{equation}}

\newcommand{\kms}{{\rm km~s^{-1}}}

\newcommand{\summnu}{\Sigma m_\nu}

\newcommand{\kmaxeff}{k_{\rm max,eff}}

\newcommand{\vtheta}{\mathbf{\theta}}

\newcommand{\vk}{\mathbf{k}}

\newcommand{\vC}{\mathbf{C}}

\newcommand{\lya}{Ly$\alpha$}
\newcommand{\lyb}{Ly$\beta$}
\newcommand{\lyaf}{Ly$\alpha$ forest}

\newcommand{\mnras}{{\em Mon. Not. Roy. Astron. Soc. }}
\newcommand{\apjl}{{\em Astrophys. J. Let. }}
\newcommand{\apjs}{ApJS}
\newcommand{\jcap}{JCAP}

\newcommand{\aap}{{\em Astron. Astrophys. }}
\newcommand{\aj}{AJ }

\def\prd{{\em Phys. Rev. }{\bf D }}

\begin{document}

\title{DESI and other Dark Energy experiments in the era of neutrino mass
measurements} 

\author{Andreu Font-Ribera\footnote{author list alphabetized}}
\email{afont@lbl.gov}
\affiliation{Institute of Theoretical Physics, University of Zurich,
        Winterthurerstrasse 190, 8057 Zurich, Switzerland}
\affiliation{Lawrence Berkeley National Laboratory,
        1 Cyclotron Road, Berkeley, CA 94720, USA}
\author{Patrick McDonald}
\email{PVMcDonald@lbl.gov}
\affiliation{Lawrence Berkeley National Laboratory, 1 Cyclotron Road, Berkeley,
             CA 94720, USA}
\author{Nick Mostek}
\affiliation{Lawrence Berkeley National Laboratory, 1 Cyclotron Road, Berkeley,
             CA 94720, USA}
\author{Beth A. Reid}
\email{BAReid@lbl.gov}
\affiliation{Lawrence Berkeley National Laboratory, 1 Cyclotron Road, Berkeley,
             CA 94720, USA}
\author{Hee-Jong Seo}
\email{hee-jongseo@lbl.gov}
\affiliation{Lawrence Berkeley National Laboratory, 1 Cyclotron Road, Berkeley,
             CA 94720, USA}
\author{An\v{z}e Slosar}
\email{anze@bnl.gov}
\affiliation{Brookhaven National Laboratory, Upton, NY 11973, USA}

\date{\today}

\begin{abstract}

We present Fisher matrix projections for future cosmological parameter 
measurements, including neutrino masses, Dark Energy, curvature, modified 
gravity, the inflationary perturbation spectrum, non-Gaussianity, and dark 
radiation. We focus on DESI and generally redshift surveys (BOSS, HETDEX, 
eBOSS, Euclid, and WFIRST), but also include CMB (Planck) and weak 
gravitational lensing (DES and LSST) constraints.  The goal is to present a 
consistent set of projections, for concrete experiments, which are otherwise 
scattered throughout many papers and proposals.  We include neutrino mass as a 
free parameter in most projections, as it will inevitably be relevant -- DESI 
and other experiments can measure the sum of neutrino masses to $\sim 0.02$ eV 
or better, while the minimum possible sum is $\sim 0.06$ eV. 
We note that constraints on Dark Energy are significantly degraded by the 
presence of neutrino mass uncertainty, especially when using galaxy clustering  
only as a probe of the BAO distance scale (because this introduces additional 
uncertainty in the background evolution after the CMB epoch). Using broadband
galaxy power
becomes relatively more powerful, and bigger gains are achieved by combining 
lensing survey constraints with redshift survey constraints.
We do not try to be especially 
innovative, e.g., with complex treatments of potential systematic errors -- 
these projections are intended as a straightforward baseline for comparison to 
more detailed analyses.
 
\end{abstract}


\maketitle

\section{Introduction}

The Fisher matrix formalism is a standard tool for forecasting the
statistical ability of future experiments to measure cosmological
parameters of interest \cite{2006astro.ph..9591A,
  Bassett:2009uv,Cooray:1999rv, Giannantonio:2011ya,
  Huterer:2005ez,2012JCAP...11..052H,2012PhRvD..86b3526J,
  Kitching:2009yr,More:2012xa,2012arXiv1201.2434W}.  The Fisher matrix
is the expectation value of the second derivative matrix of log
likelihood with respect to parameters of interest
\begin{equation}
  F_{AB}= -\left< \frac{ \partial^2 \ln L}{\partial A \partial B} \right>,
\end{equation}
where $L$ is likelihood, $A$ and $B$ are parameters and the average is
over all possible realizations of the data assuming a certain fiducial
model. In the limit of Gaussian likelihood, the Fisher information matrix
can be thought of as the inverse of a typical covariance matrix. In
particular, in the limit of Gaussian likelihood surface, 
$F_{AA}^{-1/2}$ is the expected error on the parameter $A$ assuming
the values of all other parameters are known, while
$\sigma(A)=(F^{-1})_{AA}^{1/2}$ is the marginalized error on the
parameter $A$. The Cram\'{e}r-Rao bound also stipulates that no
unbiased estimator of $A$  performs better than the Fisher matrix
error $\sigma(A)$ and therefore Fisher information is
the lower limit on the error obtainable from a given data set using an
optimal estimator.  

Fisher matrix forecasts generally should not be taken literally for
poorly constrained parameters where the likelihood will often be
non-Gaussian \cite{Wolz:2012sr, Khedekar:2012sh,2013arXiv1304.2321B,
2013JCAP...06..020C}, but even then the
sense that the constraint is poor is generally preserved. When used
thoughtfully, Fisher matrices allow us to map out the progression of
parameter measurements we can expect from future experiments.
Unfortunately, results of these exercises are generally scattered
throughout the literature (or worse, non-public proposals), often with 
slightly different assumptions
and methodologies that make direct comparisons difficult.  The main
purpose of this paper is to make predictions for a suite of models
and planned experiments with a consistent set of assumptions.

We do not attempt to be particularly innovative, but
one thing we emphasize is that our basic cosmological model assumes that the 
neutrino mass is not
known. Since the effect of massive neutrinos is important given the
accuracy of the data we consider and since it is unlikely that it will be
measured in terrestrial experiments in the next decade, it is a
logical necessity to include it as a free parameter in all forecasts.

We also note that no projection method can ever give an unambiguously fair
comparison of experiments, as it is impossible to anticipate and model
all possible sources of systematic errors (see e.g. 
\cite{Pullen:2012rd,Ross:2012qm}). It is also difficult to
forecast our advances in theoretical problems in the next decade (for
example, our understanding of non-linear matter clustering,
redshift-space distortions and biasing of cosmic tracers for the large
scale structure probes; similar issues exist for all other
probes). Therefore, we sometimes give pessimistic and optimistic forecasts
(e.g., quoting BAO-only errors for a galaxy redshift survey is essentially 
a very pessimistic use of the survey).

The rest of this paper is structured as follows: In \S \ref{sec:params}
we define our cosmological parameters and fiducial model, and discuss general
methodology. 
In \S \ref{sec:cosm-micr-backgr} we discuss our 
Cosmic Microwave Background, i.e., Planck, treatment.
In \S \ref{sec:redshift-surveys} we discuss spectroscopic, i.e., redshift 
surveys like DESI.
In \S \ref{sec:photometric-surveys} we discuss photometric, i.e., 
gravitational lensing-oriented surveys.
In \S \ref{sec:results} we give the main results on parameter constraints. 
Finally in \S \ref{sec:conclusions} we give some discussion and conclusions.  
(In the Appendix we give some more traditional FoM numbers without free 
neutrino mass, and discuss the issue of overlapping lensing and redshift 
surveys.)  

This is intended to be a technical reference paper -- see, e.g., 
\cite{2012arXiv1201.2434W} for a much higher ratio of pedagogy to tables.

\section{Parameters, fiducial model and general methodology\label{sec:params}}

In \S \ref{sec:results} we give constraint projections for this baseline 
model first, and then several scenarios with added parameters.

\subsection{Baseline model parameters}

Our baseline model is flat $\Lambda$CDM with massive neutrinos.  This
model is specified by eight parameters which are listed, together with
their fiducial values, in Table \ref{tab:fid}.   
\begin{table}
  \caption{Parameterization of the cosmological model and
    parameter values for the fiducial model.  See text for discussion of
    symbol meanings. Eight
    parameters in the upper part of the table are always free (except when we
fix the neutrino mass in the Appendix). Parameters
    in the second half of the table are extensions of the simplest model
    discussed below.  }
  \label{tab:fid}
  \centering
  \begin{tabular}{ccp{9.5cm}}
    Parameter & Value & Description \\
\hline
$\omega_b$ & $0.02214$ & Physical baryon density $\omega_b=\Omega_b h^2$ \\
$\omega_m$ & $0.1414$  & Physical matter density $\omega_m=\Omega_m h^2$
(including neutrinos which are non-relativistic at $z=0$)\\
$\theta_s$ & $0.59680$ deg & Angular size of sound horizon at the
surface of last scattering (a substitute for, e.g., Hubble's constant) 
 \\
$\summnu$ & $0.06$ eV & Sum of neutrino masses (we assume they are degenerate)
\\
$A_s$ & $2.198\times 10^{-9}$ & Amplitude of the primordial power
spectrum at $k=0.05 \iMpc$ (for the numerical Fisher matrix we actually 
use $\log_{10}A_s$)
\\
$n_s$ &  $0.9608$ & Spectral index of primordial matter fluctuations, i.e., 
 $P_{\rm primordial}(k) \propto k^{n_s}$\\
$\tau$ & $0.092$ & Optical depth to the last scattering surface
assuming instantaneous reionization.\\
$T/S$ & $0$ & Ratio of tensor to scalar perturbations (we assume 
inflationary tensor fluctuation's spectral index $n_t=-\frac{1}{8}T/S$) \\
\hline
$w_0$ & $-1$ & pressure/density ratio for Dark Energy at the present time \\
$w^\prime$ & $0$ & Rate of change of Dark Energy equation of state in the 
formula $w(a)=w_0+(1-a)w^\prime$ \\
$\Omega_k$ & 0 & Curvature of the homogeneous model \\
$\Delta \gamma$ & 0 & Modification of the growth factor $d\ln D/d \ln
a = f_{\rm GR}(a) \Omega_m(a)^{\Delta \gamma}$\\
$G_9$ & 1 & Arbitrary normalization multiplier applied to the linear 
perturbations at $z<9$, i.e., all our observables except the CMB.\\
$\alpha_s$ & 0 & Running of the spectral index $\alpha_s=d \log n_s/d \log
k$ with pivot scale $k=0.05 \iMpc $ \\
$\fNL$ & 0 & normalization of local model quadratic non-Gaussianity of the 
initial perturbations \\
$N_\nu$ & 3.046 & Effective number of neutrino species
($N_\nu>3.046 \rightarrow$ dark radiation). \\
\hline
  \end{tabular}
\end{table}
Parameter symbols have their
conventional meanings. Capital $\Omega$s are densities
of various components expressed as a fraction of critical density
today. Small case $\omega$s correspond to physical density $\omega_x =
\Omega_x h^2$ (for the component $x$), where $h$ is the dimensionless reduced 
Hubble's
parameter today $h=H_0/(100 {\rm km~s^{-1} Mpc^{-1}})$. Matter density contains
contribution from baryons, dark matter and massive neutrinos 
\begin{equation}
\omega_m=\omega_b+\omega_{\rm cdm}+\omega_\nu
\end{equation}
(this is used only where massive neutrinos are non-relativistic).
Parameter $\theta_s$ is angular size of sound
horizon at the surface of last scattering, i.e.  
\begin{equation}
\theta_s = r(a_\star)^{-1} s,
\end{equation} where 
\begin{equation}
s=\int_0^{t_\star} \frac{c_s(t)}{a} dt
=\int_0^{a_\star} \frac{c_s(a)}{a^2 H(a)} da ~.
\label{eq:s}
\end{equation}
Here 
\begin{equation}
c_s(a)=\frac{c}{\sqrt{3 \left[1+\frac{\rho_b(a)}{\rho_\gamma(a)}
\right]}} 
\end{equation}
is the sound speed of cosmic plasma 
($c$ is the speed of light and $\rho_b/\rho_\gamma$ is the ratio of
baryon to photon energy density), $r$ is the comoving
angular diameter distance, and $a_\star$ is the scale factor at the redshift 
of decoupling as given in a fitting formula in \cite{2005ASPC..339..215H}:
\begin{equation}
a_\star^{-1}=1047.5\left[1+0.00124~\omega_b^{-0.738}\right]
\left[1+b_1 \omega_m^{b_2}\right]
\end{equation}
with
\begin{equation}
b_1=0.0783~\omega_b^{-0.238}\left[1+39.5~\omega_b^{0.763}\right]^{-1}
\end{equation}
and
\begin{equation}
b_2=0.560\left[1+21.1~\omega_b^{1.81}\right]^{-1} ~.
\end{equation}

Our standard fiducial parameter values follow Planck
results, specifically the P+WP+highL+BAO column of Table 5 of
\cite{2013arXiv1303.5076P}. As mentioned before, in addition to the
conventional 6 parameters of the minimal cosmological model, we 
also
always vary the neutrino mass and the amount of tensor modes.  
Varying
T/S is largely irrelevant to anything else in the paper because the
T/S measurement is completely dominated by Planck and essentially
uncorrelated with anything else -- the error is always
$\sigma_{T/S}=0.006$, so we do not print it in tables (note that this error
is surely optimistic due to lack of consideration of foregrounds 
\cite{2011MNRAS.418.1498A}, and note that it does depend on the fiducial
value of $T/S$, e.g., we find $\sigma_{T/S}=0.026$ for fiducial 
$T/S=0.1$).  

\subsection{Extended model parameters}

Our first extension beyond the baseline model is to the Dark Energy Task Force 
(DETF) Figure of Merit
(FoM) scenario \cite{2006astro.ph..9591A}, except with the DETF definition 
modified to include marginalization over neutrino mass. 
As usual, we define the equation of dark-energy $w(a)=p(a)/\rho(a)$ where $p$ 
and $\rho$ are the Dark Energy pressure and density. A cosmological constant
is equivalent to $w(a)=-1$.
Linear dependence of the equation of state on expansion factor is allowed 
by introducing parameters $w_0$ and $w^\prime$ in $w(a)=w_0+(1-a)w^\prime$.
The DETF FoM was originally defined as the inverse of the area inside the 
$w_0-w^\prime$ 95\% confidence constraint interval, but we follow the 
subsequently generally adopted modified normalization convention
that the FoM is simply 
$\left(\sigma_{w_p} \sigma_{w^\prime}\right)^{-1}$ where
$w(z) = w_p + (a_p-a)w^\prime$ and $a_p$ is chosen to make the errors on $w_p$
and $w^\prime$ independent.      
We follow the DETF standard of allowing curvature, parameterized by 
$\Omega_k$, to be free in this scenario (i.e., marginalized over when computing
the FoM). 
While we believe that it is more useful to compute FoMs with free neutrino 
mass, in the Appendix we give results following the original DETF convention
of fixing the neutrino mass, to show the difference and allow comparison with
past calculations.

Going beyond the DETF FoM model, our next extension is to modify gravity
following a model similar to but not exactly that of
\cite{2009arXiv0901.0721A}. Rather than defining
$d\ln D/d\ln a=\Omega_m^\gamma(a)$ with $\gamma$ as a free parameter, we 
define $d\ln D/d\ln a=f_{\rm GR}(a) \Omega_m^{\Delta \gamma}(a)$, where
$\Delta \gamma$ is the free parameter, with a fiducial value of zero, and
$f_{\rm GR}(a)$ is $d\ln D/d\ln a$ computed given the background 
evolution and assuming GR. This is of course exactly equivalent to the usual
parameterization if $f_{\rm GR}(a)$ is exactly described by  
$\Omega_m^\gamma(a)$ with unvarying $\gamma$, but allows for any variation in 
$\gamma$ within GR to be properly propagated (we originally implemented this 
form because neutrinos certainly modify $\gamma$ in principle, but in practice
it did not make a noticeable difference -- note that with neutrinos 
$\Omega_m$ here is defined to only include CDM and baryons). 
Similarly following \cite{2009arXiv0901.0721A}, we include a parameter 
representing
a multiplicative offset of the amplitude of perturbations, $G_9$ 
(\cite{2009arXiv0901.0721A} called it $G_0$), relative to the
GR-predicted amplitude at $z=9$ (applied to the $z<9$ power, to decouple the
low redshift amplitude from CMB measurements), i.e., 
for every use of the power spectrum other than the CMB, we multiply it by 
$G_9^2$ (as pointed out by \cite{2009PhRvD..79f3519L}, 
equation (29) of \cite{2009arXiv0901.0721A} and the text that follows it could 
be 
interpreted as defining $G_9$ in a way that deviated from 1 even within GR --
clearly this would be a bad thing, although the definition of parameters list
at the start of \S III of \cite{2009arXiv0901.0721A} suggests that they 
really intended the definition to be the one we are using here, which does not 
have this problem). 
We include $w_0$, $w^\prime$, and $\Omega_k$ as free parameters in the 
modified gravity scenario, as the main point is to see how well
these things can be distinguished (generally a realistic modified gravity 
model would contain its own background evolution modifications, but these will
be degenerate with changes in a Dark Energy equation of state). 

We add a running of the inflationary perturbation spectral index, 
$\alpha_s=d\ln n_s /d\ln k$ to the baseline 
model as a single parameter extension, i.e., in that case
\begin{equation}
P_{\rm primordial}(k) = A_s \left(\frac{k}{k_\star}\right)^{n_s + 
\frac{1}{2}\alpha_s
\ln\left(k/k_\star\right)}
\end{equation}
where $k_\star=0.05 \iMpc$.

Another single-parameter extension describes non-Gaussianity, $\fNL$, 
parameterizing the usual local model:
\begin{equation}
\Phi = \phi +\fNL \left(\phi^2- \left<\phi^2\right>\right)
\end{equation}
where $\Phi$ is proportional to the initial potential fluctuations and 
$\phi$ is the underlying Gaussian initial field. 

Finally, we consider a single parameter extension allowing for 
``dark radiation'', i.e., a contribution to the relativistic energy density
of the Universe which otherwise does nothing. As is traditional, we call the
parameter for this $N_\nu$, measuring the amount of radiation in units of the
amount contributed by a massless standard model neutrino, but it should be
kept in mind that a measurement of this parameter differing from the 
standard model value 3.046 would not necessarily literally imply extra 
neutrinos, only extra radiation of some kind. 

\subsection{Fisher matrix parameter errors}

Through this work we assume Gaussian likelihoods and propagate
experimental designs into the Fisher matrices for the intermediate
products of individual experiments, such as Fisher matrices for power
spectrum measurements, or BAO distance scale parameters. These
intermediate results are in turn used to form Fisher matrices for
cosmological parameters for individual experiments. 
Except as otherwise discussed, we assume experimental errors are independent
and combine experiments by adding their Fisher matrices. 

For a typical vector of measured quantities, $\mathbf{O}$, for which we can 
assume the likelihood function is Gaussian, with vector of means, 
$\bar{\mathbf{O}}(\vtheta)$, that is predictable given 
parameters $\vtheta$, and covariance $\vC$ which we can assume is independent
of the parameters, the Fisher matrix is:
\begin{equation}
F_{ij} = \frac{\partial \bar{\mathbf{O}}^T}{\partial \theta_i}\vC^{-1}
\frac{\partial \bar{\mathbf{O}}}{\partial \theta_j}~.
\label{eq:simpleFisher}
\end{equation}
While in general the covariance matrix does depend on parameters, this 
dependence becomes a sub-dominant part of the likelihood function once the
parameters are sufficiently precisely determined
\cite{2009A&A...502..721E,2006MNRAS.366..983K}, i.e., essentially the same
limit in which the Fisher matrix is an accurate estimator of errors to begin
with. \cite{2009A&A...502..721E} show that it is important to compute the 
covariance for a model sufficiently close to the best fit, and all Fisher 
matrix calculations implicitly assume that this is done.
This equation is used repeatedly throughout the paper, e.g., the observable 
could be
a measurement of the BAO distance scale, or the CMB $C_\ell$'s, or galaxy
band powers, etc.

The predictions for the linear perturbations in cosmological models
are performed using \texttt{CMBFAST} \cite{1996ApJ...469..437S}, for
historical reasons, but we have checked that results using CAMB
\cite{2000ApJ...538..473L} are virtually identical.

\section{Cosmic Microwave Background: Planck}
\label{sec:cosm-micr-backgr}
We include the Planck CMB satellite as a baseline experiment in all 
projections. Without it our 
interpretation of low redshift measurements would be dominated by 
constraints on strongly degenerate directions that are actually irrelevant 
in global constraints. Planck
constraints are expected to improve with future releases including
polarization, so we continue to project results using a Fisher matrix,
following, e.g., \cite{2002PhRvD..65b3003H}.
We assume a usable fraction of the sky $f_{\rm sky}=0.7$. We assume 3
channels can be effectively used for cosmological measurements, 100,
143, and 217 GHz, with FWHM resolution $\theta_i = 9.65, 7.25$, and 4.99 arcmin,
temperature noise 
$\Delta T/T_i=2.5, 2.2$, and $4.8 \times 10^{-6}$ per resolution
element 
and polarization noise $\Delta P/T_i=
6.7, 4.0$, and $9.8 \times 10^{-6}$ \cite{2006astro.ph..4069T}
(i.e., noise is in units of the mean temperature).  
We use
all $\ell$ up to 2000 for temperature and 2500 for polarization.

To compute the Fisher matrix we use equation (\ref{eq:simpleFisher}) with the 
observables being the autocorrelations of temperature, and $E$ and $B$ modes of
polarization, and the cross-correlation between temperature and $E$ mode 
polarization, i.e., $C^{TT}_\ell$, $C^{EE}_\ell$, $C^{TE}_\ell$, and 
$C^{BB}_\ell$, 
at each multipole $\ell$.

Defining $\Delta C^{\alpha\beta}_\ell = 
\hat{C}^{\alpha\beta}_\ell -C^{\alpha\beta}_\ell$, where 
$\hat{C}^{\alpha\beta}_\ell$ is the estimated value and
$C^{\alpha\beta}_\ell$ the true value, with $\alpha$ and $\beta$ equal 
$T$, $E$, $B$, the 
covariance matrix for the observables is given by
\begin{equation}
\left<\Delta C^{\alpha\beta}_\ell \Delta C^{\delta\gamma}_\ell \right>
\simeq \left[\left(2 \ell +1\right) f_{\rm sky}\right]^{-1} 
\left( C^{\alpha\delta}_\ell C^{\beta\gamma}_\ell +C^{\alpha\gamma}_\ell
C^{\beta \delta}_\ell   \right)~,
\label{eq:cmbcovariance}
\end{equation}
e.g., 
\begin{equation}
\left<\left(\Delta C^{TT}_\ell\right)^2 \right>
\simeq \left[\left(\ell +\frac{1}{2}\right) f_{\rm sky}\right]^{-1} 
\left( C^{TT}_\ell\right)^2 ~.
\end{equation}
Note that for the CMB we always sum over integer $\ell$, not aggregated
$\Delta \ell$ bands as discussed below for photometric surveys. We assume
that different $\ell$ are uncorrelated, which they will not be with
$f_{\rm sky}<1$, but this will not affect the results as long as models
have no fine structure in $\ell$.

$C^{TT}_\ell$, $C^{EE}_\ell$, and
and $C^{BB}_\ell$ contain a noise contribution which we compute using 
\begin{equation}
N_\ell^{-1} = 
\sum_i \left[\left(\frac{\Delta T}{T}\right)_i \theta_i 
e^{\ell (\ell+1) \theta_i^2/ 16 \ln{2}} \right]^{-2}~,
\end{equation}
where $\theta_i$ is in radians \cite{2002PhRvD..65b3003H}.
$\Delta P=\Delta E$ or $\Delta B$ can be substituted for 
$\Delta T$ to compute $EE$ and $BB$ noise power.

For equivalent parameter spaces, our Planck projections agree well with, e.g., 
\cite{2004IJTP...43..599B,2012JCAP...04..027H,2010PhRvD..82l3504G,
2010JPhCS.259a2004M}.

We do not include CMB lensing
\cite{2013PhLB..718.1186O,2012MNRAS.425.1170H,2009PhRvD..79f5033D}, 
which could provide additional constraints
which will probably be qualitatively similar to the galaxy lensing surveys
discussed below, but with much different systematics. 

\subsection{Planck projections vs. reality}

Obviously our overall Planck projections are stronger than the published 
results, because they include polarization. This should change as new results 
are published. Our resolution numbers accurately reflect the achieved ones
\cite{2013arXiv1303.5075P}, and we are consistent with the 
Planck power spectrum paper \cite{2013arXiv1303.5075P} in using the
100, 143, and 217 GHz channels. For $\ell>50$, \cite{2013arXiv1303.5075P} 
use only 58\% of the sky at 100 GHz, and 37\% at 143 and 217 GHz, so it 
appears that we are optimistic in using 70\%, although there is some suggestion
in \cite{2013arXiv1303.5075P} that they expect the fractions to improve in 
future releases. For $\ell<50$ \cite{2013arXiv1303.5075P} uses 87\% of the
sky so we are a bit pessimistic there. The achieved noise in the published 15 
month data set is very 
similar to projections in the 143 and 217 GHz channels, but somewhat worse in
the 100 GHz channel 
(\url{http://www.sciops.esa.int/wikiSI/planckpla/index.php?title=HFI_performance_summary&instance=Planck_Public_PLA}), 
which should nevertheless almost achieve
the goal performance that we use in projections by the end of the 30 month 
extended mission.

One of the most important remaining uncertainties is whether or not 
low-$\ell$ Planck polarization 
measurements will be sufficiently clean to achieve the $\tau$ error
we project, which determines the error on the CMB measurement of the power 
spectrum amplitude, which is compared in turn to lower redshift amplitude
measurements from redshift-space distortions or lensing to determine things
like neutrino mass. 

\section{Spectroscopic surveys}
\label{sec:redshift-surveys}

In this section we first describe how we compute projections for redshift 
surveys,
including galaxy clustering, quasar clustering, and correlations of \lyaf\ 
absorption in quasar spectra. Then we describe the specific redshift surveys
we include. 

\subsection{Galaxy and quasar clustering}

Galaxies and quasars are point tracers of the underlying cosmic
structure. The physics of how they trace the dark matter fluctuations
is well understood based on arguments about locality of galaxy
formation \cite{2009JCAP...08..020M,2012PhRvD..86h3540B,
2006PhRvD..74j3512M}
and heuristic understanding that astrophysical
objects form in the peaks (halos) of the primordial density fields
\cite{2013PhRvD..87d3505D}. On
very large scales bias is scale independent and redshift-space distortions are
described by linear perturbation theory \cite{KAISE87}. 
Beyond-linear perturbative 
corrections can be used on intermediate scales before perturbation theory 
breaks down entirely on small scales 
\cite{2009ApJ...691..569J,2011JCAP...11..039S,2013MNRAS.433..209N,
2012PhRvD..86d3508W,2012PhRvD..85h3509C,2011PhRvD..83h3518M,
2013PhRvD..87l3510S,2012JCAP...11..009V,2012JCAP...11..014O,
2008PhRvD..78j3512S,2012JCAP...07..051B}. 

The basic model for galaxy clustering (or quasar clustering -- when quasars
are treated as clustering objects, as opposed to probes of the \lyaf, there is 
no fundamental difference between them and galaxies) is simply linear bias
and a shot noise contribution, i.e., 
\begin{equation}
P_{i j}(k,\mu)=(b_i+f \mu^2)(b_j+f \mu^2) P_{\rm mass}(k) + 
\bar{n}_i^{-1}\delta^{\rm K}_{ij}~,
\label{eq:galaxypower}
\end{equation} 
where $b_i$ is the bias for tracer type $i$, $f$ is the 
growth rate, i.e., $f\equiv d\ln D/d\ln a$, $\mu$ is the cosine of the angle
between the wavevector and our line of sight, $P_{\rm mass}(k)$ is the 
linear theory mass power spectrum, and $\bar{n}$ is the number density.
We generally assume fiducial biases follow constant
$b(z) D(z)$, where $D(z)$ is the linear growth factor normalized
by $D(z=0) \equiv 1$. 

\subsubsection{BAO}

Isolating the BAO feature gives the most robust, but pessimistic, view of 
the information that
one can recover from galaxy clustering measurements, since BAO can be
measured even in the presence of large unknown systematic effects 
(very generally
these will not change the BAO scale \cite{2010ApJ...720.1650S}).
To compute isolated galaxy BAO errors we use a lightly modified version of the
code that accompanies
\cite{2007ApJ...665...14S}, assuming 50\% reconstruction, i.e.,
reduction of the BAO damping scale of \cite{2007ApJ...665...14S} by a
factor 0.5, except at very low number density, where we degrade
reconstruction based on \cite{2010arXiv1004.0250W}. This method has held up
well under close scrutiny \cite{2012PhRvD..85j3523S,2012MNRAS.419.2949N,
2009PhRvD..80l3503T}.
We generally quote errors on the transverse and radial BAO scales as errors on 
$D_A(z)/s$ and $H(z) s$, respectively, where $s$ is the BAO length scale. For
galaxy and quasar clustering these errors always have nearly identical 
correlation coefficient 0.4. 

To understand what we mean by ``50\%'' reconstruction (and the details of our
broadband calculations below, even though they don't use the same code), one 
has to understand
how the computation in the code of \cite{2007ApJ...665...14S} works. 
Conceptually at least (i.e., before some approximations they make for purely
technical reasons) they start with the idea that the observable in the 
Fisher matrix calculation is
the BAO-only part of the power spectrum as damped by non-linear evolution 
in the form of Lagrangian displacements,
specifically:
\begin{equation}
P_{\rm BAO, nl}(k,\mu) = P_{\rm BAO, lin}(k,\mu)
\exp\left(-\frac{k_\perp^2 \Sigma_\perp^2}{2}-
\frac{k_\parallel^2 \Sigma_\parallel^2}{2} \right)
\label{eq:SEsignal}
\end{equation}
where $P_{\rm BAO,lin}(k,\mu)$ is includes the usual bias and RSD factors.
The Lagrangian displacement distances are estimated to be 
$\Sigma_\perp=9.4 \left(\sigma_8(z)/0.9\right) \hMpc$ and 
$\Sigma_\parallel = \left(1+f(z)\right) \Sigma_\perp$. These damping factors, 
along with the RSD factor, are taken outside the Fisher matrix derivatives
to avoid using their structure to measure distance, which means they have the
effect of modifying the covariance used in the Fisher matrix by 
a factor of their inverse. What we mean by ``50\% reconstruction'' is that
$\Sigma_\parallel$ and $\Sigma_\perp$ are both multiplied by a factor 0.5
relative to the above unreconstructed values. We very roughly estimate a 
degradation in this reconstruction due to shot-noise following 
\cite{2010arXiv1004.0250W}. The reconstruction multiplier used,
$r(\bar{n}P)$, is obtained by interpolating over the table defined
by the vectors $r=(1.0, 0.9, 0.8, 0.70, 0.6, 0.55, 0.52, 0.5)$,
$x=(0.2, 0.3, 0.5, 1.0, 2.0, 3.0, 6.0, 10.0)$ where 
$x\equiv \bar{n}P(k=0.14\ihMpc,\mu=0.6)/0.1734$. For $x>10$, $r=0.5$, while
for $x<0.2$ $r=1$, i.e., at low number density there is no reconstruction, 
while at high density the factor is the traditional 0.5.
Note that the covariance effectively used in the BAO Fisher calculation 
(before including the factors pulled outside the derivatives) is still 
computed using the full linear theory power spectrum with shot-noise, i.e., 
equation (\ref{eq:galaxypower}). 

Our primary modification of the code of \cite{2007ApJ...665...14S} is to allow
multiple galaxy populations probing the same volume of space to be combined.
We do this optimally by summing their contribution to the signal-to-noise 
mode-by-mode, i.e., $\left(\bar{n} P\right)_{\rm combined}(k,\mu)=
\sum_i \bar{n}_i P_i(k,\mu)$.

Once we have estimated the covariance matrix of 
$D_A(z)/s$ and $H(z) s$ using the code of \cite{2007ApJ...665...14S}, we 
propagate these 
constraints into more fundamental
parameter constraints using the usual Fisher matrix equation 
(\ref{eq:simpleFisher}). The observables are $D_A(z)/s$ and $H(z) s$ with 
dependence of $s$ on parameters included through equation \ref{eq:s}.

We also quote errors on an isotropic dilation factor 
$R/s$, defined as the error one would measure on a single parameter 
that rescales radial and transverse directions by equal amounts. 
Note that this is used only when a simpler condensation of the information
in the $H-D_A$ covariance matrix is desired, e.g., for plotting basic 
experimental power vs. redshift. The $H-D_A$ constraints are used for all 
constraints on more fundamental parameters. 
But to be more explicit, the fractional change in $R$ for which we project 
errors, $\delta R/R \simeq \delta \ln R$, is defined by
\begin{equation}
D_A = \left(1+\delta \ln R\right) D_{A,{\rm fid}}
\end{equation} 
and
\begin{equation}
H = \left(1+\delta\ln R\right)^{-1} H_{\rm fid}
\end{equation} 
where $D_{A,{\rm fid}}(z)$ and $H_{\rm fid}(z)$ are the angular diameter 
distance and Hubble parameter in a fiducial Universe. 
The effective sensitivity of $R$ to
$H$ and $D_A$ depends on the experimental 
scenario. For example, the simplest cases are easy to understand: for a purely
transverse measurement (e.g., photometric survey) $R=D_A$, while for a 
purely radial measurement (e.g., something closer to the \lyaf, although it is
not purely radial) $R=H^{-1}$ (or, if one is concerned about nonequivalent 
units, $R=H^{-1} H_{\rm fid} D_{A,{\rm fid}}$). For intermediate cases like 
typical galaxy clustering the appropriate combination of $H$ and $D_A$ can 
always be determined given the covariance matrix between them.  
Note that this error on $R$ is clearly not in general
equivalent to an error on the specific combination of parameters that determine
the volume element,
$(D_A^2 H^{-1})^{1/3}$, as is easy to see by
considering the purely radial or
purely transverse examples. While those cases give a perfectly well-defined
error on $R$, the error on 
$(D_A^2 H^{-1})^{1/3}$ is formally infinite, because one of the two parts is
unconstrained. Of course, one can add the assumption that $D_A$ and $H^{-1}$
vary proportionally, but then saying ``$(D_A^2 H^{-1})^{1/3}$'' just becomes 
an oblique
way to say $R$ -- the powers applied to $D_A$ and $H$ have no real meaning. 
$(D_A^2 H^{-1})^{1/3}$ would be directly measured by 
something that was really sensitive to volume, e.g., counts of a source with 
known physical number density). 

\subsubsection{Broadband}

Going beyond BAO, we use ``broadband'' galaxy power, i.e. measurements
of the power spectrum as a function of redshift, wavenumber and angle
with respect to the line of sight.  This treatment automatically
recovers all available information, i.e. not just the shape of the
isotropic power spectrum, but also redshift-space distortions,
Alcock-Paczynski \cite{1979Natur.281..358A}, and of course also the
BAO information. 
Discussing isolated redshift-space distortions 
as is often done, or the monopole power spectrum alone (which may sometimes
be used for neutrino mass constraints),  may be useful for pedagogical
reasons, but generally once we go beyond BAO there is no clear
systematic advantage in any subset of the broadband information, so it
makes sense to just use all of it.

Bias uncertainty is modeled by a free parameter in 
each redshift bin, generally of width $\Delta z=0.1$, for each type of galaxy.  
Our results are not sensitive to the redshift bin width, maybe surprisingly.
For example, we show an explicit comparison of some cases in 
Table \ref{tableoverlap}, and have 
checked other cases (e.g., the neutrino mass constraints we show are identical
to two significant digits between $\Delta z=0.1$ and $\Delta z=0.2$ bins). 
We believe the reason for this is that the extra freedom allowed by, say,
splitting an already fairly narrow bin in half, i.e., for the bias in one half 
to go up while the bias
in the other goes down, still summing to what would have been the bias for the
coarser bin, is not generally at all degenerate with cosmological 
parameters, because cosmological models generally do not predict this kind of
rapid, anti-correlated change in relevant quantities like $f(z)$.   

We compute the broadband Fisher matrix using the usual generic equation  
(\ref{eq:simpleFisher}),  evaluated by taking
numerical derivatives of $P_{ij}(k_\parallel,k_\perp)$ with 
respect to all parameters. 
To include all geometric effects appropriately, the observable 
band power measurements are written in observable coordinates, i.e., 
radial distance is measured in $\kms$ and transverse distance in degrees, i.e.,
\begin{equation}
P^{\rm obs}_{ij}(k^{\rm obs}_\parallel,k^{\rm obs}_\perp)=
a~ H(a)~ r^{-2}(a)~ P^{\rm com}_{ij}\left[a~ H(a)~ k^{\rm obs}_\parallel,~
r^{-1}(a)~ k^{\rm obs}_\perp\right]
\end{equation}
where obs stands for ``observed'' and com stands for ``comoving'' 
(recall that $D_A(a) \equiv a~ r(a)$). Band power measurements are labeled
by $k^{\rm obs}_\parallel$ and $k^{\rm obs}_\perp$, which are held fixed under
numerical derivatives with respect to parameters. 

The covariance matrix of band power errors,
$\Delta P_{ij} = \hat{P}_{ij}-P_{ij}$, where 
$\hat{P}_{ij}(k_\parallel,k_\perp)$ is the estimate and 
$P_{ij}$ is the true power, 
with $i$ and $j$ labeling potentially 
multiple tracers of LSS in the same volume of space, is
\begin{equation}
\left< \Delta P_{ij} \Delta P_{mn}\right>=
\frac{2 \pi^2}{V k^2 \Delta k \Delta \mu}
\left(P_{im} P_{jn}+P_{in} P_{jm}\right) ~,
\label{eq:Pcovariance}
\end{equation}
where $V$ is the volume of the survey and $\Delta k$ and $\Delta \mu$ are
the bin widths (this formula is really only correct in the small-bin limit, 
and in practice we make the bins fine enough that the Fisher 
calculation is effectively an integral -- note that as defined here that 
integral only covers $0<\mu<1$). 
Equation (\ref{eq:Pcovariance}) is valid for all combinations of $i$, $j$, $m$,
and $n$  (e.g., even if some are equal). 
Recall that $P_{i j}(k,\mu)=(b_i+f \mu^2)(b_j+f \mu^2) P_{\rm mass}(k) + 
\bar{n}_i^{-1}\delta^{\rm K}_{ij}$, so shot noise enters the errors through 
terms where $i$ or $j$ are equal to $m$ or $n$. The prefactor in 
Equation (\ref{eq:Pcovariance}) accounts for sample variance due to finite 
volume. Different bands of $k$ and $\mu$ are assumed to be independent, which,
as usual, is not strictly true for a finite volume survey but will be 
irrelevant as long as the theoretical power spectrum does not have fine 
structure in $k$ (surveys with narrow strips may violate this condition with
respect to the BAO wiggles, but not the large surveys we are most interested
in). 

We use broadband power up to some quoted $\kmaxeff$.
At $k>\kmaxeff$ we continue to use BAO information as usual
(to be clear, we compute the usual BAO fisher
matrix, but cut modes with $k<\kmaxeff$ out of the integration,
since they are already included in the broadband calculation).
We use two simple choices of $\kmaxeff$, 0.1 and 0.2 $\ihMpc$. These should
not be taken literally as a scale up to which we think linear theory will be
sufficient for an analysis of future high precision data.
Deviations will clearly be present even at $0.1~\ihMpc$.
These cutoffs are just intended to 
give an idea of the sensitivity of results to the effective scale 
where information is recovered after 
making corrections for non-linearity, presumably including some marginalization
over beyond-linear bias parameters (information at higher $k$ might be
used to constrain these parameters -- this is why we write 
$\kmaxeff$, where ``eff'' stands for ``effective'', instead of simply
$k_{\rm max}$). It will be a major program of the next
decade to figure out exactly how to do this fitting in practice for a high
precision survey
like DESI -- how well we can do this will determine how well we can measure
parameters. 
The value 0.1-0.2 $\ihMpc$ is motivated by, e.g., the finding of
\cite{2012JCAP...02..010O} that a Taylor series representation of redshift 
space distortions could be summed to high precision up to $k\sim 0.2\ihMpc$, 
after which it appeared that the power spectrum had become deeply 
non-linear, i.e., information is probably hopelessly scrambled, and many 
other similar findings (e.g., \cite{2012JCAP...11..009V,
2012JCAP...11..029G,2013PhRvD..87h3509T, 
2013PhRvD..87l3510S,2012JCAP...11..014O,
2013arXiv1307.2906V,2013JCAP...02..025L}). 

As an additional measure to be sure we are not making unreasonable predictions
using the non-linear regime, we use the same information damping factors 
from \cite{2007ApJ...665...14S} that we use for BAO for the broadband signal,
i.e., the exponential factor in equation (\ref{eq:SEsignal}). 
This is well-motivated from a theoretical point of view, i.e., the damping
is related to the propagator of \cite{2008PhRvD..77b3533C}, which suppresses 
all linear theory information, not just BAO.
We also include the same reconstruction factor,
as there is no logical reason why
similar methods could not be used to recover non-BAO information, although
this has not been worked through yet.
The logic for also using 
$\kmaxeff$ for broadband power (i.e., not relying on these damping
factors as our only cutoff), and only using BAO beyond that, is that, while
for BAO we largely only need to worry about the statistical effect of damping
the signal relative to effective noise power coming from higher order terms, 
to use the broadband power this effective noise must actually be predicted, 
which is generally harder. 
To avoid using
the scale of this damping as a new standard ruler, we again pull it outside
the Fisher matrix derivatives, effectively multiplying the power used to
compute the covariance by its inverse (of course the RSD factor is not 
pulled out of the derivatives, as here it is part of the signal that we
are interested in). 
To reiterate: the damping factors are applied as 
an additional limitation on the higher $k$ power spectrum, in addition to 
$\kmaxeff$, and the reconstruction factor is included
to make the broadband treatment consistent with the isolated
BAO treatment (i.e., note that if we did not do this the BAO information at 
$k<\kmaxeff$ would not be the same between isolated BAO and broadband cases,
which clearly does not make sense, although we could in principle adjust it as 
an additional step). 

\subsubsection{Isolated redshift-space distortions}

Frequently, galaxy constraints beyond BAO are described as a measurement of 
a single redshift-space distortion (RSD) amplitude as a function of $z$, e.g., 
``$f(z) \sigma_8(z)$''. While it is always nice to have a one dimensional 
scalar function to make simple plots, we do not use this method for our main
results
because it requires us to either ignore or marginalize away the uncertainty in 
geometry and the primordial power spectrum shape. It is easy enough, although
harder to visualize, to just include all of the broadband information. 
However, since it has come to be expected, we do quote isolated RSD errors 
as a function of redshift for BOSS and DESI, calculated using exactly the
code described
in \cite{2009JCAP...10..007M}. This calculation makes the assumption that both
the geometry and power spectrum shape are fixed by external constraints.
This is probably the only scenario in which an isolated RSD measurement is a 
useful thing
to quote, but we do not claim it applies in our cases. 
Specifically, if we assume the $k$-dependence of $P_{\rm mass}(k)$  is known,
the formula $P_{\rm red}(k,\mu)=(b+f \mu^2)^2 P_{\rm mass}(k)$ decomposes into 
a function of two parameters, $b \sigma$ and $f \sigma$, where 
$\sigma(z)\propto P_{\rm mass}^{1/2}(z,k)$ is the rms normalization of the
linear mass density fluctuations as a function of $z$. In tables, we identify
the maximum $k$ used to compute the error on $f \sigma$ by labeling it 
$f \sigma_k$, e.g., $f \sigma_{0.1}$ means the error calculation included
information up to $\kmaxeff=0.1 \ihMpc$. These fractional errors are equivalent
to what one usually sees quoted as an error on ``$f \sigma_8$'', i.e., 
$\sigma_8$ here is intended only loosely as a parameter normalizing 
the power spectrum, not to mean that you necessarily have a direct measurement
of fluctuations on the scale of $8 \hMpc$ radius spheres. We need to make
the scale of sensitivity more explicit, because we have more than one.
As in the broadband case, we always
include the information damping factors of \cite{2007ApJ...665...14S},
with reconstruction.

\subsection{LyaF}

Spectroscopic surveys designed for galaxy redshift surveys can often
also probe large-scale structure using the \lyaf\
\cite{2006ApJS..163...80M,2011JCAP...09..001S},
i.e., the Lyman-$\alpha$ absorption by
neutral gas in the intergalactic medium in the spectra of high redshift quasars
(or maybe at faint enough magnitudes Lyman-break galaxies
\cite{2007PhRvD..76f3009M}).

We model the three dimensional power spectrum of \lya\ fluctuations using 
the analytic formula of \cite{2003ApJ...585...34M}:
\begin{equation}
P_F(k,\mu,z) = b_F(z)^2 (1+\beta_F(z) \mu^2)^2 P_{mass}(k,z) D(k,\mu,z) ~,
\label{lyaf_power}
\end{equation}
where $b_F$ is the linear bias parameter, $\beta_F$ the redshift space
distortion parameter and $D(k,\mu,z)$ 
is a non-linear correction calibrated
from simulations, i.e.,  
\begin{equation}
D(k,\mu)\equiv \exp\left(\left[\frac{k}{k_{\rm NL}}\right]^{\alpha_{\rm NL}}
-\left[\frac{k}{k_P}\right]^{\alpha_P}-
\left[\frac{k_\parallel}{k_V(k)}\right]^{\alpha_V}\right)~,
\end{equation}
where $k_V(k)=k_{V0}(1+k/k^\prime_V)^{\alpha^\prime_V}$. The first term in
the exponential represents non-linear growth of real space power 
(at the central model of
\cite{2003ApJ...585...34M}, $k_{\rm NL}=6.4\ihMpc$ and 
$\alpha_{\rm NL}=0.569$), the 2nd term represents pressure smoothing of 
small-scale structure ($k_P\sim 15.3\ihMpc$ and $\alpha_P\sim 2.01$), and
finally the 3rd term represents Fingers-of-God-type suppression of radial 
power ($k_{V0}\sim 1.22 \ihMpc$, $\alpha_V\sim 1.5$, 
$k^\prime_V\sim 0.923\ihMpc$, and $\alpha^\prime_V\sim 0.451$).
Table I of \cite{2003ApJ...585...34M} gives parameter dependence of $b_F$,
$\beta_F$, and fitting parameters of $D$.

Similar to galaxies, the \lyaf\ can be viewed
pessimistically as a probe of the BAO feature, or more optimistically using 
the broadband and smaller scale power spectrum.

\subsubsection{BAO}
 
The BAO distance scale has recently been measured in the three dimensional 
correlation of 
the \lyaf\ in nearby quasar lines of sight from the BOSS survey 
\cite{2013A&A...552A..96B,2013JCAP...04..026S}. 
Using one third of the final BOSS area, the authors were able to measure 
BAO distance scale at redshift $z=2.4$ with an 
uncertainty of 2\% \cite{2013JCAP...04..026S}.

The parameters of equation (\ref{lyaf_power}) given by 
\cite{2003ApJ...585...34M} 
are only valid near $z\sim 2.25$. For BAO error 
estimates, which we want to make well away from this redshift and do not
require us to use
detailed parameter dependence, we simply use the central model parameters 
(the Planck model happens to have nearly exactly the same amplitude and
slope of the power spectrum relevant to the \lyaf, although the WMAP model was
lower) with the power
spectrum additionally multiplied by a factor $((1+z)/3.2)^{3.8}$ 
to match the
evolution of the 1D power with redshift \cite{2006ApJS..163...80M}. 
Except as otherwise noted, we use the method of \cite{2007PhRvD..76f3009M} to 
estimate the obtainable errors (a similar method was derived by
\cite{2011MNRAS.415.2257M}). 

\cite{2007PhRvD..76f3009M} derived the three dimensional flux power from a 
hypothetical 3D Fourier transform of a \lyaf\ data set to be
\begin{equation}
P_F^{\rm 3D, obs}(\vk)= P_F^{\rm 3D}(\vk) +
P_F^{\rm 1D}(k_\parallel)P_w^{\rm 2D}+P_N^{\rm eff}
\label{eq:lyafobs3D}
\end{equation}
(we say ``hypothetical'' because we generally would not do the data analysis 
with
a literal 3D Fourier transform, but any near-optimal analysis should obtain
similar results). Here  $P_F^{\rm 3D}(\vk)$ is the true 3D flux power spectrum
that one would measure with infinite sampling, $P_F^{\rm 1D}(k_\parallel)$ 
is the one dimensional power spectrum along single lines of sight, 
$P^{\rm 2D}_w$ is the power spectrum of the weighted quasar sampling function, 
and $P_N^{\rm eff}$ is the weighted pixel noise power. 
\cite{2007PhRvD..76f3009M} derived that 
\begin{equation}
P^{\rm 2D}_w=\frac{I_2}{I_1^2 L_q}
\end{equation}
and
\begin{equation}
P_N^{\rm eff}=\frac{I_3 l_p}{I_1^2 L_q}
\end{equation}
where $L_q$ is the length of the forest in a quasar spectrum, $l_p$ is the 
pixel width
\begin{equation}
I_1=\int dm \frac{dn_q}{dm} w(m)
\end{equation}
where $dn_q/dm$ is the luminosity function of observed quasars as a function 
of magnitude m, $w(m)$ is the weight as a function of $m$, 
\begin{equation}
I_2=\int dm \frac{dn_q}{dm} w^2(m)
\end{equation}
and
\begin{equation}
I_3=\int dm \frac{dn_q}{dm}\sigma_N^2(m) w^2(m)
\end{equation}
where $\sigma_N(m)$ is the pixel noise, which will generally be a function
of magnitude. The weight function is
\begin{equation}
w(m) = \frac{P_S/P_N(m)}{1+P_S/P_N(m)}
\end{equation}
where $P_S$ is, following \cite{1994ApJ...426...23F}, taken to be the signal
power at some typical wavenumber, which we take to be $k=0.07\ihMpc$, 
$\mu=0.5$, and $P_N(m)=\sigma^2_N(m)l_p /I_1 L_q$ (this must be determined
iteratively because it both determines and depends on the weights). All
of this is discussed in more detail in \cite{2007PhRvD..76f3009M}. It was a 
guess in \cite{2007PhRvD..76f3009M}
that $k_\perp$ must be restricted to be less than the Nyquist frequency
corresponding to the typical separation between quasars, but quickly after we
realized that this definitely must be correct
because it is the only way to recover the correct 1D Fisher matrix in the 
limit of infinitely sparse quasars. 

Given equation (\ref{eq:lyafobs3D}), the \lyaf\ Fisher matrix calculations 
proceed similarly to the galaxy calculations, i.e., we evaluate the basic 
Fisher matrix equation (\ref{eq:simpleFisher}) with $P_F^{\rm 3D, obs}(\vk)$
as the observable, and compute the covariance matrix using
$P_F^{\rm 3D, obs}(\vk)$ in equation 
(\ref{eq:Pcovariance}). 
 
In contrast to past projections which often used the rest wavelength range 
$1041<\lambda<1185$\AA\ (following \cite{2006ApJS..163...80M}), we expand the 
range to include the \lyb\ forest and
move slightly closer to the quasar, 
$985<\lambda<1200$\AA, reflecting our increasing confidence that we understand
the relevant issues well enough to measure BAO across this range
\cite{2013JCAP...09..016I}. 
Gains from this enhancement of effective
number density (and cross-correlations with quasars below) are substantial
because the measurement is quite sparse, i.e.,
in what for galaxies we would call the shot-noise limited regime.

We isolate the BAO signal by subtracting a smoothed version of the power 
spectrum from the wiggly one and then using the residual wiggles in our Fisher
matrix derivatives. We follow the procedure of \cite{2007ApJ...665...14S} in
dividing the noise contribution to the power errors by the 
$(1+\beta \mu^2)^2$ RSD factor rather than 
including this factor in the derivative term, which would lead to artificial 
(i.e.,
non-BAO-distance) breaking of the degeneracy between radial and transverse
distance errors (this was not done in \cite{2007PhRvD..76f3009M}, leading to
some underestimation of the degeneracy between $H(z)$ and $D_A(z)$). We
also include the \cite{2007ApJ...665...14S} damping factors, with no 
reconstruction. We have tested that our approach agrees with 
\cite{2007ApJ...665...14S} to percent level given matching assumed data 
sets. To be clear, the primary difference between the \lyaf\ BAO Fisher 
matrix calculation and the galaxy version is  
the need to evaluate the integrals 
over the quasar luminosity function and spectrograph noise distribution to 
determine the signal to noise level as a function of redshift, with another
difference being that we compute the error on BAO distance through direct 
Fisher matrix derivatives of the wiggles-only power spectrum rather than the 
procedure of \cite{2007ApJ...665...14S} of averaging over a cosine squared
approximation for the derivatives (but again, we have checked that these
methods agree remarkably precisely).

\subsubsection{Broadband and 1D power}

The correlation of \lya\ absorption in quasar spectra can provide other 
cosmological information beyond BAO. Several studies have already constrained
cosmological parameters from the line of sight power spectrum 
\cite{2006ApJS..163...80M,2013A&A...559A..85P,2011MNRAS.413.1717B,
2009JCAP...08..030A,2006JCAP...10..014S,2006MNRAS.370L..51V,
2006JCAP...06..019G,2005PhRvD..71j3515S,2005PhRvD..71f3534V,Gratton:2007tb}, 
and one can also 
obtain valuable information from the full shape of the three-dimensional 
clustering \cite{2003ApJ...585...34M}. In the projections below we 
distinguish between \lyaf\ BAO measurements and broadband
measurements that include the one dimensional power spectrum 
measurement. 

For interpreting broadband measurements, we need the parameter dependence of 
Table I of \cite{2003ApJ...585...34M}, 
i.e., $b_F$ and $\beta_F$, along with the fitting parameters of $D(k,\mu)$,
depend on the amplitude and slope of the linear power spectrum, 
temperature-density relation
\cite{2001ApJ...562...52M,2012ApJ...757L..30R}, and mean level of absorption
\cite{2008ApJ...681..831F}, 
all of which are
varied in our Fisher matrix calculations. 
To help constrain these parameters, we include the 1D power spectrum 
that could be measured from $\sim 100$ (existing) high resolution spectra
\cite{2000ApJ...543....1M,2004MNRAS.347..355K}.

The constraints from the \lyaf\ are difficult to predict accurately, because
they require careful simulation work to achieve
\cite{2005ApJ...635..761M,2005MNRAS.360.1471M,2006MNRAS.365..231V,
2007MNRAS.374..196R} -- 
more careful than the community has
been able to muster so far. The numbers we give are
intended to be a good central value guess, i.e., 
while there is uncertainty, it is at least
as likely that we could do better as worse, basically because we intentionally
leave a lot of information for ``contingency.'' 
For these projections we continue to use the rest wavelength range 
$1041<\lambda<1185$\AA, although the \lyb\ forest region should provide 
valuable complementary information. 
We do not include the bispectrum or any other statistics besides the power 
spectrum, which are known to be powerful for breaking IGM model degeneracies 
(e.g., \cite{2003MNRAS.344..776M,2004MNRAS.347L..26V}). 
We do not use cross-correlations with quasar density. 
Finally, we only use the redshift range 2-2.7. The original reason for this
was the limited range of applicability of the parameters of
\cite{2003ApJ...585...34M}, but it has the effect of reserving the large 
amount of higher redshift information to help allow for expansion in the 
modeling uncertainty.   

\subsection{LyaF-quasar cross-correlation}

The cross-correlation of quasars with the \lyaf\ 
\cite{2013JCAP...05..018F} provides a complementary 
measurement of BAO at high redshift. We use a high-noise approximation to 
combine separately computed constraints from \lyaf\ and quasars into one. 
In general, if we have multiple observable tracers, $o_i$, of the mass density 
field, $\delta_m$, of the form 
$o_i = c_i \delta_m +\epsilon_i$ where $c_i$ is the generalized bias
(including the RSD factor) and $\epsilon_i$ is the noise for tracer $i$ where
$\left<\left|\epsilon_i\right|^2\right> \equiv N_i$, and we assume the noise 
is uncorrelated (a generally good but not always perfect assumption
\cite{2010PhRvD..82d3515H,2012PhRvD..86j3513H,2009PhRvL.103i1303S}), 
it is easy to show
that the optimally weighted estimate of 
$\delta_m$ has
noise variance $\left(\sum_i c_i^2 N_i^{-1}\right)^{-1}$, where this applies
mode-by-mode in Fourier space. 
This is equivalent, for galaxies where 
$N_i = \bar{n}_i^{-1}$, to the statement that we can simply add 
$\bar{n}_i P_i$ to 
find the signal-to-noise ratio for the optimal combined tracer, 
where $P_i\equiv c_i^2 P_m$ (as mentioned above, this is how we do 
multiple-tracer BAO calculations, including full $k$ and $\mu$ dependence). 
The rms fractional error on a combined BAO measurement can then be approximated
by 
$\sigma_{\rm BAO} = \sigma_{\rm BAO, V} \left[1+ 
\left(\sum_i P_i N_i^{-1}\right)^{-1}\right]$ 
where $\sigma_{\rm BAO, V}$ is the fractional
error we would find from the given volume in the zero noise limit and
$P_i(k,\mu)N_i^{-1}(k,\mu)$ is evaluated at the typical $k$ and $\mu$ of the
BAO feature (remember that for the \lyaf\ $N$ does depend on 
$k_\parallel \equiv k \mu$). 
If we have BAO measurements from the individual tracers, which obey
$\sigma_{\rm BAO,i} = \sigma_{\rm BAO, V} \left[1+ N_i P_i^{-1}\right]$, we
can re-write $P_i N_i^{-1}$ in terms of $\sigma_{\rm BAO,i}$ and in the high
noise limit obtain 
$\sigma_{\rm BAO} \overset{N/S\rightarrow \infty}{=} 
\left(\sum_i \sigma_{\rm BAO,i}^{-1}\right)^{-1}$,
i.e., the combined fractional BAO error is given by the inverse sum of 
individual fractional BAO errors, not by the
inverse quadrature sum as it would be for measurements in different volumes
(the same approach could be used to derive the error without the high noise
approximation, it would just produce a more complicated-looking equation). 
This inverse sum property makes the addition of a subdominant
tracer like the quasars surprisingly valuable, compared to our usual intuition
based on inverse quadrature sums.  
  
We will justify the high noise approximation for DESI
below. It is certainly possible to do a full multiple-tracer Fisher 
matrix
calculation for \lyaf\ and quasars, like we do for ELG, LRG, and QSO tracers
at lower redshift, 
but this approximation should be sufficient and is easier given
the available code.  
We use cross-correlations with quasars only for BAO measurements, not for
broadband, although generally they should add information there too. 

\subsection{BOSS}

BOSS \cite{2013AJ....145...10D} is a 10000 sq. deg. survey that is almost 
completed. Analyzers have chosen a certain redshift binning
for the data \cite{2012MNRAS.426.2719R}, but we give the continuous numbers we 
have been 
using for Fisher matrix projections in Table \ref{tableBOSS}.

\begin{table}
\caption{
Basic numbers for BOSS. 
$P_{0.2,0}\equiv P(k=0.2\ihMpc, \mu=0)$,
$P_{0.14,0.6}\equiv P(k=0.14\ihMpc, \mu=0.6)$.
$f(z) \sigma_k(z)$ is what is often called $f(z) \sigma_8(z)$ --
more precisely, $\frac{\sigma_{f\sigma_k }}{f \sigma_k}$
is the fractional error on the normalization of 
$f(z) P^{1/2}(k,z)$, assuming known shape of the power spectrum and 
known geometry, using $\kmaxeff=k \ihMpc$.
}
\label{tableBOSS} 
\begin{tabular}{lccccccccc}
\hline
\hline
$z$ & $\frac{\sigma_{D_A/s}}{D_A/s}$ & $\frac{\sigma_{Hs}}{Hs}$ & 
$\frac{\sigma_{R/s}}{R/s}$ &
$\bar{n}P_{0.2,0}$ & $\bar{n}P_{0.14,0.6}$ & $V$ & 
$\frac{dN_{LRG}}{dz~ d{\rm deg}^2}$ & 
$\frac{\sigma_{f\sigma_{0.1} }}{f \sigma_{0.1}}$
& $\frac{\sigma_{f \sigma_{0.2}}}{f \sigma_{0.2}}$ \\
 & \% & \% & \% &
 &  & $\left(h^{-1}{\rm Gpc}\right)^3$ & & \% & \% 
 \\
\hline
0.05 & 10.64 & 19.44 & 7.34 & 1.79 & 4.08 & 0.03 &   8 & 49.99 & 24.75 \\ 
0.15 & 4.07 & 7.46 & 2.81 & 1.81 & 4.16 & 0.16 &  50 & 18.88 & 9.25 \\ 
0.25 & 2.53 & 4.64 & 1.75 & 1.83 & 4.24 & 0.40 & 125 & 11.68 & 5.66 \\ 
0.35 & 1.86 & 3.42 & 1.29 & 1.86 & 4.32 & 0.70 & 222 & 8.64 & 4.13 \\ 
0.45 & 1.50 & 2.74 & 1.03 & 1.88 & 4.38 & 1.04 & 332 & 7.02 & 3.32 \\ 
0.55 & 1.27 & 2.32 & 0.88 & 1.90 & 4.43 & 1.39 & 447 & 6.06 & 2.84 \\ 
0.65 & 1.60 & 2.68 & 1.07 & 0.71 & 1.65 & 1.73 & 208 & 6.23 & 3.30 \\ 
0.75 & 6.00 & 9.01 & 3.87 & 0.09 & 0.20 & 2.05 &  30 & 13.73 & 10.42 \\ 
\hline
\hline
\end{tabular}
\end{table}

We use $b_{\rm LRG}(z) D(z) = 1.7$.
For the \lyaf\ calculations we use the luminosity function of 
\cite{2006AJ....131.2788J}, for magnitude $g<22$, multiplied by 0.73,
in order to match the observed number density of quasars 
\cite{2013JCAP...04..026S}.

The published analyses of the 
first BOSS data release (DR9) \cite{2012MNRAS.427.3435A,2012MNRAS.426.2719R} 
give us
the opportunity to evaluate the Fisher matrix projections relative to achieved
reality, which we do in the next two subsections, for BAO and RSD.
DR9 covered an effective area of 3275 square 
degrees and these analyses focused only on the high redshift sample, 
CMASS, which dominates the redshift 
distribution above $z \sim 0.45$; the formal redshift cuts were 
$0.43 < z < 0.7$ and did not include any LOWZ targets.  
The number densities were very similar to our assumption (by design).
\cite{2012MNRAS.426.2719R} found clustering amplitude corresponding to
$b \sigma_8(z=0.57)= 1.23$, or $b(z=0.57)=1.97$ for our model with
$\sigma_8(z=0.57)=0.624$, corresponding to $b(z) D(z)=1.48$, in contrast to
1.7 that we assume in the Fisher matrix calculations.

\subsubsection{BOSS BAO projections vs. reality}

To match the CMASS sample 
presented in \cite{2012MNRAS.427.3435A}, we combine the three redshift 
bins in Table~\ref{tableBOSS} at $z=0.45$, 0.55, and 0.65, halving the volume 
for $z=0.45$ to 
account for the observed redshift range of $0.43 < z < 0.7$ (with 
much-suppressed number density at the low $z$ end). We combine 
the errors by simply taking the inverse of the square root of the sum of 
inverse squares of fractional errors.
The combined errors are rescaled to 
account for the effective area of 3275 square degrees. For the dilation 
factor, $R/s$, this gives
projections of 1.85\% error without reconstruction and  1.08\% with 
reconstruction for the DR9 CMASS sample. Using the bias measured from the
data as mentioned above, $b D=1.48$, instead of $b D=1.7$,
and a slightly more exact match to the number density distribution,
we derive 1.94\% and 1.15\% (this $\sim$6\% change in distance error is so 
small that we continue to use the traditional $b D=1.7$ for projections).
This is to be compared with the 
average of the mocks in Fig. 13 of \cite{2012MNRAS.427.3435A}, which is 
approximately 2.6\% and 1.8\% before and after reconstruction, respectively, 
and therefore the mock results in \cite{2012MNRAS.427.3435A} are 1.34 (before 
reconstruction) and 1.57 (after reconstruction) times the Fisher matrix 
projections. 
The actual measured results from the data are better than expected,
1.7\% both before and after reconstruction,
but if we believe the mocks accurately represent the statistics of the 
measurement these differences must be just statistical fluctuations, 
and in any case the post-reconstruction
ratio of error measured from data to Fisher estimate is still 1.48.

We can only speculate about the reasons for this discrepancy, 
i.e, $\sim 34\%$ 
before reconstruction and $\sim 57\%$ after reconstruction. 
First, note that \cite{2012MNRAS.427.3435A} find practically identical errors 
using the power spectrum or correlation function, so a difference between 
these two approaches cannot be the explanation.
\cite{2012MNRAS.427.3435A} used a spherically averaged power 
spectrum while our Fisher matrix assumes an anisotropic power spectrum when 
deriving an isotropic error. According to \cite{2009ApJ...700..479T} (their 
eq. [9]),    
using a spherically averaged power spectrum would return a slightly worse error 
than using an anisotropic power spectrum and then projecting two dimensional 
errors in ${D_A/s}$ and $Hs$ 
on $R/s$; for the fiducial value of $f$, we only expect $\sim 7\%$ difference 
in error.
A non-Gaussian aspect could exist in a likelihood curve of the dilation 
factor in the real survey, however 
\cite{2012MNRAS.427.3435A} estimates the likelihood to be fairly Gaussian.
Meanwhile we expect that the sample 
variance on the errors of the dilation scale from the finite number of mocks
is not big enough to explain this 
discrepancy; we expect $\sim 3\%$ for the 600 mocks 
(i.e., $1/\sqrt{600\times 2}$). 
The level of a Finger-of-God in the CMASS sample introduces only a small effect in the Fisher errors
and therefore cannot explain the $\sim 34\%$ discrepancy.  
It is possible that small scale power due to nonlinear structure growth and 
bias 
could have increased an effective shot noise level relative to the underlying 
BAO signal, although an amount of power large enough to increase the errors
this much should be obvious in the band-power measurement.
Finally, the power spectrum/correlation function
estimators used in \cite{2012MNRAS.427.3435A} are not precisely optimal, 
although we would not expect the effect to be this large.

The discrepancy between CMASS DR9 and the Fisher formalism increases after 
reconstruction.
We note that the
Fisher matrix errors have not been as rigorously tested for the reconstructed 
field as the original field (although see \cite{2012MNRAS.419.2949N}, which
did do careful tests with a reconstructed field, although only for the 
mass density). 
The dependence of the noise properties of the reconstructed field 
on the details of the reconstruction needs detailed tests in the future. 
For example, the reconstruction in  
\cite{2012MNRAS.427.3435A,2012MNRAS.426.2719R} 
adapted conventions to restore isotropy on large scales, which were not used in
many of the tests of the method, and may not have well understood noise 
properties.

We note that the geometry of DR9 was significantly stripey, i.e., not a nice 
compact (e.g., square) 3275 sq. deg. \cite{2012MNRAS.427.3435A}, 
which could lead to an unavoidable degradation in the
measurement relative to the Fisher matrix (basically because the $k$-band
correlation length could become too large to resolve the BAO wiggles), could 
degrade reconstruction, and
could exacerbate any sub-optimality in the analysis. The next generation of 
analyses will be quite compact, so it will be interesting to see if the 
discrepancy is reduced. In fact, while this paper was under review, BOSS
published an analysis of the 8500 sq. deg. DR11 data set 
\cite{2013arXiv1312.4877A}. Their mock-based mean error estimates, from 
their Table 4, are $\sim 27$\% above our projections for this volume,
with this factor essentially identical pre- and post-reconstruction. This 
does suggest that a substantial part of the problem with DR9 reconstruction
was the pathological geometry. By the same token, the relatively modest 
improvement in the pre-reconstruction results suggests that the remaining 
discrepancy is not likely to be related to geometry. However, it is small 
enough to be more likely explained by combinations of the other reasons  
discussed above.

The bottom line is: 
the errors predicted by BOSS collaboration mocks are somewhat ($\sim 27$\% in
DR11)
larger than standard Fisher matrix projections for them, and we
aren't entirely sure why, considering 
the many tests that have been done. 
However, we have already identified $\sim$10\% worth of {\it fixable} 
sub-optimality above, so overall this does not seem like a big problem for
our projections for future surveys. 
There are three logical possibilities that
should be explored for the remainder: the Fisher projections are overly 
optimistic somehow, the
analysis of the mocks and data is noticeably sub-optimal, or some imperfection 
in the mocks leads
to even an optimal analysis producing incorrectly large errors.

\subsubsection{BOSS redshift-space distortion projections vs. reality}

Predictions for RSD are generally less certain than for BAO
(really we only project a range of possibilities, for different $\kmaxeff$),
but it is still useful to see how they compare to the real analysis of
\cite{2012MNRAS.426.2719R}. 
From Table~\ref{tableBOSS}, we estimate the expected fractional error on 
$f\sigma_{0.1}$ ($f\sigma_{0.2}$) to be 
4.0 (2.0)\% for 10000 sq. deg., or 7.0 (3.4)\% for DR9.  

Model 2 in \cite{2012MNRAS.426.2719R}, which treats $f\sigma_8$ as a free 
parameter and marginalizes over uncertainty in the linear matter power spectrum
and flat $\Lambda$CDM distance-redshift relation, finds 
$f\sigma_8 = 0.415 \pm 0.033$, which is an 8.0\% fractional error relative to
the measured value, or  6.9\% of our fiducial $f\sigma_8 = 0.48$  
(there is some statistical error in this kind of percent error, of order the 
error itself, i.e., if the measured value fluctuates low, a percent error
based on it will be larger, and vice versa).
Table 2 of \cite{2012MNRAS.426.2719R} shows that the uncertainty remaining 
after including CMB constraints on the linear matter power spectrum or 
$\Lambda$CDM distance-redshift relation does not contribute to the DR9 error 
budget for measuring $f\sigma_8$, so we can directly compare with the Fisher 
projections above.  Table 2 of \cite{2012MNRAS.426.2719R} also shows that the 
uncertainty would be reduced to 7.0\% (6.0\% of our fiducial value) if 
a Finger-of-God nuisance parameter were held fixed.  Since the DR9 analysis was
performed in configuration space and the Fisher analysis in Fourier space, an
attempt to compare at equal $k_{\rm max}$ can only be approximate.  
\cite{2011MNRAS.417.1913R} found an approximate mapping between minimum 
configuration scale $s_{\rm min}$ and an equivalent  
$k_{\rm max}$, which suggests for the DR9 analysis 
$k_{\rm max} \sim 0.14\ihMpc$.
The bottom line seems to be that the DR9 results are consistent with the 
projections assuming $\kmaxeff \sim 0.1 \ihMpc$.
This is consistent with the idea that fits will go to somewhat larger $k$
than $\kmaxeff$, but lose some information to marginalization over nuisance
nuisance parameters describing nonlinear 
effects such as Fingers-of-God, and some to non-Gaussian errors that enhance 
the data covariance matrix above the naive Fisher prediction.  
We conclude that the scheme adopted in this 
work gives a reasonable
estimate of constraints that are achievable today with $\kmaxeff = 0.1$; we are
optimistic that future theoretical improvements will further enhance the 
constraining power of future surveys.

\subsection{eBOSS}

eBOSS is a proposed extension of BOSS that would cover 7500 sq. deg., 6000
sq. deg. focused entirely on quasars and LRGs at slightly higher redshift than
BOSS LRGs, and another 1500 sq. deg. that adds ELGs similar to those discussed
below for DESI (in addition to LRGs and quasars as in the 6000 sq. deg.). 
Table \ref{tableeBOSS} shows basic numbers for eBOSS \cite{eBOSS}.
eBOSS will also target quasars at $z > 2.15$ over the 7500 sq. deg.,
and re-observe some BOSS quasars to obtain better signal-to-noise in the
spectra, in order
to improve the \lya\ BAO measurement from BOSS by $\sim 25\%$. 
For simplicity, 
we do not consider the \lya\ part of eBOSS in our analysis. 

\begin{table}
\caption{Basic numbers for eBOSS.
The redshift range is covered twice, first showing the 1500 sq. deg. that
will include ELGs, and then the 6000 sq. deg. that will not.}
\label{tableeBOSS} 
\begin{tabular}{lccccccccc}
\hline
\hline
$z$ & $ \frac{\sigma_{D_A/s}}{D_A/s}$ & $\frac{\sigma_{Hs}}{Hs}$ &
$\frac{\sigma_{R/s}}{R/s}$ &
$\bar{n}P_{0.2,0}$ & $\bar{n}P_{0.14,0.6}$ & $V$ &
$\frac{dN_{ELG}}{dz~ d{\rm deg}^2}$ & $\frac{dN_{LRG}}{dz ~d{\rm deg}^2}$ &
$\frac{dN_{QSO}}{dz~ d{\rm deg}^2}$ \\
 & \% & \% & \% &
 &  & $\left(h^{-1}{\rm Gpc}\right)^3$ &
 \\
\hline
0.55 & 32.45 & 48.38 & 20.91 & 0.05 & 0.11 & 0.21 &   0 &  11 &   0 \\
0.65 & 3.63 & 6.15 & 2.44 & 0.95 & 2.28 & 0.26 & 156 & 240 &   0 \\ 
0.75 & 3.42 & 5.53 & 2.26 & 0.83 & 2.17 & 0.31 & 619 & 139 &   0 \\ 
0.85 & 4.02 & 6.09 & 2.59 & 0.50 & 1.32 & 0.35 & 506 &  76 &   0 \\ 
0.95 & 10.60 & 13.71 & 6.40 & 0.11 & 0.29 & 0.39 & 159 &   8 &   0 \\ 
1.05 & 21.65 & 29.28 & 13.39 & 0.05 & 0.11 & 0.43 &   0 &   0 &  44 \\ 
1.15 & 22.11 & 29.85 & 13.67 & 0.04 & 0.10 & 0.46 &   0 &   0 &  44 \\ 
1.25 & 19.07 & 25.83 & 11.80 & 0.05 & 0.12 & 0.49 &   0 &   0 &  53 \\ 
1.35 & 19.30 & 26.14 & 11.95 & 0.05 & 0.11 & 0.51 &   0 &   0 &  53 \\ 
1.45 & 16.98 & 23.06 & 10.53 & 0.05 & 0.12 & 0.53 &   0 &   0 &  62 \\ 
1.55 & 17.09 & 23.23 & 10.60 & 0.05 & 0.12 & 0.55 &   0 &   0 &  62 \\ 
1.65 & 18.46 & 25.07 & 11.45 & 0.04 & 0.10 & 0.57 &   0 &   0 &  57 \\ 
1.75 & 18.52 & 25.19 & 11.50 & 0.04 & 0.10 & 0.58 &   0 &   0 &  57 \\ 
1.85 & 20.09 & 27.32 & 12.47 & 0.04 & 0.09 & 0.59 &   0 &   0 &  52 \\ 
1.95 & 20.11 & 27.39 & 12.49 & 0.04 & 0.09 & 0.60 &   0 &   0 &  52 \\ 
2.05 & 21.94 & 29.90 & 13.64 & 0.03 & 0.08 & 0.60 &   0 &   0 &  47 \\ 
2.15 & 21.93 & 29.93 & 13.64 & 0.03 & 0.08 & 0.61 &   0 &   0 &  47 \\ 
\hline
0.55 & 16.22 & 24.19 & 10.45 & 0.05 & 0.11 & 0.83 &   0 &  11 &   0 \\ 
0.65 & 1.94 & 3.28 & 1.30 & 0.82 & 1.90 & 1.04 &   0 & 240 &   0 \\ 
0.75 & 2.52 & 4.02 & 1.66 & 0.40 & 0.93 & 1.23 &   0 & 139 &   0 \\ 
0.85 & 3.76 & 5.73 & 2.44 & 0.19 & 0.44 & 1.41 &   0 &  76 &   0 \\ 
0.95 & 28.68 & 42.75 & 18.51 & 0.02 & 0.04 & 1.57 &   0 &   8 &   0 \\ 
1.05 & 10.83 & 14.64 & 6.69 & 0.05 & 0.11 & 1.72 &   0 &   0 &  44 \\ 
1.15 & 11.05 & 14.93 & 6.83 & 0.04 & 0.10 & 1.85 &   0 &   0 &  44 \\ 
1.25 & 9.53 & 12.91 & 5.90 & 0.05 & 0.12 & 1.96 &   0 &   0 &  53 \\ 
1.35 & 9.65 & 13.07 & 5.97 & 0.05 & 0.11 & 2.06 &   0 &   0 &  53 \\ 
1.45 & 8.49 & 11.53 & 5.26 & 0.05 & 0.12 & 2.14 &   0 &   0 &  62 \\ 
1.55 & 8.55 & 11.61 & 5.30 & 0.05 & 0.12 & 2.21 &   0 &   0 &  62 \\ 
1.65 & 9.23 & 12.54 & 5.72 & 0.04 & 0.10 & 2.27 &   0 &   0 &  57 \\ 
1.75 & 9.26 & 12.59 & 5.75 & 0.04 & 0.10 & 2.31 &   0 &   0 &  57 \\ 
1.85 & 10.04 & 13.66 & 6.24 & 0.04 & 0.09 & 2.35 &   0 &   0 &  52 \\ 
1.95 & 10.06 & 13.69 & 6.25 & 0.04 & 0.09 & 2.38 &   0 &   0 &  52 \\ 
2.05 & 10.97 & 14.95 & 6.82 & 0.03 & 0.08 & 2.40 &   0 &   0 &  47 \\ 
2.15 & 10.96 & 14.97 & 6.82 & 0.03 & 0.08 & 2.42 &   0 &   0 &  47 \\ 
\hline
\hline
\end{tabular}
\end{table}

We assume that we can add the constraints from the two areas as if they are
independent (in the usual way of this kind of Fisher matrix, this will be 
correct up to survey edge effects). 

\subsection{HETDEX}

For HETDEX (\url{http://hetdex.org}), 
we do not have a complete redshift distribution, only the 
number 0.8 million galaxies, area 420 sq. deg., and a redshift range 
$1.9<z<3.5$
\cite{2013arXiv1306.4157C}, 
so we use a fixed $\frac{dN}{dz~ d{\rm deg}^2}=1190$.
We use bias $b(z)D(z)=0.89$ \cite{2013arXiv1306.4157C}.
Table \ref{tableHETDEX} shows the basic HETDEX numbers.
\begin{table}
\caption{
Basic numbers for HETDEX, covering 420 sq. deg. }
\label{tableHETDEX}
\begin{tabular}{lccccccc}
\hline
\hline
$z$ & $ \frac{\sigma_{D_A/s}}{D_A/s}$ & $\frac{\sigma_{Hs}}{Hs}$ &
$\frac{\sigma_{R/s}}{R/s}$ &
$\bar{n}P_{0.2,0}$ & $\bar{n}P_{0.14,0.6}$ & $V$ &
$\frac{dN}{dz~ d{\rm deg}^2}$ \\
 & \% & \% & \% &
 &  & $\left(h^{-1}{\rm Gpc}\right)^3$ &
 \\
\hline
1.95 & 5.51 & 7.93 & 3.49 & 0.48 & 1.21 & 0.17 & 1190 \\ 
2.05 & 5.50 & 7.90 & 3.48 & 0.48 & 1.19 & 0.17 & 1190 \\ 
2.15 & 5.49 & 7.88 & 3.47 & 0.48 & 1.18 & 0.17 & 1190 \\ 
2.25 & 5.47 & 7.86 & 3.46 & 0.47 & 1.16 & 0.17 & 1190 \\ 
2.35 & 5.46 & 7.84 & 3.46 & 0.47 & 1.15 & 0.17 & 1190 \\ 
2.45 & 5.45 & 7.82 & 3.45 & 0.47 & 1.14 & 0.17 & 1190 \\ 
2.55 & 5.44 & 7.81 & 3.45 & 0.47 & 1.14 & 0.17 & 1190 \\ 
2.65 & 5.44 & 7.81 & 3.44 & 0.47 & 1.13 & 0.17 & 1190 \\ 
2.75 & 5.43 & 7.80 & 3.44 & 0.48 & 1.13 & 0.17 & 1190 \\ 
2.85 & 5.42 & 7.80 & 3.43 & 0.48 & 1.13 & 0.17 & 1190 \\ 
2.95 & 5.41 & 7.79 & 3.43 & 0.48 & 1.13 & 0.17 & 1190 \\ 
3.05 & 5.40 & 7.79 & 3.43 & 0.48 & 1.13 & 0.17 & 1190 \\ 
3.15 & 5.40 & 7.79 & 3.43 & 0.49 & 1.13 & 0.17 & 1190 \\ 
3.25 & 5.39 & 7.79 & 3.43 & 0.49 & 1.14 & 0.16 & 1190 \\ 
3.35 & 5.39 & 7.80 & 3.42 & 0.49 & 1.14 & 0.16 & 1190 \\ 
3.45 & 5.38 & 7.80 & 3.42 & 0.50 & 1.14 & 0.16 & 1190 \\ 
\hline
\hline
\end{tabular}
\end{table}

In the interest of limiting the length of our main results tables, we only 
include a limited set of cases using HETDEX.

\subsection{DESI}

DESI (short for Dark Energy Spectroscopic Instrument 
\cite{2013arXiv1308.0847L}) is a 
galaxy and quasar redshift survey likely to run on the Mayall 4 
meter telescope at Kitt Peak National
Observatory near Tucson, AZ, over an approximately five year period from
2018 to 2022.  The baseline area is 14000 sq. deg.
We also consider the possibility that the spectrograph could then be moved to 
the
twin Blanco telescope in Chile to cover another $\sim 10000$ sq. deg.
We take the BigBOSS numbers 
to represent 
DESI, although this is not set in stone 
(see \cite{2011arXiv1106.1706S,2012SPIE.8446E..0QM}, but the numbers
we actually use are revised ones presented in 
\cite{2012AAS...21933513M, 2013arXiv1308.0847L}). 
DESI will target 3 types of objects: 
Luminous Red
Galaxies (LRGs) are bright, highly biased red objects that are easy to
target from spectroscopic data (since the galaxy target selection for
BOSS is not exactly the same as for the SDSS I and
II LRGs, they are distinguished as CMASS and LOWZ galaxies 
in BOSS analysis papers; however, at the level of this paper they
are essentially the same class of objects, which we will call LRGs).
A second class of objects are Emission Line Galaxies (ELGs)
\cite{2013ApJ...767...89M}, which require a higher
resolution spectrograph to type and redshift, since this is only possible
if the OII doublet is resolved. ELGs are considerably less biased than LRGs.
For ELGs we use
$b_{\rm ELG}(z) D(z) = 0.84$  \cite{2013ApJ...767...89M}.
Finally, we use quasars as tracers of cosmic structure.  Quasars are
difficult to target photometrically, especially in the redshift range
$2<z<3.5$, but can be very efficiently targeted using
variability. They are very highly biased, but are limited by their
limited number density, which is considerably lower than that of LRGs
and ELGs. For quasars we use
$b_{\rm QSO}(z) D(z) = 1.2$, loosely based on \cite{2009ApJ...697.1634R}.

Numbers we use for Fisher matrix projections are given in Table 
\ref{tableDESI}.

\begin{table}
\caption{
Basic numbers for DESI, covering 14000 sq. deg. }
\label{tableDESI} 
\begin{tabular}{lccccccccccc}
\hline
\hline
$z$ & $ \frac{\sigma_{D_A/s}}{D_A/s}$ & $\frac{\sigma_{Hs}}{Hs}$ & 
$\frac{\sigma_{R/s}}{R/s}$ &
$\bar{n}P_{0.2,0}$ & $\bar{n}P_{0.14,0.6}$ & $V$ & 
$\frac{dN_{ELG}}{dz~ d{\rm deg}^2}$ & $\frac{dN_{LRG}}{dz ~d{\rm deg}^2}$ &
$\frac{dN_{QSO}}{dz~ d{\rm deg}^2}$ &
$\frac{\sigma_{f\sigma_{0.1} }}{f \sigma_{0.1}}$
& $\frac{\sigma_{f \sigma_{0.2}}}{f \sigma_{0.2}}$ \\
 & \% & \% & \% &
 &  & $\left(h^{-1}{\rm Gpc}\right)^3$ & & & & \% & \%
 \\
\hline
0.15 & 2.78 & 5.34 & 1.95 & 5.24 & 13.79 & 0.23 & 376 &  50 &   8 & 7.51 & 3.60 \\ 
0.25 & 1.87 & 3.51 & 1.30 & 3.24 & 8.19 & 0.56 & 347 & 125 &  23 & 5.24 & 2.55 \\ 
0.35 & 1.45 & 2.69 & 1.00 & 2.58 & 6.35 & 0.99 & 291 & 222 &  31 & 4.44 & 2.17 \\ 
0.45 & 1.19 & 2.20 & 0.82 & 2.36 & 5.74 & 1.46 & 285 & 332 &  31 & 3.92 & 1.91 \\ 
0.55 & 1.01 & 1.85 & 0.70 & 2.42 & 5.90 & 1.94 & 431 & 448 &  32 & 3.31 & 1.60 \\ 
0.65 & 0.87 & 1.60 & 0.60 & 2.58 & 6.34 & 2.42 & 722 & 563 &  34 & 2.80 & 1.34 \\ 
0.75 & 0.77 & 1.41 & 0.53 & 2.77 & 6.85 & 2.87 & 1112 & 675 &  37 & 2.47 & 1.18 \\ 
0.85 & 0.76 & 1.35 & 0.52 & 2.05 & 5.17 & 3.29 & 1333 & 471 &  44 & 2.34 & 1.11 \\ 
0.95 & 0.88 & 1.42 & 0.58 & 1.03 & 2.76 & 3.67 & 1401 &  91 &  50 & 2.34 & 1.13 \\ 
1.05 & 0.91 & 1.41 & 0.59 & 0.82 & 2.24 & 4.01 & 1469 &  11 &  56 & 2.32 & 1.12 \\ 
1.15 & 0.91 & 1.38 & 0.58 & 0.75 & 2.05 & 4.31 & 1483 &   0 &  62 & 2.30 & 1.12 \\ 
1.25 & 0.91 & 1.36 & 0.58 & 0.69 & 1.86 & 4.57 & 1421 &   0 &  69 & 2.32 & 1.14 \\ 
1.35 & 1.00 & 1.46 & 0.64 & 0.53 & 1.42 & 4.80 & 1120 &   0 &  75 & 2.45 & 1.26 \\ 
1.45 & 1.17 & 1.66 & 0.74 & 0.38 & 1.00 & 4.99 & 775 &   0 &  81 & 2.71 & 1.47 \\ 
1.55 & 1.50 & 2.04 & 0.93 & 0.25 & 0.63 & 5.15 & 460 &   0 &  83 & 3.22 & 1.89 \\ 
1.65 & 2.36 & 3.15 & 1.45 & 0.13 & 0.33 & 5.29 & 179 &   0 &  80 & 4.63 & 3.06 \\ 
1.75 & 3.62 & 4.87 & 2.23 & 0.08 & 0.19 & 5.40 &  49 &   0 &  77 & 7.17 & 5.14 \\ 
1.85 & 4.79 & 6.55 & 2.98 & 0.06 & 0.13 & 5.49 &   0 &   0 &  74 & 10.26 & 7.66 \\ 
\hline
\hline
\end{tabular}
\end{table}

\begin{figure}[tb]
\centering
\includegraphics[height=4.5in]{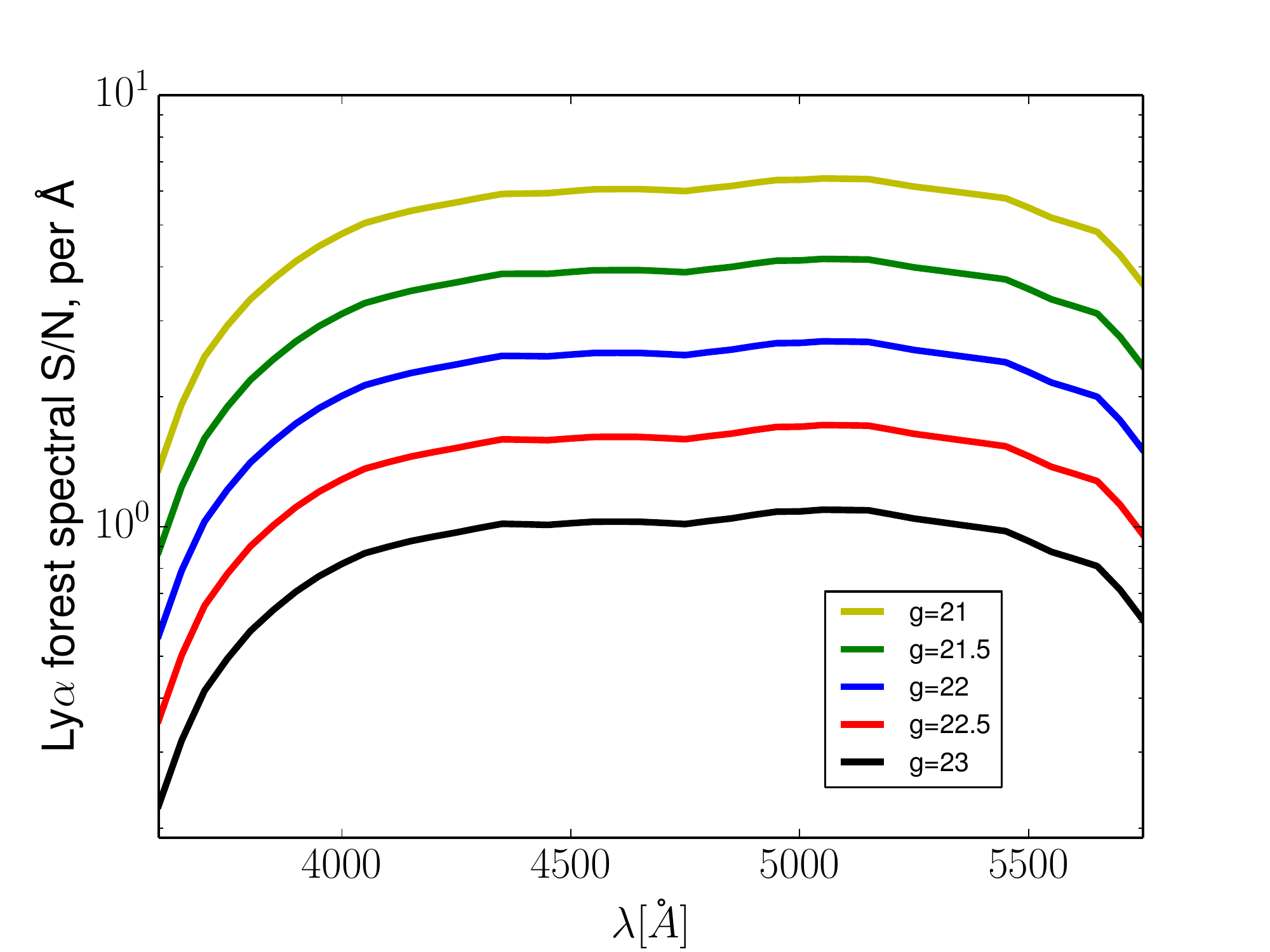}
\caption{
Signal-to-noise ratio per \AA\ used for DESI quasar spectra (detector noise,
not absorption noise), for different 
$g$ magnitudes, accounting for mean \lyaf\ absorption. 
(We only use the blue DESI spectrograph -- we could squeeze out a little 
more BAO information at $z\gtrsim 3.7$ by including the red spectrograph.)
}
\label{fig_qSN}
\end{figure}

The quasar luminosity function use for the \lyaf\ calculation follows 
\cite{2013A&A...551A..29P}, for magnitude $g<23$, with a 0.8 reduction in 
numbers to allow for targeting inefficiency. 
The spectral signal-to-noise ratio that we use, computed using the 
{\it BBspecsim} code \cite{2012SPIE.8446E..0QM}, is shown in 
Figure \ref{fig_qSN}.
Note that we always absorb BOSS and eBOSS into DESI, as they will be physically
overlapping (it was not necessary to combine BOSS and eBOSS because they
cover distinct redshift ranges).

At the heart of the \lyaf\ data at $z\sim 2.5$, 
$\bar{n}P(k=0.14\ihMpc, \mu=0.6)= 0.09$ for DESI quasar 
clustering, while scaling from the \lyaf\ BAO errors gives $\sim 0.22$ for the
\lyaf\, i.e., 
any corrections to the high-noise limit when combining the two will be 
modest, although we are on the borderline of applicability for this  
approximation.
Note that we have not included reconstruction of the non-linear damping of
the BAO feature, which might produce a small ($\sim 10-15$\%) improvement. 

\subsection{Euclid}

We only include the redshift survey from Euclid, assuming 15000 sq. deg. and
a total of 50 million galaxies, based on \cite{2011arXiv1110.3193L}
(note that
there is some uncertainty in the expected number of galaxies). 
Adding Euclid lensing
should be qualitatively similar to LSST. 
Numbers we use for Euclid Fisher matrix projections
are given in Table \ref{tableEuclid}.
\begin{table}
\caption{
Basic numbers for Euclid (50 million total galaxies), covering 15000 sq.
deg. Note that the number densities here may be optimistic by a factor 
$\sim 2$ \cite{2013arXiv1305.5422S}.}
\label{tableEuclid} 
\begin{tabular}{lccccccc}
\hline
\hline
$z$ & $ \frac{\sigma_{D_A/s}}{D_A/s}$ & $\frac{\sigma_{Hs}}{Hs}$ & 
$\frac{\sigma_{R/s}}{R/s}$ &
$\bar{n}P_{0.2,0}$ & $\bar{n}P_{0.14,0.6}$ & $V$ & 
$\frac{dN}{dz~ d{\rm deg}^2}$ \\
 & \% & \% & \% &
 &  & $\left(h^{-1}{\rm Gpc}\right)^3$ & 
 \\
\hline
0.65 & 1.23 & 1.89 & 0.79 & 0.75 & 2.24 & 2.59 & 1100 \\ 
0.75 & 0.83 & 1.42 & 0.56 & 1.69 & 5.03 & 3.07 & 2950 \\ 
0.85 & 0.74 & 1.27 & 0.50 & 1.90 & 5.60 & 3.52 & 3800 \\ 
0.95 & 0.71 & 1.19 & 0.48 & 1.75 & 5.11 & 3.93 & 3900 \\ 
1.05 & 0.70 & 1.14 & 0.46 & 1.55 & 4.48 & 4.29 & 3775 \\ 
1.15 & 0.70 & 1.12 & 0.46 & 1.35 & 3.85 & 4.62 & 3525 \\ 
1.25 & 0.70 & 1.10 & 0.46 & 1.17 & 3.31 & 4.90 & 3250 \\ 
1.35 & 0.73 & 1.11 & 0.47 & 0.98 & 2.74 & 5.14 & 2850 \\ 
1.45 & 0.78 & 1.16 & 0.50 & 0.78 & 2.15 & 5.35 & 2350 \\ 
1.55 & 0.87 & 1.24 & 0.55 & 0.59 & 1.62 & 5.52 & 1850 \\ 
1.65 & 1.01 & 1.40 & 0.63 & 0.43 & 1.16 & 5.66 & 1375 \\ 
1.75 & 1.23 & 1.64 & 0.75 & 0.30 & 0.80 & 5.78 & 975 \\ 
1.85 & 1.61 & 2.07 & 0.97 & 0.20 & 0.52 & 5.88 & 650 \\ 
1.95 & 2.32 & 2.90 & 1.38 & 0.12 & 0.31 & 5.95 & 400 \\ 
2.05 & 5.32 & 6.39 & 3.11 & 0.04 & 0.12 & 6.01 & 150 \\ 
\hline
\hline
\end{tabular}
\end{table}
We assume fiducial $b(z)D(z)=0.76$. For simplicity, we assume Euclid does
not overlap with DESI. To the extent that there is some overlap, there will be 
some degradation of combined constraints relative to what we quote (we do not
see any case where this should critically change one's basic picture of how 
well parameters can be measured).

Note that the WFIRST-AFTA report \cite{2013arXiv1305.5422S} appears to suggest
that these Euclid numbers are very optimistic. They forecast that Euclid will
find factors of 8, 16, and 30 lower number density, at $z=1.1$, 1.5, and
1.9, respectively, than they forecast for WFIRST, which correspond to factors
0.38, 0.49, and 0.49 smaller number density than we use in this paper -- 
this would obviously lead to some degradation of our projections for Euclid.  
To be clear: we continue to use the relatively optimistic ``official'' Euclid
numbers, with 50 million total galaxies as shown in Table \ref{tableEuclid}. 
The WFIRST report suggests that these numbers that we use are {\it too high} 
by a factor 
$\gtrsim 2$ (the factors 8, 16, and 30 are relative to the much higher WFIRST 
densities -- we quote these numbers to be very accurate about what the WFIRST 
report says).

\subsection{WFIRST}

We implement WFIRST-2.4 following \cite{2013arXiv1305.5422S}.
Number densities come from their Table 2.2. 
The bias formula that we use for Euclid, $b(z)D(z)=0.76$,
happens to be exactly equal to the formula of
\cite{2013arXiv1305.5422S}, $b=1.5+0.4 (z-1.5)$, at $z=1.5$, and 
within 10\% over the full redshift range, so we use this also for WFIRST.
The numbers we use for Fisher projections are given in 
Table \ref{tableWFIRST}.
\begin{table}
\caption{
Basic numbers for WFIRST-2.4, covering 2000 sq.
deg. }
\label{tableWFIRST}
\begin{tabular}{lccccccc}
\hline
\hline
$z$ & $ \frac{\sigma_{D_A/s}}{D_A/s}$ & $\frac{\sigma_{Hs}}{Hs}$ &
$\frac{\sigma_{R/s}}{R/s}$ &
$\bar{n}P_{0.2,0}$ & $\bar{n}P_{0.14,0.6}$ & $V$ &
$\frac{dN}{dz~ d{\rm deg}^2}$ \\
 & \% & \% & \% &
 &  & $\left(h^{-1}{\rm Gpc}\right)^3$ &
 \\
\hline
1.05 & 1.51 & 2.72 & 1.03 & 4.37 & 12.60 & 0.57 & 10623 \\ 
1.15 & 1.43 & 2.56 & 0.98 & 4.50 & 12.85 & 0.62 & 11776 \\ 
1.25 & 1.35 & 2.42 & 0.92 & 5.00 & 14.13 & 0.65 & 13877 \\ 
1.35 & 1.29 & 2.30 & 0.88 & 5.33 & 14.90 & 0.69 & 15527 \\ 
1.45 & 1.24 & 2.21 & 0.85 & 5.58 & 15.42 & 0.71 & 16890 \\ 
1.55 & 1.23 & 2.16 & 0.84 & 5.04 & 13.79 & 0.74 & 15759 \\ 
1.65 & 1.25 & 2.15 & 0.84 & 4.15 & 11.23 & 0.76 & 13305 \\ 
1.75 & 1.28 & 2.16 & 0.86 & 3.33 & 8.94 & 0.77 & 10918 \\ 
1.85 & 1.33 & 2.19 & 0.88 & 2.61 & 6.94 & 0.78 & 8697 \\ 
1.95 & 1.41 & 2.27 & 0.93 & 1.99 & 5.25 & 0.79 & 6718 \\ 
2.05 & 2.51 & 3.52 & 1.57 & 0.47 & 1.23 & 0.80 & 1610 \\ 
2.15 & 2.60 & 3.62 & 1.62 & 0.44 & 1.14 & 0.81 & 1509 \\ 
2.25 & 2.74 & 3.78 & 1.70 & 0.40 & 1.02 & 0.81 & 1368 \\ 
2.35 & 3.02 & 4.09 & 1.86 & 0.33 & 0.85 & 0.81 & 1156 \\ 
2.45 & 3.38 & 4.52 & 2.08 & 0.28 & 0.70 & 0.81 & 960 \\ 
2.55 & 3.87 & 5.11 & 2.36 & 0.23 & 0.57 & 0.81 & 781 \\ 
2.65 & 4.52 & 5.90 & 2.75 & 0.18 & 0.45 & 0.81 & 626 \\ 
2.75 & 5.41 & 6.99 & 3.27 & 0.14 & 0.35 & 0.81 & 490 \\ 
\hline
\hline
\end{tabular}
\end{table}

In the interest of limiting the length of our main results tables, we only 
include a limited set of cases using WFIRST.

\subsection{Summary of S/N and BAO distance errors vs. redshift for redshift 
surveys}

\begin{figure}[tb]
\centering
\includegraphics[height=4.5in]{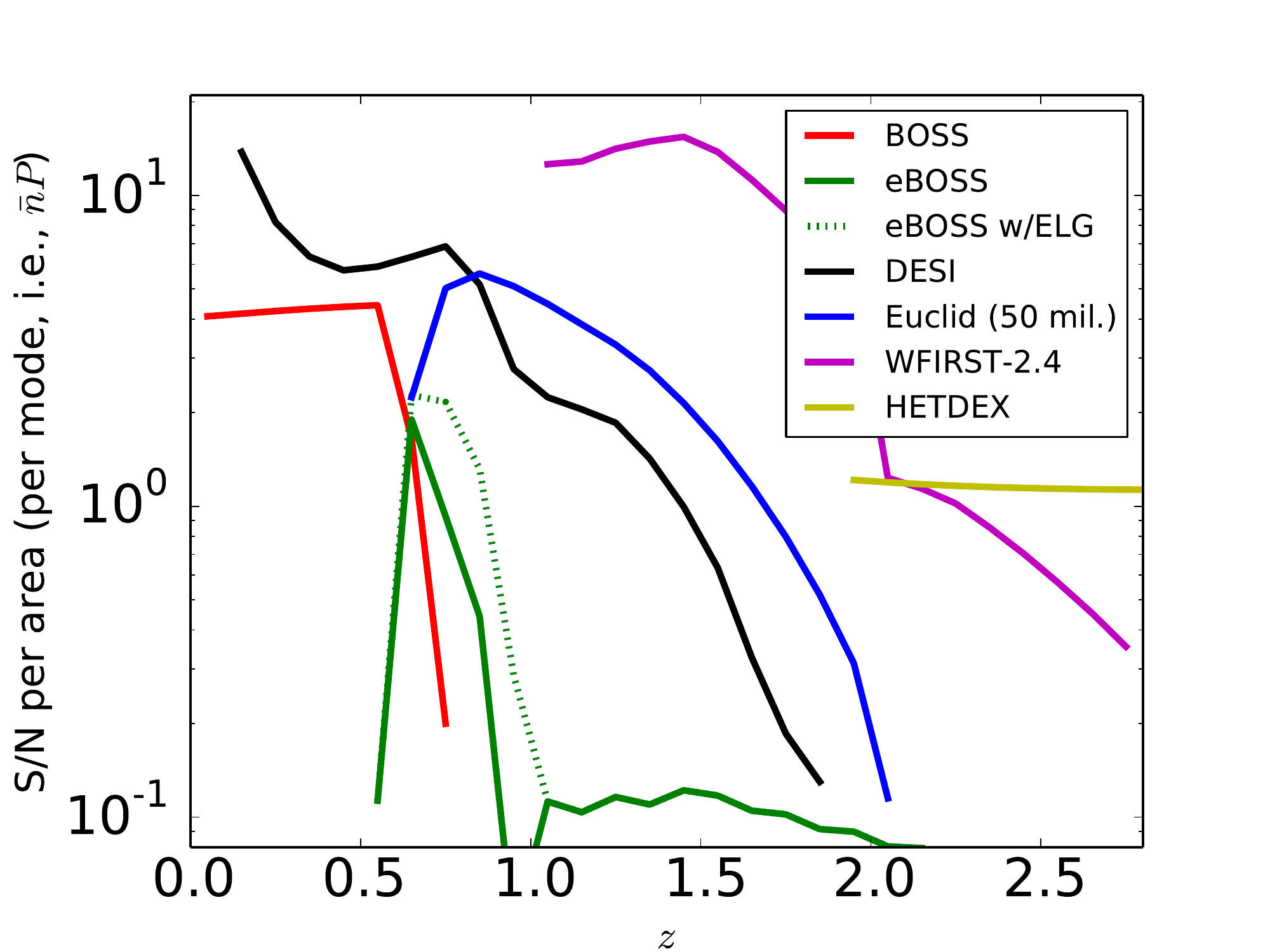}
\caption{
$\bar{n} P(k=0.14 \ihMpc, \mu=0.6)$ comparison. DESI does not include the 
\lyaf\ contribution, which would bring it to effective $\bar{n}P\sim 0.3$ 
at $z\sim 2.5$ (over a much wider area than HETDEX and WFIRST).
}
\label{fig_nP}
\end{figure}

Coincidentally, the BAO scale and non-linear scale are quite similar, so BAO
errors can summarize well the general relative constraining power
of redshift surveys and their redshift dependence. 
The signal-to-noise for typical BAO-scale modes in redshift space is shown
in Fig. \ref{fig_nP}.
We evaluate $\bar{n}P$ at $k=0.14\ihMpc$,
$\mu=0.6$, an approximate center-of-weight point for BAO measurements.
We chose the numbers 0.14 and 0.6 by looking for the point where
$\bar{n}P=1$ corresponded to the optimum in a trade-off between area and
number density at fixed total number of objects (specifically, for the full 
range 
of parameters covered by DESI LRGs and ELGs). We think this definition reflects 
the origin of the idea that
$\bar{n}P=1$ is a special point, but it should be kept in mind that 
achieving $\bar{n}P$ by this 
definition does leave a survey significantly farther away from the 
sample variance limit than the traditional definition $k=0.2\ihMpc$,
$\mu=0$.

\begin{figure}[tb]
\centering
\includegraphics[height=4.5in]{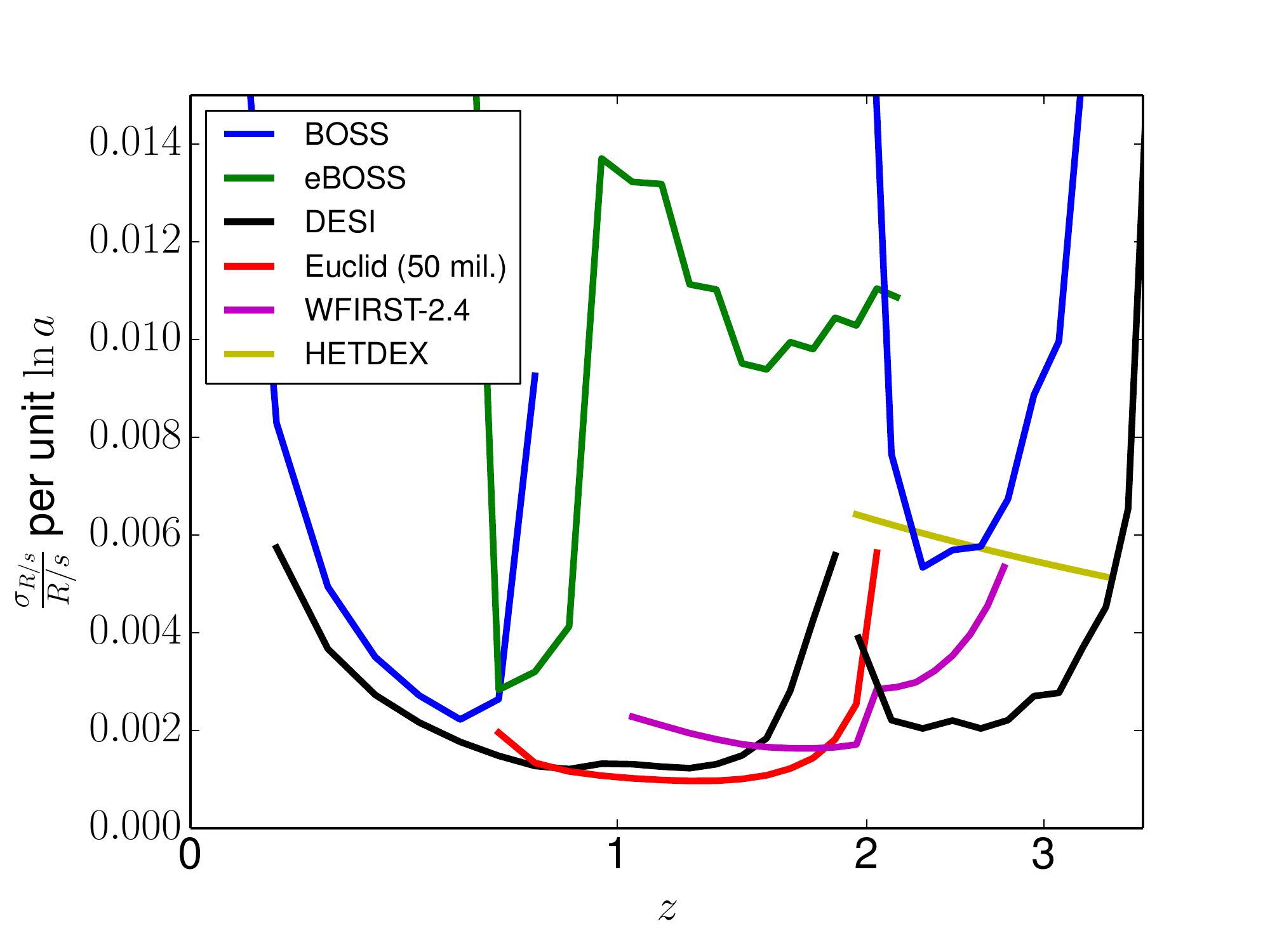}
\caption{
Fractional error on the dilation factor as a function of redshift, 
per unit $\ln a$, i.e., this is some sense a distance error {\it density} 
(plotting points aggregated over different bin widths, as is sometimes done, 
is essentially like plotting densities in different units on the same scale
-- of course, to add up information one still needs to integrate the inverse
square of the curve).
In other words, the effect of a width $\Delta z$ is removed in this plot.}
\label{fig_R}
\end{figure}

Projected BAO distance errors are shown in Fig. \ref{fig_R}.

\section{Photometric surveys}
\label{sec:photometric-surveys}

We compute photometric survey Fisher matrices in terms of 
angular cross- and auto-power spectra, $C_\ell^{xy}$, 
between objects divided up into
nominal redshift bins based on photometric redshift estimates, i.e., 
photo-$z$'s (in case it is not obvious, the fundamental reason why it is 
natural to use
different computational methods for redshift and photometric surveys is that 
redshift surveys derive most of their power from fluctuations on radial 
scales much smaller than the scale of cosmological evolution, while lensing 
surveys do not). 
Except when otherwise specified, we use all cross- and auto-correlations 
involving lensing and galaxy density in a set of tomographic bins of width
$\Delta z=0.2$ and maximum redshift $z_{\rm max}=2$, i.e., information from
photometric 
galaxy clustering and galaxy-galaxy lensing are automatically included
\cite{2012MNRAS.420.3240N}. We use $\Delta z=0.2$ after checking that there is 
little change in results in going to finer bins, while the computational 
time does increase significantly (e.g., as shown in Table \ref{tableoverlap},
changes in DETF Figure of Merit when switching to $\Delta z=0.1$ are $<10$\% 
for cases including spectroscopic
surveys and $<20$\% for cases including only lensing, and the changes in 
single-parameter errors are even smaller than this). 
Bins in $\ell$ have size $\Delta \ln \ell=0.2$.
We assume no photo-$z$ systematics, no intrinsic alignments, and no shear
calibration bias.
For shear-shear we use $\ell_{\rm max}=500$, to reflect the limitation set by 
non-linear effects \cite{2012PhRvD..86h3504Y,2009arXiv0901.0721A,
2009ApJ...701..945S, 2001ApJ...554...56C}. 
Overall our treatment of lensing is optimistic, at least relative to the 
FoMSWG 
treatment 
\cite{2009arXiv0901.0721A} (see Table \ref{tableFoMSWGFoM}), but it is beyond 
the scope of this paper to get into lensing 
systematics in detail -- these projections should be thought of as targets
for lensing researchers to work towards. 

In the rest of this section we give some more details describing our 
angular clustering treatment. 

\subsection{Angular Clustering}

Suppose we define an angular field $\delta_{i,x}(\theta)$ as the redshift
integral with weight $W_x(z)$ over a 3D field $\delta_i(z,\theta)$, i.e.,
\begin{equation}
\delta_{ix}(\theta) = \int dz~ W_x(z) ~\delta_i(z,\theta)
\end{equation}
(at this stage this is just a definition, but $\delta_i(z,\theta)$ could be 
the density of some type of galaxy, or mass density, with 
$W_x(z)$ the true redshift distribution of galaxies within some nominal
photo-$z$ bin, or lensing weight, for example).
The angular correlation function of two such fields is
\begin{equation}
\xi_{ixjy}(\Delta\theta) = 
\int dz \int dz' W_x(z) W_y(z') 
\left<\delta_i(z,\theta)\delta_j(z',\theta+\Delta \theta)\right>=
\int d\bar{z} \int d\Delta z W_x(\bar{z}+\Delta z/2) W_y(\bar{z}-\Delta z/2) 
\xi_{ij}(\bar{z},\Delta z, \Delta\theta)
\end{equation}
where note that there is no approximation or loss of generality here, i.e.,
$\bar{z}=(z+z')/2$ and $\Delta z=z-z'$ are equally good parameters of two-point
function evolving completely generally with redshift (although note that in
general the sign of $\Delta z$ does matter, i.e., we cannot use
$\left|\Delta z\right|$). 
We can now FT $\xi_{ij}(\bar{z},\Delta z, \Delta\theta)$  with respect to
$\Delta \theta$ and $\Delta z$ and plug that in to obtain
\begin{equation}
P_{xiyj}(k_\theta) =  
\int d\bar{z} \int d\Delta z W_x(\bar{z}+\Delta z/2) W_y(\bar{z}-\Delta z/2) 
\int \frac{dk_z}{2 \pi} \exp(-i k_z \Delta z) P_{ij}(\bar{z},k_z,k_\theta) 
\end{equation}
Now we make an approximation, that the kernels $W(z)$ are broad enough that
we can assume we are sensitive only to modes with small $k_z$, specifically
$k_z$ much less than $k_\theta$, so that
$P(k_z,k_\theta)\simeq P(0,k_\theta)$ (note that it would be fairly
straightforward to
include higher order terms in a Taylor expansion here).
This allows us to integrate the
exponential factor over $k_z$ to obtain $2 \pi \delta^D(\Delta z)$ and then
\begin{equation}
P_{xiyj}(k_\theta) \simeq
\int dz~ W_x(z)~W_y(z) ~
P_{ij}(z,k_z=0,k_\theta) ~.
\end{equation}
(Needless to say, this approximation breaks down as $k_\theta\rightarrow 0$,
but there are few modes there so our calculation is not sensitive to this.)
Note that the units of $P_{ij}$ are redshift times angle squared, i.e., it is
related to the usual comoving coordinate power spectrum, which we will
identify by arguments $k_\parallel$ and $k_\perp$, by
\begin{equation}
P_{ij}(z,k_z=0,k_\theta) = \frac{H(z)/c}{r^2(z)} P_{ij}(z,k_\parallel=0,
k_\perp = k_\theta /r(z) ) 
\end{equation}
The final step to obtain the usual Limber approximation equation is to
identify $k_\theta$ with $\ell+1/2$, and then $C_{xiyj}(\ell) = 
P_{xiyj}[k_\theta = l+1/2]$,
i.e.,
\begin{equation}  
C_{xiyj}(\ell) = 
\int dz~ W_x(z)~W_y(z) ~
\frac{H(z)/c}{r^2(z)} P_{ij}\left(z,k_\parallel=0,
k_\perp = \frac{\ell+1/2}{r(z)} \right)
~.
\label{eqCxiyj}
\end{equation}
This expression is often found in the literature (e.g.,
\cite{2010PhRvD..81b3503G}), but this derivation may show more clearly the
origin and units of the various factors.

\subsection{Weight functions}

\subsubsection{Density}

One simple example of the use of equation (\ref{eqCxiyj}) is clustering of two
types of galaxy, where $W(z) = n(z)/n_{tot}$ if
$n(z)=dN/dz d\theta^2 = [r^2(z) c/ H(z)] n_{\rm com}(z)$
and $n_{tot} \equiv \int dz~ n(z)$, and in the usual
linear regime picture $P_{ij}(z,0,k_\perp) = b_i(z) b_j(z) P_m(k_\perp) 
+ \delta^K_{ij} n^{-1}_{\rm com}$ ($n_{\rm com}$ is the 3D comoving coordinate
density).
Weight functions here are defined in terms of the {\it actual} redshift
distribution of galaxies in a bin. For a bin defined by
measured redshifts, the redshift errors must be accounted for.

Often we see the formula 
\begin{equation}
n(z) \propto \left(z/z_\star\right)^\alpha
\exp\left[-\left(z/z_\star\right)^\beta\right]
\label{eqnumberdensities}
\end{equation}
used for number densities.
This can integrated analytically to provide a normalization:
\begin{equation}
\int_0^\infty dz \left(\frac{z}{z_\star}\right)^\alpha 
\exp\left[-\left(\frac{z}{z_\star}\right)^\beta\right]=z_\star \beta^{-1}
\Gamma\left[\frac{\alpha+1}{\beta}\right]
\end{equation}
E.g., for $\alpha=2$, $\beta=1$, the $\gamma$ function evaluates to 2, so
\begin{equation}
\frac{dN}{dz}(z) = \frac{n_{tot}}{2 z_\star} \left(\frac{z}{z_\star}\right)^2
\exp\left[-\frac{z}{z_\star}\right]
\end{equation}

\subsubsection{Weak gravitational lensing}

For weak lensing, the weight function is \cite{2004PhRvD..70d3009H}:
\begin{equation}
W_\kappa(z_l) = \frac{3}{2}\Omega_{m,0} H_0^2 \frac{r(z_l)}{c H(z_l) a(z_l)}
\int_{z_l}^\infty dz_s \frac{r(z_s)-r(z_l)}{r(z_s)} \frac{n(z_s)}{n_{tot}}
\end{equation}

\subsection{Non-linearity}
We use the non-linear mass power spectrum for lensing-lensing
correlations \cite{2006MNRAS.366..547M,2003MNRAS.341.1311S,
1998ApJ...508L...5M} and the linear power spectrum for 
everything else.
To cut off the angular galaxy clustering before non-linear effects destroy its
usefulness, we multiply the power spectrum of the galaxy field (including
both clustering and noise) by $\exp\left[\left(\ell/\ell_c\right)^2\right]$
when computing the error covariance matrix for $g-g$ and $g-\kappa$
correlations, where $\ell_c(z)=\kmaxeff~ r(z)$, $r(z)$ is the comoving
angular diameter distance to redshift $z$, and $\kmaxeff$ is the maximum
$k$ used for the redshift survey. To be clear, this is a
``soft'' cutoff on $\ell$, increasing the noise term 
(this is why the argument of the exponential is positive, equivalent to 
suppressing
the signal term, outside the Fisher derivative) as an alternative to simply
truncating our calculation at some maximum $\ell$.  We introduced 
this Gaussian cutoff in an attempt to be slightly more realistic by 
allowing $g-\kappa$ to be sensitive to a somewhat 
smaller scale than $g-g$ (because it has only one power of the cutoff 
Gaussian instead of two), but it gives
remarkably close to exactly the same result as a sharp cutoff.

\subsection{Noise}

Noise power is $\bar{n}^{-1}$ for galaxy-galaxy auto-correlations,
$\frac{0.3^2}{2} \bar{n}^{-1}$ for $\kappa-\kappa$ auto-correlations, 
where $0.3^2/2$ is from the 
intrinsic shape noise of galaxies \cite{2012JCAP...04..034H}, where
$\bar{n}$ is the surface density (per steradian) in the tomographic bin.

\subsection{$C_\ell$ covariance matrices}

In our most standard angular calculation we divide the objects into
$\Delta z=0.2$
groupings by estimated redshift, spanning the range $0<z<2$. 
At the maximum, when we consider a lensing survey overlapping with 
a spectroscopic survey (in the Appendix), we have 4 tracers --
LRGs, ELGs, photo-galaxy density, and photo-galaxy lensing -- which gives 40
angular fields (for completeness we tried including
lensing of LRGs and ELGs, but they made no contribution).
From these we can measure 820 cross and auto correlations of the form
$\hat{C}_{\ell, ij}$.
The covariance between two of these measurements is
\begin{equation}
\left< \Delta C_{\ell, xy} \Delta C_{\ell, mn}\right>=
\left(f_{\rm sky} \Delta \ell \left(2 \ell+1\right)\right)^{-1}
\left(C_{\ell, xm} C_{\ell, yn}+C_{\ell, xn} C_{\ell, ym}\right)
\label{eq:covariance}
\end{equation}
where $\Delta C_{\ell, xy}=\hat{C}_{\ell, ij}-C_{\ell, xy}$ is the error in the
measurement as usual (this is of course the same equation as equation 
\ref{eq:cmbcovariance}, except allowing for $\ell$ binning),
$\Delta \ell$ is the width of the bin in $\ell$, $f_{\rm sky}$ is the
fraction of the sky covered by the survey, and $C_\ell$s include appropriate
noise, i.e., $N_{\ell, ij}=\delta^{\rm K}_{ij} ~\bar{n}^{-1}$ when $i=j$ 
labels a tracer of galaxy density, or 
$N_{\ell, ij}=\delta^{\rm K}_{ij}~ \frac{0.3^2}{2} \bar{n}^{-1}$ when $i=j$ 
labels a lensing convergence field.

\subsection{Photo-$z$ error distribution}

For a given estimated photo-$z$, we assume a true distribution of
galaxy redshifts following a simple
Gaussian distribution with rms width $0.05 (1+z)$. This propagates into the
calculation through the weight kernels $W_x(z)$. 
In this paper we assume
the distribution is exactly known, i.e., well-calibrated by direct redshift
measurements. 

\subsection{DES}

We include lensing, galaxy clustering, and their cross-correlations from DES,
an imaging survey covering 5000 sq. deg. 
(\url{www.darkenergysurvey.org}). 
We use $\alpha=1.25$, $\beta=2.29$, $z_\star=0.88$, and
$n_{tot}=12$ arcmin$^{-2}$ in equation (\ref{eqnumberdensities}).
For the bias of DES galaxies we adopt
$b(z) D(z)/D(0) = 0.95$. This agrees well with
halo number matching biases at $z\lesssim 1-1.5$, but is lower at higher $z$.
We include a free bias parameter for each photo-$z$ bin. In some circumstances
this is optimistic, as generally galaxies with different true redshift within
a photo-$z$ bin can have different bias (this is important if, e.g., we want
to calibrate photo-$z$ errors by cross-correlation with a redshift survey
\cite{2010MNRAS.401.1399B,2010MNRAS.405..359Z,2008ApJ...684...88N,
2008ApJ...682...39M,2006ApJ...636...21M}). 

\subsection{LSST}

Similar to DES, for LSST \cite{2009arXiv0912.0201L}, 
covering 20000 sq. deg., we use 
$\alpha=2.0$, $\beta=1.0$, $z_\star=0.3$, and
$n_{tot}=50$ arcmin$^{-2}$ in equation (\ref{eqnumberdensities}), with 
$b(z) D(z)/D(0) = 0.95$.

\section{Parameter constraint projections \label{sec:results}}

In this section we give parameter constraint projections for a large number of 
combinations of experiments and parameters. Table 
\ref{tab:experimentabbreviations} lists abbreviations for experiments that we
use in all the other tables. 

\subsection{Vanilla model (including neutrino mass!) 
\label{sec:vanilla-model-incl}}

Table \ref{tablemnu} shows constraints in our baseline model, which may be of
primary interest to readers interested in neutrino mass (see 
\cite{2012arXiv1212.6154L} for a review). 
\begin{table}
\caption{Abbreviations for experiments/sub-experiments in our tables. }
\label{tab:experimentabbreviations}
\centering
\begin{tabular}{llp{9.5cm}}
Abbreviation & Data Set \\
\hline
$P$ & Planck CMB (and a 5\% constraint on $H_0$ that only
matters in severely under-constrained cases).  \\
$BgB$ & BOSS galaxy BAO.  \\
$BlB$ & BOSS \lyaf\ and high-$z$ quasar BAO.  \\
$BgA \kmaxeff$ &  BOSS galaxy broadband to $k< \kmaxeff \ihMpc$ 
(plus BAO beyond that). \\
$DES$ & DES lensing and galaxy clustering. \\
$hdB$ & HETDEX BAO \\
$hdA \kmaxeff$ &  HETDEX broadband to $k< \kmaxeff \ihMpc$ 
(plus BAO beyond that). \\
$ebgA \kmaxeff$ & eBOSS galaxy broadband to $k< \kmaxeff \ihMpc$ (plus BAO
beyond that). \\
$BBgB$ & DESI galaxy BAO.  \\
$BBlB$ & DESI \lyaf\ and high-$z$ quasar BAO.  \\
$BBA \kmaxeff$ & DESI galaxy broadband to $k< \kmaxeff \ihMpc$ (plus BAO beyond
that). \\ 
$euB$ & Euclid BAO (for 50 million galaxies).  \\
$euA \kmaxeff$ & Euclid galaxy broadband to $k< \kmaxeff \ihMpc$ (plus BAO 
beyond that). \\
$LSST$ & LSST lensing and galaxy clustering. \\
$BlA$ & BOSS \lyaf\ broadband (including relatively small, $\sim$1D scales). \\
$l1D$ & $\sim 100$ high resolution \lyaf\ spectra. \\
$BBlA$ & DESI \lyaf\ broadband (including relatively small, $\sim$1D scales).\\
$BB24$ & 24 is appended to BB to indicate 24000 sq. deg. DESI instead of
the baseline 14000 sq. deg.  \\
$wfB$ & WFIRST BAO.  \\
$wfA \kmaxeff$ & WFIRST galaxy broadband to $k< \kmaxeff \ihMpc$ (plus BAO 
beyond that). \\
\hline
\end{tabular}
\end{table}
\begin{table}
\caption{
Neutrino mass and other basic parameter projections.
See Table \ref{tab:experimentabbreviations} for experiment codes } 
\label{tablemnu}  
\begin{tabular}{lcccccccc}
\hline
\hline
$ $ & $\omega_m$ & $\omega_b$ & $\theta_s$ & $\Sigma m_{\nu}$ & $\log_{10}(A)$ & $n_s$ & $\tau$ \\
\hline
${\rm value}$ & $0.141$ & $0.0221$ & $0.597$ & $0.0600$ & $-8.66$ & $0.961$ & $0.0920$ \\
\hline
$P$ & $0.0037$ & $0.00015$ & $0.00035$ & $0.35$ & $0.0039$ & $0.0038$ & $0.0045$ \\
$P+BgB+BlB$ & $0.00074$ & $0.00015$ & $0.00014$ & $0.10$ & $0.0038$ & $0.0038$ & $0.0044$ \\
$P+BgA0.1+BlB$ & $0.00070$ & $0.00013$ & $0.00014$ & $0.068$ & $0.0037$ & $0.0031$ & $0.0044$ \\
$P+BgA0.2+BlB$ & $0.00071$ & $0.00012$ & $0.00015$ & $0.046$ & $0.0037$ & $0.0028$ & $0.0043$ \\
$P+DES$ & $0.0013$ & $0.00013$ & $0.00017$ & $0.041$ & $0.0036$ & $0.0032$ & $0.0043$ \\
$P+BgB+BlB+DES$ & $0.00069$ & $0.00011$ & $0.00014$ & $0.030$ & $0.0035$ & $0.0027$ & $0.0043$ \\
$P+BgA0.1+BlB+DES$ & $0.00067$ & $0.00011$ & $0.00014$ & $0.029$ & $0.0035$ & $0.0027$ & $0.0042$ \\
$P+BgA0.1+BlB+ebA0.1$ & $0.00064$ & $0.00012$ & $0.00014$ & $0.052$ & $0.0037$ & $0.0029$ & $0.0043$ \\
$P+BgA0.2+BlB+ebA0.2$ & $0.00064$ & $0.00011$ & $0.00014$ & $0.036$ & $0.0037$ & $0.0027$ & $0.0043$ \\
$P+BgA0.1+BlB+ebA0.1+DES$ & $0.00062$ & $0.00011$ & $0.00014$ & $0.028$ & $0.0035$ & $0.0026$ & $0.0042$ \\
$P+hdB+BgB$ & $0.00074$ & $0.00015$ & $0.00014$ & $0.099$ & $0.0038$ & $0.0038$ & $0.0044$ \\
$P+hdA0.1+BgA0.1$ & $0.00069$ & $0.00012$ & $0.00014$ & $0.061$ & $0.0037$ & $0.0030$ & $0.0044$ \\
$P+hdA0.2+BgA0.2$ & $0.00068$ & $0.00011$ & $0.00014$ & $0.039$ & $0.0037$ & $0.0027$ & $0.0043$ \\
$P+BBgB$ & $0.00055$ & $0.00015$ & $0.00014$ & $0.090$ & $0.0038$ & $0.0038$ & $0.0044$ \\
$P+BBgB+BlB$ & $0.00055$ & $0.00015$ & $0.00014$ & $0.090$ & $0.0038$ & $0.0038$ & $0.0044$ \\
$P+BBlB+BgB$ & $0.00072$ & $0.00015$ & $0.00014$ & $0.098$ & $0.0038$ & $0.0038$ & $0.0044$ \\
$P+BBgB+BBlB$ & $0.00055$ & $0.00015$ & $0.00014$ & $0.090$ & $0.0038$ & $0.0038$ & $0.0044$ \\
$P+BBgB+BBlB+DES$ & $0.00045$ & $0.00011$ & $0.00014$ & $0.027$ & $0.0035$ & $0.0025$ & $0.0043$ \\
$P+BBgA0.1$ & $0.00044$ & $0.00011$ & $0.00014$ & $0.024$ & $0.0036$ & $0.0024$ & $0.0043$ \\
$P+BBgA0.1+BBlB$ & $0.00044$ & $0.00011$ & $0.00014$ & $0.024$ & $0.0036$ & $0.0024$ & $0.0043$ \\
$P+BBgA0.1+BBlB+DES$ & $0.00043$ & $0.00011$ & $0.00014$ & $0.021$ & $0.0034$ & $0.0024$ & $0.0041$ \\
$P+BBgA0.2+BBlB$ & $0.00042$ & $0.00010$ & $0.00014$ & $0.017$ & $0.0035$ & $0.0022$ & $0.0043$ \\
$P+BBgA0.2+BBlB+DES$ & $0.00042$ & $0.00010$ & $0.00014$ & $0.017$ & $0.0033$ & $0.0022$ & $0.0040$ \\
$P+BB24gB+BB24lB$ & $0.00052$ & $0.00015$ & $0.00014$ & $0.088$ & $0.0038$ & $0.0037$ & $0.0044$ \\
$P+BB24gA0.1+BB24lB$ & $0.00039$ & $0.00011$ & $0.00014$ & $0.020$ & $0.0035$ & $0.0023$ & $0.0043$ \\
$P+BB24gA0.1+BB24lB+DES$ & $0.00038$ & $0.00011$ & $0.00013$ & $0.019$ & $0.0033$ & $0.0023$ & $0.0040$ \\
$P+BB24gA0.2+BB24lB$ & $0.00037$ & $9.9e-05$ & $0.00014$ & $0.015$ & $0.0035$ & $0.0020$ & $0.0042$ \\
$P+BB24gA0.2+BB24lB+DES$ & $0.00037$ & $9.9e-05$ & $0.00013$ & $0.015$ & $0.0032$ & $0.0020$ & $0.0040$ \\
$P+BgB+BlB+euB$ & $0.00054$ & $0.00015$ & $0.00014$ & $0.090$ & $0.0038$ & $0.0038$ & $0.0044$ \\
$P+BgA0.1+BlB+euA0.1$ & $0.00043$ & $0.00011$ & $0.00014$ & $0.021$ & $0.0036$ & $0.0024$ & $0.0043$ \\
$P+BgA0.1+BlB+euA0.1+DES$ & $0.00043$ & $0.00011$ & $0.00014$ & $0.019$ & $0.0034$ & $0.0023$ & $0.0041$ \\
$P+BgA0.2+BlB+euA0.2$ & $0.00042$ & $0.00010$ & $0.00014$ & $0.015$ & $0.0035$ & $0.0021$ & $0.0042$ \\
$P+BgA0.2+BlB+euA0.2+DES$ & $0.00041$ & $0.00010$ & $0.00014$ & $0.015$ & $0.0033$ & $0.0021$ & $0.0040$ \\
$P+BB24gA0.1+BB24lB+euA0.1$ & $0.00036$ & $0.00010$ & $0.00013$ & $0.017$ & $0.0035$ & $0.0022$ & $0.0042$ \\
$P+BB24gA0.1+BB24lB+euA0.1+DES$ & $0.00036$ & $0.00010$ & $0.00013$ & $0.016$ & $0.0032$ & $0.0022$ & $0.0040$ \\
$P+BB24gA0.2+BB24lB+euA0.2$ & $0.00034$ & $9.6e-05$ & $0.00013$ & $0.014$ & $0.0034$ & $0.0018$ & $0.0041$ \\
$P+BB24gA0.2+BB24lB+euA0.2+DES$ & $0.00034$ & $9.6e-05$ & $0.00013$ & $0.013$ & $0.0032$ & $0.0018$ & $0.0039$ \\
$P+LSST$ & $0.00080$ & $0.00011$ & $0.00015$ & $0.020$ & $0.0030$ & $0.0029$ & $0.0036$ \\
$P+BgB+BlB+LSST$ & $0.00060$ & $0.00011$ & $0.00014$ & $0.018$ & $0.0030$ & $0.0025$ & $0.0036$ \\
$P+BBgB+BBlB+LSST$ & $0.00044$ & $0.00011$ & $0.00013$ & $0.016$ & $0.0030$ & $0.0022$ & $0.0036$ \\
$P+BBgA0.1+BBlB+LSST$ & $0.00042$ & $0.00010$ & $0.00013$ & $0.015$ & $0.0028$ & $0.0021$ & $0.0034$ \\
$P+BBgA0.2+BBlB+LSST$ & $0.00041$ & $0.00010$ & $0.00013$ & $0.014$ & $0.0026$ & $0.0020$ & $0.0032$ \\
$P+BB24gA0.1+BB24lB+LSST$ & $0.00038$ & $0.00010$ & $0.00013$ & $0.015$ & $0.0027$ & $0.0020$ & $0.0033$ \\
$P+BB24gA0.2+BB24lB+LSST$ & $0.00036$ & $9.8e-05$ & $0.00013$ & $0.013$ & $0.0025$ & $0.0018$ & $0.0031$ \\
$P+BB24gA0.1+BB24lB+euA0.1+LSST$ & $0.00035$ & $0.00010$ & $0.00013$ & $0.014$ & $0.0026$ & $0.0019$ & $0.0032$ \\
$P+BB24gA0.2+BB24lB+euA0.2+LSST$ & $0.00033$ & $9.5e-05$ & $0.00013$ & $0.011$ & $0.0024$ & $0.0016$ & $0.0030$ \\
$P+wfB+BgB$ & $0.00064$ & $0.00015$ & $0.00014$ & $0.095$ & $0.0038$ & $0.0038$ & $0.0044$ \\
$P+wfA0.1+BgA0.1$ & $0.00058$ & $0.00011$ & $0.00014$ & $0.037$ & $0.0037$ & $0.0027$ & $0.0043$ \\
$P+wfA0.2+BgA0.2$ & $0.00056$ & $0.00011$ & $0.00014$ & $0.021$ & $0.0036$ & $0.0025$ & $0.0043$ \\
$P+BgB+BlA+l1D$ & $0.00066$ & $0.00011$ & $0.00014$ & $0.053$ & $0.0037$ & $0.0032$ & $0.0044$ \\
$P+BgA0.1+BlA+l1D$ & $0.00065$ & $0.00011$ & $0.00014$ & $0.048$ & $0.0037$ & $0.0030$ & $0.0043$ \\
$P+BgA0.2+BlA+l1D$ & $0.00066$ & $0.00011$ & $0.00014$ & $0.040$ & $0.0037$ & $0.0027$ & $0.0043$ \\
$P+BBgB+BBlA+l1D$ & $0.00041$ & $0.00010$ & $0.00014$ & $0.039$ & $0.0037$ & $0.0029$ & $0.0043$ \\
$P+BBgA0.1+BBlA+l1D$ & $0.00039$ & $0.00010$ & $0.00014$ & $0.023$ & $0.0035$ & $0.0021$ & $0.0043$ \\
$P+BBgA0.2+BBlA+l1D$ & $0.00038$ & $0.00010$ & $0.00014$ & $0.017$ & $0.0035$ & $0.0019$ & $0.0042$ \\
$P+BB24gB+BB24lA+l1D$ & $0.00036$ & $0.00010$ & $0.00014$ & $0.034$ & $0.0036$ & $0.0028$ & $0.0043$ \\
$P+BB24gA0.1+BB24lA+l1D$ & $0.00035$ & $0.00010$ & $0.00013$ & $0.019$ & $0.0035$ & $0.0019$ & $0.0042$ \\
$P+BB24gA0.2+BB24lA+l1D$ & $0.00034$ & $9.8e-05$ & $0.00014$ & $0.015$ & $0.0034$ & $0.0016$ & $0.0041$ \\
$P+BB24gA0.2+BB24lA+l1D+euA0.2$ & $0.00032$ & $9.5e-05$ & $0.00013$ & $0.013$ & $0.0033$ & $0.0015$ & $0.0040$ \\
$P+BB24gA0.2+BB24lA+l1D+LSST$ & $0.00033$ & $9.7e-05$ & $0.00013$ & $0.012$ & $0.0025$ & $0.0015$ & $0.0031$ \\
$P+BB24gA0.2+BB24lA+l1D+euA0.2+LSST$ & $0.00032$ & $9.5e-05$ & $0.00013$ & $0.011$ & $0.0024$ & $0.0014$ & $0.0030$ \\

\hline
\hline
\end{tabular}
\end{table}

We see that the 14000 sq. deg. baseline DESI can measure neutrino masses 
to $0.024$ eV for $\kmaxeff=0.1\ihMpc$ or $0.017$ eV for $\kmaxeff=0.2\ihMpc$.
DES can improve the more pessimistic number to $0.021$ eV, i.e., by the end
of the DESI baseline survey the minimal neutrino mass of $0.057$ eV should
definitely be detected at $\sim 3\sigma$. Euclid's redshift survey can produce
similar measurements.
(Note that 
Euclid lensing can presumably be substituted for LSST here without a 
qualitative change in results.) 

\cite{2013NuPhS.237...50C} found qualitatively similar expectations for Euclid,
as did \cite{2013arXiv1307.2919S}.
\cite{2012JCAP...11..052H} appear to have found somewhat weaker constraints
for the Euclid redshift survey, but it is not clear that they fully include
the measurement that comes from redshift space distortions.  
\cite{2013JCAP...06..020C,2013arXiv1304.2321B} found $0.01-0.02$ eV constraints 
from Euclid using only
the photometric survey, i.e., not including the redshift survey, but 
including clusters (amounting to a largely independent form of measurement 
from our redshift survey projections).
The results of \cite{2011PhRvD..83k5023G} are not directly comparable to our 
results because 
they simultaneously varied the sum of masses of standard neutrinos and a 
separate sterile neutrino mass.  
\cite{2013PhLB..718.1186O} found qualitatively similar, although not very
directly comparable, results for a futuristic 21 cm equivalent of a redshift 
survey.
They found CMB lensing in the form of CMBpol to be somewhat more limited than
our galaxy lensing experiments, producing an error $\sim 0.05$eV.
We find \cite{2012MNRAS.425.1170H} to be a little confusing to read, 
but a bottom line appears
to be that their proposed COrE CMB lensing experiment could measure $\summnu$ 
to $\sim 0.02$ eV when combined with BAO measurements (from CMB lensing alone
they appear to be consistent with \cite{2013PhLB..718.1186O}). 
\cite{2014arXiv1402.4108W} found that future ground based CMB lensing 
experiments with $10^4-10^5$ detectors can measure $\summnu$ to
$\sim 0.02$ eV, when combined with DESI BAO measurements.  
\cite{2012PhRvD..86b3526J} found some combinations of future experiments 
reaching the 0.02-0.03 eV
level, but do not appear to have included a full redshift survey.
The LSST project book \cite{2009arXiv0912.0201L} quotes an error 
$0.03-0.07$ eV depending on their fiducial $\summnu$ value, not as good as
our $0.02$ eV projections for them, but they appear to
be using only lensing-lensing correlations for this calculation, not including
correlations involving galaxy density.
Generally, the idea that $\sim 0.02$ eV level constraints will be 
achieved by cosmological measurements in the 2020's appears very secure --
it is projected redundantly for several different kinds of probes, which are
unlikely to all fail. A constraint $\sim 0.01$ eV may be possible.

\subsubsection{Neutrino mass hierarchy}

\begin{figure}[htbp]
\centering
\includegraphics[width=0.8\textwidth]{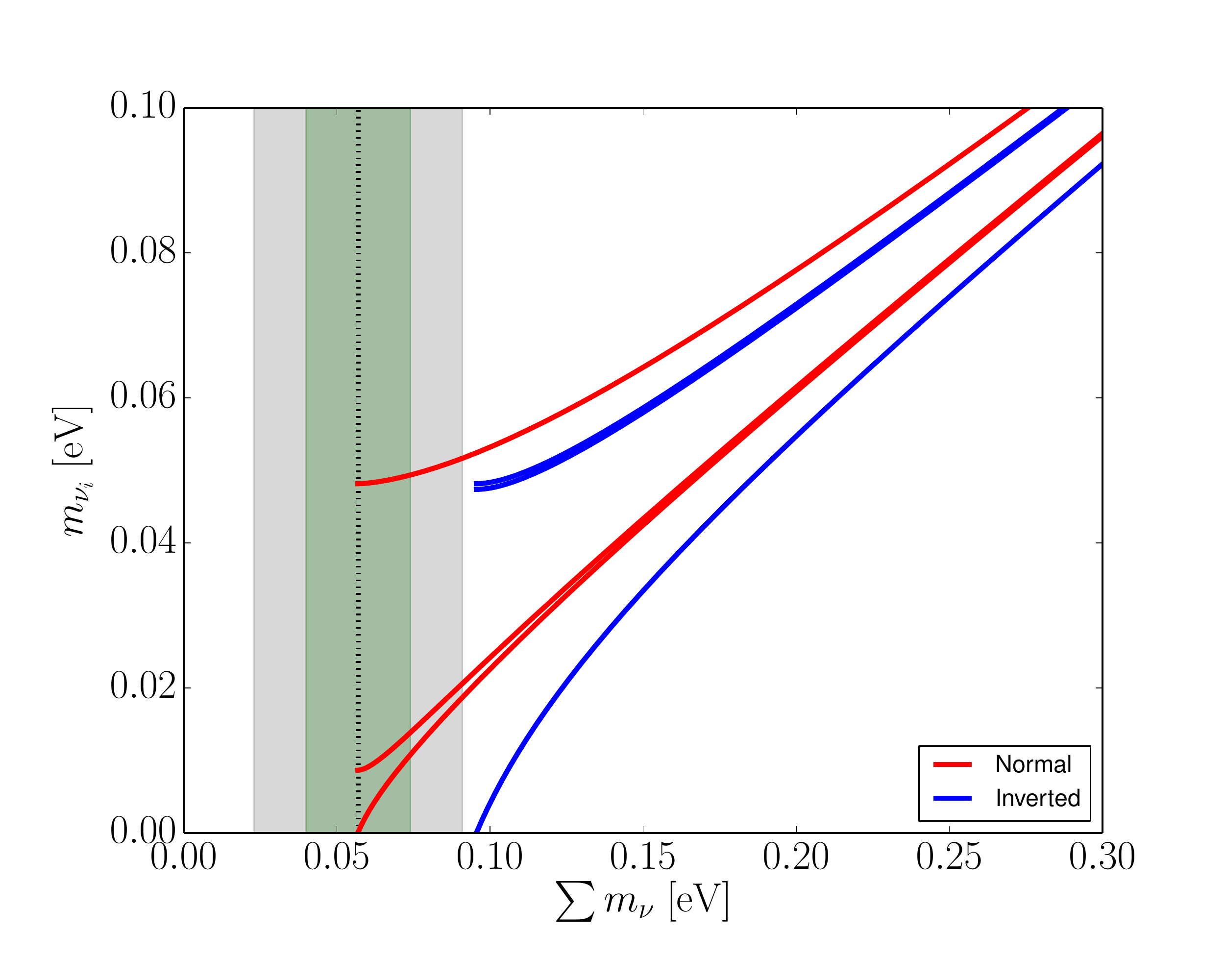}
\caption{
Inverted (blue) vs. normal (red) hierarchies, given the current
mass-squared difference measurements from 
\cite{2012PhRvD..86a0001B}.
Each line shows the mass of one of the neutrinos, plotted as a function of the
sum of masses in each case (in the inverted case the two most massive
neutrinos have almost indistinguishable mass on this plot). The green
and gray bands indicate the 1 and 2 sigma error for an experiment 
with $\sigma_{m_\nu}=0.017$eV,
assuming no prior on $\sum m_\nu$, for a fiducial model with $\sum
m_\nu=0.057$eV.
\label{fig:masscases} }
\end{figure}

There is great interest in determining the distribution of masses
between neutrino species \cite{2013arXiv1307.5487C}. Although
cosmology can, in principle, measure individual neutrino mass
eigenstates, this is unrealistic at the level of precision of experiments
discussed here
\cite{2012AAS...21933501S,2006PhRvD..73l3501S,2010JCAP...05..035J,
  2012ApJ...752L..31W,2009PhRvD..80l3509D} and therefore we are
effectively limited to determining only the sum of neutrino masses
directly.  The situation given experimental constraints is illustrated
in Figure \ref{fig:masscases}.
In the case where the hierarchy is in fact normal with sum of
masses near the minimum, i.e., sum of masses $\sim 57$ meV,
the fact that the hierarchy is normal can be proven, because the minimum total
mass in the inverted
hierarchy is $\sim 96$ meV, however, if the hierarchy is inverted, or normal
but with mass much above the minimum, there will be no possibility of 
distinguishing the two cases. 
We see that in the best case the experiments in Table \ref{tablemnu} can
hope to distinguish the hierarchy at about $3.5\sigma$ level. 

\subsection{Dark Energy Figures of Merit}
\label{sec:dark-energy-figures}
Table \ref{tableFoM} shows Dark Energy Task Force (DETF) Figures of Merit
(FoMs) \cite{2006astro.ph..9591A}, except with the DETF definition modified to 
include
marginalization over neutrino mass (to be clear, the additional parameters 
beyond our baseline are $w_0$, $w^\prime$, and $\Omega_k$). 
\begin{table}
\caption{
DETF Figures of Merit (including marginalization over $\summnu$).
See Table \ref{tab:experimentabbreviations} for survey codes. 
All lines include a 5\% $H_0$ constraint (which only matters for 
under-constrained cases).}   \label{tableFoM}  
\begin{tabular}{lrccc}
\hline
\hline
Surveys & FoM & $a_p$ & $\sigma_{w_p}$ & $\sigma_{\Omega_k}$ \\
\hline
P & 2 & 0.76 & 0.287 & 0.0085 \\
P+BgB+BlB & 12 & 0.51 & 0.095 & 0.0057 \\
P+BgA0.1+BlB & 22 & 0.60 & 0.076 & 0.0029 \\
P+BgA0.2+BlB & 25 & 0.60 & 0.075 & 0.0026 \\
P+DES & 38 & 0.71 & 0.032 & 0.0032 \\
P+BgB+BlB+DES & 71 & 0.71 & 0.028 & 0.0019 \\
P+BgA0.1+BlB+DES & 89 & 0.69 & 0.025 & 0.0019 \\
P+BgA0.1+BlB+ebA0.1 & 32 & 0.60 & 0.058 & 0.0023 \\
P+BgA0.2+BlB+ebA0.2 & 42 & 0.58 & 0.054 & 0.0020 \\
P+BgA0.1+BlB+ebA0.1+DES & 100 & 0.69 & 0.024 & 0.0018 \\
P+hdB+BgB & 13 & 0.52 & 0.094 & 0.0056 \\
P+hdA0.1+BgA0.1 & 29 & 0.60 & 0.061 & 0.0026 \\
P+hdA0.2+BgA0.2 & 45 & 0.57 & 0.045 & 0.0024 \\
P+BBgB & 23 & 0.52 & 0.070 & 0.0064 \\
P+BBgB+BlB & 27 & 0.53 & 0.063 & 0.0055 \\
P+BBlB+BgB & 17 & 0.54 & 0.082 & 0.0039 \\
P+BBgB+BBlB & 37 & 0.56 & 0.048 & 0.0036 \\
P+BBgB+BBlB+DES & 128 & 0.70 & 0.024 & 0.0013 \\
P+BBgA0.1 & 104 & 0.67 & 0.029 & 0.0013 \\
P+BBgA0.1+BBlB & 117 & 0.67 & 0.029 & 0.0012 \\
P+BBgA0.1+BBlB+DES & 186 & 0.69 & 0.019 & 0.0012 \\
P+BBgA0.2+BBlB & 199 & 0.63 & 0.024 & 0.0010 \\
P+BBgA0.2+BBlB+DES & 327 & 0.67 & 0.015 & 0.0010 \\
P+BB24gB+BB24lB & 58 & 0.56 & 0.037 & 0.0029 \\
P+BB24gA0.1+BB24lB & 182 & 0.66 & 0.023 & 0.0011 \\
P+BB24gA0.1+BB24lB+DES & 258 & 0.68 & 0.017 & 0.0010 \\
P+BB24gA0.2+BB24lB & 318 & 0.62 & 0.018 & 0.0008 \\
P+BB24gA0.2+BB24lB+DES & 459 & 0.66 & 0.013 & 0.0008 \\
P+BgB+BlB+euB & 27 & 0.54 & 0.062 & 0.0052 \\
P+BgA0.1+BlB+euA0.1 & 123 & 0.69 & 0.028 & 0.0012 \\
P+BgA0.1+BlB+euA0.1+DES & 183 & 0.71 & 0.019 & 0.0011 \\
P+BgA0.2+BlB+euA0.2 & 228 & 0.65 & 0.021 & 0.0009 \\
P+BgA0.2+BlB+euA0.2+DES & 342 & 0.68 & 0.015 & 0.0009 \\
P+BB24gA0.1+BB24lB+euA0.1 & 250 & 0.67 & 0.020 & 0.0009 \\
P+BB24gA0.1+BB24lB+euA0.1+DES & 320 & 0.68 & 0.016 & 0.0009 \\
P+BB24gA0.2+BB24lB+euA0.2 & 468 & 0.63 & 0.015 & 0.0007 \\
P+BB24gA0.2+BB24lB+euA0.2+DES & 595 & 0.66 & 0.012 & 0.0007 \\
P+LSST & 134 & 0.72 & 0.019 & 0.0019 \\
P+BgB+BlB+LSST & 176 & 0.72 & 0.017 & 0.0013 \\
P+BBgB+BBlB+LSST & 230 & 0.71 & 0.017 & 0.0011 \\
P+BBgA0.1+BBlB+LSST & 300 & 0.70 & 0.014 & 0.0010 \\
P+BBgA0.2+BBlB+LSST & 518 & 0.69 & 0.011 & 0.0009 \\
P+BB24gA0.1+BB24lB+LSST & 391 & 0.69 & 0.013 & 0.0009 \\
P+BB24gA0.2+BB24lB+LSST & 708 & 0.68 & 0.010 & 0.0008 \\
P+BB24gA0.1+BB24lB+euA0.1+LSST & 467 & 0.70 & 0.012 & 0.0008 \\
P+BB24gA0.2+BB24lB+euA0.2+LSST & 877 & 0.68 & 0.009 & 0.0007 \\
P+wfB+BgB & 19 & 0.55 & 0.077 & 0.0054 \\
P+wfA0.1+BgA0.1 & 69 & 0.68 & 0.038 & 0.0014 \\
P+wfA0.2+BgA0.2 & 116 & 0.63 & 0.031 & 0.0012 \\
\hline
\hline
\end{tabular}
\end{table}
For the common normalization convention
that we follow, the FoM is simply 
$\left(\sigma_{w_p} \sigma_{w^\prime}\right)^{-1}$ where
$w(z) = w_p + (a_p-a)w^\prime$ and $a_p$ is chosen to make the errors on $w_p$
and $w^\prime$ independent.      
Overall we find the complementarity between different 
experiments striking -- each of the major experiments contributes 
significantly and non-redundantly to building up our understanding of dark 
energy properties. 

One additional thing that this Table shows, although it requires comparison 
with the fixed neutrino mass calculations in the Appendix, 
Table \ref{tableoldFoM}, to see it, is that Dark Energy constraints are 
generally significantly degraded by the uncertainty in neutrino mass.
Maybe surprisingly, this degradation is largest when Planck is combined with
isolated BAO distance measurements alone, even though neutrinos are generally
thought of primarily as effecting the power spectrum growth and shape. 
This happens because neutrino mass also introduces an additional uncertainty 
in the background evolution after CMB decoupling.
It becomes even more
useful to include a probe sensitive to the amplitude/growth of structure, 
either broadband galaxy power or lensing or ideally both. 
This is easy to understand qualitatively:
the neutrinos act as essentially a form of modified gravity, breaking what
we usually think of as the GR relation between background evolution and 
growth of structure. 

For completeness, Table \ref{tableOmkWz} shows broader constraints in the FoM 
parameter space.
\begin{table}
\caption{
Projections in the FoM parameter space.
See Table \ref{tab:experimentabbreviations} for survey codes.}
\label{tableOmkWz}  
\begin{tabular}{lcccccccccc}
\hline
\hline
$ $ & $\omega_m$ & $\omega_b$ & $\theta_s$ & $a_p$ & $w_p$ & $w_0$ & $w'$ & $\Omega_k$ & $\Sigma m_{\nu}$ & $n_s$  \\
\hline
${\rm value}$ & $0.141$ & $0.0221$ & $0.597$ & $$ & $-1.00$ & $-1.00$ & $0.00$ & $0$ & $0.0600$ & $0.961$ \\
\hline
$P$ & $0.011$ & $0.00015$ & $0.00095$ & $0.64$ & $0.29$ & $0.66$ & $1.6$ & $0.0085$ & $1.0$ & $0.0050$ \\
$P+BgB+BlB$ & $0.0069$ & $0.00015$ & $0.00064$ & $0.80$ & $0.095$ & $0.21$ & $0.90$ & $0.0057$ & $0.66$ & $0.0045$ \\
$P+BgA0.1+BlB$ & $0.0026$ & $0.00014$ & $0.00025$ & $0.74$ & $0.076$ & $0.17$ & $0.60$ & $0.0029$ & $0.18$ & $0.0034$ \\
$P+BgA0.2+BlB$ & $0.0023$ & $0.00013$ & $0.00023$ & $0.74$ & $0.075$ & $0.16$ & $0.53$ & $0.0026$ & $0.15$ & $0.0033$ \\
$P+DES$ & $0.0017$ & $0.00014$ & $0.00019$ & $0.66$ & $0.032$ & $0.29$ & $0.84$ & $0.0032$ & $0.063$ & $0.0034$ \\
$P+BgB+BlB+DES$ & $0.0016$ & $0.00014$ & $0.00019$ & $0.67$ & $0.028$ & $0.17$ & $0.51$ & $0.0019$ & $0.054$ & $0.0034$ \\
$P+BgA0.1+BlB+DES$ & $0.0016$ & $0.00014$ & $0.00018$ & $0.68$ & $0.025$ & $0.15$ & $0.45$ & $0.0019$ & $0.052$ & $0.0033$ \\
$P+BgA0.1+BlB+ebA0.1$ & $0.0021$ & $0.00014$ & $0.00022$ & $0.73$ & $0.058$ & $0.15$ & $0.54$ & $0.0023$ & $0.13$ & $0.0034$ \\
$P+BgA0.2+BlB+ebA0.2$ & $0.0019$ & $0.00013$ & $0.00021$ & $0.75$ & $0.054$ & $0.12$ & $0.44$ & $0.0020$ & $0.11$ & $0.0033$ \\
$P+BgA0.1+BlB+ebA0.1+DES$ & $0.0016$ & $0.00014$ & $0.00018$ & $0.68$ & $0.024$ & $0.14$ & $0.42$ & $0.0018$ & $0.050$ & $0.0033$ \\
$P+hdB+BgB$ & $0.0066$ & $0.00015$ & $0.00062$ & $0.79$ & $0.094$ & $0.20$ & $0.85$ & $0.0056$ & $0.64$ & $0.0044$ \\
$P+hdA0.1+BgA0.1$ & $0.0021$ & $0.00014$ & $0.00022$ & $0.73$ & $0.061$ & $0.16$ & $0.57$ & $0.0026$ & $0.13$ & $0.0034$ \\
$P+hdA0.2+BgA0.2$ & $0.0017$ & $0.00013$ & $0.00019$ & $0.75$ & $0.045$ & $0.13$ & $0.49$ & $0.0024$ & $0.086$ & $0.0033$ \\
$P+BBgB$ & $0.0067$ & $0.00015$ & $0.00061$ & $0.79$ & $0.070$ & $0.15$ & $0.61$ & $0.0064$ & $0.64$ & $0.0044$ \\
$P+BBgB+BlB$ & $0.0059$ & $0.00015$ & $0.00055$ & $0.78$ & $0.063$ & $0.15$ & $0.60$ & $0.0055$ & $0.57$ & $0.0043$ \\
$P+BBlB+BgB$ & $0.0053$ & $0.00015$ & $0.00051$ & $0.77$ & $0.082$ & $0.18$ & $0.73$ & $0.0039$ & $0.52$ & $0.0042$ \\
$P+BBgB+BBlB$ & $0.0042$ & $0.00015$ & $0.00041$ & $0.76$ & $0.048$ & $0.14$ & $0.56$ & $0.0036$ & $0.41$ & $0.0040$ \\
$P+BBgB+BBlB+DES$ & $0.0016$ & $0.00014$ & $0.00018$ & $0.67$ & $0.024$ & $0.11$ & $0.33$ & $0.0013$ & $0.051$ & $0.0033$ \\
$P+BBgA0.1$ & $0.0015$ & $0.00013$ & $0.00018$ & $0.69$ & $0.029$ & $0.10$ & $0.33$ & $0.0013$ & $0.058$ & $0.0032$ \\
$P+BBgA0.1+BBlB$ & $0.0015$ & $0.00013$ & $0.00018$ & $0.69$ & $0.029$ & $0.095$ & $0.29$ & $0.0012$ & $0.058$ & $0.0032$ \\
$P+BBgA0.1+BBlB+DES$ & $0.0014$ & $0.00013$ & $0.00018$ & $0.68$ & $0.019$ & $0.092$ & $0.28$ & $0.0012$ & $0.044$ & $0.0031$ \\
$P+BBgA0.2+BBlB$ & $0.0012$ & $0.00011$ & $0.00017$ & $0.71$ & $0.024$ & $0.065$ & $0.21$ & $0.00098$ & $0.047$ & $0.0030$ \\
$P+BBgA0.2+BBlB+DES$ & $0.0012$ & $0.00011$ & $0.00016$ & $0.69$ & $0.015$ & $0.064$ & $0.20$ & $0.00097$ & $0.038$ & $0.0029$ \\
$P+BB24gB+BB24lB$ & $0.0035$ & $0.00015$ & $0.00035$ & $0.76$ & $0.037$ & $0.12$ & $0.46$ & $0.0029$ & $0.34$ & $0.0039$ \\
$P+BB24gA0.1+BB24lB$ & $0.0014$ & $0.00012$ & $0.00018$ & $0.70$ & $0.023$ & $0.075$ & $0.24$ & $0.0011$ & $0.050$ & $0.0031$ \\
$P+BB24gA0.1+BB24lB+DES$ & $0.0013$ & $0.00012$ & $0.00017$ & $0.68$ & $0.017$ & $0.074$ & $0.23$ & $0.0010$ & $0.041$ & $0.0030$ \\
$P+BB24gA0.2+BB24lB$ & $0.0011$ & $0.00010$ & $0.00016$ & $0.72$ & $0.018$ & $0.051$ & $0.17$ & $0.00083$ & $0.040$ & $0.0028$ \\
$P+BB24gA0.2+BB24lB+DES$ & $0.0010$ & $0.00010$ & $0.00016$ & $0.70$ & $0.013$ & $0.051$ & $0.16$ & $0.00082$ & $0.035$ & $0.0027$ \\
$P+BgB+BlB+euB$ & $0.0057$ & $0.00015$ & $0.00053$ & $0.77$ & $0.062$ & $0.15$ & $0.61$ & $0.0052$ & $0.55$ & $0.0042$ \\
$P+BgA0.1+BlB+euA0.1$ & $0.0014$ & $0.00012$ & $0.00018$ & $0.68$ & $0.028$ & $0.099$ & $0.29$ & $0.0012$ & $0.051$ & $0.0032$ \\
$P+BgA0.1+BlB+euA0.1+DES$ & $0.0014$ & $0.00012$ & $0.00017$ & $0.67$ & $0.019$ & $0.096$ & $0.28$ & $0.0011$ & $0.042$ & $0.0032$ \\
$P+BgA0.2+BlB+euA0.2$ & $0.0011$ & $0.00010$ & $0.00017$ & $0.70$ & $0.021$ & $0.064$ & $0.20$ & $0.00091$ & $0.041$ & $0.0029$ \\
$P+BgA0.2+BlB+euA0.2+DES$ & $0.0011$ & $0.00010$ & $0.00016$ & $0.68$ & $0.015$ & $0.064$ & $0.19$ & $0.00090$ & $0.035$ & $0.0029$ \\
$P+BB24gA0.1+BB24lB+euA0.1$ & $0.0012$ & $0.00011$ & $0.00017$ & $0.69$ & $0.020$ & $0.066$ & $0.20$ & $0.00093$ & $0.042$ & $0.0030$ \\
$P+BB24gA0.1+BB24lB+euA0.1+DES$ & $0.0012$ & $0.00011$ & $0.00017$ & $0.68$ & $0.016$ & $0.065$ & $0.20$ & $0.00092$ & $0.038$ & $0.0029$ \\
$P+BB24gA0.2+BB24lB+euA0.2$ & $0.00098$ & $9.7e-05$ & $0.00016$ & $0.71$ & $0.015$ & $0.044$ & $0.14$ & $0.00070$ & $0.034$ & $0.0026$ \\
$P+BB24gA0.2+BB24lB+euA0.2+DES$ & $0.00095$ & $9.6e-05$ & $0.00016$ & $0.70$ & $0.012$ & $0.044$ & $0.14$ & $0.00070$ & $0.031$ & $0.0025$ \\
$P+LSST$ & $0.0013$ & $0.00013$ & $0.00017$ & $0.66$ & $0.019$ & $0.13$ & $0.38$ & $0.0019$ & $0.035$ & $0.0032$ \\
$P+BgB+BlB+LSST$ & $0.0013$ & $0.00013$ & $0.00017$ & $0.66$ & $0.017$ & $0.11$ & $0.33$ & $0.0013$ & $0.034$ & $0.0032$ \\
$P+BBgB+BBlB+LSST$ & $0.0013$ & $0.00013$ & $0.00017$ & $0.66$ & $0.017$ & $0.089$ & $0.26$ & $0.0011$ & $0.033$ & $0.0031$ \\
$P+BBgA0.1+BBlB+LSST$ & $0.0012$ & $0.00012$ & $0.00017$ & $0.67$ & $0.014$ & $0.078$ & $0.23$ & $0.0010$ & $0.032$ & $0.0030$ \\
$P+BBgA0.2+BBlB+LSST$ & $0.0010$ & $0.00010$ & $0.00016$ & $0.68$ & $0.011$ & $0.058$ & $0.18$ & $0.00088$ & $0.029$ & $0.0028$ \\
$P+BB24gA0.1+BB24lB+LSST$ & $0.0011$ & $0.00011$ & $0.00016$ & $0.67$ & $0.013$ & $0.065$ & $0.20$ & $0.00090$ & $0.031$ & $0.0029$ \\
$P+BB24gA0.2+BB24lB+LSST$ & $0.00092$ & $0.00010$ & $0.00016$ & $0.68$ & $0.0096$ & $0.048$ & $0.15$ & $0.00076$ & $0.028$ & $0.0026$ \\
$P+BB24gA0.1+BB24lB+euA0.1+LSST$ & $0.0010$ & $0.00011$ & $0.00016$ & $0.67$ & $0.012$ & $0.059$ & $0.18$ & $0.00082$ & $0.030$ & $0.0028$ \\
$P+BB24gA0.2+BB24lB+euA0.2+LSST$ & $0.00086$ & $9.6e-05$ & $0.00015$ & $0.68$ & $0.0089$ & $0.042$ & $0.13$ & $0.00066$ & $0.026$ & $0.0024$ \\
$P+wfB+BgB$ & $0.0060$ & $0.00015$ & $0.00056$ & $0.77$ & $0.077$ & $0.18$ & $0.68$ & $0.0054$ & $0.58$ & $0.0043$ \\
$P+wfA0.1+BgA0.1$ & $0.0017$ & $0.00014$ & $0.00019$ & $0.68$ & $0.038$ & $0.13$ & $0.38$ & $0.0014$ & $0.072$ & $0.0033$ \\
$P+wfA0.2+BgA0.2$ & $0.0014$ & $0.00012$ & $0.00018$ & $0.72$ & $0.031$ & $0.085$ & $0.28$ & $0.0012$ & $0.053$ & $0.0031$ \\

\hline
\hline
\end{tabular}
\end{table}

\subsection{Modified Gravity}

Table \ref{tableMoG} shows constraints on the FoMSWG
\cite{2009arXiv0901.0721A} modified gravity 
parameters (as we interpret them, explained in \S \ref{sec:params}). 
To be clear, we add 
$\Delta \gamma$ and $G_9$ to the DETF FoM scenario in 
\S \ref{sec:dark-energy-figures}.
\begin{table}
\caption{
Projections for modified gravity parameters.
See Table \ref{tab:experimentabbreviations} for survey codes.}
\label{tableMoG}  
\begin{tabular}{lcccccccccccc}
\hline
\hline
$ $ & $\omega_m$ & $a_p$ & $w_p$ & $w_0$ & $w'$ & $\Omega_k$ & $\Sigma m_{\nu}$ & $n_s$ & $\Delta \gamma$ & $G_9$ \\
\hline
${\rm value}$ & $0.14$ & $$ & $-1.0$ & $-1.0$ & $0.0$ & $0$ & $0.060$ & $0.96$ & $0.0$ & $1.0$ \\
\hline
$P+BgA0.1+BlB$ & $0.0036$ & $0.75$ & $0.077$ & $0.18$ & $0.66$ & $0.0037$ & $0.30$ & $0.0036$ & $0.18$ & $0.14$ \\
$P+BgA0.2+BlB$ & $0.0035$ & $0.75$ & $0.075$ & $0.18$ & $0.63$ & $0.0035$ & $0.28$ & $0.0034$ & $0.10$ & $0.094$ \\
$P+DES$ & $0.0029$ & $0.65$ & $0.046$ & $0.35$ & $1.0$ & $0.0036$ & $0.22$ & $0.0035$ & $0.16$ & $0.063$ \\
$P+BgB+BlB+DES$ & $0.0023$ & $0.66$ & $0.038$ & $0.18$ & $0.51$ & $0.0029$ & $0.16$ & $0.0035$ & $0.13$ & $0.044$ \\
$P+BgA0.1+BlB+DES$ & $0.0023$ & $0.66$ & $0.033$ & $0.17$ & $0.49$ & $0.0027$ & $0.15$ & $0.0034$ & $0.067$ & $0.037$ \\
$P+BgA0.1+BlB+ebA0.1$ & $0.0029$ & $0.73$ & $0.061$ & $0.17$ & $0.60$ & $0.0030$ & $0.22$ & $0.0034$ & $0.10$ & $0.074$ \\
$P+BgA0.2+BlB+ebA0.2$ & $0.0028$ & $0.73$ & $0.057$ & $0.16$ & $0.57$ & $0.0028$ & $0.21$ & $0.0033$ & $0.070$ & $0.060$ \\
$P+BgA0.1+BlB+ebA0.1+DES$ & $0.0022$ & $0.66$ & $0.032$ & $0.16$ & $0.45$ & $0.0024$ & $0.13$ & $0.0034$ & $0.062$ & $0.034$ \\
$P+hdA0.1+BgA0.1$ & $0.0032$ & $0.73$ & $0.070$ & $0.18$ & $0.63$ & $0.0032$ & $0.26$ & $0.0035$ & $0.093$ & $0.082$ \\
$P+hdA0.2+BgA0.2$ & $0.0030$ & $0.73$ & $0.060$ & $0.17$ & $0.60$ & $0.0031$ & $0.23$ & $0.0034$ & $0.061$ & $0.066$ \\
$P+BBgB+BBlB+DES$ & $0.0020$ & $0.68$ & $0.030$ & $0.12$ & $0.35$ & $0.0018$ & $0.13$ & $0.0034$ & $0.12$ & $0.040$ \\
$P+BBgA0.1$ & $0.0023$ & $0.70$ & $0.035$ & $0.12$ & $0.37$ & $0.0021$ & $0.14$ & $0.0033$ & $0.041$ & $0.036$ \\
$P+BBgA0.1+BBlB$ & $0.0022$ & $0.71$ & $0.034$ & $0.11$ & $0.34$ & $0.0018$ & $0.14$ & $0.0032$ & $0.039$ & $0.033$ \\
$P+BBgA0.1+BBlB+DES$ & $0.0018$ & $0.68$ & $0.025$ & $0.10$ & $0.30$ & $0.0016$ & $0.10$ & $0.0032$ & $0.031$ & $0.025$ \\
$P+BBgA0.2+BBlB$ & $0.0020$ & $0.72$ & $0.029$ & $0.085$ & $0.29$ & $0.0016$ & $0.13$ & $0.0031$ & $0.025$ & $0.030$ \\
$P+BBgA0.2+BBlB+DES$ & $0.0016$ & $0.68$ & $0.021$ & $0.078$ & $0.24$ & $0.0014$ & $0.096$ & $0.0031$ & $0.022$ & $0.024$ \\
$P+BB24gA0.1+BB24lB$ & $0.0019$ & $0.71$ & $0.027$ & $0.083$ & $0.28$ & $0.0015$ & $0.11$ & $0.0032$ & $0.030$ & $0.026$ \\
$P+BB24gA0.1+BB24lB+DES$ & $0.0017$ & $0.69$ & $0.022$ & $0.080$ & $0.25$ & $0.0014$ & $0.090$ & $0.0031$ & $0.025$ & $0.022$ \\
$P+BB24gA0.2+BB24lB$ & $0.0016$ & $0.73$ & $0.023$ & $0.066$ & $0.23$ & $0.0013$ & $0.11$ & $0.0029$ & $0.019$ & $0.024$ \\
$P+BB24gA0.2+BB24lB+DES$ & $0.0014$ & $0.69$ & $0.018$ & $0.062$ & $0.19$ & $0.0012$ & $0.084$ & $0.0029$ & $0.017$ & $0.020$ \\
$P+BgA0.1+BlB+euA0.1$ & $0.0021$ & $0.68$ & $0.036$ & $0.12$ & $0.36$ & $0.0018$ & $0.13$ & $0.0032$ & $0.052$ & $0.032$ \\
$P+BgA0.1+BlB+euA0.1+DES$ & $0.0018$ & $0.66$ & $0.027$ & $0.11$ & $0.33$ & $0.0016$ & $0.10$ & $0.0032$ & $0.043$ & $0.025$ \\
$P+BgA0.2+BlB+euA0.2$ & $0.0018$ & $0.68$ & $0.031$ & $0.11$ & $0.32$ & $0.0015$ & $0.12$ & $0.0030$ & $0.037$ & $0.028$ \\
$P+BgA0.2+BlB+euA0.2+DES$ & $0.0015$ & $0.65$ & $0.023$ & $0.099$ & $0.28$ & $0.0013$ & $0.092$ & $0.0030$ & $0.033$ & $0.023$ \\
$P+BB24gA0.1+BB24lB+euA0.1$ & $0.0017$ & $0.70$ & $0.024$ & $0.075$ & $0.24$ & $0.0014$ & $0.095$ & $0.0031$ & $0.026$ & $0.022$ \\
$P+BB24gA0.1+BB24lB+euA0.1+DES$ & $0.0016$ & $0.69$ & $0.020$ & $0.073$ & $0.22$ & $0.0013$ & $0.082$ & $0.0031$ & $0.023$ & $0.019$ \\
$P+BB24gA0.2+BB24lB+euA0.2$ & $0.0014$ & $0.71$ & $0.019$ & $0.059$ & $0.19$ & $0.0011$ & $0.085$ & $0.0027$ & $0.017$ & $0.019$ \\
$P+BB24gA0.2+BB24lB+euA0.2+DES$ & $0.0012$ & $0.69$ & $0.016$ & $0.056$ & $0.17$ & $0.00099$ & $0.072$ & $0.0027$ & $0.016$ & $0.017$ \\
$P+LSST$ & $0.0014$ & $0.65$ & $0.028$ & $0.16$ & $0.44$ & $0.0020$ & $0.070$ & $0.0033$ & $0.056$ & $0.017$ \\
$P+BgB+BlB+LSST$ & $0.0014$ & $0.65$ & $0.022$ & $0.13$ & $0.35$ & $0.0015$ & $0.063$ & $0.0033$ & $0.050$ & $0.015$ \\
$P+BBgB+BBlB+LSST$ & $0.0014$ & $0.65$ & $0.021$ & $0.095$ & $0.27$ & $0.0012$ & $0.059$ & $0.0033$ & $0.046$ & $0.013$ \\
$P+BBgA0.1+BBlB+LSST$ & $0.0013$ & $0.66$ & $0.017$ & $0.085$ & $0.24$ & $0.0012$ & $0.055$ & $0.0032$ & $0.024$ & $0.012$ \\
$P+BBgA0.2+BBlB+LSST$ & $0.0011$ & $0.66$ & $0.014$ & $0.069$ & $0.20$ & $0.0010$ & $0.052$ & $0.0030$ & $0.018$ & $0.011$ \\
$P+BB24gA0.1+BB24lB+LSST$ & $0.0012$ & $0.66$ & $0.016$ & $0.070$ & $0.20$ & $0.0010$ & $0.053$ & $0.0031$ & $0.020$ & $0.011$ \\
$P+BB24gA0.2+BB24lB+LSST$ & $0.0010$ & $0.66$ & $0.013$ & $0.056$ & $0.16$ & $0.00089$ & $0.049$ & $0.0029$ & $0.014$ & $0.011$ \\
$P+BB24gA0.1+BB24lB+euA0.1+LSST$ & $0.0011$ & $0.66$ & $0.015$ & $0.065$ & $0.19$ & $0.00096$ & $0.051$ & $0.0030$ & $0.018$ & $0.011$ \\
$P+BB24gA0.2+BB24lB+euA0.2+LSST$ & $0.00095$ & $0.66$ & $0.012$ & $0.051$ & $0.15$ & $0.00078$ & $0.046$ & $0.0027$ & $0.013$ & $0.010$ \\
$P+wfA0.1+BgA0.1$ & $0.0026$ & $0.68$ & $0.049$ & $0.15$ & $0.46$ & $0.0023$ & $0.19$ & $0.0034$ & $0.072$ & $0.049$ \\
$P+wfA0.2+BgA0.2$ & $0.0024$ & $0.69$ & $0.045$ & $0.14$ & $0.41$ & $0.0020$ & $0.17$ & $0.0032$ & $0.050$ & $0.042$ \\
$P+BgB+BlA+l1D$ & $0.0046$ & $0.79$ & $0.083$ & $0.19$ & $0.79$ & $0.0031$ & $0.42$ & $0.0038$ & $1.5$ & $0.10$ \\
$P+BgA0.1+BlA+l1D$ & $0.0032$ & $0.76$ & $0.073$ & $0.16$ & $0.59$ & $0.0027$ & $0.26$ & $0.0035$ & $0.066$ & $0.062$ \\
$P+BgA0.2+BlA+l1D$ & $0.0031$ & $0.75$ & $0.065$ & $0.16$ & $0.56$ & $0.0026$ & $0.24$ & $0.0034$ & $0.049$ & $0.054$ \\
$P+BBgB+BBlA+l1D$ & $0.0025$ & $0.73$ & $0.039$ & $0.12$ & $0.43$ & $0.0019$ & $0.21$ & $0.0034$ & $1.5$ & $0.053$ \\
$P+BBgA0.1+BBlA+l1D$ & $0.0019$ & $0.71$ & $0.033$ & $0.097$ & $0.32$ & $0.0015$ & $0.12$ & $0.0030$ & $0.031$ & $0.025$ \\
$P+BBgA0.2+BBlA+l1D$ & $0.0017$ & $0.73$ & $0.028$ & $0.080$ & $0.28$ & $0.0013$ & $0.12$ & $0.0028$ & $0.021$ & $0.023$ \\
$P+BB24gB+BB24lA+l1D$ & $0.0021$ & $0.74$ & $0.031$ & $0.096$ & $0.34$ & $0.0016$ & $0.17$ & $0.0033$ & $1.5$ & $0.044$ \\
$P+BB24gA0.1+BB24lA+l1D$ & $0.0016$ & $0.72$ & $0.026$ & $0.076$ & $0.25$ & $0.0012$ & $0.10$ & $0.0028$ & $0.024$ & $0.020$ \\
$P+BB24gA0.2+BB24lA+l1D$ & $0.0014$ & $0.73$ & $0.022$ & $0.063$ & $0.22$ & $0.0011$ & $0.094$ & $0.0025$ & $0.017$ & $0.018$ \\
$P+BB24gA0.2+BB24lA+l1D+euA0.2$ & $0.0012$ & $0.71$ & $0.018$ & $0.056$ & $0.19$ & $0.00091$ & $0.076$ & $0.0023$ & $0.015$ & $0.016$ \\
$P+BB24gA0.2+BB24lA+l1D+LSST$ & $0.00092$ & $0.66$ & $0.012$ & $0.053$ & $0.15$ & $0.00082$ & $0.043$ & $0.0024$ & $0.014$ & $0.0091$ \\
$P+BB24gA0.2+BB24lA+l1D+euA0.2+LSST$ & $0.00086$ & $0.66$ & $0.012$ & $0.049$ & $0.14$ & $0.00072$ & $0.041$ & $0.0023$ & $0.013$ & $0.0087$ \\

\hline
\hline
\end{tabular}
\end{table}

One thing we note is that neutrino mass measurements are substantially 
degraded by 
including the MoG parameters, indicating that the $\summnu$ constraint is 
driven measuring the low-$z$ structure amplitude relative to the CMB, more than
the scale dependence of the neutrino power suppression, which would not be
degenerate with these MoG parameters.

\subsection{Inflation}

We consider two ways to probe inflation: measurement of the scale dependence
of the initial perturbations, and their level of Gaussianity.
 
\subsubsection{Running of the spectral index}

Table \ref{tablealphas} shows constraints with free running of the spectral 
index. 
\begin{table}
\caption{
Projections including running of the spectral index.
See Table \ref{tab:experimentabbreviations} for survey codes.}
\label{tablealphas} 
\begin{tabular}{lccccccc}
\hline
\hline
$ $ & $\omega_m$ & $\omega_b$ & $\theta_s$ & $\Sigma m_{\nu}$ & $\log_{10}(A)$ & $n_s$ & $\alpha_s$ \\
\hline
${\rm value}$ & $0.141$ & $0.0221$ & $0.597$ & $0.0600$ & $-8.66$ & $0.961$ & $0.00$  \\
\hline
$P$ & $0.0037$ & $0.00017$ & $0.00036$ & $0.35$ & $0.0043$ & $0.0038$ & $0.0054$ \\
$P+BgB+BlB$ & $0.00074$ & $0.00017$ & $0.00014$ & $0.10$ & $0.0042$ & $0.0038$ & $0.0054$ \\
$P+BgA0.1+BlB$ & $0.00070$ & $0.00015$ & $0.00014$ & $0.069$ & $0.0042$ & $0.0031$ & $0.0053$ \\
$P+BgA0.2+BlB$ & $0.00072$ & $0.00013$ & $0.00015$ & $0.046$ & $0.0041$ & $0.0028$ & $0.0050$ \\
$P+DES$ & $0.0013$ & $0.00014$ & $0.00017$ & $0.041$ & $0.0041$ & $0.0032$ & $0.0049$ \\
$P+BgB+BlB+DES$ & $0.00070$ & $0.00013$ & $0.00014$ & $0.030$ & $0.0041$ & $0.0027$ & $0.0049$ \\
$P+BgA0.1+BlB+DES$ & $0.00068$ & $0.00013$ & $0.00014$ & $0.029$ & $0.0041$ & $0.0027$ & $0.0049$ \\
$P+BgA0.1+BlB+ebA0.1$ & $0.00064$ & $0.00014$ & $0.00014$ & $0.053$ & $0.0041$ & $0.0029$ & $0.0053$ \\
$P+BgA0.2+BlB+ebA0.2$ & $0.00065$ & $0.00013$ & $0.00014$ & $0.036$ & $0.0040$ & $0.0027$ & $0.0049$ \\
$P+BgA0.1+BlB+ebA0.1+DES$ & $0.00062$ & $0.00013$ & $0.00014$ & $0.028$ & $0.0040$ & $0.0026$ & $0.0048$ \\
$P+hdB+BgB$ & $0.00074$ & $0.00017$ & $0.00014$ & $0.10$ & $0.0042$ & $0.0038$ & $0.0054$ \\
$P+hdA0.1+BgA0.1$ & $0.00069$ & $0.00015$ & $0.00014$ & $0.061$ & $0.0041$ & $0.0030$ & $0.0053$ \\
$P+hdA0.2+BgA0.2$ & $0.00069$ & $0.00013$ & $0.00014$ & $0.039$ & $0.0040$ & $0.0027$ & $0.0050$ \\
$P+BBgB$ & $0.00056$ & $0.00017$ & $0.00014$ & $0.090$ & $0.0042$ & $0.0038$ & $0.0054$ \\
$P+BBgB+BlB$ & $0.00056$ & $0.00017$ & $0.00014$ & $0.090$ & $0.0042$ & $0.0038$ & $0.0054$ \\
$P+BBlB+BgB$ & $0.00073$ & $0.00017$ & $0.00014$ & $0.099$ & $0.0042$ & $0.0038$ & $0.0054$ \\
$P+BBgB+BBlB$ & $0.00056$ & $0.00017$ & $0.00014$ & $0.090$ & $0.0042$ & $0.0038$ & $0.0054$ \\
$P+BBgB+BBlB+DES$ & $0.00047$ & $0.00013$ & $0.00014$ & $0.027$ & $0.0040$ & $0.0025$ & $0.0049$ \\
$P+BBgA0.1$ & $0.00045$ & $0.00013$ & $0.00014$ & $0.025$ & $0.0040$ & $0.0024$ & $0.0051$ \\
$P+BBgA0.1+BBlB$ & $0.00045$ & $0.00013$ & $0.00014$ & $0.025$ & $0.0040$ & $0.0024$ & $0.0051$ \\
$P+BBgA0.1+BBlB+DES$ & $0.00044$ & $0.00013$ & $0.00014$ & $0.022$ & $0.0040$ & $0.0024$ & $0.0046$ \\
$P+BBgA0.2+BBlB$ & $0.00043$ & $0.00011$ & $0.00014$ & $0.017$ & $0.0037$ & $0.0022$ & $0.0040$ \\
$P+BBgA0.2+BBlB+DES$ & $0.00043$ & $0.00011$ & $0.00014$ & $0.017$ & $0.0036$ & $0.0022$ & $0.0038$ \\
$P+BB24gB+BB24lB$ & $0.00053$ & $0.00016$ & $0.00014$ & $0.088$ & $0.0042$ & $0.0037$ & $0.0054$ \\
$P+BB24gA0.1+BB24lB$ & $0.00039$ & $0.00013$ & $0.00014$ & $0.021$ & $0.0040$ & $0.0023$ & $0.0050$ \\
$P+BB24gA0.1+BB24lB+DES$ & $0.00039$ & $0.00012$ & $0.00013$ & $0.020$ & $0.0040$ & $0.0023$ & $0.0045$ \\
$P+BB24gA0.2+BB24lB$ & $0.00037$ & $0.00010$ & $0.00014$ & $0.015$ & $0.0036$ & $0.0021$ & $0.0036$ \\
$P+BB24gA0.2+BB24lB+DES$ & $0.00037$ & $0.00010$ & $0.00014$ & $0.015$ & $0.0035$ & $0.0021$ & $0.0035$ \\
$P+BgB+BlB+euB$ & $0.00055$ & $0.00017$ & $0.00014$ & $0.090$ & $0.0042$ & $0.0038$ & $0.0054$ \\
$P+BgA0.1+BlB+euA0.1$ & $0.00044$ & $0.00013$ & $0.00014$ & $0.022$ & $0.0040$ & $0.0024$ & $0.0050$ \\
$P+BgA0.1+BlB+euA0.1+DES$ & $0.00044$ & $0.00012$ & $0.00014$ & $0.020$ & $0.0040$ & $0.0023$ & $0.0046$ \\
$P+BgA0.2+BlB+euA0.2$ & $0.00043$ & $0.00011$ & $0.00014$ & $0.016$ & $0.0037$ & $0.0022$ & $0.0037$ \\
$P+BgA0.2+BlB+euA0.2+DES$ & $0.00042$ & $0.00011$ & $0.00014$ & $0.015$ & $0.0035$ & $0.0022$ & $0.0036$ \\
$P+BB24gA0.1+BB24lB+euA0.1$ & $0.00037$ & $0.00013$ & $0.00013$ & $0.018$ & $0.0039$ & $0.0022$ & $0.0047$ \\
$P+BB24gA0.1+BB24lB+euA0.1+DES$ & $0.00037$ & $0.00012$ & $0.00013$ & $0.018$ & $0.0039$ & $0.0022$ & $0.0043$ \\
$P+BB24gA0.2+BB24lB+euA0.2$ & $0.00034$ & $9.9e-05$ & $0.00013$ & $0.014$ & $0.0035$ & $0.0019$ & $0.0031$ \\
$P+BB24gA0.2+BB24lB+euA0.2+DES$ & $0.00034$ & $9.9e-05$ & $0.00013$ & $0.014$ & $0.0034$ & $0.0019$ & $0.0030$ \\
$P+LSST$ & $0.00084$ & $0.00012$ & $0.00015$ & $0.023$ & $0.0039$ & $0.0030$ & $0.0038$ \\
$P+BgB+BlB+LSST$ & $0.00062$ & $0.00012$ & $0.00014$ & $0.020$ & $0.0039$ & $0.0025$ & $0.0038$ \\
$P+BBgB+BBlB+LSST$ & $0.00045$ & $0.00012$ & $0.00013$ & $0.018$ & $0.0038$ & $0.0023$ & $0.0037$ \\
$P+BBgA0.1+BBlB+LSST$ & $0.00043$ & $0.00012$ & $0.00013$ & $0.018$ & $0.0038$ & $0.0022$ & $0.0036$ \\
$P+BBgA0.2+BBlB+LSST$ & $0.00042$ & $0.00011$ & $0.00014$ & $0.015$ & $0.0033$ & $0.0022$ & $0.0033$ \\
$P+BB24gA0.1+BB24lB+LSST$ & $0.00039$ & $0.00012$ & $0.00013$ & $0.017$ & $0.0038$ & $0.0021$ & $0.0035$ \\
$P+BB24gA0.2+BB24lB+LSST$ & $0.00037$ & $0.00010$ & $0.00013$ & $0.014$ & $0.0031$ & $0.0020$ & $0.0030$ \\
$P+BB24gA0.1+BB24lB+euA0.1+LSST$ & $0.00036$ & $0.00011$ & $0.00013$ & $0.016$ & $0.0037$ & $0.0020$ & $0.0034$ \\
$P+BB24gA0.2+BB24lB+euA0.2+LSST$ & $0.00034$ & $9.9e-05$ & $0.00013$ & $0.012$ & $0.0029$ & $0.0019$ & $0.0028$ \\
$P+wfB+BgB$ & $0.00065$ & $0.00017$ & $0.00014$ & $0.096$ & $0.0042$ & $0.0038$ & $0.0054$ \\
$P+wfA0.1+BgA0.1$ & $0.00058$ & $0.00014$ & $0.00014$ & $0.037$ & $0.0041$ & $0.0027$ & $0.0053$ \\
$P+wfA0.2+BgA0.2$ & $0.00057$ & $0.00012$ & $0.00014$ & $0.021$ & $0.0039$ & $0.0025$ & $0.0045$ \\
$P+BgB+BlA+l1D$ & $0.00067$ & $0.00011$ & $0.00014$ & $0.068$ & $0.0041$ & $0.0033$ & $0.0038$ \\
$P+BgA0.1+BlA+l1D$ & $0.00065$ & $0.00011$ & $0.00014$ & $0.056$ & $0.0040$ & $0.0030$ & $0.0034$ \\
$P+BgA0.2+BlA+l1D$ & $0.00066$ & $0.00011$ & $0.00014$ & $0.043$ & $0.0039$ & $0.0027$ & $0.0031$ \\
$P+BBgB+BBlA+l1D$ & $0.00041$ & $0.00010$ & $0.00014$ & $0.047$ & $0.0039$ & $0.0029$ & $0.0027$ \\
$P+BBgA0.1+BBlA+l1D$ & $0.00040$ & $0.00010$ & $0.00014$ & $0.023$ & $0.0036$ & $0.0023$ & $0.0021$ \\
$P+BBgA0.2+BBlA+l1D$ & $0.00038$ & $0.00010$ & $0.00014$ & $0.017$ & $0.0036$ & $0.0022$ & $0.0020$ \\
$P+BB24gB+BB24lA+l1D$ & $0.00036$ & $0.00010$ & $0.00014$ & $0.040$ & $0.0038$ & $0.0028$ & $0.0023$ \\
$P+BB24gA0.1+BB24lA+l1D$ & $0.00035$ & $0.00010$ & $0.00014$ & $0.019$ & $0.0036$ & $0.0022$ & $0.0018$ \\
$P+BB24gA0.2+BB24lA+l1D$ & $0.00034$ & $9.8e-05$ & $0.00014$ & $0.015$ & $0.0035$ & $0.0020$ & $0.0017$ \\
$P+BB24gA0.2+BB24lA+l1D+euA0.2$ & $0.00032$ & $9.6e-05$ & $0.00013$ & $0.014$ & $0.0034$ & $0.0019$ & $0.0016$ \\
$P+BB24gA0.2+BB24lA+l1D+LSST$ & $0.00034$ & $9.8e-05$ & $0.00013$ & $0.012$ & $0.0026$ & $0.0019$ & $0.0016$ \\
$P+BB24gA0.2+BB24lA+l1D+euA0.2+LSST$ & $0.00032$ & $9.5e-05$ & $0.00013$ & $0.011$ & $0.0025$ & $0.0018$ & $0.0016$ \\

\hline
\hline
\end{tabular}
\end{table}

We first note that previous tables, like this one, have shown that galaxy 
surveys can improve the constraint on $n_s$ by about a factor of 2 relative
to Planck alone (similar constraints have been projected for CMBpol 
\cite{2010PhRvD..82l3504G}). 
We see that the \lyaf\ broadband constraint is important
for achieving the tightest constraints on $\alpha_s$, i.e., 
starting from a Planck-only baseline of $0.0054$,
aggressive galaxy clustering
from DESI and Euclid, combined with LSST, can reach an error $0.0028$, 
while 0.0016 can be achieved with the \lyaf\ broadband power,
a critical improvement in the quest to find
expected deviations from a pure power law at the $10^{-3}$ level. 
The statistical power of the DESI \lyaf\ data will also help to control
systematics and extensions of the gas model, and might allow further
improvements if external data sets that are limiting the measurement are
improved. The \lyaf\ is especially powerful for constraining $\alpha_s$ because
it is sensitive to the power spectrum on smaller scales than our other probes, 
giving a longer lever arm to measure its scale dependence.

\subsubsection{Non-Gaussianity ($\fNL$)} 

In this section we do not attempt to be as comprehensive as others. We simply
give the estimated constraining power for the main redshift surveys, based on 
the power spectrum, for the local model of non-Gaussianity, following 
\cite{2008PhRvD..78l3519M,2009JCAP...10..007M}.
Recall that when inflation produces curvature perturbations that are non-linear
in a Gaussian field, $\phi$ (which has the usual power spectrum that we have in
the Gaussian case), i.e., 
\begin{equation}
\Phi = \phi +\fNL \left(\phi^2- \left<\phi^2\right>\right)
\end{equation}
it can be shown in many ways \cite{2008ApJ...687...12D,
2008ApJ...677L..77M,2008PhRvD..78l3519M,
2008PhRvD..78l3507A,2008JCAP...08..031S}
that the linear regime bias model for galaxy 
clustering must be extended to
\begin{equation}
\delta_g = b_\delta \delta + b_\phi \fNL \phi +\epsilon +...
\end{equation}
where $b_\delta$ is the usual linear bias present in the Gaussian case and 
$b_\phi$ is the new ``potential bias'', which becomes significant on large
scales, i.e., low $k$,  because $\phi_\vk \propto k^{-2} \delta_\vk$.
\cite{2008JCAP...08..031S} use the halo model to estimate
\begin{equation}
b_\phi \simeq 3.372 (b_\delta-1)~.
\end{equation}
This is not an exact calculation but should be good enough to roughly 
estimate the expected detection significance in a given scenario. 
To make projections, we use exactly the code of \cite{2009JCAP...10..007M}
so we refer the reader there for details. 

Table \ref{tablefNL} shows the projections. 
\begin{table}
\caption{
Estimated constraints on $\fNL$ for the local model of non-Gaussianity.}
\label{tablefNL} 
\begin{tabular}{lc}
\hline
\hline
Survey & $\sigma_{\fNL}$  \\
\hline
BOSS & 23 \\
BOSS+eBOSS & 11 \\
DESI & 3.8 \\
BOSS+Euclid & 6.7 \\
\hline
\hline
\end{tabular}
\end{table}
The results are weakly sensitive to redshift binning (which sets the 
interval
of free bias parameters), e.g., the DESI number would be 3.4 instead of 3.8 
for $\Delta z=0.2$
instead of the usual 0.1. The DESI high-$z$ quasars ($z>1.9$) add what might be 
considered a surprising amount to the constraint -- it would be 4.7 instead of
3.8 without them (we do not use the \lyaf\ here because we suspect that 
things like radiation background fluctuations will make it difficult to use
the very largest scales for this kind of measurement 
\cite{2005MNRAS.360.1471M}).

The $\sigma_{\fNL} \sim 5$ constraints possible with DESI or Euclid are
comparable to Planck constraints \cite{2013arXiv1303.5084P}. 
It may be possible to do better using the bispectrum 
\cite{2012MNRAS.425.2903S,2011JCAP...04..006B,2010JCAP...07..002N,
2010AdAst2010E..73L,2009PhRvD..80l3002S,2009ApJ...703.1230J}. 

\subsection{Dark radiation ($N_\nu$)}

Table \ref{tableNnu} shows constraints on extra radiation beyond the standard
amount contributed by photons and neutrinos, parameterized as is traditional
by an effective number of neutrino species $N_\nu$ \cite{2012PhRvD..85b3001S,
2013PhRvD..87h3008H,2013arXiv1303.5379B,2012JCAP...12..027H}.
\begin{table}
\caption{
Projections including $N_\nu$.
See Table \ref{tab:experimentabbreviations} for survey codes.
}
\label{tableNnu} 
\begin{tabular}{lccccccc}
\hline
\hline
$ $ & $\omega_m$ & $\omega_b$ & $\theta_s$ & $\Sigma m_{\nu}$ & $N_{\nu,l}$ & $\log_{10}(A)$ & $n_s$  \\
\hline
${\rm value}$ & $0.141$ & $0.0221$ & $0.597$ & $0.0600$ & $3.05$ & $-8.66$ & $0.961$  \\
\hline
$P$ & $0.0050$ & $0.00023$ & $0.00042$ & $0.35$ & $0.18$ & $0.0049$ & $0.0082$ \\
$P+BgB+BlB$ & $0.0033$ & $0.00023$ & $0.00026$ & $0.12$ & $0.18$ & $0.0049$ & $0.0081$ \\
$P+BgA0.1+BlB$ & $0.0031$ & $0.00020$ & $0.00025$ & $0.086$ & $0.18$ & $0.0048$ & $0.0073$ \\
$P+BgA0.2+BlB$ & $0.0025$ & $0.00019$ & $0.00022$ & $0.061$ & $0.15$ & $0.0045$ & $0.0061$ \\
$P+DES$ & $0.0020$ & $0.00019$ & $0.00020$ & $0.048$ & $0.12$ & $0.0038$ & $0.0059$ \\
$P+BgB+BlB+DES$ & $0.0019$ & $0.00016$ & $0.00020$ & $0.045$ & $0.11$ & $0.0037$ & $0.0048$ \\
$P+BgA0.1+BlB+DES$ & $0.0019$ & $0.00016$ & $0.00019$ & $0.043$ & $0.11$ & $0.0037$ & $0.0048$ \\
$P+BgA0.1+BlB+ebA0.1$ & $0.0029$ & $0.00019$ & $0.00024$ & $0.068$ & $0.17$ & $0.0048$ & $0.0068$ \\
$P+BgA0.2+BlB+ebA0.2$ & $0.0022$ & $0.00018$ & $0.00020$ & $0.048$ & $0.13$ & $0.0044$ & $0.0055$ \\
$P+BgA0.1+BlB+ebA0.1+DES$ & $0.0018$ & $0.00015$ & $0.00019$ & $0.041$ & $0.11$ & $0.0037$ & $0.0046$ \\
$P+hdB+BgB$ & $0.0033$ & $0.00023$ & $0.00026$ & $0.12$ & $0.18$ & $0.0049$ & $0.0081$ \\
$P+hdA0.1+BgA0.1$ & $0.0030$ & $0.00020$ & $0.00025$ & $0.078$ & $0.17$ & $0.0048$ & $0.0071$ \\
$P+hdA0.2+BgA0.2$ & $0.0023$ & $0.00018$ & $0.00021$ & $0.053$ & $0.14$ & $0.0045$ & $0.0058$ \\
$P+BBgB$ & $0.0032$ & $0.00023$ & $0.00026$ & $0.11$ & $0.18$ & $0.0049$ & $0.0081$ \\
$P+BBgB+BlB$ & $0.0032$ & $0.00023$ & $0.00026$ & $0.11$ & $0.18$ & $0.0049$ & $0.0081$ \\
$P+BBlB+BgB$ & $0.0033$ & $0.00023$ & $0.00026$ & $0.12$ & $0.18$ & $0.0049$ & $0.0081$ \\
$P+BBgB+BBlB$ & $0.0032$ & $0.00023$ & $0.00026$ & $0.11$ & $0.18$ & $0.0049$ & $0.0081$ \\
$P+BBgB+BBlB+DES$ & $0.0019$ & $0.00015$ & $0.00019$ & $0.044$ & $0.11$ & $0.0037$ & $0.0046$ \\
$P+BBgA0.1$ & $0.0022$ & $0.00016$ & $0.00020$ & $0.036$ & $0.13$ & $0.0046$ & $0.0048$ \\
$P+BBgA0.1+BBlB$ & $0.0022$ & $0.00016$ & $0.00020$ & $0.036$ & $0.13$ & $0.0046$ & $0.0048$ \\
$P+BBgA0.1+BBlB+DES$ & $0.0015$ & $0.00014$ & $0.00018$ & $0.029$ & $0.086$ & $0.0036$ & $0.0036$ \\
$P+BBgA0.2+BBlB$ & $0.0014$ & $0.00014$ & $0.00017$ & $0.024$ & $0.084$ & $0.0042$ & $0.0031$ \\
$P+BBgA0.2+BBlB+DES$ & $0.0012$ & $0.00013$ & $0.00017$ & $0.021$ & $0.070$ & $0.0035$ & $0.0027$ \\
$P+BB24gB+BB24lB$ & $0.0032$ & $0.00023$ & $0.00026$ & $0.11$ & $0.18$ & $0.0049$ & $0.0081$ \\
$P+BB24gA0.1+BB24lB$ & $0.0020$ & $0.00015$ & $0.00019$ & $0.031$ & $0.11$ & $0.0045$ & $0.0041$ \\
$P+BB24gA0.1+BB24lB+DES$ & $0.0014$ & $0.00013$ & $0.00017$ & $0.026$ & $0.079$ & $0.0035$ & $0.0032$ \\
$P+BB24gA0.2+BB24lB$ & $0.0013$ & $0.00014$ & $0.00017$ & $0.022$ & $0.074$ & $0.0041$ & $0.0025$ \\
$P+BB24gA0.2+BB24lB+DES$ & $0.0011$ & $0.00013$ & $0.00016$ & $0.019$ & $0.063$ & $0.0035$ & $0.0023$ \\
$P+BgB+BlB+euB$ & $0.0032$ & $0.00023$ & $0.00026$ & $0.11$ & $0.18$ & $0.0049$ & $0.0081$ \\
$P+BgA0.1+BlB+euA0.1$ & $0.0021$ & $0.00015$ & $0.00020$ & $0.031$ & $0.12$ & $0.0045$ & $0.0044$ \\
$P+BgA0.1+BlB+euA0.1+DES$ & $0.0014$ & $0.00013$ & $0.00017$ & $0.026$ & $0.081$ & $0.0035$ & $0.0033$ \\
$P+BgA0.2+BlB+euA0.2$ & $0.0013$ & $0.00014$ & $0.00017$ & $0.022$ & $0.077$ & $0.0041$ & $0.0028$ \\
$P+BgA0.2+BlB+euA0.2+DES$ & $0.0011$ & $0.00013$ & $0.00016$ & $0.019$ & $0.065$ & $0.0035$ & $0.0025$ \\
$P+BB24gA0.1+BB24lB+euA0.1$ & $0.0017$ & $0.00014$ & $0.00018$ & $0.026$ & $0.097$ & $0.0044$ & $0.0035$ \\
$P+BB24gA0.1+BB24lB+euA0.1+DES$ & $0.0013$ & $0.00013$ & $0.00017$ & $0.022$ & $0.072$ & $0.0035$ & $0.0028$ \\
$P+BB24gA0.2+BB24lB+euA0.2$ & $0.0012$ & $0.00013$ & $0.00016$ & $0.019$ & $0.065$ & $0.0040$ & $0.0021$ \\
$P+BB24gA0.2+BB24lB+euA0.2+DES$ & $0.0010$ & $0.00012$ & $0.00016$ & $0.017$ & $0.057$ & $0.0034$ & $0.0019$ \\
$P+LSST$ & $0.00096$ & $0.00014$ & $0.00016$ & $0.020$ & $0.063$ & $0.0031$ & $0.0038$ \\
$P+BgB+BlB+LSST$ & $0.00094$ & $0.00012$ & $0.00016$ & $0.018$ & $0.055$ & $0.0030$ & $0.0029$ \\
$P+BBgB+BBlB+LSST$ & $0.00093$ & $0.00012$ & $0.00016$ & $0.017$ & $0.051$ & $0.0030$ & $0.0023$ \\
$P+BBgA0.1+BBlB+LSST$ & $0.00090$ & $0.00012$ & $0.00016$ & $0.016$ & $0.050$ & $0.0029$ & $0.0023$ \\
$P+BBgA0.2+BBlB+LSST$ & $0.00086$ & $0.00012$ & $0.00016$ & $0.014$ & $0.049$ & $0.0027$ & $0.0021$ \\
$P+BB24gA0.1+BB24lB+LSST$ & $0.00089$ & $0.00011$ & $0.00016$ & $0.015$ & $0.049$ & $0.0028$ & $0.0021$ \\
$P+BB24gA0.2+BB24lB+LSST$ & $0.00083$ & $0.00011$ & $0.00016$ & $0.013$ & $0.046$ & $0.0025$ & $0.0019$ \\
$P+BB24gA0.1+BB24lB+euA0.1+LSST$ & $0.00086$ & $0.00011$ & $0.00016$ & $0.014$ & $0.047$ & $0.0027$ & $0.0020$ \\
$P+BB24gA0.2+BB24lB+euA0.2+LSST$ & $0.00079$ & $0.00011$ & $0.00015$ & $0.012$ & $0.043$ & $0.0024$ & $0.0017$ \\
$P+wfB+BgB$ & $0.0033$ & $0.00023$ & $0.00026$ & $0.12$ & $0.18$ & $0.0049$ & $0.0081$ \\
$P+wfA0.1+BgA0.1$ & $0.0026$ & $0.00018$ & $0.00022$ & $0.050$ & $0.15$ & $0.0047$ & $0.0060$ \\
$P+wfA0.2+BgA0.2$ & $0.0017$ & $0.00016$ & $0.00018$ & $0.030$ & $0.11$ & $0.0043$ & $0.0042$ \\
$P+BgB+BlA+l1D$ & $0.0026$ & $0.00021$ & $0.00023$ & $0.11$ & $0.15$ & $0.0042$ & $0.0073$ \\
$P+BgA0.1+BlA+l1D$ & $0.0022$ & $0.00019$ & $0.00021$ & $0.083$ & $0.13$ & $0.0042$ & $0.0059$ \\
$P+BgA0.2+BlA+l1D$ & $0.0019$ & $0.00017$ & $0.00019$ & $0.059$ & $0.11$ & $0.0041$ & $0.0047$ \\
$P+BBgB+BBlA+l1D$ & $0.0020$ & $0.00018$ & $0.00020$ & $0.082$ & $0.11$ & $0.0041$ & $0.0056$ \\
$P+BBgA0.1+BBlA+l1D$ & $0.0013$ & $0.00013$ & $0.00017$ & $0.031$ & $0.070$ & $0.0039$ & $0.0025$ \\
$P+BBgA0.2+BBlA+l1D$ & $0.0012$ & $0.00013$ & $0.00016$ & $0.022$ & $0.063$ & $0.0038$ & $0.0021$ \\
$P+BB24gB+BB24lA+l1D$ & $0.0018$ & $0.00017$ & $0.00018$ & $0.071$ & $0.099$ & $0.0040$ & $0.0048$ \\
$P+BB24gA0.1+BB24lA+l1D$ & $0.0012$ & $0.00013$ & $0.00016$ & $0.026$ & $0.062$ & $0.0038$ & $0.0020$ \\
$P+BB24gA0.2+BB24lA+l1D$ & $0.0011$ & $0.00013$ & $0.00016$ & $0.019$ & $0.057$ & $0.0037$ & $0.0017$ \\
$P+BB24gA0.2+BB24lA+l1D+euA0.2$ & $0.0010$ & $0.00012$ & $0.00016$ & $0.017$ & $0.054$ & $0.0036$ & $0.0016$ \\
$P+BB24gA0.2+BB24lA+l1D+LSST$ & $0.00081$ & $0.00011$ & $0.00015$ & $0.013$ & $0.043$ & $0.0025$ & $0.0015$ \\
$P+BB24gA0.2+BB24lA+l1D+euA0.2+LSST$ & $0.00078$ & $0.00011$ & $0.00015$ & $0.011$ & $0.041$ & $0.0024$ & $0.0014$ \\

\hline
\hline
\end{tabular}
\end{table}
DESI-era LSS measurements should be able to verify if recent hints for extra 
radiation are correct
\cite{2013arXiv1307.7715W}. If they are, a new quest will begin to measure 
the properties of the radiation in detail -- if not, we can continue to search 
at higher precision. A constraint $\sigma_{N_{\rm eff}}\sim 0.044$
has been projected for
CMB lensing from CMBPol \cite{2010PhRvD..82l3504G,2010JPhCS.259a2004M}.

\section{Discussion and Conclusions \label{sec:conclusions}}

Our choice of experiments is undoubtedly a somewhat subjective combination of
predictability of results, likelihood to happen, and simply author capabilities
and interest. In the future we might hope to add 
CMB lensing \cite{2014arXiv1402.4108W,2013PhLB..718.1186O,
2012MNRAS.425.1170H,2009PhRvD..79f5033D},
higher resolution CMB experiments \cite{2014arXiv1402.4108W,
2012ApJ...749...90H},
21 cm intensity mapping surveys
\cite{2008PhRvL.100i1303C,2013AJ....145...65P,2010PhRvD..81j3527M,
2010PhRvD..82j3501T,
2010PhRvD..81f2001M,2008PhRvL.100i1303C}, etc.
Our choice of cosmological models to project was also limited largely by author
capability. We could hope to add projections for, 
e.g., other models of non-Gaussianity \cite{2010AdAst2010E..89D,
2010PhRvD..82d3531T,2010PhRvD..81b3006D,2010JCAP...10..022W,
2010PhRvD..82j6009B,2011JCAP...01..006B,2011JCAP...08..003L,
2011JCAP...11..009S,2012JCAP...03..002W,2012PhRvD..85h3002S,
2012JCAP...08..033S,2013MNRAS.429.1774B,2013PhRvL.110m1301N,
2009PhRvD..80l6018B,2009JCAP...12..022S,2009JCAP...01..042K,
2009JCAP...01..026K},
warm dark matter 
\cite{2013arXiv1306.2314V,2008PhRvL.100d1304V,2006PhRvL..97s1303S} or 
primordial black holes
contributing to the power spectrum 
\cite{2003ApJ...594L..71A,2010PhRvD..81j4019C} 
(for the latter two we would really like to have a better \lyaf\ 
implementation). 
In general there is more information in all of the data sets than we 
include here, e.g., obtainable by measuring higher order statistics like the 
bispectrum \cite{2011MNRAS.410.2730K}.
There is also a lot of work to be done to systematically quantify sensitivity 
to various potential systematic errors \cite{2009MNRAS.399.1074H}.

This is intended primarily as a reference paper, i.e., when you wonder how
well we hope to measure parameters in a given scenario, you can look it up 
in the appropriate table.  However, 
some of the initial take-home points, not all new to this paper but in
any case quantified and highlighted, might be:
\begin{itemize}
\item Redshift surveys like DESI, with help from Planck and possibly
  lensing surveys, will measure the sum of neutrino masses to $\sim
  0.01-0.02$ eV in the 2020's. This will give a strong detection of the 
minimum possible sum of masses $\sim 0.06$ eV, however, the mass hierarchy
will only be distinguishable with luck, if the true sum of masses is right at 
the minimum. 
\item Because it will inevitably be relevant, we should always
  include neutrino mass uncertainty in projections for other
  parameters. This introduces new uncertainty in the background evolution 
after the CMB epoch, in addition to the more commonly discussed change in 
evolution of the power spectrum, which increases the value of adding 
redshift space distortion and/or lensing measurements to Dark Energy 
measurements based on BAO distance measurements alone. The combination
of growth of structure and background evolution constraints helps to break
the degeneracy between Dark Energy and neutrino mass, although in the end the
Dark Energy constraints are always degraded by the neutrino mass uncertainty. 
\item Redshift surveys and lensing surveys are highly complementary as
  Dark Energy probes. (Somewhat surprisingly, however, overlapping the
  volume of the two such surveys so that cross-correlations of galaxy density
with lensing can be used to calibrate bias does not 
significantly improve fundamental parameter measurements,
as discussed in the Appendix, \ref{sec:overl-redsh-phot}.)
\item The ``Stage IV'' redshift surveys are fairly comparable in their 
constraining power, and comparable to LSST when using broadband power
constraints, although WFIRST's limited area leaves it a little bit 
under-powered statistically.
The Dark Energy equation of state will be constrained to 1-2\% in the 
Stage IV epoch. 
\item The curvature parameter $\Omega_k$ will be constrained to better than
$\pm 0.001$.
\item Allowing for deviations from GR significantly degrades neutrino
  mass constraints, highlighting the fact that these constraints at
  their most powerful come from measuring the amplitude of fluctuations
  at low redshift relative to the CMB (with
  redshift-space distortions for spectroscopic surveys or lensing for
  photometric surveys), more than measuring the effect of neutrinos on
  the shape of the power spectrum.
\item Of the two modified gravity parameters, gravitational lensing is 
relatively more powerful for measuring $G_9$, the parameter that sets the 
overall normalization of the late-time perturbations relative to the CMB, 
while redshift space
distortions are relatively better at measuring $\Delta \gamma$, deviations of 
the late-time growth rate from GR (presumably because RSDs are directly 
sensitive to the growth
rate at a given redshift, while lensing measures a broadly averaged 
normalization of fluctuations). The most powerful test of GR will be obtained
by putting the two probes together.  
\item Using the \lyaf\ enhanced by cross-correlation with quasar
  density, DESI will be able to make very precise BAO measurements at
  $z>2$, one of its unique advantages over a survey like Euclid (the
measurement at $z>2$ will also be better than WFIRST's).
\item \lyaf\ power spectrum measurements are potentially our best
  probe of the running of the inflationary spectral index, which can be 
constrained to better than $\pm 0.002$, while
  galaxy clustering can help to improve constraints 
  and the spectral index itself, also to $\sim \pm 0.002$.
\item One place redshift surveys are not likely to improve on Planck is 
measuring
the local non-Gaussianity parameter, $\fNL$, although they can achieve 
comparable precision, $~\pm 4$ 
(using the power spectrum alone -- with the bispectrum
they might be able to do better). 
\item Dark radiation in the Universe will be measured at a level equivalent to 
$\sim 0.05$ times the
contribution of a standard model neutrino species. 
\end{itemize}

We thank Yan-Chuan Cai for invaluable spot-checks of the lensing/redshift
survey overlap calculations, and Zhaoming Ma for helpful conversations and
the DES numbers. We thank Uro\v{s} Seljak for helpful comments.  

\bibliography{cosmo,cosmo_preprints}

\begin{thebibliography}{197}%
\makeatletter
\providecommand \@ifxundefined [1]{%
 \@ifx{#1\undefined}
}%
\providecommand \@ifnum [1]{%
 \ifnum #1\expandafter \@firstoftwo
 \else \expandafter \@secondoftwo
 \fi
}%
\providecommand \@ifx [1]{%
 \ifx #1\expandafter \@firstoftwo
 \else \expandafter \@secondoftwo
 \fi
}%
\providecommand \natexlab [1]{#1}%
\providecommand \enquote  [1]{``#1''}%
\providecommand \bibnamefont  [1]{#1}%
\providecommand \bibfnamefont [1]{#1}%
\providecommand \citenamefont [1]{#1}%
\providecommand \href@noop [0]{\@secondoftwo}%
\providecommand \href [0]{\begingroup \@sanitize@url \@href}%
\providecommand \@href[1]{\@@startlink{#1}\@@href}%
\providecommand \@@href[1]{\endgroup#1\@@endlink}%
\providecommand \@sanitize@url [0]{\catcode `\\12\catcode `\$12\catcode
  `\&12\catcode `\#12\catcode `\^12\catcode `\_12\catcode `\%12\relax}%
\providecommand \@@startlink[1]{}%
\providecommand \@@endlink[0]{}%
\providecommand \url  [0]{\begingroup\@sanitize@url \@url }%
\providecommand \@url [1]{\endgroup\@href {#1}{\urlprefix }}%
\providecommand \urlprefix  [0]{URL }%
\providecommand \Eprint [0]{\href }%
\providecommand \doibase [0]{http://dx.doi.org/}%
\providecommand \selectlanguage [0]{\@gobble}%
\providecommand \bibinfo  [0]{\@secondoftwo}%
\providecommand \bibfield  [0]{\@secondoftwo}%
\providecommand \translation [1]{[#1]}%
\providecommand \BibitemOpen [0]{}%
\providecommand \bibitemStop [0]{}%
\providecommand \bibitemNoStop [0]{.\EOS\space}%
\providecommand \EOS [0]{\spacefactor3000\relax}%
\providecommand \BibitemShut  [1]{\csname bibitem#1\endcsname}%
\let\auto@bib@innerbib\@empty
\bibitem [{\citenamefont {{Albrecht}}\ \emph {et~al.}(2006)\citenamefont
  {{Albrecht}}, \citenamefont {{Bernstein}}, \citenamefont {{Cahn}},
  \citenamefont {{Freedman}}, \citenamefont {{Hewitt}}, \citenamefont {{Hu}},
  \citenamefont {{Huth}}, \citenamefont {{Kamionkowski}}, \citenamefont
  {{Kolb}}, \citenamefont {{Knox}}, \citenamefont {{Mather}}, \citenamefont
  {{Staggs}},\ and\ \citenamefont {{Suntzeff}}}]{2006astro.ph..9591A}%
  \BibitemOpen
  \bibfield  {author} {\bibinfo {author} {\bibfnamefont {A.}~\bibnamefont
  {{Albrecht}}}, \bibinfo {author} {\bibfnamefont {G.}~\bibnamefont
  {{Bernstein}}}, \bibinfo {author} {\bibfnamefont {R.}~\bibnamefont {{Cahn}}},
  \bibinfo {author} {\bibfnamefont {W.~L.}\ \bibnamefont {{Freedman}}},
  \bibinfo {author} {\bibfnamefont {J.}~\bibnamefont {{Hewitt}}}, \bibinfo
  {author} {\bibfnamefont {W.}~\bibnamefont {{Hu}}}, \bibinfo {author}
  {\bibfnamefont {J.}~\bibnamefont {{Huth}}}, \bibinfo {author} {\bibfnamefont
  {M.}~\bibnamefont {{Kamionkowski}}}, \bibinfo {author} {\bibfnamefont
  {E.~W.}\ \bibnamefont {{Kolb}}}, \bibinfo {author} {\bibfnamefont
  {L.}~\bibnamefont {{Knox}}}, \bibinfo {author} {\bibfnamefont {J.~C.}\
  \bibnamefont {{Mather}}}, \bibinfo {author} {\bibfnamefont {S.}~\bibnamefont
  {{Staggs}}}, \ and\ \bibinfo {author} {\bibfnamefont {N.~B.}\ \bibnamefont
  {{Suntzeff}}},\ }\href@noop {} {\bibfield  {journal} {\bibinfo  {journal}
  {ArXiv Astrophysics e-prints}\ } (\bibinfo {year} {2006})},\ \Eprint
  {http://arxiv.org/abs/astro-ph/0609591} {astro-ph/0609591} \BibitemShut
  {NoStop}%
\bibitem [{\citenamefont {Bassett}\ \emph {et~al.}(2011)\citenamefont
  {Bassett}, \citenamefont {Fantaye}, \citenamefont {Hlozek},\ and\
  \citenamefont {Kotze}}]{Bassett:2009uv}%
  \BibitemOpen
  \bibfield  {author} {\bibinfo {author} {\bibfnamefont {B.~A.}\ \bibnamefont
  {Bassett}}, \bibinfo {author} {\bibfnamefont {Y.}~\bibnamefont {Fantaye}},
  \bibinfo {author} {\bibfnamefont {R.}~\bibnamefont {Hlozek}}, \ and\ \bibinfo
  {author} {\bibfnamefont {J.}~\bibnamefont {Kotze}},\ }\href {\doibase
  10.1142/S0218271811020548} {\bibfield  {journal} {\bibinfo  {journal}
  {Int.J.Mod.Phys.}\ }\textbf {\bibinfo {volume} {D20}},\ \bibinfo {pages}
  {2559} (\bibinfo {year} {2011})},\ \Eprint {http://arxiv.org/abs/0906.0993}
  {arXiv:0906.0993 [astro-ph.CO]} \BibitemShut {NoStop}%
\bibitem [{\citenamefont {Cooray}(1999)}]{Cooray:1999rv}%
  \BibitemOpen
  \bibfield  {author} {\bibinfo {author} {\bibfnamefont {A.~R.}\ \bibnamefont
  {Cooray}},\ }\href@noop {} {\bibfield  {journal} {\bibinfo  {journal}
  {Astron.Astrophys.}\ }\textbf {\bibinfo {volume} {348}},\ \bibinfo {pages}
  {31} (\bibinfo {year} {1999})},\ \Eprint
  {http://arxiv.org/abs/astro-ph/9904246} {arXiv:astro-ph/9904246 [astro-ph]}
  \BibitemShut {NoStop}%
\bibitem [{\citenamefont {Giannantonio}\ \emph {et~al.}(2012)\citenamefont
  {Giannantonio}, \citenamefont {Porciani}, \citenamefont {Carron},
  \citenamefont {Amara},\ and\ \citenamefont
  {Pillepich}}]{Giannantonio:2011ya}%
  \BibitemOpen
  \bibfield  {author} {\bibinfo {author} {\bibfnamefont {T.}~\bibnamefont
  {Giannantonio}}, \bibinfo {author} {\bibfnamefont {C.}~\bibnamefont
  {Porciani}}, \bibinfo {author} {\bibfnamefont {J.}~\bibnamefont {Carron}},
  \bibinfo {author} {\bibfnamefont {A.}~\bibnamefont {Amara}}, \ and\ \bibinfo
  {author} {\bibfnamefont {A.}~\bibnamefont {Pillepich}},\ }\href {\doibase
  10.1111/j.1365-2966.2012.20604.x} {\bibfield  {journal} {\bibinfo  {journal}
  {Mon.Not.Roy.Astron.Soc.}\ }\textbf {\bibinfo {volume} {422}},\ \bibinfo
  {pages} {2854} (\bibinfo {year} {2012})},\ \Eprint
  {http://arxiv.org/abs/1109.0958} {arXiv:1109.0958 [astro-ph.CO]} \BibitemShut
  {NoStop}%
\bibitem [{\citenamefont {Huterer}\ \emph {et~al.}(2006)\citenamefont
  {Huterer}, \citenamefont {Takada}, \citenamefont {Bernstein},\ and\
  \citenamefont {Jain}}]{Huterer:2005ez}%
  \BibitemOpen
  \bibfield  {author} {\bibinfo {author} {\bibfnamefont {D.}~\bibnamefont
  {Huterer}}, \bibinfo {author} {\bibfnamefont {M.}~\bibnamefont {Takada}},
  \bibinfo {author} {\bibfnamefont {G.}~\bibnamefont {Bernstein}}, \ and\
  \bibinfo {author} {\bibfnamefont {B.}~\bibnamefont {Jain}},\ }\href {\doibase
  10.1111/j.1365-2966.2005.09782.x} {\bibfield  {journal} {\bibinfo  {journal}
  {Mon.Not.Roy.Astron.Soc.}\ }\textbf {\bibinfo {volume} {366}},\ \bibinfo
  {pages} {101} (\bibinfo {year} {2006})},\ \Eprint
  {http://arxiv.org/abs/astro-ph/0506030} {arXiv:astro-ph/0506030 [astro-ph]}
  \BibitemShut {NoStop}%
\bibitem [{\citenamefont {{Hamann}}\ \emph {et~al.}(2012)\citenamefont
  {{Hamann}}, \citenamefont {{Hannestad}},\ and\ \citenamefont
  {{Wong}}}]{2012JCAP...11..052H}%
  \BibitemOpen
  \bibfield  {author} {\bibinfo {author} {\bibfnamefont {J.}~\bibnamefont
  {{Hamann}}}, \bibinfo {author} {\bibfnamefont {S.}~\bibnamefont
  {{Hannestad}}}, \ and\ \bibinfo {author} {\bibfnamefont {Y.~Y.~Y.}\
  \bibnamefont {{Wong}}},\ }\href {\doibase 10.1088/1475-7516/2012/11/052}
  {\bibfield  {journal} {\bibinfo  {journal} {\jcap}\ }\textbf {\bibinfo
  {volume} {11}},\ \bibinfo {eid} {052} (\bibinfo {year} {2012})},\ \Eprint
  {http://arxiv.org/abs/1209.1043} {arXiv:1209.1043 [astro-ph.CO]} \BibitemShut
  {NoStop}%
\bibitem [{\citenamefont {{Joudaki}}\ and\ \citenamefont
  {{Kaplinghat}}(2012)}]{2012PhRvD..86b3526J}%
  \BibitemOpen
  \bibfield  {author} {\bibinfo {author} {\bibfnamefont {S.}~\bibnamefont
  {{Joudaki}}}\ and\ \bibinfo {author} {\bibfnamefont {M.}~\bibnamefont
  {{Kaplinghat}}},\ }\href {\doibase 10.1103/PhysRevD.86.023526} {\bibfield
  {journal} {\bibinfo  {journal} {\prd}\ }\textbf {\bibinfo {volume} {86}},\
  \bibinfo {eid} {023526} (\bibinfo {year} {2012})},\ \Eprint
  {http://arxiv.org/abs/1106.0299} {arXiv:1106.0299 [astro-ph.CO]} \BibitemShut
  {NoStop}%
\bibitem [{\citenamefont {{Kitching}}\ and\ \citenamefont
  {{Amara}}(2009)}]{Kitching:2009yr}%
  \BibitemOpen
  \bibfield  {author} {\bibinfo {author} {\bibfnamefont {T.~D.}\ \bibnamefont
  {{Kitching}}}\ and\ \bibinfo {author} {\bibfnamefont {A.}~\bibnamefont
  {{Amara}}},\ }\href {\doibase 10.1111/j.1365-2966.2009.15263.x} {\bibfield
  {journal} {\bibinfo  {journal} {\mnras}\ }\textbf {\bibinfo {volume} {398}},\
  \bibinfo {pages} {2134} (\bibinfo {year} {2009})},\ \Eprint
  {http://arxiv.org/abs/0905.3383} {arXiv:0905.3383 [astro-ph.CO]} \BibitemShut
  {NoStop}%
\bibitem [{\citenamefont {More}\ \emph {et~al.}(2013)\citenamefont {More},
  \citenamefont {van~den Bosch}, \citenamefont {Cacciato}, \citenamefont
  {More}, \citenamefont {Mo} \emph {et~al.}}]{More:2012xa}%
  \BibitemOpen
  \bibfield  {author} {\bibinfo {author} {\bibfnamefont {S.}~\bibnamefont
  {More}}, \bibinfo {author} {\bibfnamefont {F.}~\bibnamefont {van~den Bosch}},
  \bibinfo {author} {\bibfnamefont {M.}~\bibnamefont {Cacciato}}, \bibinfo
  {author} {\bibfnamefont {A.}~\bibnamefont {More}}, \bibinfo {author}
  {\bibfnamefont {H.}~\bibnamefont {Mo}},  \emph {et~al.},\ }\href {\doibase
  10.1093/mnras/sts697} {\bibfield  {journal} {\bibinfo  {journal}
  {Mon.Not.Roy.Astron.Soc.}\ }\textbf {\bibinfo {volume} {430}},\ \bibinfo
  {pages} {747} (\bibinfo {year} {2013})},\ \Eprint
  {http://arxiv.org/abs/1207.0004} {arXiv:1207.0004 [astro-ph.CO]} \BibitemShut
  {NoStop}%
\bibitem [{\citenamefont {{Weinberg}}\ \emph {et~al.}(2012)\citenamefont
  {{Weinberg}}, \citenamefont {{Mortonson}}, \citenamefont {{Eisenstein}},
  \citenamefont {{Hirata}}, \citenamefont {{Riess}},\ and\ \citenamefont
  {{Rozo}}}]{2012arXiv1201.2434W}%
  \BibitemOpen
  \bibfield  {author} {\bibinfo {author} {\bibfnamefont {D.~H.}\ \bibnamefont
  {{Weinberg}}}, \bibinfo {author} {\bibfnamefont {M.~J.}\ \bibnamefont
  {{Mortonson}}}, \bibinfo {author} {\bibfnamefont {D.~J.}\ \bibnamefont
  {{Eisenstein}}}, \bibinfo {author} {\bibfnamefont {C.}~\bibnamefont
  {{Hirata}}}, \bibinfo {author} {\bibfnamefont {A.~G.}\ \bibnamefont
  {{Riess}}}, \ and\ \bibinfo {author} {\bibfnamefont {E.}~\bibnamefont
  {{Rozo}}},\ }\href@noop {} {\bibfield  {journal} {\bibinfo  {journal} {ArXiv
  e-prints}\ } (\bibinfo {year} {2012})},\ \Eprint
  {http://arxiv.org/abs/1201.2434} {arXiv:1201.2434 [astro-ph.CO]} \BibitemShut
  {NoStop}%
\bibitem [{\citenamefont {Wolz}\ \emph {et~al.}(2012)\citenamefont {Wolz},
  \citenamefont {Kilbinger}, \citenamefont {Weller},\ and\ \citenamefont
  {Giannantonio}}]{Wolz:2012sr}%
  \BibitemOpen
  \bibfield  {author} {\bibinfo {author} {\bibfnamefont {L.}~\bibnamefont
  {Wolz}}, \bibinfo {author} {\bibfnamefont {M.}~\bibnamefont {Kilbinger}},
  \bibinfo {author} {\bibfnamefont {J.}~\bibnamefont {Weller}}, \ and\ \bibinfo
  {author} {\bibfnamefont {T.}~\bibnamefont {Giannantonio}},\ }\href {\doibase
  10.1088/1475-7516/2012/09/009} {\bibfield  {journal} {\bibinfo  {journal}
  {JCAP}\ }\textbf {\bibinfo {volume} {1209}},\ \bibinfo {pages} {009}
  (\bibinfo {year} {2012})},\ \Eprint {http://arxiv.org/abs/1205.3984}
  {arXiv:1205.3984 [astro-ph.CO]} \BibitemShut {NoStop}%
\bibitem [{\citenamefont {Khedekar}\ and\ \citenamefont
  {Majumdar}(2013)}]{Khedekar:2012sh}%
  \BibitemOpen
  \bibfield  {author} {\bibinfo {author} {\bibfnamefont {S.}~\bibnamefont
  {Khedekar}}\ and\ \bibinfo {author} {\bibfnamefont {S.}~\bibnamefont
  {Majumdar}},\ }\href {\doibase 10.1088/1475-7516/2013/02/030} {\bibfield
  {journal} {\bibinfo  {journal} {JCAP}\ }\textbf {\bibinfo {volume} {1302}},\
  \bibinfo {pages} {030} (\bibinfo {year} {2013})},\ \Eprint
  {http://arxiv.org/abs/1210.5586} {arXiv:1210.5586 [astro-ph.CO]} \BibitemShut
  {NoStop}%
\bibitem [{\citenamefont {{Basse}}\ \emph {et~al.}(2013)\citenamefont
  {{Basse}}, \citenamefont {{Eggers Bjaelde}}, \citenamefont {{Hamann}},
  \citenamefont {{Hannestad}},\ and\ \citenamefont
  {{Wong}}}]{2013arXiv1304.2321B}%
  \BibitemOpen
  \bibfield  {author} {\bibinfo {author} {\bibfnamefont {T.}~\bibnamefont
  {{Basse}}}, \bibinfo {author} {\bibfnamefont {O.}~\bibnamefont {{Eggers
  Bjaelde}}}, \bibinfo {author} {\bibfnamefont {J.}~\bibnamefont {{Hamann}}},
  \bibinfo {author} {\bibfnamefont {S.}~\bibnamefont {{Hannestad}}}, \ and\
  \bibinfo {author} {\bibfnamefont {Y.~Y.~Y.}\ \bibnamefont {{Wong}}},\
  }\href@noop {} {\bibfield  {journal} {\bibinfo  {journal} {ArXiv e-prints}\ }
  (\bibinfo {year} {2013})},\ \Eprint {http://arxiv.org/abs/1304.2321}
  {arXiv:1304.2321} \BibitemShut {NoStop}%
\bibitem [{\citenamefont {{Costanzi Alunno Cerbolini}}\ \emph
  {et~al.}(2013)\citenamefont {{Costanzi Alunno Cerbolini}}, \citenamefont
  {{Sartoris}}, \citenamefont {{Xia}}, \citenamefont {{Biviano}}, \citenamefont
  {{Borgani}},\ and\ \citenamefont {{Viel}}}]{2013JCAP...06..020C}%
  \BibitemOpen
  \bibfield  {author} {\bibinfo {author} {\bibfnamefont {M.}~\bibnamefont
  {{Costanzi Alunno Cerbolini}}}, \bibinfo {author} {\bibfnamefont
  {B.}~\bibnamefont {{Sartoris}}}, \bibinfo {author} {\bibfnamefont {J.-Q.}\
  \bibnamefont {{Xia}}}, \bibinfo {author} {\bibfnamefont {A.}~\bibnamefont
  {{Biviano}}}, \bibinfo {author} {\bibfnamefont {S.}~\bibnamefont
  {{Borgani}}}, \ and\ \bibinfo {author} {\bibfnamefont {M.}~\bibnamefont
  {{Viel}}},\ }\href {\doibase 10.1088/1475-7516/2013/06/020} {\bibfield
  {journal} {\bibinfo  {journal} {\jcap}\ }\textbf {\bibinfo {volume} {6}},\
  \bibinfo {eid} {020} (\bibinfo {year} {2013})},\ \Eprint
  {http://arxiv.org/abs/1303.4550} {arXiv:1303.4550 [astro-ph.CO]} \BibitemShut
  {NoStop}%
\bibitem [{\citenamefont {Pullen}\ and\ \citenamefont
  {Hirata}(2012)}]{Pullen:2012rd}%
  \BibitemOpen
  \bibfield  {author} {\bibinfo {author} {\bibfnamefont {A.~R.}\ \bibnamefont
  {Pullen}}\ and\ \bibinfo {author} {\bibfnamefont {C.~M.}\ \bibnamefont
  {Hirata}},\ }\href@noop {} {\bibfield  {journal} {\bibinfo  {journal} {ArXiv
  e-prints}\ } (\bibinfo {year} {2012})},\ \Eprint
  {http://arxiv.org/abs/1212.4500} {arXiv:1212.4500 [astro-ph.CO]} \BibitemShut
  {NoStop}%
\bibitem [{\citenamefont {Ross}\ \emph {et~al.}(2012)\citenamefont {Ross} \emph
  {et~al.}}]{Ross:2012qm}%
  \BibitemOpen
  \bibfield  {author} {\bibinfo {author} {\bibfnamefont {A.~J.}\ \bibnamefont
  {Ross}} \emph {et~al.} (\bibinfo {collaboration} {BOSS Collaboration}),\
  }\href {\doibase 10.1111/j.1365-2966.2012.21235.x} {\bibfield  {journal}
  {\bibinfo  {journal} {Mon.Not.Roy.Astron.Soc.}\ }\textbf {\bibinfo {volume}
  {424}},\ \bibinfo {pages} {564} (\bibinfo {year} {2012})},\ \Eprint
  {http://arxiv.org/abs/1203.6499} {arXiv:1203.6499 [astro-ph.CO]} \BibitemShut
  {NoStop}%
\bibitem [{\citenamefont {{Hu}}(2005)}]{2005ASPC..339..215H}%
  \BibitemOpen
  \bibfield  {author} {\bibinfo {author} {\bibfnamefont {W.}~\bibnamefont
  {{Hu}}},\ }in\ \href@noop {} {\emph {\bibinfo {booktitle} {ASP Conf. Ser.
  339: Observing Dark Energy}}}\ (\bibinfo {year} {2005})\ pp.\ \bibinfo
  {pages} {215--+}\BibitemShut {NoStop}%
\bibitem [{\citenamefont {{Planck Collaboration}}\ \emph
  {et~al.}(2013)\citenamefont {{Planck Collaboration}}, \citenamefont {{Ade}},
  \citenamefont {{Aghanim}}, \citenamefont {{Armitage-Caplan}}, \citenamefont
  {{Arnaud}}, \citenamefont {{Ashdown}}, \citenamefont {{Atrio-Barandela}},
  \citenamefont {{Aumont}}, \citenamefont {{Baccigalupi}}, \citenamefont
  {{Banday}},\ and\ \citenamefont {et~al.}}]{2013arXiv1303.5076P}%
  \BibitemOpen
  \bibfield  {author} {\bibinfo {author} {\bibnamefont {{Planck
  Collaboration}}}, \bibinfo {author} {\bibfnamefont {P.~A.~R.}\ \bibnamefont
  {{Ade}}}, \bibinfo {author} {\bibfnamefont {N.}~\bibnamefont {{Aghanim}}},
  \bibinfo {author} {\bibfnamefont {C.}~\bibnamefont {{Armitage-Caplan}}},
  \bibinfo {author} {\bibfnamefont {M.}~\bibnamefont {{Arnaud}}}, \bibinfo
  {author} {\bibfnamefont {M.}~\bibnamefont {{Ashdown}}}, \bibinfo {author}
  {\bibfnamefont {F.}~\bibnamefont {{Atrio-Barandela}}}, \bibinfo {author}
  {\bibfnamefont {J.}~\bibnamefont {{Aumont}}}, \bibinfo {author}
  {\bibfnamefont {C.}~\bibnamefont {{Baccigalupi}}}, \bibinfo {author}
  {\bibfnamefont {A.~J.}\ \bibnamefont {{Banday}}}, \ and\ \bibinfo {author}
  {\bibnamefont {et~al.}},\ }\href@noop {} {\bibfield  {journal} {\bibinfo
  {journal} {ArXiv e-prints}\ } (\bibinfo {year} {2013})},\ \Eprint
  {http://arxiv.org/abs/1303.5076} {arXiv:1303.5076 [astro-ph.CO]} \BibitemShut
  {NoStop}%
\bibitem [{\citenamefont {{Armitage-Caplan}}\ \emph {et~al.}(2011)\citenamefont
  {{Armitage-Caplan}}, \citenamefont {{Dunkley}}, \citenamefont {{Eriksen}},\
  and\ \citenamefont {{Dickinson}}}]{2011MNRAS.418.1498A}%
  \BibitemOpen
  \bibfield  {author} {\bibinfo {author} {\bibfnamefont {C.}~\bibnamefont
  {{Armitage-Caplan}}}, \bibinfo {author} {\bibfnamefont {J.}~\bibnamefont
  {{Dunkley}}}, \bibinfo {author} {\bibfnamefont {H.~K.}\ \bibnamefont
  {{Eriksen}}}, \ and\ \bibinfo {author} {\bibfnamefont {C.}~\bibnamefont
  {{Dickinson}}},\ }\href {\doibase 10.1111/j.1365-2966.2011.19307.x}
  {\bibfield  {journal} {\bibinfo  {journal} {\mnras}\ }\textbf {\bibinfo
  {volume} {418}},\ \bibinfo {pages} {1498} (\bibinfo {year} {2011})},\ \Eprint
  {http://arxiv.org/abs/1103.2554} {arXiv:1103.2554 [astro-ph.CO]} \BibitemShut
  {NoStop}%
\bibitem [{\citenamefont {{Albrecht}}\ \emph {et~al.}(2009)\citenamefont
  {{Albrecht}}, \citenamefont {{Amendola}}, \citenamefont {{Bernstein}},
  \citenamefont {{Clowe}}, \citenamefont {{Eisenstein}}, \citenamefont
  {{Guzzo}}, \citenamefont {{Hirata}}, \citenamefont {{Huterer}}, \citenamefont
  {{Kirshner}}, \citenamefont {{Kolb}},\ and\ \citenamefont
  {{Nichol}}}]{2009arXiv0901.0721A}%
  \BibitemOpen
  \bibfield  {author} {\bibinfo {author} {\bibfnamefont {A.}~\bibnamefont
  {{Albrecht}}}, \bibinfo {author} {\bibfnamefont {L.}~\bibnamefont
  {{Amendola}}}, \bibinfo {author} {\bibfnamefont {G.}~\bibnamefont
  {{Bernstein}}}, \bibinfo {author} {\bibfnamefont {D.}~\bibnamefont
  {{Clowe}}}, \bibinfo {author} {\bibfnamefont {D.}~\bibnamefont
  {{Eisenstein}}}, \bibinfo {author} {\bibfnamefont {L.}~\bibnamefont
  {{Guzzo}}}, \bibinfo {author} {\bibfnamefont {C.}~\bibnamefont {{Hirata}}},
  \bibinfo {author} {\bibfnamefont {D.}~\bibnamefont {{Huterer}}}, \bibinfo
  {author} {\bibfnamefont {R.}~\bibnamefont {{Kirshner}}}, \bibinfo {author}
  {\bibfnamefont {E.}~\bibnamefont {{Kolb}}}, \ and\ \bibinfo {author}
  {\bibfnamefont {R.}~\bibnamefont {{Nichol}}},\ }\href@noop {} {\bibfield
  {journal} {\bibinfo  {journal} {ArXiv e-prints}\ } (\bibinfo {year}
  {2009})},\ \Eprint {http://arxiv.org/abs/0901.0721} {arXiv:0901.0721}
  \BibitemShut {NoStop}%
\bibitem [{\citenamefont {{Linder}}(2009)}]{2009PhRvD..79f3519L}%
  \BibitemOpen
  \bibfield  {author} {\bibinfo {author} {\bibfnamefont {E.~V.}\ \bibnamefont
  {{Linder}}},\ }\href {\doibase 10.1103/PhysRevD.79.063519} {\bibfield
  {journal} {\bibinfo  {journal} {\prd}\ }\textbf {\bibinfo {volume} {79}},\
  \bibinfo {eid} {063519} (\bibinfo {year} {2009})},\ \Eprint
  {http://arxiv.org/abs/0901.0918} {arXiv:0901.0918 [astro-ph.CO]} \BibitemShut
  {NoStop}%
\bibitem [{\citenamefont {{Eifler}}\ \emph {et~al.}(2009)\citenamefont
  {{Eifler}}, \citenamefont {{Schneider}},\ and\ \citenamefont
  {{Hartlap}}}]{2009A&A...502..721E}%
  \BibitemOpen
  \bibfield  {author} {\bibinfo {author} {\bibfnamefont {T.}~\bibnamefont
  {{Eifler}}}, \bibinfo {author} {\bibfnamefont {P.}~\bibnamefont
  {{Schneider}}}, \ and\ \bibinfo {author} {\bibfnamefont {J.}~\bibnamefont
  {{Hartlap}}},\ }\href {\doibase 10.1051/0004-6361/200811276} {\bibfield
  {journal} {\bibinfo  {journal} {\aap}\ }\textbf {\bibinfo {volume} {502}},\
  \bibinfo {pages} {721} (\bibinfo {year} {2009})},\ \Eprint
  {http://arxiv.org/abs/0810.4254} {arXiv:0810.4254} \BibitemShut {NoStop}%
\bibitem [{\citenamefont {{Kilbinger}}\ and\ \citenamefont
  {{Munshi}}(2006)}]{2006MNRAS.366..983K}%
  \BibitemOpen
  \bibfield  {author} {\bibinfo {author} {\bibfnamefont {M.}~\bibnamefont
  {{Kilbinger}}}\ and\ \bibinfo {author} {\bibfnamefont {D.}~\bibnamefont
  {{Munshi}}},\ }\href {\doibase 10.1111/j.1365-2966.2005.09857.x} {\bibfield
  {journal} {\bibinfo  {journal} {\mnras}\ }\textbf {\bibinfo {volume} {366}},\
  \bibinfo {pages} {983} (\bibinfo {year} {2006})},\ \Eprint
  {http://arxiv.org/abs/astro-ph/0509548} {astro-ph/0509548} \BibitemShut
  {NoStop}%
\bibitem [{\citenamefont {{Seljak}}\ and\ \citenamefont
  {{Zaldarriaga}}(1996)}]{1996ApJ...469..437S}%
  \BibitemOpen
  \bibfield  {author} {\bibinfo {author} {\bibfnamefont {U.}~\bibnamefont
  {{Seljak}}}\ and\ \bibinfo {author} {\bibfnamefont {M.}~\bibnamefont
  {{Zaldarriaga}}},\ }\href@noop {} {\bibfield  {journal} {\bibinfo  {journal}
  {\apj}\ }\textbf {\bibinfo {volume} {469}},\ \bibinfo {pages} {437} (\bibinfo
  {year} {1996})}\BibitemShut {NoStop}%
\bibitem [{\citenamefont {{Lewis}}\ \emph {et~al.}(2000)\citenamefont
  {{Lewis}}, \citenamefont {{Challinor}},\ and\ \citenamefont
  {{Lasenby}}}]{2000ApJ...538..473L}%
  \BibitemOpen
  \bibfield  {author} {\bibinfo {author} {\bibfnamefont {A.}~\bibnamefont
  {{Lewis}}}, \bibinfo {author} {\bibfnamefont {A.}~\bibnamefont
  {{Challinor}}}, \ and\ \bibinfo {author} {\bibfnamefont {A.}~\bibnamefont
  {{Lasenby}}},\ }\href@noop {} {\bibfield  {journal} {\bibinfo  {journal}
  {\apj}\ }\textbf {\bibinfo {volume} {538}},\ \bibinfo {pages} {473} (\bibinfo
  {year} {2000})}\BibitemShut {NoStop}%
\bibitem [{\citenamefont {{Hu}}(2002)}]{2002PhRvD..65b3003H}%
  \BibitemOpen
  \bibfield  {author} {\bibinfo {author} {\bibfnamefont {W.}~\bibnamefont
  {{Hu}}},\ }\href {\doibase 10.1103/PhysRevD.65.023003} {\bibfield  {journal}
  {\bibinfo  {journal} {\prd}\ }\textbf {\bibinfo {volume} {65}},\ \bibinfo
  {eid} {023003} (\bibinfo {year} {2002})},\ \Eprint
  {http://arxiv.org/abs/arXiv:astro-ph/0108090} {arXiv:astro-ph/0108090}
  \BibitemShut {NoStop}%
\bibitem [{\citenamefont {{The Planck
  Collaboration}}(2006)}]{2006astro.ph..4069T}%
  \BibitemOpen
  \bibfield  {author} {\bibinfo {author} {\bibnamefont {{The Planck
  Collaboration}}},\ }\href@noop {} {\bibfield  {journal} {\bibinfo  {journal}
  {ArXiv Astrophysics e-prints}\ } (\bibinfo {year} {2006})},\ \Eprint
  {http://arxiv.org/abs/arXiv:astro-ph/0604069} {arXiv:astro-ph/0604069}
  \BibitemShut {NoStop}%
\bibitem [{\citenamefont {{Bond}}\ \emph {et~al.}(2004)\citenamefont {{Bond}},
  \citenamefont {{Contaldi}}, \citenamefont {{Lewis}},\ and\ \citenamefont
  {{Pogosyan}}}]{2004IJTP...43..599B}%
  \BibitemOpen
  \bibfield  {author} {\bibinfo {author} {\bibfnamefont {J.~R.}\ \bibnamefont
  {{Bond}}}, \bibinfo {author} {\bibfnamefont {C.}~\bibnamefont {{Contaldi}}},
  \bibinfo {author} {\bibfnamefont {A.}~\bibnamefont {{Lewis}}}, \ and\
  \bibinfo {author} {\bibfnamefont {D.}~\bibnamefont {{Pogosyan}}},\ }\href
  {\doibase 10.1023/B:IJTP.0000048167.24074.ef} {\bibfield  {journal} {\bibinfo
   {journal} {International Journal of Theoretical Physics}\ }\textbf {\bibinfo
  {volume} {43}},\ \bibinfo {pages} {599} (\bibinfo {year} {2004})},\ \Eprint
  {http://arxiv.org/abs/arXiv:astro-ph/0406195} {arXiv:astro-ph/0406195}
  \BibitemShut {NoStop}%
\bibitem [{\citenamefont {{Howlett}}\ \emph {et~al.}(2012)\citenamefont
  {{Howlett}}, \citenamefont {{Lewis}}, \citenamefont {{Hall}},\ and\
  \citenamefont {{Challinor}}}]{2012JCAP...04..027H}%
  \BibitemOpen
  \bibfield  {author} {\bibinfo {author} {\bibfnamefont {C.}~\bibnamefont
  {{Howlett}}}, \bibinfo {author} {\bibfnamefont {A.}~\bibnamefont {{Lewis}}},
  \bibinfo {author} {\bibfnamefont {A.}~\bibnamefont {{Hall}}}, \ and\ \bibinfo
  {author} {\bibfnamefont {A.}~\bibnamefont {{Challinor}}},\ }\href {\doibase
  10.1088/1475-7516/2012/04/027} {\bibfield  {journal} {\bibinfo  {journal}
  {\jcap}\ }\textbf {\bibinfo {volume} {4}},\ \bibinfo {eid} {027} (\bibinfo
  {year} {2012})},\ \Eprint {http://arxiv.org/abs/1201.3654} {arXiv:1201.3654
  [astro-ph.CO]} \BibitemShut {NoStop}%
\bibitem [{\citenamefont {{Galli}}\ \emph {et~al.}(2010)\citenamefont
  {{Galli}}, \citenamefont {{Martinelli}}, \citenamefont {{Melchiorri}},
  \citenamefont {{Pagano}}, \citenamefont {{Sherwin}},\ and\ \citenamefont
  {{Spergel}}}]{2010PhRvD..82l3504G}%
  \BibitemOpen
  \bibfield  {author} {\bibinfo {author} {\bibfnamefont {S.}~\bibnamefont
  {{Galli}}}, \bibinfo {author} {\bibfnamefont {M.}~\bibnamefont
  {{Martinelli}}}, \bibinfo {author} {\bibfnamefont {A.}~\bibnamefont
  {{Melchiorri}}}, \bibinfo {author} {\bibfnamefont {L.}~\bibnamefont
  {{Pagano}}}, \bibinfo {author} {\bibfnamefont {B.~D.}\ \bibnamefont
  {{Sherwin}}}, \ and\ \bibinfo {author} {\bibfnamefont {D.~N.}\ \bibnamefont
  {{Spergel}}},\ }\href {\doibase 10.1103/PhysRevD.82.123504} {\bibfield
  {journal} {\bibinfo  {journal} {\prd}\ }\textbf {\bibinfo {volume} {82}},\
  \bibinfo {eid} {123504} (\bibinfo {year} {2010})},\ \Eprint
  {http://arxiv.org/abs/1005.3808} {arXiv:1005.3808 [astro-ph.CO]} \BibitemShut
  {NoStop}%
\bibitem [{\citenamefont {{Melchiorri}}\ \emph {et~al.}(2010)\citenamefont
  {{Melchiorri}}, \citenamefont {{Galli}}, \citenamefont {{Martinelli}},\ and\
  \citenamefont {{Pagano}}}]{2010JPhCS.259a2004M}%
  \BibitemOpen
  \bibfield  {author} {\bibinfo {author} {\bibfnamefont {A.}~\bibnamefont
  {{Melchiorri}}}, \bibinfo {author} {\bibfnamefont {S.}~\bibnamefont
  {{Galli}}}, \bibinfo {author} {\bibfnamefont {M.}~\bibnamefont
  {{Martinelli}}}, \ and\ \bibinfo {author} {\bibfnamefont {L.}~\bibnamefont
  {{Pagano}}},\ }\href {\doibase 10.1088/1742-6596/259/1/012004} {\bibfield
  {journal} {\bibinfo  {journal} {Journal of Physics Conference Series}\
  }\textbf {\bibinfo {volume} {259}},\ \bibinfo {eid} {012004} (\bibinfo {year}
  {2010})}\BibitemShut {NoStop}%
\bibitem [{\citenamefont {{Oyama}}\ \emph {et~al.}(2013)\citenamefont
  {{Oyama}}, \citenamefont {{Shimizu}},\ and\ \citenamefont
  {{Kohri}}}]{2013PhLB..718.1186O}%
  \BibitemOpen
  \bibfield  {author} {\bibinfo {author} {\bibfnamefont {Y.}~\bibnamefont
  {{Oyama}}}, \bibinfo {author} {\bibfnamefont {A.}~\bibnamefont {{Shimizu}}},
  \ and\ \bibinfo {author} {\bibfnamefont {K.}~\bibnamefont {{Kohri}}},\ }\href
  {\doibase 10.1016/j.physletb.2012.12.053} {\bibfield  {journal} {\bibinfo
  {journal} {Physics Letters B}\ }\textbf {\bibinfo {volume} {718}},\ \bibinfo
  {pages} {1186} (\bibinfo {year} {2013})},\ \Eprint
  {http://arxiv.org/abs/1205.5223} {arXiv:1205.5223 [astro-ph.CO]} \BibitemShut
  {NoStop}%
\bibitem [{\citenamefont {{Hall}}\ and\ \citenamefont
  {{Challinor}}(2012)}]{2012MNRAS.425.1170H}%
  \BibitemOpen
  \bibfield  {author} {\bibinfo {author} {\bibfnamefont {A.~C.}\ \bibnamefont
  {{Hall}}}\ and\ \bibinfo {author} {\bibfnamefont {A.}~\bibnamefont
  {{Challinor}}},\ }\href {\doibase 10.1111/j.1365-2966.2012.21493.x}
  {\bibfield  {journal} {\bibinfo  {journal} {\mnras}\ }\textbf {\bibinfo
  {volume} {425}},\ \bibinfo {pages} {1170} (\bibinfo {year} {2012})},\ \Eprint
  {http://arxiv.org/abs/1205.6172} {arXiv:1205.6172 [astro-ph.CO]} \BibitemShut
  {NoStop}%
\bibitem [{\citenamefont {{de Putter}}\ \emph {et~al.}(2009)\citenamefont {{de
  Putter}}, \citenamefont {{Zahn}},\ and\ \citenamefont
  {{Linder}}}]{2009PhRvD..79f5033D}%
  \BibitemOpen
  \bibfield  {author} {\bibinfo {author} {\bibfnamefont {R.}~\bibnamefont {{de
  Putter}}}, \bibinfo {author} {\bibfnamefont {O.}~\bibnamefont {{Zahn}}}, \
  and\ \bibinfo {author} {\bibfnamefont {E.~V.}\ \bibnamefont {{Linder}}},\
  }\href {\doibase 10.1103/PhysRevD.79.065033} {\bibfield  {journal} {\bibinfo
  {journal} {\prd}\ }\textbf {\bibinfo {volume} {79}},\ \bibinfo {eid} {065033}
  (\bibinfo {year} {2009})},\ \Eprint {http://arxiv.org/abs/0901.0916}
  {arXiv:0901.0916 [astro-ph.CO]} \BibitemShut {NoStop}%
\bibitem [{\citenamefont {{Planck collaboration}}\ \emph
  {et~al.}(2013)\citenamefont {{Planck collaboration}}, \citenamefont {{Ade}},
  \citenamefont {{Aghanim}}, \citenamefont {{Armitage-Caplan}}, \citenamefont
  {{Arnaud}}, \citenamefont {{Ashdown}}, \citenamefont {{Atrio-Barandela}},
  \citenamefont {{Aumont}}, \citenamefont {{Baccigalupi}}, \citenamefont
  {{Banday}},\ and\ \citenamefont {et~al.}}]{2013arXiv1303.5075P}%
  \BibitemOpen
  \bibfield  {author} {\bibinfo {author} {\bibnamefont {{Planck
  collaboration}}}, \bibinfo {author} {\bibfnamefont {P.~A.~R.}\ \bibnamefont
  {{Ade}}}, \bibinfo {author} {\bibfnamefont {N.}~\bibnamefont {{Aghanim}}},
  \bibinfo {author} {\bibfnamefont {C.}~\bibnamefont {{Armitage-Caplan}}},
  \bibinfo {author} {\bibfnamefont {M.}~\bibnamefont {{Arnaud}}}, \bibinfo
  {author} {\bibfnamefont {M.}~\bibnamefont {{Ashdown}}}, \bibinfo {author}
  {\bibfnamefont {F.}~\bibnamefont {{Atrio-Barandela}}}, \bibinfo {author}
  {\bibfnamefont {J.}~\bibnamefont {{Aumont}}}, \bibinfo {author}
  {\bibfnamefont {C.}~\bibnamefont {{Baccigalupi}}}, \bibinfo {author}
  {\bibfnamefont {A.~J.}\ \bibnamefont {{Banday}}}, \ and\ \bibinfo {author}
  {\bibnamefont {et~al.}},\ }\href@noop {} {\bibfield  {journal} {\bibinfo
  {journal} {ArXiv e-prints}\ } (\bibinfo {year} {2013})},\ \Eprint
  {http://arxiv.org/abs/1303.5075} {arXiv:1303.5075 [astro-ph.CO]} \BibitemShut
  {NoStop}%
\bibitem [{\citenamefont {{McDonald}}\ and\ \citenamefont
  {{Roy}}(2009)}]{2009JCAP...08..020M}%
  \BibitemOpen
  \bibfield  {author} {\bibinfo {author} {\bibfnamefont {P.}~\bibnamefont
  {{McDonald}}}\ and\ \bibinfo {author} {\bibfnamefont {A.}~\bibnamefont
  {{Roy}}},\ }\href {\doibase 10.1088/1475-7516/2009/08/020} {\bibfield
  {journal} {\bibinfo  {journal} {\jcap}\ }\textbf {\bibinfo {volume} {8}},\
  \bibinfo {eid} {020} (\bibinfo {year} {2009})},\ \Eprint
  {http://arxiv.org/abs/0902.0991} {arXiv:0902.0991 [astro-ph.CO]} \BibitemShut
  {NoStop}%
\bibitem [{\citenamefont {{Baldauf}}\ \emph {et~al.}(2012)\citenamefont
  {{Baldauf}}, \citenamefont {{Seljak}}, \citenamefont {{Desjacques}},\ and\
  \citenamefont {{McDonald}}}]{2012PhRvD..86h3540B}%
  \BibitemOpen
  \bibfield  {author} {\bibinfo {author} {\bibfnamefont {T.}~\bibnamefont
  {{Baldauf}}}, \bibinfo {author} {\bibfnamefont {U.}~\bibnamefont {{Seljak}}},
  \bibinfo {author} {\bibfnamefont {V.}~\bibnamefont {{Desjacques}}}, \ and\
  \bibinfo {author} {\bibfnamefont {P.}~\bibnamefont {{McDonald}}},\ }\href
  {\doibase 10.1103/PhysRevD.86.083540} {\bibfield  {journal} {\bibinfo
  {journal} {\prd}\ }\textbf {\bibinfo {volume} {86}},\ \bibinfo {eid} {083540}
  (\bibinfo {year} {2012})},\ \Eprint {http://arxiv.org/abs/1201.4827}
  {arXiv:1201.4827 [astro-ph.CO]} \BibitemShut {NoStop}%
\bibitem [{\citenamefont {{McDonald}}(2006)}]{2006PhRvD..74j3512M}%
  \BibitemOpen
  \bibfield  {author} {\bibinfo {author} {\bibfnamefont {P.}~\bibnamefont
  {{McDonald}}},\ }\href {\doibase 10.1103/PhysRevD.74.103512} {\bibfield
  {journal} {\bibinfo  {journal} {\prd}\ }\textbf {\bibinfo {volume} {74}},\
  \bibinfo {pages} {103512} (\bibinfo {year} {2006})}\BibitemShut {NoStop}%
\bibitem [{\citenamefont {{Desjacques}}(2013)}]{2013PhRvD..87d3505D}%
  \BibitemOpen
  \bibfield  {author} {\bibinfo {author} {\bibfnamefont {V.}~\bibnamefont
  {{Desjacques}}},\ }\href {\doibase 10.1103/PhysRevD.87.043505} {\bibfield
  {journal} {\bibinfo  {journal} {\prd}\ }\textbf {\bibinfo {volume} {87}},\
  \bibinfo {eid} {043505} (\bibinfo {year} {2013})},\ \Eprint
  {http://arxiv.org/abs/1211.4128} {arXiv:1211.4128 [astro-ph.CO]} \BibitemShut
  {NoStop}%
\bibitem [{\citenamefont {{Kaiser}}(1987)}]{KAISE87}%
  \BibitemOpen
  \bibfield  {author} {\bibinfo {author} {\bibfnamefont {N.}~\bibnamefont
  {{Kaiser}}},\ }\href@noop {} {\bibfield  {journal} {\bibinfo  {journal}
  {\mnras}\ }\textbf {\bibinfo {volume} {227}},\ \bibinfo {pages} {1} (\bibinfo
  {year} {1987})}\BibitemShut {NoStop}%
\bibitem [{\citenamefont {{Jeong}}\ and\ \citenamefont
  {{Komatsu}}(2009{\natexlab{a}})}]{2009ApJ...691..569J}%
  \BibitemOpen
  \bibfield  {author} {\bibinfo {author} {\bibfnamefont {D.}~\bibnamefont
  {{Jeong}}}\ and\ \bibinfo {author} {\bibfnamefont {E.}~\bibnamefont
  {{Komatsu}}},\ }\href {\doibase 10.1088/0004-637X/691/1/569} {\bibfield
  {journal} {\bibinfo  {journal} {\apj}\ }\textbf {\bibinfo {volume} {691}},\
  \bibinfo {pages} {569} (\bibinfo {year} {2009}{\natexlab{a}})},\ \Eprint
  {http://arxiv.org/abs/0805.2632} {arXiv:0805.2632} \BibitemShut {NoStop}%
\bibitem [{\citenamefont {{Seljak}}\ and\ \citenamefont
  {{McDonald}}(2011)}]{2011JCAP...11..039S}%
  \BibitemOpen
  \bibfield  {author} {\bibinfo {author} {\bibfnamefont {U.}~\bibnamefont
  {{Seljak}}}\ and\ \bibinfo {author} {\bibfnamefont {P.}~\bibnamefont
  {{McDonald}}},\ }\href {\doibase 10.1088/1475-7516/2011/11/039} {\bibfield
  {journal} {\bibinfo  {journal} {\jcap}\ }\textbf {\bibinfo {volume} {11}},\
  \bibinfo {eid} {039} (\bibinfo {year} {2011})},\ \Eprint
  {http://arxiv.org/abs/1109.1888} {arXiv:1109.1888 [astro-ph.CO]} \BibitemShut
  {NoStop}%
\bibitem [{\citenamefont {{Nishizawa}}\ \emph {et~al.}(2013)\citenamefont
  {{Nishizawa}}, \citenamefont {{Takada}},\ and\ \citenamefont
  {{Nishimichi}}}]{2013MNRAS.433..209N}%
  \BibitemOpen
  \bibfield  {author} {\bibinfo {author} {\bibfnamefont {A.~J.}\ \bibnamefont
  {{Nishizawa}}}, \bibinfo {author} {\bibfnamefont {M.}~\bibnamefont
  {{Takada}}}, \ and\ \bibinfo {author} {\bibfnamefont {T.}~\bibnamefont
  {{Nishimichi}}},\ }\href {\doibase 10.1093/mnras/stt716} {\bibfield
  {journal} {\bibinfo  {journal} {\mnras}\ }\textbf {\bibinfo {volume} {433}},\
  \bibinfo {pages} {209} (\bibinfo {year} {2013})},\ \Eprint
  {http://arxiv.org/abs/1212.4025} {arXiv:1212.4025 [astro-ph.CO]} \BibitemShut
  {NoStop}%
\bibitem [{\citenamefont {{Wang}}\ and\ \citenamefont
  {{Szalay}}(2012)}]{2012PhRvD..86d3508W}%
  \BibitemOpen
  \bibfield  {author} {\bibinfo {author} {\bibfnamefont {X.}~\bibnamefont
  {{Wang}}}\ and\ \bibinfo {author} {\bibfnamefont {A.}~\bibnamefont
  {{Szalay}}},\ }\href {\doibase 10.1103/PhysRevD.86.043508} {\bibfield
  {journal} {\bibinfo  {journal} {\prd}\ }\textbf {\bibinfo {volume} {86}},\
  \bibinfo {eid} {043508} (\bibinfo {year} {2012})},\ \Eprint
  {http://arxiv.org/abs/1204.0019} {arXiv:1204.0019 [astro-ph.CO]} \BibitemShut
  {NoStop}%
\bibitem [{\citenamefont {{Chan}}\ \emph {et~al.}(2012)\citenamefont {{Chan}},
  \citenamefont {{Scoccimarro}},\ and\ \citenamefont
  {{Sheth}}}]{2012PhRvD..85h3509C}%
  \BibitemOpen
  \bibfield  {author} {\bibinfo {author} {\bibfnamefont {K.~C.}\ \bibnamefont
  {{Chan}}}, \bibinfo {author} {\bibfnamefont {R.}~\bibnamefont
  {{Scoccimarro}}}, \ and\ \bibinfo {author} {\bibfnamefont {R.~K.}\
  \bibnamefont {{Sheth}}},\ }\href {\doibase 10.1103/PhysRevD.85.083509}
  {\bibfield  {journal} {\bibinfo  {journal} {\prd}\ }\textbf {\bibinfo
  {volume} {85}},\ \bibinfo {eid} {083509} (\bibinfo {year} {2012})},\ \Eprint
  {http://arxiv.org/abs/1201.3614} {arXiv:1201.3614 [astro-ph.CO]} \BibitemShut
  {NoStop}%
\bibitem [{\citenamefont {{Matsubara}}(2011)}]{2011PhRvD..83h3518M}%
  \BibitemOpen
  \bibfield  {author} {\bibinfo {author} {\bibfnamefont {T.}~\bibnamefont
  {{Matsubara}}},\ }\href {\doibase 10.1103/PhysRevD.83.083518} {\bibfield
  {journal} {\bibinfo  {journal} {\prd}\ }\textbf {\bibinfo {volume} {83}},\
  \bibinfo {eid} {083518} (\bibinfo {year} {2011})},\ \Eprint
  {http://arxiv.org/abs/1102.4619} {arXiv:1102.4619 [astro-ph.CO]} \BibitemShut
  {NoStop}%
\bibitem [{\citenamefont {{Song}}\ \emph {et~al.}(2013)\citenamefont {{Song}},
  \citenamefont {{Nishimichi}}, \citenamefont {{Taruya}},\ and\ \citenamefont
  {{Kayo}}}]{2013PhRvD..87l3510S}%
  \BibitemOpen
  \bibfield  {author} {\bibinfo {author} {\bibfnamefont {Y.-S.}\ \bibnamefont
  {{Song}}}, \bibinfo {author} {\bibfnamefont {T.}~\bibnamefont
  {{Nishimichi}}}, \bibinfo {author} {\bibfnamefont {A.}~\bibnamefont
  {{Taruya}}}, \ and\ \bibinfo {author} {\bibfnamefont {I.}~\bibnamefont
  {{Kayo}}},\ }\href {\doibase 10.1103/PhysRevD.87.123510} {\bibfield
  {journal} {\bibinfo  {journal} {\prd}\ }\textbf {\bibinfo {volume} {87}},\
  \bibinfo {eid} {123510} (\bibinfo {year} {2013})},\ \Eprint
  {http://arxiv.org/abs/1301.3133} {arXiv:1301.3133 [astro-ph.CO]} \BibitemShut
  {NoStop}%
\bibitem [{\citenamefont {{Vlah}}\ \emph {et~al.}(2012)\citenamefont {{Vlah}},
  \citenamefont {{Seljak}}, \citenamefont {{McDonald}}, \citenamefont
  {{Okumura}},\ and\ \citenamefont {{Baldauf}}}]{2012JCAP...11..009V}%
  \BibitemOpen
  \bibfield  {author} {\bibinfo {author} {\bibfnamefont {Z.}~\bibnamefont
  {{Vlah}}}, \bibinfo {author} {\bibfnamefont {U.}~\bibnamefont {{Seljak}}},
  \bibinfo {author} {\bibfnamefont {P.}~\bibnamefont {{McDonald}}}, \bibinfo
  {author} {\bibfnamefont {T.}~\bibnamefont {{Okumura}}}, \ and\ \bibinfo
  {author} {\bibfnamefont {T.}~\bibnamefont {{Baldauf}}},\ }\href {\doibase
  10.1088/1475-7516/2012/11/009} {\bibfield  {journal} {\bibinfo  {journal}
  {\jcap}\ }\textbf {\bibinfo {volume} {11}},\ \bibinfo {eid} {009} (\bibinfo
  {year} {2012})},\ \Eprint {http://arxiv.org/abs/1207.0839} {arXiv:1207.0839
  [astro-ph.CO]} \BibitemShut {NoStop}%
\bibitem [{\citenamefont {{Okumura}}\ \emph
  {et~al.}(2012{\natexlab{a}})\citenamefont {{Okumura}}, \citenamefont
  {{Seljak}},\ and\ \citenamefont {{Desjacques}}}]{2012JCAP...11..014O}%
  \BibitemOpen
  \bibfield  {author} {\bibinfo {author} {\bibfnamefont {T.}~\bibnamefont
  {{Okumura}}}, \bibinfo {author} {\bibfnamefont {U.}~\bibnamefont {{Seljak}}},
  \ and\ \bibinfo {author} {\bibfnamefont {V.}~\bibnamefont {{Desjacques}}},\
  }\href {\doibase 10.1088/1475-7516/2012/11/014} {\bibfield  {journal}
  {\bibinfo  {journal} {\jcap}\ }\textbf {\bibinfo {volume} {11}},\ \bibinfo
  {eid} {014} (\bibinfo {year} {2012}{\natexlab{a}})},\ \Eprint
  {http://arxiv.org/abs/1206.4070} {arXiv:1206.4070 [astro-ph.CO]} \BibitemShut
  {NoStop}%
\bibitem [{\citenamefont {{Shaw}}\ and\ \citenamefont
  {{Lewis}}(2008)}]{2008PhRvD..78j3512S}%
  \BibitemOpen
  \bibfield  {author} {\bibinfo {author} {\bibfnamefont {J.~R.}\ \bibnamefont
  {{Shaw}}}\ and\ \bibinfo {author} {\bibfnamefont {A.}~\bibnamefont
  {{Lewis}}},\ }\href {\doibase 10.1103/PhysRevD.78.103512} {\bibfield
  {journal} {\bibinfo  {journal} {\prd}\ }\textbf {\bibinfo {volume} {78}},\
  \bibinfo {pages} {103512} (\bibinfo {year} {2008})},\ \Eprint
  {http://arxiv.org/abs/0808.1724} {arXiv:0808.1724} \BibitemShut {NoStop}%
\bibitem [{\citenamefont {{Baumann}}\ \emph {et~al.}(2012)\citenamefont
  {{Baumann}}, \citenamefont {{Nicolis}}, \citenamefont {{Senatore}},\ and\
  \citenamefont {{Zaldarriaga}}}]{2012JCAP...07..051B}%
  \BibitemOpen
  \bibfield  {author} {\bibinfo {author} {\bibfnamefont {D.}~\bibnamefont
  {{Baumann}}}, \bibinfo {author} {\bibfnamefont {A.}~\bibnamefont
  {{Nicolis}}}, \bibinfo {author} {\bibfnamefont {L.}~\bibnamefont
  {{Senatore}}}, \ and\ \bibinfo {author} {\bibfnamefont {M.}~\bibnamefont
  {{Zaldarriaga}}},\ }\href {\doibase 10.1088/1475-7516/2012/07/051} {\bibfield
   {journal} {\bibinfo  {journal} {\jcap}\ }\textbf {\bibinfo {volume} {7}},\
  \bibinfo {eid} {051} (\bibinfo {year} {2012})},\ \Eprint
  {http://arxiv.org/abs/1004.2488} {arXiv:1004.2488 [astro-ph.CO]} \BibitemShut
  {NoStop}%
\bibitem [{\citenamefont {{Seo}}\ \emph {et~al.}(2010)\citenamefont {{Seo}},
  \citenamefont {{Eckel}}, \citenamefont {{Eisenstein}}, \citenamefont
  {{Mehta}}, \citenamefont {{Metchnik}}, \citenamefont {{Padmanabhan}},
  \citenamefont {{Pinto}}, \citenamefont {{Takahashi}}, \citenamefont
  {{White}},\ and\ \citenamefont {{Xu}}}]{2010ApJ...720.1650S}%
  \BibitemOpen
  \bibfield  {author} {\bibinfo {author} {\bibfnamefont {H.-J.}\ \bibnamefont
  {{Seo}}}, \bibinfo {author} {\bibfnamefont {J.}~\bibnamefont {{Eckel}}},
  \bibinfo {author} {\bibfnamefont {D.~J.}\ \bibnamefont {{Eisenstein}}},
  \bibinfo {author} {\bibfnamefont {K.}~\bibnamefont {{Mehta}}}, \bibinfo
  {author} {\bibfnamefont {M.}~\bibnamefont {{Metchnik}}}, \bibinfo {author}
  {\bibfnamefont {N.}~\bibnamefont {{Padmanabhan}}}, \bibinfo {author}
  {\bibfnamefont {P.}~\bibnamefont {{Pinto}}}, \bibinfo {author} {\bibfnamefont
  {R.}~\bibnamefont {{Takahashi}}}, \bibinfo {author} {\bibfnamefont
  {M.}~\bibnamefont {{White}}}, \ and\ \bibinfo {author} {\bibfnamefont
  {X.}~\bibnamefont {{Xu}}},\ }\href {\doibase 10.1088/0004-637X/720/2/1650}
  {\bibfield  {journal} {\bibinfo  {journal} {\apj}\ }\textbf {\bibinfo
  {volume} {720}},\ \bibinfo {pages} {1650} (\bibinfo {year} {2010})},\ \Eprint
  {http://arxiv.org/abs/0910.5005} {arXiv:0910.5005 [astro-ph.CO]} \BibitemShut
  {NoStop}%
\bibitem [{\citenamefont {{Seo}}\ and\ \citenamefont
  {{Eisenstein}}(2007)}]{2007ApJ...665...14S}%
  \BibitemOpen
  \bibfield  {author} {\bibinfo {author} {\bibfnamefont {H.-J.}\ \bibnamefont
  {{Seo}}}\ and\ \bibinfo {author} {\bibfnamefont {D.~J.}\ \bibnamefont
  {{Eisenstein}}},\ }\href {\doibase 10.1086/519549} {\bibfield  {journal}
  {\bibinfo  {journal} {\apj}\ }\textbf {\bibinfo {volume} {665}},\ \bibinfo
  {pages} {14} (\bibinfo {year} {2007})},\ \Eprint
  {http://arxiv.org/abs/arXiv:astro-ph/0701079} {arXiv:astro-ph/0701079}
  \BibitemShut {NoStop}%
\bibitem [{\citenamefont {{White}}(2010)}]{2010arXiv1004.0250W}%
  \BibitemOpen
  \bibfield  {author} {\bibinfo {author} {\bibfnamefont {M.}~\bibnamefont
  {{White}}},\ }\href@noop {} {\bibfield  {journal} {\bibinfo  {journal} {ArXiv
  e-prints}\ } (\bibinfo {year} {2010})},\ \Eprint
  {http://arxiv.org/abs/1004.0250} {arXiv:1004.0250 [astro-ph.CO]} \BibitemShut
  {NoStop}%
\bibitem [{\citenamefont {{Sherwin}}\ and\ \citenamefont
  {{Zaldarriaga}}(2012)}]{2012PhRvD..85j3523S}%
  \BibitemOpen
  \bibfield  {author} {\bibinfo {author} {\bibfnamefont {B.~D.}\ \bibnamefont
  {{Sherwin}}}\ and\ \bibinfo {author} {\bibfnamefont {M.}~\bibnamefont
  {{Zaldarriaga}}},\ }\href {\doibase 10.1103/PhysRevD.85.103523} {\bibfield
  {journal} {\bibinfo  {journal} {\prd}\ }\textbf {\bibinfo {volume} {85}},\
  \bibinfo {eid} {103523} (\bibinfo {year} {2012})},\ \Eprint
  {http://arxiv.org/abs/1202.3998} {arXiv:1202.3998 [astro-ph.CO]} \BibitemShut
  {NoStop}%
\bibitem [{\citenamefont {{Ngan}}\ \emph {et~al.}(2012)\citenamefont {{Ngan}},
  \citenamefont {{Harnois-D{\'e}raps}}, \citenamefont {{Pen}}, \citenamefont
  {{McDonald}},\ and\ \citenamefont {{MacDonald}}}]{2012MNRAS.419.2949N}%
  \BibitemOpen
  \bibfield  {author} {\bibinfo {author} {\bibfnamefont {W.}~\bibnamefont
  {{Ngan}}}, \bibinfo {author} {\bibfnamefont {J.}~\bibnamefont
  {{Harnois-D{\'e}raps}}}, \bibinfo {author} {\bibfnamefont {U.-L.}\
  \bibnamefont {{Pen}}}, \bibinfo {author} {\bibfnamefont {P.}~\bibnamefont
  {{McDonald}}}, \ and\ \bibinfo {author} {\bibfnamefont {I.}~\bibnamefont
  {{MacDonald}}},\ }\href {\doibase 10.1111/j.1365-2966.2011.19936.x}
  {\bibfield  {journal} {\bibinfo  {journal} {\mnras}\ }\textbf {\bibinfo
  {volume} {419}},\ \bibinfo {pages} {2949} (\bibinfo {year} {2012})},\ \Eprint
  {http://arxiv.org/abs/1106.5548} {arXiv:1106.5548 [astro-ph.CO]} \BibitemShut
  {NoStop}%
\bibitem [{\citenamefont {{Taruya}}\ \emph {et~al.}(2009)\citenamefont
  {{Taruya}}, \citenamefont {{Nishimichi}}, \citenamefont {{Saito}},\ and\
  \citenamefont {{Hiramatsu}}}]{2009PhRvD..80l3503T}%
  \BibitemOpen
  \bibfield  {author} {\bibinfo {author} {\bibfnamefont {A.}~\bibnamefont
  {{Taruya}}}, \bibinfo {author} {\bibfnamefont {T.}~\bibnamefont
  {{Nishimichi}}}, \bibinfo {author} {\bibfnamefont {S.}~\bibnamefont
  {{Saito}}}, \ and\ \bibinfo {author} {\bibfnamefont {T.}~\bibnamefont
  {{Hiramatsu}}},\ }\href {\doibase 10.1103/PhysRevD.80.123503} {\bibfield
  {journal} {\bibinfo  {journal} {\prd}\ }\textbf {\bibinfo {volume} {80}},\
  \bibinfo {eid} {123503} (\bibinfo {year} {2009})},\ \Eprint
  {http://arxiv.org/abs/0906.0507} {arXiv:0906.0507 [astro-ph.CO]} \BibitemShut
  {NoStop}%
\bibitem [{\citenamefont {{Alcock}}\ and\ \citenamefont
  {{Paczynski}}(1979)}]{1979Natur.281..358A}%
  \BibitemOpen
  \bibfield  {author} {\bibinfo {author} {\bibfnamefont {C.}~\bibnamefont
  {{Alcock}}}\ and\ \bibinfo {author} {\bibfnamefont {B.}~\bibnamefont
  {{Paczynski}}},\ }\href@noop {} {\bibfield  {journal} {\bibinfo  {journal}
  {\nat}\ }\textbf {\bibinfo {volume} {281}},\ \bibinfo {pages} {358} (\bibinfo
  {year} {1979})}\BibitemShut {NoStop}%
\bibitem [{\citenamefont {{Okumura}}\ \emph
  {et~al.}(2012{\natexlab{b}})\citenamefont {{Okumura}}, \citenamefont
  {{Seljak}}, \citenamefont {{McDonald}},\ and\ \citenamefont
  {{Desjacques}}}]{2012JCAP...02..010O}%
  \BibitemOpen
  \bibfield  {author} {\bibinfo {author} {\bibfnamefont {T.}~\bibnamefont
  {{Okumura}}}, \bibinfo {author} {\bibfnamefont {U.}~\bibnamefont {{Seljak}}},
  \bibinfo {author} {\bibfnamefont {P.}~\bibnamefont {{McDonald}}}, \ and\
  \bibinfo {author} {\bibfnamefont {V.}~\bibnamefont {{Desjacques}}},\ }\href
  {\doibase 10.1088/1475-7516/2012/02/010} {\bibfield  {journal} {\bibinfo
  {journal} {\jcap}\ }\textbf {\bibinfo {volume} {2}},\ \bibinfo {eid} {010}
  (\bibinfo {year} {2012}{\natexlab{b}})},\ \Eprint
  {http://arxiv.org/abs/1109.1609} {arXiv:1109.1609 [astro-ph.CO]} \BibitemShut
  {NoStop}%
\bibitem [{\citenamefont {{Gil-Mar{\'{\i}}n}}\ \emph
  {et~al.}(2012)\citenamefont {{Gil-Mar{\'{\i}}n}}, \citenamefont {{Wagner}},
  \citenamefont {{Verde}}, \citenamefont {{Porciani}},\ and\ \citenamefont
  {{Jimenez}}}]{2012JCAP...11..029G}%
  \BibitemOpen
  \bibfield  {author} {\bibinfo {author} {\bibfnamefont {H.}~\bibnamefont
  {{Gil-Mar{\'{\i}}n}}}, \bibinfo {author} {\bibfnamefont {C.}~\bibnamefont
  {{Wagner}}}, \bibinfo {author} {\bibfnamefont {L.}~\bibnamefont {{Verde}}},
  \bibinfo {author} {\bibfnamefont {C.}~\bibnamefont {{Porciani}}}, \ and\
  \bibinfo {author} {\bibfnamefont {R.}~\bibnamefont {{Jimenez}}},\ }\href
  {\doibase 10.1088/1475-7516/2012/11/029} {\bibfield  {journal} {\bibinfo
  {journal} {\jcap}\ }\textbf {\bibinfo {volume} {11}},\ \bibinfo {eid} {029}
  (\bibinfo {year} {2012})},\ \Eprint {http://arxiv.org/abs/1209.3771}
  {arXiv:1209.3771 [astro-ph.CO]} \BibitemShut {NoStop}%
\bibitem [{\citenamefont {{Taruya}}\ \emph {et~al.}(2013)\citenamefont
  {{Taruya}}, \citenamefont {{Nishimichi}},\ and\ \citenamefont
  {{Bernardeau}}}]{2013PhRvD..87h3509T}%
  \BibitemOpen
  \bibfield  {author} {\bibinfo {author} {\bibfnamefont {A.}~\bibnamefont
  {{Taruya}}}, \bibinfo {author} {\bibfnamefont {T.}~\bibnamefont
  {{Nishimichi}}}, \ and\ \bibinfo {author} {\bibfnamefont {F.}~\bibnamefont
  {{Bernardeau}}},\ }\href {\doibase 10.1103/PhysRevD.87.083509} {\bibfield
  {journal} {\bibinfo  {journal} {\prd}\ }\textbf {\bibinfo {volume} {87}},\
  \bibinfo {eid} {083509} (\bibinfo {year} {2013})},\ \Eprint
  {http://arxiv.org/abs/1301.3624} {arXiv:1301.3624 [astro-ph.CO]} \BibitemShut
  {NoStop}%
\bibitem [{\citenamefont {{Vallinotto}}\ and\ \citenamefont
  {{Linder}}(2013)}]{2013arXiv1307.2906V}%
  \BibitemOpen
  \bibfield  {author} {\bibinfo {author} {\bibfnamefont {A.}~\bibnamefont
  {{Vallinotto}}}\ and\ \bibinfo {author} {\bibfnamefont {E.~V.}\ \bibnamefont
  {{Linder}}},\ }\href@noop {} {\bibfield  {journal} {\bibinfo  {journal}
  {ArXiv e-prints}\ } (\bibinfo {year} {2013})},\ \Eprint
  {http://arxiv.org/abs/1307.2906} {arXiv:1307.2906 [astro-ph.CO]} \BibitemShut
  {NoStop}%
\bibitem [{\citenamefont {{Linder}}\ and\ \citenamefont
  {{Samsing}}(2013)}]{2013JCAP...02..025L}%
  \BibitemOpen
  \bibfield  {author} {\bibinfo {author} {\bibfnamefont {E.~V.}\ \bibnamefont
  {{Linder}}}\ and\ \bibinfo {author} {\bibfnamefont {J.}~\bibnamefont
  {{Samsing}}},\ }\href {\doibase 10.1088/1475-7516/2013/02/025} {\bibfield
  {journal} {\bibinfo  {journal} {\jcap}\ }\textbf {\bibinfo {volume} {2}},\
  \bibinfo {eid} {025} (\bibinfo {year} {2013})},\ \Eprint
  {http://arxiv.org/abs/1211.2274} {arXiv:1211.2274 [astro-ph.CO]} \BibitemShut
  {NoStop}%
\bibitem [{\citenamefont {{Crocce}}\ and\ \citenamefont
  {{Scoccimarro}}(2008)}]{2008PhRvD..77b3533C}%
  \BibitemOpen
  \bibfield  {author} {\bibinfo {author} {\bibfnamefont {M.}~\bibnamefont
  {{Crocce}}}\ and\ \bibinfo {author} {\bibfnamefont {R.}~\bibnamefont
  {{Scoccimarro}}},\ }\href {\doibase 10.1103/PhysRevD.77.023533} {\bibfield
  {journal} {\bibinfo  {journal} {\prd}\ }\textbf {\bibinfo {volume} {77}},\
  \bibinfo {pages} {023533} (\bibinfo {year} {2008})},\ \Eprint
  {http://arxiv.org/abs/arXiv:0704.2783} {arXiv:0704.2783} \BibitemShut
  {NoStop}%
\bibitem [{\citenamefont {{McDonald}}\ and\ \citenamefont
  {{Seljak}}(2009)}]{2009JCAP...10..007M}%
  \BibitemOpen
  \bibfield  {author} {\bibinfo {author} {\bibfnamefont {P.}~\bibnamefont
  {{McDonald}}}\ and\ \bibinfo {author} {\bibfnamefont {U.}~\bibnamefont
  {{Seljak}}},\ }\href {\doibase 10.1088/1475-7516/2009/10/007} {\bibfield
  {journal} {\bibinfo  {journal} {\jcap}\ }\textbf {\bibinfo {volume} {10}},\
  \bibinfo {eid} {007} (\bibinfo {year} {2009})},\ \Eprint
  {http://arxiv.org/abs/0810.0323} {arXiv:0810.0323} \BibitemShut {NoStop}%
\bibitem [{\citenamefont {{McDonald}}\ \emph
  {et~al.}(2006{\natexlab{a}})\citenamefont {{McDonald}}, \citenamefont
  {{Seljak}}, \citenamefont {{Burles}}, \citenamefont {{Schlegel}},
  \citenamefont {{Weinberg}}, \citenamefont {{Cen}}, \citenamefont {{Shih}},
  \citenamefont {{Schaye}}, \citenamefont {{Schneider}}, \citenamefont
  {{Bahcall}}, \citenamefont {{Briggs}}, \citenamefont {{Brinkmann}},
  \citenamefont {{Brunner}}, \citenamefont {{Fukugita}}, \citenamefont
  {{Gunn}}, \citenamefont {{Ivezi{\'c}}}, \citenamefont {{Kent}}, \citenamefont
  {{Lupton}},\ and\ \citenamefont {{Vanden Berk}}}]{2006ApJS..163...80M}%
  \BibitemOpen
  \bibfield  {author} {\bibinfo {author} {\bibfnamefont {P.}~\bibnamefont
  {{McDonald}}}, \bibinfo {author} {\bibfnamefont {U.}~\bibnamefont
  {{Seljak}}}, \bibinfo {author} {\bibfnamefont {S.}~\bibnamefont {{Burles}}},
  \bibinfo {author} {\bibfnamefont {D.~J.}\ \bibnamefont {{Schlegel}}},
  \bibinfo {author} {\bibfnamefont {D.~H.}\ \bibnamefont {{Weinberg}}},
  \bibinfo {author} {\bibfnamefont {R.}~\bibnamefont {{Cen}}}, \bibinfo
  {author} {\bibfnamefont {D.}~\bibnamefont {{Shih}}}, \bibinfo {author}
  {\bibfnamefont {J.}~\bibnamefont {{Schaye}}}, \bibinfo {author}
  {\bibfnamefont {D.~P.}\ \bibnamefont {{Schneider}}}, \bibinfo {author}
  {\bibfnamefont {N.~A.}\ \bibnamefont {{Bahcall}}}, \bibinfo {author}
  {\bibfnamefont {J.~W.}\ \bibnamefont {{Briggs}}}, \bibinfo {author}
  {\bibfnamefont {J.}~\bibnamefont {{Brinkmann}}}, \bibinfo {author}
  {\bibfnamefont {R.~J.}\ \bibnamefont {{Brunner}}}, \bibinfo {author}
  {\bibfnamefont {M.}~\bibnamefont {{Fukugita}}}, \bibinfo {author}
  {\bibfnamefont {J.~E.}\ \bibnamefont {{Gunn}}}, \bibinfo {author}
  {\bibfnamefont {{\v Z}.}~\bibnamefont {{Ivezi{\'c}}}}, \bibinfo {author}
  {\bibfnamefont {S.}~\bibnamefont {{Kent}}}, \bibinfo {author} {\bibfnamefont
  {R.~H.}\ \bibnamefont {{Lupton}}}, \ and\ \bibinfo {author} {\bibfnamefont
  {D.~E.}\ \bibnamefont {{Vanden Berk}}},\ }\href {\doibase 10.1086/444361}
  {\bibfield  {journal} {\bibinfo  {journal} {\apjs}\ }\textbf {\bibinfo
  {volume} {163}},\ \bibinfo {pages} {80} (\bibinfo {year}
  {2006}{\natexlab{a}})},\ \Eprint
  {http://arxiv.org/abs/arXiv:astro-ph/0405013} {arXiv:astro-ph/0405013}
  \BibitemShut {NoStop}%
\bibitem [{\citenamefont {{Slosar}}\ \emph {et~al.}(2011)\citenamefont
  {{Slosar}}, \citenamefont {{Font-Ribera}}, \citenamefont {{Pieri}},
  \citenamefont {{Rich}}, \citenamefont {{Le Goff}}, \citenamefont {{Aubourg}},
  \citenamefont {{Brinkmann}}, \citenamefont {{Busca}}, \citenamefont
  {{Carithers}}, \citenamefont {{Charlassier}}, \citenamefont {{Cort{\^e}s}},
  \citenamefont {{Croft}}, \citenamefont {{Dawson}}, \citenamefont
  {{Eisenstein}}, \citenamefont {{Hamilton}}, \citenamefont {{Ho}},
  \citenamefont {{Lee}}, \citenamefont {{Lupton}}, \citenamefont {{McDonald}},
  \citenamefont {{Medolin}}, \citenamefont {{Muna}}, \citenamefont
  {{Miralda-Escud{\'e}}}, \citenamefont {{Myers}}, \citenamefont {{Nichol}},
  \citenamefont {{Palanque-Delabrouille}}, \citenamefont {{P{\^a}ris}},
  \citenamefont {{Petitjean}}, \citenamefont {{Pi{\v s}kur}}, \citenamefont
  {{Rollinde}}, \citenamefont {{Ross}}, \citenamefont {{Schlegel}},
  \citenamefont {{Schneider}}, \citenamefont {{Sheldon}}, \citenamefont
  {{Weaver}}, \citenamefont {{Weinberg}}, \citenamefont {{Yeche}},\ and\
  \citenamefont {{York}}}]{2011JCAP...09..001S}%
  \BibitemOpen
  \bibfield  {author} {\bibinfo {author} {\bibfnamefont {A.}~\bibnamefont
  {{Slosar}}}, \bibinfo {author} {\bibfnamefont {A.}~\bibnamefont
  {{Font-Ribera}}}, \bibinfo {author} {\bibfnamefont {M.~M.}\ \bibnamefont
  {{Pieri}}}, \bibinfo {author} {\bibfnamefont {J.}~\bibnamefont {{Rich}}},
  \bibinfo {author} {\bibfnamefont {J.-M.}\ \bibnamefont {{Le Goff}}}, \bibinfo
  {author} {\bibfnamefont {{\'E}.}~\bibnamefont {{Aubourg}}}, \bibinfo {author}
  {\bibfnamefont {J.}~\bibnamefont {{Brinkmann}}}, \bibinfo {author}
  {\bibfnamefont {N.}~\bibnamefont {{Busca}}}, \bibinfo {author} {\bibfnamefont
  {B.}~\bibnamefont {{Carithers}}}, \bibinfo {author} {\bibfnamefont
  {R.}~\bibnamefont {{Charlassier}}}, \bibinfo {author} {\bibfnamefont
  {M.}~\bibnamefont {{Cort{\^e}s}}}, \bibinfo {author} {\bibfnamefont
  {R.}~\bibnamefont {{Croft}}}, \bibinfo {author} {\bibfnamefont {K.~S.}\
  \bibnamefont {{Dawson}}}, \bibinfo {author} {\bibfnamefont {D.}~\bibnamefont
  {{Eisenstein}}}, \bibinfo {author} {\bibfnamefont {J.-C.}\ \bibnamefont
  {{Hamilton}}}, \bibinfo {author} {\bibfnamefont {S.}~\bibnamefont {{Ho}}},
  \bibinfo {author} {\bibfnamefont {K.-G.}\ \bibnamefont {{Lee}}}, \bibinfo
  {author} {\bibfnamefont {R.}~\bibnamefont {{Lupton}}}, \bibinfo {author}
  {\bibfnamefont {P.}~\bibnamefont {{McDonald}}}, \bibinfo {author}
  {\bibfnamefont {B.}~\bibnamefont {{Medolin}}}, \bibinfo {author}
  {\bibfnamefont {D.}~\bibnamefont {{Muna}}}, \bibinfo {author} {\bibfnamefont
  {J.}~\bibnamefont {{Miralda-Escud{\'e}}}}, \bibinfo {author} {\bibfnamefont
  {A.~D.}\ \bibnamefont {{Myers}}}, \bibinfo {author} {\bibfnamefont {R.~C.}\
  \bibnamefont {{Nichol}}}, \bibinfo {author} {\bibfnamefont {N.}~\bibnamefont
  {{Palanque-Delabrouille}}}, \bibinfo {author} {\bibfnamefont
  {I.}~\bibnamefont {{P{\^a}ris}}}, \bibinfo {author} {\bibfnamefont
  {P.}~\bibnamefont {{Petitjean}}}, \bibinfo {author} {\bibfnamefont
  {Y.}~\bibnamefont {{Pi{\v s}kur}}}, \bibinfo {author} {\bibfnamefont
  {E.}~\bibnamefont {{Rollinde}}}, \bibinfo {author} {\bibfnamefont {N.~P.}\
  \bibnamefont {{Ross}}}, \bibinfo {author} {\bibfnamefont {D.~J.}\
  \bibnamefont {{Schlegel}}}, \bibinfo {author} {\bibfnamefont {D.~P.}\
  \bibnamefont {{Schneider}}}, \bibinfo {author} {\bibfnamefont
  {E.}~\bibnamefont {{Sheldon}}}, \bibinfo {author} {\bibfnamefont {B.~A.}\
  \bibnamefont {{Weaver}}}, \bibinfo {author} {\bibfnamefont {D.~H.}\
  \bibnamefont {{Weinberg}}}, \bibinfo {author} {\bibfnamefont
  {C.}~\bibnamefont {{Yeche}}}, \ and\ \bibinfo {author} {\bibfnamefont
  {D.~G.}\ \bibnamefont {{York}}},\ }\href {\doibase
  10.1088/1475-7516/2011/09/001} {\bibfield  {journal} {\bibinfo  {journal}
  {\jcap}\ }\textbf {\bibinfo {volume} {9}},\ \bibinfo {pages} {1} (\bibinfo
  {year} {2011})},\ \Eprint {http://arxiv.org/abs/1104.5244} {arXiv:1104.5244
  [astro-ph.CO]} \BibitemShut {NoStop}%
\bibitem [{\citenamefont {{McDonald}}\ and\ \citenamefont
  {{Eisenstein}}(2007)}]{2007PhRvD..76f3009M}%
  \BibitemOpen
  \bibfield  {author} {\bibinfo {author} {\bibfnamefont {P.}~\bibnamefont
  {{McDonald}}}\ and\ \bibinfo {author} {\bibfnamefont {D.~J.}\ \bibnamefont
  {{Eisenstein}}},\ }\href {\doibase 10.1103/PhysRevD.76.063009} {\bibfield
  {journal} {\bibinfo  {journal} {\prd}\ }\textbf {\bibinfo {volume} {76}},\
  \bibinfo {pages} {063009} (\bibinfo {year} {2007})},\ \Eprint
  {http://arxiv.org/abs/arXiv:astro-ph/0607122} {arXiv:astro-ph/0607122}
  \BibitemShut {NoStop}%
\bibitem [{\citenamefont {{McDonald}}(2003)}]{2003ApJ...585...34M}%
  \BibitemOpen
  \bibfield  {author} {\bibinfo {author} {\bibfnamefont {P.}~\bibnamefont
  {{McDonald}}},\ }\href {\doibase 10.1086/345945} {\bibfield  {journal}
  {\bibinfo  {journal} {\apj}\ }\textbf {\bibinfo {volume} {585}},\ \bibinfo
  {pages} {34} (\bibinfo {year} {2003})},\ \Eprint
  {http://arxiv.org/abs/arXiv:astro-ph/0108064} {arXiv:astro-ph/0108064}
  \BibitemShut {NoStop}%
\bibitem [{\citenamefont {{Busca}}\ \emph {et~al.}(2013)\citenamefont
  {{Busca}}, \citenamefont {{Delubac}}, \citenamefont {{Rich}}, \citenamefont
  {{Bailey}}, \citenamefont {{Font-Ribera}}, \citenamefont {{Kirkby}},
  \citenamefont {{Le Goff}}, \citenamefont {{Pieri}}, \citenamefont {{Slosar}},
  \citenamefont {{Aubourg}}, \citenamefont {{Bautista}}, \citenamefont
  {{Bizyaev}}, \citenamefont {{Blomqvist}}, \citenamefont {{Bolton}},
  \citenamefont {{Bovy}}, \citenamefont {{Brewington}}, \citenamefont
  {{Borde}}, \citenamefont {{Brinkmann}}, \citenamefont {{Carithers}},
  \citenamefont {{Croft}}, \citenamefont {{Dawson}}, \citenamefont {{Ebelke}},
  \citenamefont {{Eisenstein}}, \citenamefont {{Hamilton}}, \citenamefont
  {{Ho}}, \citenamefont {{Hogg}}, \citenamefont {{Honscheid}}, \citenamefont
  {{Lee}}, \citenamefont {{Lundgren}}, \citenamefont {{Malanushenko}},
  \citenamefont {{Malanushenko}}, \citenamefont {{Margala}}, \citenamefont
  {{Maraston}}, \citenamefont {{Mehta}}, \citenamefont {{Miralda-Escud{\'e}}},
  \citenamefont {{Myers}}, \citenamefont {{Nichol}}, \citenamefont
  {{Noterdaeme}}, \citenamefont {{Olmstead}}, \citenamefont {{Oravetz}},
  \citenamefont {{Palanque-Delabrouille}}, \citenamefont {{Pan}}, \citenamefont
  {{P{\^a}ris}}, \citenamefont {{Percival}}, \citenamefont {{Petitjean}},
  \citenamefont {{Roe}}, \citenamefont {{Rollinde}}, \citenamefont {{Ross}},
  \citenamefont {{Rossi}}, \citenamefont {{Schlegel}}, \citenamefont
  {{Schneider}}, \citenamefont {{Shelden}}, \citenamefont {{Sheldon}},
  \citenamefont {{Simmons}}, \citenamefont {{Snedden}}, \citenamefont
  {{Tinker}}, \citenamefont {{Viel}}, \citenamefont {{Weaver}}, \citenamefont
  {{Weinberg}}, \citenamefont {{White}}, \citenamefont {{Y{\`e}che}},\ and\
  \citenamefont {{York}}}]{2013A&A...552A..96B}%
  \BibitemOpen
  \bibfield  {author} {\bibinfo {author} {\bibfnamefont {N.~G.}\ \bibnamefont
  {{Busca}}}, \bibinfo {author} {\bibfnamefont {T.}~\bibnamefont {{Delubac}}},
  \bibinfo {author} {\bibfnamefont {J.}~\bibnamefont {{Rich}}}, \bibinfo
  {author} {\bibfnamefont {S.}~\bibnamefont {{Bailey}}}, \bibinfo {author}
  {\bibfnamefont {A.}~\bibnamefont {{Font-Ribera}}}, \bibinfo {author}
  {\bibfnamefont {D.}~\bibnamefont {{Kirkby}}}, \bibinfo {author}
  {\bibfnamefont {J.-M.}\ \bibnamefont {{Le Goff}}}, \bibinfo {author}
  {\bibfnamefont {M.~M.}\ \bibnamefont {{Pieri}}}, \bibinfo {author}
  {\bibfnamefont {A.}~\bibnamefont {{Slosar}}}, \bibinfo {author}
  {\bibfnamefont {{\'E}.}~\bibnamefont {{Aubourg}}}, \bibinfo {author}
  {\bibfnamefont {J.~E.}\ \bibnamefont {{Bautista}}}, \bibinfo {author}
  {\bibfnamefont {D.}~\bibnamefont {{Bizyaev}}}, \bibinfo {author}
  {\bibfnamefont {M.}~\bibnamefont {{Blomqvist}}}, \bibinfo {author}
  {\bibfnamefont {A.~S.}\ \bibnamefont {{Bolton}}}, \bibinfo {author}
  {\bibfnamefont {J.}~\bibnamefont {{Bovy}}}, \bibinfo {author} {\bibfnamefont
  {H.}~\bibnamefont {{Brewington}}}, \bibinfo {author} {\bibfnamefont
  {A.}~\bibnamefont {{Borde}}}, \bibinfo {author} {\bibfnamefont
  {J.}~\bibnamefont {{Brinkmann}}}, \bibinfo {author} {\bibfnamefont
  {B.}~\bibnamefont {{Carithers}}}, \bibinfo {author} {\bibfnamefont
  {R.~A.~C.}\ \bibnamefont {{Croft}}}, \bibinfo {author} {\bibfnamefont
  {K.~S.}\ \bibnamefont {{Dawson}}}, \bibinfo {author} {\bibfnamefont
  {G.}~\bibnamefont {{Ebelke}}}, \bibinfo {author} {\bibfnamefont {D.~J.}\
  \bibnamefont {{Eisenstein}}}, \bibinfo {author} {\bibfnamefont {J.-C.}\
  \bibnamefont {{Hamilton}}}, \bibinfo {author} {\bibfnamefont
  {S.}~\bibnamefont {{Ho}}}, \bibinfo {author} {\bibfnamefont {D.~W.}\
  \bibnamefont {{Hogg}}}, \bibinfo {author} {\bibfnamefont {K.}~\bibnamefont
  {{Honscheid}}}, \bibinfo {author} {\bibfnamefont {K.-G.}\ \bibnamefont
  {{Lee}}}, \bibinfo {author} {\bibfnamefont {B.}~\bibnamefont {{Lundgren}}},
  \bibinfo {author} {\bibfnamefont {E.}~\bibnamefont {{Malanushenko}}},
  \bibinfo {author} {\bibfnamefont {V.}~\bibnamefont {{Malanushenko}}},
  \bibinfo {author} {\bibfnamefont {D.}~\bibnamefont {{Margala}}}, \bibinfo
  {author} {\bibfnamefont {C.}~\bibnamefont {{Maraston}}}, \bibinfo {author}
  {\bibfnamefont {K.}~\bibnamefont {{Mehta}}}, \bibinfo {author} {\bibfnamefont
  {J.}~\bibnamefont {{Miralda-Escud{\'e}}}}, \bibinfo {author} {\bibfnamefont
  {A.~D.}\ \bibnamefont {{Myers}}}, \bibinfo {author} {\bibfnamefont {R.~C.}\
  \bibnamefont {{Nichol}}}, \bibinfo {author} {\bibfnamefont {P.}~\bibnamefont
  {{Noterdaeme}}}, \bibinfo {author} {\bibfnamefont {M.~D.}\ \bibnamefont
  {{Olmstead}}}, \bibinfo {author} {\bibfnamefont {D.}~\bibnamefont
  {{Oravetz}}}, \bibinfo {author} {\bibfnamefont {N.}~\bibnamefont
  {{Palanque-Delabrouille}}}, \bibinfo {author} {\bibfnamefont
  {K.}~\bibnamefont {{Pan}}}, \bibinfo {author} {\bibfnamefont
  {I.}~\bibnamefont {{P{\^a}ris}}}, \bibinfo {author} {\bibfnamefont {W.~J.}\
  \bibnamefont {{Percival}}}, \bibinfo {author} {\bibfnamefont
  {P.}~\bibnamefont {{Petitjean}}}, \bibinfo {author} {\bibfnamefont {N.~A.}\
  \bibnamefont {{Roe}}}, \bibinfo {author} {\bibfnamefont {E.}~\bibnamefont
  {{Rollinde}}}, \bibinfo {author} {\bibfnamefont {N.~P.}\ \bibnamefont
  {{Ross}}}, \bibinfo {author} {\bibfnamefont {G.}~\bibnamefont {{Rossi}}},
  \bibinfo {author} {\bibfnamefont {D.~J.}\ \bibnamefont {{Schlegel}}},
  \bibinfo {author} {\bibfnamefont {D.~P.}\ \bibnamefont {{Schneider}}},
  \bibinfo {author} {\bibfnamefont {A.}~\bibnamefont {{Shelden}}}, \bibinfo
  {author} {\bibfnamefont {E.~S.}\ \bibnamefont {{Sheldon}}}, \bibinfo {author}
  {\bibfnamefont {A.}~\bibnamefont {{Simmons}}}, \bibinfo {author}
  {\bibfnamefont {S.}~\bibnamefont {{Snedden}}}, \bibinfo {author}
  {\bibfnamefont {J.~L.}\ \bibnamefont {{Tinker}}}, \bibinfo {author}
  {\bibfnamefont {M.}~\bibnamefont {{Viel}}}, \bibinfo {author} {\bibfnamefont
  {B.~A.}\ \bibnamefont {{Weaver}}}, \bibinfo {author} {\bibfnamefont {D.~H.}\
  \bibnamefont {{Weinberg}}}, \bibinfo {author} {\bibfnamefont
  {M.}~\bibnamefont {{White}}}, \bibinfo {author} {\bibfnamefont
  {C.}~\bibnamefont {{Y{\`e}che}}}, \ and\ \bibinfo {author} {\bibfnamefont
  {D.~G.}\ \bibnamefont {{York}}},\ }\href {\doibase
  10.1051/0004-6361/201220724} {\bibfield  {journal} {\bibinfo  {journal}
  {\aap}\ }\textbf {\bibinfo {volume} {552}},\ \bibinfo {eid} {A96} (\bibinfo
  {year} {2013})},\ \Eprint {http://arxiv.org/abs/1211.2616} {arXiv:1211.2616
  [astro-ph.CO]} \BibitemShut {NoStop}%
\bibitem [{\citenamefont {{Slosar}}\ \emph {et~al.}(2013)\citenamefont
  {{Slosar}}, \citenamefont {{Ir{\v s}i{\v c}}}, \citenamefont {{Kirkby}},
  \citenamefont {{Bailey}}, \citenamefont {{Busca}}, \citenamefont {{Delubac}},
  \citenamefont {{Rich}}, \citenamefont {{Aubourg}}, \citenamefont
  {{Bautista}}, \citenamefont {{Bhardwaj}}, \citenamefont {{Blomqvist}},
  \citenamefont {{Bolton}}, \citenamefont {{Bovy}}, \citenamefont
  {{Brownstein}}, \citenamefont {{Carithers}}, \citenamefont {{Croft}},
  \citenamefont {{Dawson}}, \citenamefont {{Font-Ribera}}, \citenamefont {{Le
  Goff}}, \citenamefont {{Ho}}, \citenamefont {{Honscheid}}, \citenamefont
  {{Lee}}, \citenamefont {{Margala}}, \citenamefont {{McDonald}}, \citenamefont
  {{Medolin}}, \citenamefont {{Miralda-Escud{\'e}}}, \citenamefont {{Myers}},
  \citenamefont {{Nichol}}, \citenamefont {{Noterdaeme}}, \citenamefont
  {{Palanque-Delabrouille}}, \citenamefont {{P{\^a}ris}}, \citenamefont
  {{Petitjean}}, \citenamefont {{Pieri}}, \citenamefont {{Pi{\v s}kur}},
  \citenamefont {{Roe}}, \citenamefont {{Ross}}, \citenamefont {{Rossi}},
  \citenamefont {{Schlegel}}, \citenamefont {{Schneider}}, \citenamefont
  {{Suzuki}}, \citenamefont {{Sheldon}}, \citenamefont {{Seljak}},
  \citenamefont {{Viel}}, \citenamefont {{Weinberg}},\ and\ \citenamefont
  {{Y{\`e}che}}}]{2013JCAP...04..026S}%
  \BibitemOpen
  \bibfield  {author} {\bibinfo {author} {\bibfnamefont {A.}~\bibnamefont
  {{Slosar}}}, \bibinfo {author} {\bibfnamefont {V.}~\bibnamefont {{Ir{\v
  s}i{\v c}}}}, \bibinfo {author} {\bibfnamefont {D.}~\bibnamefont {{Kirkby}}},
  \bibinfo {author} {\bibfnamefont {S.}~\bibnamefont {{Bailey}}}, \bibinfo
  {author} {\bibfnamefont {N.~G.}\ \bibnamefont {{Busca}}}, \bibinfo {author}
  {\bibfnamefont {T.}~\bibnamefont {{Delubac}}}, \bibinfo {author}
  {\bibfnamefont {J.}~\bibnamefont {{Rich}}}, \bibinfo {author} {\bibfnamefont
  {{\'E}.}~\bibnamefont {{Aubourg}}}, \bibinfo {author} {\bibfnamefont {J.~E.}\
  \bibnamefont {{Bautista}}}, \bibinfo {author} {\bibfnamefont
  {V.}~\bibnamefont {{Bhardwaj}}}, \bibinfo {author} {\bibfnamefont
  {M.}~\bibnamefont {{Blomqvist}}}, \bibinfo {author} {\bibfnamefont {A.~S.}\
  \bibnamefont {{Bolton}}}, \bibinfo {author} {\bibfnamefont {J.}~\bibnamefont
  {{Bovy}}}, \bibinfo {author} {\bibfnamefont {J.}~\bibnamefont
  {{Brownstein}}}, \bibinfo {author} {\bibfnamefont {B.}~\bibnamefont
  {{Carithers}}}, \bibinfo {author} {\bibfnamefont {R.~A.~C.}\ \bibnamefont
  {{Croft}}}, \bibinfo {author} {\bibfnamefont {K.~S.}\ \bibnamefont
  {{Dawson}}}, \bibinfo {author} {\bibfnamefont {A.}~\bibnamefont
  {{Font-Ribera}}}, \bibinfo {author} {\bibfnamefont {J.-M.}\ \bibnamefont {{Le
  Goff}}}, \bibinfo {author} {\bibfnamefont {S.}~\bibnamefont {{Ho}}}, \bibinfo
  {author} {\bibfnamefont {K.}~\bibnamefont {{Honscheid}}}, \bibinfo {author}
  {\bibfnamefont {K.-G.}\ \bibnamefont {{Lee}}}, \bibinfo {author}
  {\bibfnamefont {D.}~\bibnamefont {{Margala}}}, \bibinfo {author}
  {\bibfnamefont {P.}~\bibnamefont {{McDonald}}}, \bibinfo {author}
  {\bibfnamefont {B.}~\bibnamefont {{Medolin}}}, \bibinfo {author}
  {\bibfnamefont {J.}~\bibnamefont {{Miralda-Escud{\'e}}}}, \bibinfo {author}
  {\bibfnamefont {A.~D.}\ \bibnamefont {{Myers}}}, \bibinfo {author}
  {\bibfnamefont {R.~C.}\ \bibnamefont {{Nichol}}}, \bibinfo {author}
  {\bibfnamefont {P.}~\bibnamefont {{Noterdaeme}}}, \bibinfo {author}
  {\bibfnamefont {N.}~\bibnamefont {{Palanque-Delabrouille}}}, \bibinfo
  {author} {\bibfnamefont {I.}~\bibnamefont {{P{\^a}ris}}}, \bibinfo {author}
  {\bibfnamefont {P.}~\bibnamefont {{Petitjean}}}, \bibinfo {author}
  {\bibfnamefont {M.~M.}\ \bibnamefont {{Pieri}}}, \bibinfo {author}
  {\bibfnamefont {Y.}~\bibnamefont {{Pi{\v s}kur}}}, \bibinfo {author}
  {\bibfnamefont {N.~A.}\ \bibnamefont {{Roe}}}, \bibinfo {author}
  {\bibfnamefont {N.~P.}\ \bibnamefont {{Ross}}}, \bibinfo {author}
  {\bibfnamefont {G.}~\bibnamefont {{Rossi}}}, \bibinfo {author} {\bibfnamefont
  {D.~J.}\ \bibnamefont {{Schlegel}}}, \bibinfo {author} {\bibfnamefont
  {D.~P.}\ \bibnamefont {{Schneider}}}, \bibinfo {author} {\bibfnamefont
  {N.}~\bibnamefont {{Suzuki}}}, \bibinfo {author} {\bibfnamefont {E.~S.}\
  \bibnamefont {{Sheldon}}}, \bibinfo {author} {\bibfnamefont {U.}~\bibnamefont
  {{Seljak}}}, \bibinfo {author} {\bibfnamefont {M.}~\bibnamefont {{Viel}}},
  \bibinfo {author} {\bibfnamefont {D.~H.}\ \bibnamefont {{Weinberg}}}, \ and\
  \bibinfo {author} {\bibfnamefont {C.}~\bibnamefont {{Y{\`e}che}}},\ }\href
  {\doibase 10.1088/1475-7516/2013/04/026} {\bibfield  {journal} {\bibinfo
  {journal} {\jcap}\ }\textbf {\bibinfo {volume} {4}},\ \bibinfo {eid} {026}
  (\bibinfo {year} {2013})},\ \Eprint {http://arxiv.org/abs/1301.3459}
  {arXiv:1301.3459 [astro-ph.CO]} \BibitemShut {NoStop}%
\bibitem [{\citenamefont {{McQuinn}}\ and\ \citenamefont
  {{White}}(2011)}]{2011MNRAS.415.2257M}%
  \BibitemOpen
  \bibfield  {author} {\bibinfo {author} {\bibfnamefont {M.}~\bibnamefont
  {{McQuinn}}}\ and\ \bibinfo {author} {\bibfnamefont {M.}~\bibnamefont
  {{White}}},\ }\href {\doibase 10.1111/j.1365-2966.2011.18855.x} {\bibfield
  {journal} {\bibinfo  {journal} {\mnras}\ }\textbf {\bibinfo {volume} {415}},\
  \bibinfo {pages} {2257} (\bibinfo {year} {2011})},\ \Eprint
  {http://arxiv.org/abs/1102.1752} {arXiv:1102.1752 [astro-ph.CO]} \BibitemShut
  {NoStop}%
\bibitem [{\citenamefont {{Feldman}}\ \emph {et~al.}(1994)\citenamefont
  {{Feldman}}, \citenamefont {{Kaiser}},\ and\ \citenamefont
  {{Peacock}}}]{1994ApJ...426...23F}%
  \BibitemOpen
  \bibfield  {author} {\bibinfo {author} {\bibfnamefont {H.~A.}\ \bibnamefont
  {{Feldman}}}, \bibinfo {author} {\bibfnamefont {N.}~\bibnamefont {{Kaiser}}},
  \ and\ \bibinfo {author} {\bibfnamefont {J.~A.}\ \bibnamefont {{Peacock}}},\
  }\href {\doibase 10.1086/174036} {\bibfield  {journal} {\bibinfo  {journal}
  {\apj}\ }\textbf {\bibinfo {volume} {426}},\ \bibinfo {pages} {23} (\bibinfo
  {year} {1994})}\BibitemShut {NoStop}%
\bibitem [{\citenamefont {{Ir{\v s}i{\v c}}}\ \emph {et~al.}(2013)\citenamefont
  {{Ir{\v s}i{\v c}}}, \citenamefont {{Slosar}}, \citenamefont {{Bailey}},
  \citenamefont {{Eisenstein}}, \citenamefont {{Font-Ribera}}, \citenamefont
  {{Le Goff}}, \citenamefont {{Lundgren}}, \citenamefont {{McDonald}},
  \citenamefont {{O'Connell}}, \citenamefont {{Palanque-Delabrouille}},
  \citenamefont {{Petitjean}}, \citenamefont {{Rich}}, \citenamefont {{Rossi}},
  \citenamefont {{Schneider}}, \citenamefont {{Sheldon}},\ and\ \citenamefont
  {{Y{\`e}che}}}]{2013JCAP...09..016I}%
  \BibitemOpen
  \bibfield  {author} {\bibinfo {author} {\bibfnamefont {V.}~\bibnamefont
  {{Ir{\v s}i{\v c}}}}, \bibinfo {author} {\bibfnamefont {A.}~\bibnamefont
  {{Slosar}}}, \bibinfo {author} {\bibfnamefont {S.}~\bibnamefont {{Bailey}}},
  \bibinfo {author} {\bibfnamefont {D.~J.}\ \bibnamefont {{Eisenstein}}},
  \bibinfo {author} {\bibfnamefont {A.}~\bibnamefont {{Font-Ribera}}}, \bibinfo
  {author} {\bibfnamefont {J.-M.}\ \bibnamefont {{Le Goff}}}, \bibinfo {author}
  {\bibfnamefont {B.}~\bibnamefont {{Lundgren}}}, \bibinfo {author}
  {\bibfnamefont {P.}~\bibnamefont {{McDonald}}}, \bibinfo {author}
  {\bibfnamefont {R.}~\bibnamefont {{O'Connell}}}, \bibinfo {author}
  {\bibfnamefont {N.}~\bibnamefont {{Palanque-Delabrouille}}}, \bibinfo
  {author} {\bibfnamefont {P.}~\bibnamefont {{Petitjean}}}, \bibinfo {author}
  {\bibfnamefont {J.}~\bibnamefont {{Rich}}}, \bibinfo {author} {\bibfnamefont
  {G.}~\bibnamefont {{Rossi}}}, \bibinfo {author} {\bibfnamefont {D.~P.}\
  \bibnamefont {{Schneider}}}, \bibinfo {author} {\bibfnamefont {E.~S.}\
  \bibnamefont {{Sheldon}}}, \ and\ \bibinfo {author} {\bibfnamefont
  {C.}~\bibnamefont {{Y{\`e}che}}},\ }\href {\doibase
  10.1088/1475-7516/2013/09/016} {\bibfield  {journal} {\bibinfo  {journal}
  {\jcap}\ }\textbf {\bibinfo {volume} {9}},\ \bibinfo {eid} {016} (\bibinfo
  {year} {2013})},\ \Eprint {http://arxiv.org/abs/1307.3403} {arXiv:1307.3403
  [astro-ph.CO]} \BibitemShut {NoStop}%
\bibitem [{\citenamefont {{Palanque-Delabrouille}}\ \emph
  {et~al.}(2013{\natexlab{a}})\citenamefont {{Palanque-Delabrouille}},
  \citenamefont {{Y{\`e}che}}, \citenamefont {{Borde}}, \citenamefont {{Le
  Goff}}, \citenamefont {{Rossi}}, \citenamefont {{Viel}}, \citenamefont
  {{Aubourg}}, \citenamefont {{Bailey}}, \citenamefont {{Bautista}},
  \citenamefont {{Blomqvist}}, \citenamefont {{Bolton}}, \citenamefont
  {{Bolton}}, \citenamefont {{Busca}}, \citenamefont {{Carithers}},
  \citenamefont {{Croft}}, \citenamefont {{Dawson}}, \citenamefont {{Delubac}},
  \citenamefont {{Font-Ribera}}, \citenamefont {{Ho}}, \citenamefont
  {{Kirkby}}, \citenamefont {{Lee}}, \citenamefont {{Margala}}, \citenamefont
  {{Miralda-Escud{\'e}}}, \citenamefont {{Muna}}, \citenamefont {{Myers}},
  \citenamefont {{Noterdaeme}}, \citenamefont {{P{\^a}ris}}, \citenamefont
  {{Petitjean}}, \citenamefont {{Pieri}}, \citenamefont {{Rich}}, \citenamefont
  {{Rollinde}}, \citenamefont {{Ross}}, \citenamefont {{Schlegel}},
  \citenamefont {{Schneider}}, \citenamefont {{Slosar}},\ and\ \citenamefont
  {{Weinberg}}}]{2013A&A...559A..85P}%
  \BibitemOpen
  \bibfield  {author} {\bibinfo {author} {\bibfnamefont {N.}~\bibnamefont
  {{Palanque-Delabrouille}}}, \bibinfo {author} {\bibfnamefont
  {C.}~\bibnamefont {{Y{\`e}che}}}, \bibinfo {author} {\bibfnamefont
  {A.}~\bibnamefont {{Borde}}}, \bibinfo {author} {\bibfnamefont {J.-M.}\
  \bibnamefont {{Le Goff}}}, \bibinfo {author} {\bibfnamefont {G.}~\bibnamefont
  {{Rossi}}}, \bibinfo {author} {\bibfnamefont {M.}~\bibnamefont {{Viel}}},
  \bibinfo {author} {\bibfnamefont {{\'E}.}~\bibnamefont {{Aubourg}}}, \bibinfo
  {author} {\bibfnamefont {S.}~\bibnamefont {{Bailey}}}, \bibinfo {author}
  {\bibfnamefont {J.}~\bibnamefont {{Bautista}}}, \bibinfo {author}
  {\bibfnamefont {M.}~\bibnamefont {{Blomqvist}}}, \bibinfo {author}
  {\bibfnamefont {A.}~\bibnamefont {{Bolton}}}, \bibinfo {author}
  {\bibfnamefont {J.~S.}\ \bibnamefont {{Bolton}}}, \bibinfo {author}
  {\bibfnamefont {N.~G.}\ \bibnamefont {{Busca}}}, \bibinfo {author}
  {\bibfnamefont {B.}~\bibnamefont {{Carithers}}}, \bibinfo {author}
  {\bibfnamefont {R.~A.~C.}\ \bibnamefont {{Croft}}}, \bibinfo {author}
  {\bibfnamefont {K.~S.}\ \bibnamefont {{Dawson}}}, \bibinfo {author}
  {\bibfnamefont {T.}~\bibnamefont {{Delubac}}}, \bibinfo {author}
  {\bibfnamefont {A.}~\bibnamefont {{Font-Ribera}}}, \bibinfo {author}
  {\bibfnamefont {S.}~\bibnamefont {{Ho}}}, \bibinfo {author} {\bibfnamefont
  {D.}~\bibnamefont {{Kirkby}}}, \bibinfo {author} {\bibfnamefont {K.-G.}\
  \bibnamefont {{Lee}}}, \bibinfo {author} {\bibfnamefont {D.}~\bibnamefont
  {{Margala}}}, \bibinfo {author} {\bibfnamefont {J.}~\bibnamefont
  {{Miralda-Escud{\'e}}}}, \bibinfo {author} {\bibfnamefont {D.}~\bibnamefont
  {{Muna}}}, \bibinfo {author} {\bibfnamefont {A.~D.}\ \bibnamefont {{Myers}}},
  \bibinfo {author} {\bibfnamefont {P.}~\bibnamefont {{Noterdaeme}}}, \bibinfo
  {author} {\bibfnamefont {I.}~\bibnamefont {{P{\^a}ris}}}, \bibinfo {author}
  {\bibfnamefont {P.}~\bibnamefont {{Petitjean}}}, \bibinfo {author}
  {\bibfnamefont {M.~M.}\ \bibnamefont {{Pieri}}}, \bibinfo {author}
  {\bibfnamefont {J.}~\bibnamefont {{Rich}}}, \bibinfo {author} {\bibfnamefont
  {E.}~\bibnamefont {{Rollinde}}}, \bibinfo {author} {\bibfnamefont {N.~P.}\
  \bibnamefont {{Ross}}}, \bibinfo {author} {\bibfnamefont {D.~J.}\
  \bibnamefont {{Schlegel}}}, \bibinfo {author} {\bibfnamefont {D.~P.}\
  \bibnamefont {{Schneider}}}, \bibinfo {author} {\bibfnamefont
  {A.}~\bibnamefont {{Slosar}}}, \ and\ \bibinfo {author} {\bibfnamefont
  {D.~H.}\ \bibnamefont {{Weinberg}}},\ }\href {\doibase
  10.1051/0004-6361/201322130} {\bibfield  {journal} {\bibinfo  {journal}
  {\aap}\ }\textbf {\bibinfo {volume} {559}},\ \bibinfo {eid} {A85} (\bibinfo
  {year} {2013}{\natexlab{a}})},\ \Eprint {http://arxiv.org/abs/1306.5896}
  {arXiv:1306.5896 [astro-ph.CO]} \BibitemShut {NoStop}%
\bibitem [{\citenamefont {{Bird}}\ \emph {et~al.}(2011)\citenamefont {{Bird}},
  \citenamefont {{Peiris}}, \citenamefont {{Viel}},\ and\ \citenamefont
  {{Verde}}}]{2011MNRAS.413.1717B}%
  \BibitemOpen
  \bibfield  {author} {\bibinfo {author} {\bibfnamefont {S.}~\bibnamefont
  {{Bird}}}, \bibinfo {author} {\bibfnamefont {H.~V.}\ \bibnamefont
  {{Peiris}}}, \bibinfo {author} {\bibfnamefont {M.}~\bibnamefont {{Viel}}}, \
  and\ \bibinfo {author} {\bibfnamefont {L.}~\bibnamefont {{Verde}}},\ }\href
  {\doibase 10.1111/j.1365-2966.2011.18245.x} {\bibfield  {journal} {\bibinfo
  {journal} {\mnras}\ }\textbf {\bibinfo {volume} {413}},\ \bibinfo {pages}
  {1717} (\bibinfo {year} {2011})},\ \Eprint {http://arxiv.org/abs/1010.1519}
  {arXiv:1010.1519 [astro-ph.CO]} \BibitemShut {NoStop}%
\bibitem [{\citenamefont {{Afshordi}}\ \emph {et~al.}(2009)\citenamefont
  {{Afshordi}}, \citenamefont {{Geshnizjani}},\ and\ \citenamefont
  {{Khoury}}}]{2009JCAP...08..030A}%
  \BibitemOpen
  \bibfield  {author} {\bibinfo {author} {\bibfnamefont {N.}~\bibnamefont
  {{Afshordi}}}, \bibinfo {author} {\bibfnamefont {G.}~\bibnamefont
  {{Geshnizjani}}}, \ and\ \bibinfo {author} {\bibfnamefont {J.}~\bibnamefont
  {{Khoury}}},\ }\href {\doibase 10.1088/1475-7516/2009/08/030} {\bibfield
  {journal} {\bibinfo  {journal} {\jcap}\ }\textbf {\bibinfo {volume} {8}},\
  \bibinfo {eid} {030} (\bibinfo {year} {2009})},\ \Eprint
  {http://arxiv.org/abs/0812.2244} {arXiv:0812.2244} \BibitemShut {NoStop}%
\bibitem [{\citenamefont {{Seljak}}\ \emph
  {et~al.}(2006{\natexlab{a}})\citenamefont {{Seljak}}, \citenamefont
  {{Slosar}},\ and\ \citenamefont {{McDonald}}}]{2006JCAP...10..014S}%
  \BibitemOpen
  \bibfield  {author} {\bibinfo {author} {\bibfnamefont {U.}~\bibnamefont
  {{Seljak}}}, \bibinfo {author} {\bibfnamefont {A.}~\bibnamefont {{Slosar}}},
  \ and\ \bibinfo {author} {\bibfnamefont {P.}~\bibnamefont {{McDonald}}},\
  }\href {\doibase 10.1088/1475-7516/2006/10/014} {\bibfield  {journal}
  {\bibinfo  {journal} {Journal of Cosmology and Astro-Particle Physics}\
  }\textbf {\bibinfo {volume} {10}},\ \bibinfo {pages} {14} (\bibinfo {year}
  {2006}{\natexlab{a}})}\BibitemShut {NoStop}%
\bibitem [{\citenamefont {{Viel}}\ \emph {et~al.}(2006)\citenamefont {{Viel}},
  \citenamefont {{Haehnelt}},\ and\ \citenamefont
  {{Lewis}}}]{2006MNRAS.370L..51V}%
  \BibitemOpen
  \bibfield  {author} {\bibinfo {author} {\bibfnamefont {M.}~\bibnamefont
  {{Viel}}}, \bibinfo {author} {\bibfnamefont {M.~G.}\ \bibnamefont
  {{Haehnelt}}}, \ and\ \bibinfo {author} {\bibfnamefont {A.}~\bibnamefont
  {{Lewis}}},\ }\href {\doibase 10.1111/j.1745-3933.2006.00187.x} {\bibfield
  {journal} {\bibinfo  {journal} {\mnras}\ }\textbf {\bibinfo {volume} {370}},\
  \bibinfo {pages} {L51} (\bibinfo {year} {2006})},\ \Eprint
  {http://arxiv.org/abs/arXiv:astro-ph/0604310} {arXiv:astro-ph/0604310}
  \BibitemShut {NoStop}%
\bibitem [{\citenamefont {{Goobar}}\ \emph {et~al.}(2006)\citenamefont
  {{Goobar}}, \citenamefont {{Hannestad}}, \citenamefont {{M{\"o}rtsell}},\
  and\ \citenamefont {{Tu}}}]{2006JCAP...06..019G}%
  \BibitemOpen
  \bibfield  {author} {\bibinfo {author} {\bibfnamefont {A.}~\bibnamefont
  {{Goobar}}}, \bibinfo {author} {\bibfnamefont {S.}~\bibnamefont
  {{Hannestad}}}, \bibinfo {author} {\bibfnamefont {E.}~\bibnamefont
  {{M{\"o}rtsell}}}, \ and\ \bibinfo {author} {\bibfnamefont {H.}~\bibnamefont
  {{Tu}}},\ }\href {\doibase 10.1088/1475-7516/2006/06/019} {\bibfield
  {journal} {\bibinfo  {journal} {\jcap}\ }\textbf {\bibinfo {volume} {6}},\
  \bibinfo {eid} {019} (\bibinfo {year} {2006})},\ \Eprint
  {http://arxiv.org/abs/arXiv:astro-ph/0602155} {arXiv:astro-ph/0602155}
  \BibitemShut {NoStop}%
\bibitem [{\citenamefont {{Seljak}}\ \emph {et~al.}(2005)\citenamefont
  {{Seljak}}, \citenamefont {{Makarov}}, \citenamefont {{McDonald}},
  \citenamefont {{Anderson}}, \citenamefont {{Bahcall}}, \citenamefont
  {{Brinkmann}}, \citenamefont {{Burles}}, \citenamefont {{Cen}}, \citenamefont
  {{Doi}}, \citenamefont {{Gunn}}, \citenamefont {{Ivezi{\' c}}}, \citenamefont
  {{Kent}}, \citenamefont {{Loveday}}, \citenamefont {{Lupton}}, \citenamefont
  {{Munn}}, \citenamefont {{Nichol}}, \citenamefont {{Ostriker}}, \citenamefont
  {{Schlegel}}, \citenamefont {{Schneider}}, \citenamefont {{Tegmark}},
  \citenamefont {{Berk}}, \citenamefont {{Weinberg}},\ and\ \citenamefont
  {{York}}}]{2005PhRvD..71j3515S}%
  \BibitemOpen
  \bibfield  {author} {\bibinfo {author} {\bibfnamefont {U.}~\bibnamefont
  {{Seljak}}}, \bibinfo {author} {\bibfnamefont {A.}~\bibnamefont {{Makarov}}},
  \bibinfo {author} {\bibfnamefont {P.}~\bibnamefont {{McDonald}}}, \bibinfo
  {author} {\bibfnamefont {S.~F.}\ \bibnamefont {{Anderson}}}, \bibinfo
  {author} {\bibfnamefont {N.~A.}\ \bibnamefont {{Bahcall}}}, \bibinfo {author}
  {\bibfnamefont {J.}~\bibnamefont {{Brinkmann}}}, \bibinfo {author}
  {\bibfnamefont {S.}~\bibnamefont {{Burles}}}, \bibinfo {author}
  {\bibfnamefont {R.}~\bibnamefont {{Cen}}}, \bibinfo {author} {\bibfnamefont
  {M.}~\bibnamefont {{Doi}}}, \bibinfo {author} {\bibfnamefont {J.~E.}\
  \bibnamefont {{Gunn}}}, \bibinfo {author} {\bibfnamefont {{\v
  Z}.}~\bibnamefont {{Ivezi{\' c}}}}, \bibinfo {author} {\bibfnamefont
  {S.}~\bibnamefont {{Kent}}}, \bibinfo {author} {\bibfnamefont
  {J.}~\bibnamefont {{Loveday}}}, \bibinfo {author} {\bibfnamefont {R.~H.}\
  \bibnamefont {{Lupton}}}, \bibinfo {author} {\bibfnamefont {J.~A.}\
  \bibnamefont {{Munn}}}, \bibinfo {author} {\bibfnamefont {R.~C.}\
  \bibnamefont {{Nichol}}}, \bibinfo {author} {\bibfnamefont {J.~P.}\
  \bibnamefont {{Ostriker}}}, \bibinfo {author} {\bibfnamefont {D.~J.}\
  \bibnamefont {{Schlegel}}}, \bibinfo {author} {\bibfnamefont {D.~P.}\
  \bibnamefont {{Schneider}}}, \bibinfo {author} {\bibfnamefont
  {M.}~\bibnamefont {{Tegmark}}}, \bibinfo {author} {\bibfnamefont {D.~E.}\
  \bibnamefont {{Berk}}}, \bibinfo {author} {\bibfnamefont {D.~H.}\
  \bibnamefont {{Weinberg}}}, \ and\ \bibinfo {author} {\bibfnamefont {D.~G.}\
  \bibnamefont {{York}}},\ }\href@noop {} {\bibfield  {journal} {\bibinfo
  {journal} {\prd}\ }\textbf {\bibinfo {volume} {71}},\ \bibinfo {pages}
  {103515} (\bibinfo {year} {2005})}\BibitemShut {NoStop}%
\bibitem [{\citenamefont {{Viel}}\ \emph {et~al.}(2005)\citenamefont {{Viel}},
  \citenamefont {{Lesgourgues}}, \citenamefont {{Haehnelt}}, \citenamefont
  {{Matarrese}},\ and\ \citenamefont {{Riotto}}}]{2005PhRvD..71f3534V}%
  \BibitemOpen
  \bibfield  {author} {\bibinfo {author} {\bibfnamefont {M.}~\bibnamefont
  {{Viel}}}, \bibinfo {author} {\bibfnamefont {J.}~\bibnamefont
  {{Lesgourgues}}}, \bibinfo {author} {\bibfnamefont {M.~G.}\ \bibnamefont
  {{Haehnelt}}}, \bibinfo {author} {\bibfnamefont {S.}~\bibnamefont
  {{Matarrese}}}, \ and\ \bibinfo {author} {\bibfnamefont {A.}~\bibnamefont
  {{Riotto}}},\ }\href {\doibase 10.1103/PhysRevD.71.063534} {\bibfield
  {journal} {\bibinfo  {journal} {\prd}\ }\textbf {\bibinfo {volume} {71}},\
  \bibinfo {pages} {063534} (\bibinfo {year} {2005})}\BibitemShut {NoStop}%
\bibitem [{\citenamefont {Gratton}\ \emph {et~al.}(2008)\citenamefont
  {Gratton}, \citenamefont {Lewis},\ and\ \citenamefont
  {Efstathiou}}]{Gratton:2007tb}%
  \BibitemOpen
  \bibfield  {author} {\bibinfo {author} {\bibfnamefont {S.}~\bibnamefont
  {Gratton}}, \bibinfo {author} {\bibfnamefont {A.}~\bibnamefont {Lewis}}, \
  and\ \bibinfo {author} {\bibfnamefont {G.}~\bibnamefont {Efstathiou}},\
  }\href {\doibase 10.1103/PhysRevD.77.083507} {\bibfield  {journal} {\bibinfo
  {journal} {Phys.Rev.}\ }\textbf {\bibinfo {volume} {D77}},\ \bibinfo {pages}
  {083507} (\bibinfo {year} {2008})},\ \Eprint {http://arxiv.org/abs/0705.3100}
  {arXiv:0705.3100 [astro-ph]} \BibitemShut {NoStop}%
\bibitem [{\citenamefont {{McDonald}}\ \emph {et~al.}(2001)\citenamefont
  {{McDonald}}, \citenamefont {{Miralda-Escud{\' e}}}, \citenamefont {{Rauch}},
  \citenamefont {{Sargent}}, \citenamefont {{Barlow}},\ and\ \citenamefont
  {{Cen}}}]{2001ApJ...562...52M}%
  \BibitemOpen
  \bibfield  {author} {\bibinfo {author} {\bibfnamefont {P.}~\bibnamefont
  {{McDonald}}}, \bibinfo {author} {\bibfnamefont {J.}~\bibnamefont
  {{Miralda-Escud{\' e}}}}, \bibinfo {author} {\bibfnamefont {M.}~\bibnamefont
  {{Rauch}}}, \bibinfo {author} {\bibfnamefont {W.~L.~W.}\ \bibnamefont
  {{Sargent}}}, \bibinfo {author} {\bibfnamefont {T.~A.}\ \bibnamefont
  {{Barlow}}}, \ and\ \bibinfo {author} {\bibfnamefont {R.}~\bibnamefont
  {{Cen}}},\ }\href@noop {} {\bibfield  {journal} {\bibinfo  {journal} {\apj}\
  }\textbf {\bibinfo {volume} {562}},\ \bibinfo {pages} {52} (\bibinfo {year}
  {2001})}\BibitemShut {NoStop}%
\bibitem [{\citenamefont {{Rudie}}\ \emph {et~al.}(2012)\citenamefont
  {{Rudie}}, \citenamefont {{Steidel}},\ and\ \citenamefont
  {{Pettini}}}]{2012ApJ...757L..30R}%
  \BibitemOpen
  \bibfield  {author} {\bibinfo {author} {\bibfnamefont {G.~C.}\ \bibnamefont
  {{Rudie}}}, \bibinfo {author} {\bibfnamefont {C.~C.}\ \bibnamefont
  {{Steidel}}}, \ and\ \bibinfo {author} {\bibfnamefont {M.}~\bibnamefont
  {{Pettini}}},\ }\href {\doibase 10.1088/2041-8205/757/2/L30} {\bibfield
  {journal} {\bibinfo  {journal} {\apjl}\ }\textbf {\bibinfo {volume} {757}},\
  \bibinfo {eid} {L30} (\bibinfo {year} {2012})},\ \Eprint
  {http://arxiv.org/abs/1209.0005} {arXiv:1209.0005 [astro-ph.CO]} \BibitemShut
  {NoStop}%
\bibitem [{\citenamefont {{Faucher-Gigu{\`e}re}}\ \emph
  {et~al.}(2008)\citenamefont {{Faucher-Gigu{\`e}re}}, \citenamefont
  {{Prochaska}}, \citenamefont {{Lidz}}, \citenamefont {{Hernquist}},\ and\
  \citenamefont {{Zaldarriaga}}}]{2008ApJ...681..831F}%
  \BibitemOpen
  \bibfield  {author} {\bibinfo {author} {\bibfnamefont {C.-A.}\ \bibnamefont
  {{Faucher-Gigu{\`e}re}}}, \bibinfo {author} {\bibfnamefont {J.~X.}\
  \bibnamefont {{Prochaska}}}, \bibinfo {author} {\bibfnamefont
  {A.}~\bibnamefont {{Lidz}}}, \bibinfo {author} {\bibfnamefont
  {L.}~\bibnamefont {{Hernquist}}}, \ and\ \bibinfo {author} {\bibfnamefont
  {M.}~\bibnamefont {{Zaldarriaga}}},\ }\href {\doibase 10.1086/588648}
  {\bibfield  {journal} {\bibinfo  {journal} {\apj}\ }\textbf {\bibinfo
  {volume} {681}},\ \bibinfo {pages} {831} (\bibinfo {year} {2008})},\ \Eprint
  {http://arxiv.org/abs/0709.2382} {arXiv:0709.2382} \BibitemShut {NoStop}%
\bibitem [{\citenamefont {{McDonald}}\ \emph {et~al.}(2000)\citenamefont
  {{McDonald}}, \citenamefont {{Miralda-Escud{\' e}}}, \citenamefont {{Rauch}},
  \citenamefont {{Sargent}}, \citenamefont {{Barlow}}, \citenamefont {{Cen}},\
  and\ \citenamefont {{Ostriker}}}]{2000ApJ...543....1M}%
  \BibitemOpen
  \bibfield  {author} {\bibinfo {author} {\bibfnamefont {P.}~\bibnamefont
  {{McDonald}}}, \bibinfo {author} {\bibfnamefont {J.}~\bibnamefont
  {{Miralda-Escud{\' e}}}}, \bibinfo {author} {\bibfnamefont {M.}~\bibnamefont
  {{Rauch}}}, \bibinfo {author} {\bibfnamefont {W.~L.~W.}\ \bibnamefont
  {{Sargent}}}, \bibinfo {author} {\bibfnamefont {T.~A.}\ \bibnamefont
  {{Barlow}}}, \bibinfo {author} {\bibfnamefont {R.}~\bibnamefont {{Cen}}}, \
  and\ \bibinfo {author} {\bibfnamefont {J.~P.}\ \bibnamefont {{Ostriker}}},\
  }\href@noop {} {\bibfield  {journal} {\bibinfo  {journal} {\apj}\ }\textbf
  {\bibinfo {volume} {543}},\ \bibinfo {pages} {1} (\bibinfo {year}
  {2000})}\BibitemShut {NoStop}%
\bibitem [{\citenamefont {{Kim}}\ \emph {et~al.}(2004)\citenamefont {{Kim}},
  \citenamefont {{Viel}}, \citenamefont {{Haehnelt}}, \citenamefont
  {{Carswell}},\ and\ \citenamefont {{Cristiani}}}]{2004MNRAS.347..355K}%
  \BibitemOpen
  \bibfield  {author} {\bibinfo {author} {\bibfnamefont {T.}~\bibnamefont
  {{Kim}}}, \bibinfo {author} {\bibfnamefont {M.}~\bibnamefont {{Viel}}},
  \bibinfo {author} {\bibfnamefont {M.~G.}\ \bibnamefont {{Haehnelt}}},
  \bibinfo {author} {\bibfnamefont {R.~F.}\ \bibnamefont {{Carswell}}}, \ and\
  \bibinfo {author} {\bibfnamefont {S.}~\bibnamefont {{Cristiani}}},\ }\href
  {\doibase 10.1111/j.1365-2966.2004.07221.x} {\bibfield  {journal} {\bibinfo
  {journal} {\mnras}\ }\textbf {\bibinfo {volume} {347}},\ \bibinfo {pages}
  {355} (\bibinfo {year} {2004})},\ \Eprint
  {http://arxiv.org/abs/arXiv:astro-ph/0308103} {arXiv:astro-ph/0308103}
  \BibitemShut {NoStop}%
\bibitem [{\citenamefont {{McDonald}}\ \emph
  {et~al.}(2005{\natexlab{a}})\citenamefont {{McDonald}}, \citenamefont
  {{Seljak}}, \citenamefont {{Cen}}, \citenamefont {{Shih}}, \citenamefont
  {{Weinberg}}, \citenamefont {{Burles}}, \citenamefont {{Schneider}},
  \citenamefont {{Schlegel}}, \citenamefont {{Bahcall}}, \citenamefont
  {{Briggs}}, \citenamefont {{Brinkmann}}, \citenamefont {{Fukugita}},
  \citenamefont {{Ivezi{\'c}}}, \citenamefont {{Kent}},\ and\ \citenamefont
  {{Vanden Berk}}}]{2005ApJ...635..761M}%
  \BibitemOpen
  \bibfield  {author} {\bibinfo {author} {\bibfnamefont {P.}~\bibnamefont
  {{McDonald}}}, \bibinfo {author} {\bibfnamefont {U.}~\bibnamefont
  {{Seljak}}}, \bibinfo {author} {\bibfnamefont {R.}~\bibnamefont {{Cen}}},
  \bibinfo {author} {\bibfnamefont {D.}~\bibnamefont {{Shih}}}, \bibinfo
  {author} {\bibfnamefont {D.~H.}\ \bibnamefont {{Weinberg}}}, \bibinfo
  {author} {\bibfnamefont {S.}~\bibnamefont {{Burles}}}, \bibinfo {author}
  {\bibfnamefont {D.~P.}\ \bibnamefont {{Schneider}}}, \bibinfo {author}
  {\bibfnamefont {D.~J.}\ \bibnamefont {{Schlegel}}}, \bibinfo {author}
  {\bibfnamefont {N.~A.}\ \bibnamefont {{Bahcall}}}, \bibinfo {author}
  {\bibfnamefont {J.~W.}\ \bibnamefont {{Briggs}}}, \bibinfo {author}
  {\bibfnamefont {J.}~\bibnamefont {{Brinkmann}}}, \bibinfo {author}
  {\bibfnamefont {M.}~\bibnamefont {{Fukugita}}}, \bibinfo {author}
  {\bibfnamefont {{\v Z}.}~\bibnamefont {{Ivezi{\'c}}}}, \bibinfo {author}
  {\bibfnamefont {S.}~\bibnamefont {{Kent}}}, \ and\ \bibinfo {author}
  {\bibfnamefont {D.~E.}\ \bibnamefont {{Vanden Berk}}},\ }\href {\doibase
  10.1086/497563} {\bibfield  {journal} {\bibinfo  {journal} {\apj}\ }\textbf
  {\bibinfo {volume} {635}},\ \bibinfo {pages} {761} (\bibinfo {year}
  {2005}{\natexlab{a}})}\BibitemShut {NoStop}%
\bibitem [{\citenamefont {{McDonald}}\ \emph
  {et~al.}(2005{\natexlab{b}})\citenamefont {{McDonald}}, \citenamefont
  {{Seljak}}, \citenamefont {{Cen}}, \citenamefont {{Bode}},\ and\
  \citenamefont {{Ostriker}}}]{2005MNRAS.360.1471M}%
  \BibitemOpen
  \bibfield  {author} {\bibinfo {author} {\bibfnamefont {P.}~\bibnamefont
  {{McDonald}}}, \bibinfo {author} {\bibfnamefont {U.}~\bibnamefont
  {{Seljak}}}, \bibinfo {author} {\bibfnamefont {R.}~\bibnamefont {{Cen}}},
  \bibinfo {author} {\bibfnamefont {P.}~\bibnamefont {{Bode}}}, \ and\ \bibinfo
  {author} {\bibfnamefont {J.~P.}\ \bibnamefont {{Ostriker}}},\ }\href
  {\doibase 10.1111/j.1365-2966.2005.09141.x} {\bibfield  {journal} {\bibinfo
  {journal} {\mnras}\ }\textbf {\bibinfo {volume} {360}},\ \bibinfo {pages}
  {1471} (\bibinfo {year} {2005}{\natexlab{b}})}\BibitemShut {NoStop}%
\bibitem [{\citenamefont {{Viel}}\ and\ \citenamefont
  {{Haehnelt}}(2006)}]{2006MNRAS.365..231V}%
  \BibitemOpen
  \bibfield  {author} {\bibinfo {author} {\bibfnamefont {M.}~\bibnamefont
  {{Viel}}}\ and\ \bibinfo {author} {\bibfnamefont {M.~G.}\ \bibnamefont
  {{Haehnelt}}},\ }\href {\doibase 10.1111/j.1365-2966.2005.09703.x} {\bibfield
   {journal} {\bibinfo  {journal} {\mnras}\ }\textbf {\bibinfo {volume}
  {365}},\ \bibinfo {pages} {231} (\bibinfo {year} {2006})}\BibitemShut
  {NoStop}%
\bibitem [{\citenamefont {{Regan}}\ \emph {et~al.}(2007)\citenamefont
  {{Regan}}, \citenamefont {{Haehnelt}},\ and\ \citenamefont
  {{Viel}}}]{2007MNRAS.374..196R}%
  \BibitemOpen
  \bibfield  {author} {\bibinfo {author} {\bibfnamefont {J.~A.}\ \bibnamefont
  {{Regan}}}, \bibinfo {author} {\bibfnamefont {M.~G.}\ \bibnamefont
  {{Haehnelt}}}, \ and\ \bibinfo {author} {\bibfnamefont {M.}~\bibnamefont
  {{Viel}}},\ }\href {\doibase 10.1111/j.1365-2966.2006.11132.x} {\bibfield
  {journal} {\bibinfo  {journal} {\mnras}\ }\textbf {\bibinfo {volume} {374}},\
  \bibinfo {pages} {196} (\bibinfo {year} {2007})},\ \Eprint
  {http://arxiv.org/abs/arXiv:astro-ph/0606638} {arXiv:astro-ph/0606638}
  \BibitemShut {NoStop}%
\bibitem [{\citenamefont {{Mandelbaum}}\ \emph {et~al.}(2003)\citenamefont
  {{Mandelbaum}}, \citenamefont {{McDonald}}, \citenamefont {{Seljak}},\ and\
  \citenamefont {{Cen}}}]{2003MNRAS.344..776M}%
  \BibitemOpen
  \bibfield  {author} {\bibinfo {author} {\bibfnamefont {R.}~\bibnamefont
  {{Mandelbaum}}}, \bibinfo {author} {\bibfnamefont {P.}~\bibnamefont
  {{McDonald}}}, \bibinfo {author} {\bibfnamefont {U.}~\bibnamefont
  {{Seljak}}}, \ and\ \bibinfo {author} {\bibfnamefont {R.}~\bibnamefont
  {{Cen}}},\ }\href@noop {} {\bibfield  {journal} {\bibinfo  {journal}
  {\mnras}\ }\textbf {\bibinfo {volume} {344}},\ \bibinfo {pages} {776}
  (\bibinfo {year} {2003})}\BibitemShut {NoStop}%
\bibitem [{\citenamefont {{Viel}}\ \emph {et~al.}(2004)\citenamefont {{Viel}},
  \citenamefont {{Matarrese}}, \citenamefont {{Heavens}}, \citenamefont
  {{Haehnelt}}, \citenamefont {{Kim}}, \citenamefont {{Springel}},\ and\
  \citenamefont {{Hernquist}}}]{2004MNRAS.347L..26V}%
  \BibitemOpen
  \bibfield  {author} {\bibinfo {author} {\bibfnamefont {M.}~\bibnamefont
  {{Viel}}}, \bibinfo {author} {\bibfnamefont {S.}~\bibnamefont {{Matarrese}}},
  \bibinfo {author} {\bibfnamefont {A.}~\bibnamefont {{Heavens}}}, \bibinfo
  {author} {\bibfnamefont {M.~G.}\ \bibnamefont {{Haehnelt}}}, \bibinfo
  {author} {\bibfnamefont {T.-S.}\ \bibnamefont {{Kim}}}, \bibinfo {author}
  {\bibfnamefont {V.}~\bibnamefont {{Springel}}}, \ and\ \bibinfo {author}
  {\bibfnamefont {L.}~\bibnamefont {{Hernquist}}},\ }\href@noop {} {\bibfield
  {journal} {\bibinfo  {journal} {\mnras}\ }\textbf {\bibinfo {volume} {347}},\
  \bibinfo {pages} {L26} (\bibinfo {year} {2004})}\BibitemShut {NoStop}%
\bibitem [{\citenamefont {{Font-Ribera}}\ \emph {et~al.}(2013)\citenamefont
  {{Font-Ribera}}, \citenamefont {{Arnau}}, \citenamefont
  {{Miralda-Escud{\'e}}}, \citenamefont {{Rollinde}}, \citenamefont
  {{Brinkmann}}, \citenamefont {{Brownstein}}, \citenamefont {{Lee}},
  \citenamefont {{Myers}}, \citenamefont {{Palanque-Delabrouille}},
  \citenamefont {{P{\^a}ris}}, \citenamefont {{Petitjean}}, \citenamefont
  {{Rich}}, \citenamefont {{Ross}}, \citenamefont {{Schneider}},\ and\
  \citenamefont {{White}}}]{2013JCAP...05..018F}%
  \BibitemOpen
  \bibfield  {author} {\bibinfo {author} {\bibfnamefont {A.}~\bibnamefont
  {{Font-Ribera}}}, \bibinfo {author} {\bibfnamefont {E.}~\bibnamefont
  {{Arnau}}}, \bibinfo {author} {\bibfnamefont {J.}~\bibnamefont
  {{Miralda-Escud{\'e}}}}, \bibinfo {author} {\bibfnamefont {E.}~\bibnamefont
  {{Rollinde}}}, \bibinfo {author} {\bibfnamefont {J.}~\bibnamefont
  {{Brinkmann}}}, \bibinfo {author} {\bibfnamefont {J.~R.}\ \bibnamefont
  {{Brownstein}}}, \bibinfo {author} {\bibfnamefont {K.-G.}\ \bibnamefont
  {{Lee}}}, \bibinfo {author} {\bibfnamefont {A.~D.}\ \bibnamefont {{Myers}}},
  \bibinfo {author} {\bibfnamefont {N.}~\bibnamefont
  {{Palanque-Delabrouille}}}, \bibinfo {author} {\bibfnamefont
  {I.}~\bibnamefont {{P{\^a}ris}}}, \bibinfo {author} {\bibfnamefont
  {P.}~\bibnamefont {{Petitjean}}}, \bibinfo {author} {\bibfnamefont
  {J.}~\bibnamefont {{Rich}}}, \bibinfo {author} {\bibfnamefont {N.~P.}\
  \bibnamefont {{Ross}}}, \bibinfo {author} {\bibfnamefont {D.~P.}\
  \bibnamefont {{Schneider}}}, \ and\ \bibinfo {author} {\bibfnamefont
  {M.}~\bibnamefont {{White}}},\ }\href {\doibase
  10.1088/1475-7516/2013/05/018} {\bibfield  {journal} {\bibinfo  {journal}
  {\jcap}\ }\textbf {\bibinfo {volume} {5}},\ \bibinfo {eid} {018} (\bibinfo
  {year} {2013})},\ \Eprint {http://arxiv.org/abs/1303.1937} {arXiv:1303.1937
  [astro-ph.CO]} \BibitemShut {NoStop}%
\bibitem [{\citenamefont {{Hamaus}}\ \emph {et~al.}(2010)\citenamefont
  {{Hamaus}}, \citenamefont {{Seljak}}, \citenamefont {{Desjacques}},
  \citenamefont {{Smith}},\ and\ \citenamefont
  {{Baldauf}}}]{2010PhRvD..82d3515H}%
  \BibitemOpen
  \bibfield  {author} {\bibinfo {author} {\bibfnamefont {N.}~\bibnamefont
  {{Hamaus}}}, \bibinfo {author} {\bibfnamefont {U.}~\bibnamefont {{Seljak}}},
  \bibinfo {author} {\bibfnamefont {V.}~\bibnamefont {{Desjacques}}}, \bibinfo
  {author} {\bibfnamefont {R.~E.}\ \bibnamefont {{Smith}}}, \ and\ \bibinfo
  {author} {\bibfnamefont {T.}~\bibnamefont {{Baldauf}}},\ }\href {\doibase
  10.1103/PhysRevD.82.043515} {\bibfield  {journal} {\bibinfo  {journal}
  {\prd}\ }\textbf {\bibinfo {volume} {82}},\ \bibinfo {pages} {043515}
  (\bibinfo {year} {2010})},\ \Eprint {http://arxiv.org/abs/1004.5377}
  {arXiv:1004.5377 [astro-ph.CO]} \BibitemShut {NoStop}%
\bibitem [{\citenamefont {{Hamaus}}\ \emph {et~al.}(2012)\citenamefont
  {{Hamaus}}, \citenamefont {{Seljak}},\ and\ \citenamefont
  {{Desjacques}}}]{2012PhRvD..86j3513H}%
  \BibitemOpen
  \bibfield  {author} {\bibinfo {author} {\bibfnamefont {N.}~\bibnamefont
  {{Hamaus}}}, \bibinfo {author} {\bibfnamefont {U.}~\bibnamefont {{Seljak}}},
  \ and\ \bibinfo {author} {\bibfnamefont {V.}~\bibnamefont {{Desjacques}}},\
  }\href {\doibase 10.1103/PhysRevD.86.103513} {\bibfield  {journal} {\bibinfo
  {journal} {\prd}\ }\textbf {\bibinfo {volume} {86}},\ \bibinfo {eid} {103513}
  (\bibinfo {year} {2012})},\ \Eprint {http://arxiv.org/abs/1207.1102}
  {arXiv:1207.1102 [astro-ph.CO]} \BibitemShut {NoStop}%
\bibitem [{\citenamefont {{Seljak}}\ \emph {et~al.}(2009)\citenamefont
  {{Seljak}}, \citenamefont {{Hamaus}},\ and\ \citenamefont
  {{Desjacques}}}]{2009PhRvL.103i1303S}%
  \BibitemOpen
  \bibfield  {author} {\bibinfo {author} {\bibfnamefont {U.}~\bibnamefont
  {{Seljak}}}, \bibinfo {author} {\bibfnamefont {N.}~\bibnamefont {{Hamaus}}},
  \ and\ \bibinfo {author} {\bibfnamefont {V.}~\bibnamefont {{Desjacques}}},\
  }\href {\doibase 10.1103/PhysRevLett.103.091303} {\bibfield  {journal}
  {\bibinfo  {journal} {Physical Review Letters}\ }\textbf {\bibinfo {volume}
  {103}},\ \bibinfo {eid} {091303} (\bibinfo {year} {2009})},\ \Eprint
  {http://arxiv.org/abs/0904.2963} {arXiv:0904.2963 [astro-ph.CO]} \BibitemShut
  {NoStop}%
\bibitem [{\citenamefont {{Dawson}}\ \emph {et~al.}(2013)\citenamefont
  {{Dawson}}, \citenamefont {{Schlegel}}, \citenamefont {{Ahn}}, \citenamefont
  {{Anderson}}, \citenamefont {{Aubourg}}, \citenamefont {{Bailey}},
  \citenamefont {{Barkhouser}}, \citenamefont {{Bautista}}, \citenamefont
  {{Beifiori}}, \citenamefont {{Berlind}}, \citenamefont {{Bhardwaj}},
  \citenamefont {{Bizyaev}}, \citenamefont {{Blake}}, \citenamefont
  {{Blanton}}, \citenamefont {{Blomqvist}}, \citenamefont {{Bolton}},
  \citenamefont {{Borde}}, \citenamefont {{Bovy}}, \citenamefont {{Brandt}},
  \citenamefont {{Brewington}}, \citenamefont {{Brinkmann}}, \citenamefont
  {{Brown}}, \citenamefont {{Brownstein}}, \citenamefont {{Bundy}},
  \citenamefont {{Busca}}, \citenamefont {{Carithers}}, \citenamefont
  {{Carnero}}, \citenamefont {{Carr}}, \citenamefont {{Chen}}, \citenamefont
  {{Comparat}}, \citenamefont {{Connolly}}, \citenamefont {{Cope}},
  \citenamefont {{Croft}}, \citenamefont {{Cuesta}}, \citenamefont {{da
  Costa}}, \citenamefont {{Davenport}}, \citenamefont {{Delubac}},
  \citenamefont {{de Putter}}, \citenamefont {{Dhital}}, \citenamefont
  {{Ealet}}, \citenamefont {{Ebelke}}, \citenamefont {{Eisenstein}},
  \citenamefont {{Escoffier}}, \citenamefont {{Fan}}, \citenamefont {{Filiz
  Ak}}, \citenamefont {{Finley}}, \citenamefont {{Font-Ribera}}, \citenamefont
  {{G{\'e}nova-Santos}}, \citenamefont {{Gunn}}, \citenamefont {{Guo}},
  \citenamefont {{Haggard}}, \citenamefont {{Hall}}, \citenamefont
  {{Hamilton}}, \citenamefont {{Harris}}, \citenamefont {{Harris}},
  \citenamefont {{Ho}}, \citenamefont {{Hogg}}, \citenamefont {{Holder}},
  \citenamefont {{Honscheid}}, \citenamefont {{Huehnerhoff}}, \citenamefont
  {{Jordan}}, \citenamefont {{Jordan}}, \citenamefont {{Kauffmann}},
  \citenamefont {{Kazin}}, \citenamefont {{Kirkby}}, \citenamefont {{Klaene}},
  \citenamefont {{Kneib}}, \citenamefont {{Le Goff}}, \citenamefont {{Lee}},
  \citenamefont {{Long}}, \citenamefont {{Loomis}}, \citenamefont {{Lundgren}},
  \citenamefont {{Lupton}}, \citenamefont {{Maia}}, \citenamefont {{Makler}},
  \citenamefont {{Malanushenko}}, \citenamefont {{Malanushenko}}, \citenamefont
  {{Mandelbaum}}, \citenamefont {{Manera}}, \citenamefont {{Maraston}},
  \citenamefont {{Margala}}, \citenamefont {{Masters}}, \citenamefont
  {{McBride}}, \citenamefont {{McDonald}}, \citenamefont {{McGreer}},
  \citenamefont {{McMahon}}, \citenamefont {{Mena}}, \citenamefont
  {{Miralda-Escud{\'e}}}, \citenamefont {{Montero-Dorta}}, \citenamefont
  {{Montesano}}, \citenamefont {{Muna}}, \citenamefont {{Myers}}, \citenamefont
  {{Naugle}}, \citenamefont {{Nichol}}, \citenamefont {{Noterdaeme}},
  \citenamefont {{Nuza}}, \citenamefont {{Olmstead}}, \citenamefont
  {{Oravetz}}, \citenamefont {{Oravetz}}, \citenamefont {{Owen}}, \citenamefont
  {{Padmanabhan}}, \citenamefont {{Palanque-Delabrouille}}, \citenamefont
  {{Pan}}, \citenamefont {{Parejko}}, \citenamefont {{P{\^a}ris}},
  \citenamefont {{Percival}}, \citenamefont {{P{\'e}rez-Fournon}},
  \citenamefont {{P{\'e}rez-R{\`a}fols}}, \citenamefont {{Petitjean}},
  \citenamefont {{Pfaffenberger}}, \citenamefont {{Pforr}}, \citenamefont
  {{Pieri}}, \citenamefont {{Prada}}, \citenamefont {{Price-Whelan}},
  \citenamefont {{Raddick}}, \citenamefont {{Rebolo}}, \citenamefont {{Rich}},
  \citenamefont {{Richards}}, \citenamefont {{Rockosi}}, \citenamefont {{Roe}},
  \citenamefont {{Ross}}, \citenamefont {{Ross}}, \citenamefont {{Rossi}},
  \citenamefont {{Rubi{\~n}o-Martin}}, \citenamefont {{Samushia}},
  \citenamefont {{S{\'a}nchez}}, \citenamefont {{Sayres}}, \citenamefont
  {{Schmidt}}, \citenamefont {{Schneider}}, \citenamefont {{Sc{\'o}ccola}},
  \citenamefont {{Seo}}, \citenamefont {{Shelden}}, \citenamefont {{Sheldon}},
  \citenamefont {{Shen}}, \citenamefont {{Shu}}, \citenamefont {{Slosar}},
  \citenamefont {{Smee}}, \citenamefont {{Snedden}}, \citenamefont
  {{Stauffer}}, \citenamefont {{Steele}}, \citenamefont {{Strauss}},
  \citenamefont {{Streblyanska}}, \citenamefont {{Suzuki}}, \citenamefont
  {{Swanson}}, \citenamefont {{Tal}}, \citenamefont {{Tanaka}}, \citenamefont
  {{Thomas}}, \citenamefont {{Tinker}}, \citenamefont {{Tojeiro}},
  \citenamefont {{Tremonti}}, \citenamefont {{Vargas Maga{\~n}a}},
  \citenamefont {{Verde}}, \citenamefont {{Viel}}, \citenamefont {{Wake}},
  \citenamefont {{Watson}}, \citenamefont {{Weaver}}, \citenamefont
  {{Weinberg}}, \citenamefont {{Weiner}}, \citenamefont {{West}}, \citenamefont
  {{White}}, \citenamefont {{Wood-Vasey}}, \citenamefont {{Yeche}},
  \citenamefont {{Zehavi}}, \citenamefont {{Zhao}},\ and\ \citenamefont
  {{Zheng}}}]{2013AJ....145...10D}%
  \BibitemOpen
  \bibfield  {author} {\bibinfo {author} {\bibfnamefont {K.~S.}\ \bibnamefont
  {{Dawson}}}, \bibinfo {author} {\bibfnamefont {D.~J.}\ \bibnamefont
  {{Schlegel}}}, \bibinfo {author} {\bibfnamefont {C.~P.}\ \bibnamefont
  {{Ahn}}}, \bibinfo {author} {\bibfnamefont {S.~F.}\ \bibnamefont
  {{Anderson}}}, \bibinfo {author} {\bibfnamefont {{\'E}.}~\bibnamefont
  {{Aubourg}}}, \bibinfo {author} {\bibfnamefont {S.}~\bibnamefont {{Bailey}}},
  \bibinfo {author} {\bibfnamefont {R.~H.}\ \bibnamefont {{Barkhouser}}},
  \bibinfo {author} {\bibfnamefont {J.~E.}\ \bibnamefont {{Bautista}}},
  \bibinfo {author} {\bibfnamefont {A.}~\bibnamefont {{Beifiori}}}, \bibinfo
  {author} {\bibfnamefont {A.~A.}\ \bibnamefont {{Berlind}}}, \bibinfo {author}
  {\bibfnamefont {V.}~\bibnamefont {{Bhardwaj}}}, \bibinfo {author}
  {\bibfnamefont {D.}~\bibnamefont {{Bizyaev}}}, \bibinfo {author}
  {\bibfnamefont {C.~H.}\ \bibnamefont {{Blake}}}, \bibinfo {author}
  {\bibfnamefont {M.~R.}\ \bibnamefont {{Blanton}}}, \bibinfo {author}
  {\bibfnamefont {M.}~\bibnamefont {{Blomqvist}}}, \bibinfo {author}
  {\bibfnamefont {A.~S.}\ \bibnamefont {{Bolton}}}, \bibinfo {author}
  {\bibfnamefont {A.}~\bibnamefont {{Borde}}}, \bibinfo {author} {\bibfnamefont
  {J.}~\bibnamefont {{Bovy}}}, \bibinfo {author} {\bibfnamefont {W.~N.}\
  \bibnamefont {{Brandt}}}, \bibinfo {author} {\bibfnamefont {H.}~\bibnamefont
  {{Brewington}}}, \bibinfo {author} {\bibfnamefont {J.}~\bibnamefont
  {{Brinkmann}}}, \bibinfo {author} {\bibfnamefont {P.~J.}\ \bibnamefont
  {{Brown}}}, \bibinfo {author} {\bibfnamefont {J.~R.}\ \bibnamefont
  {{Brownstein}}}, \bibinfo {author} {\bibfnamefont {K.}~\bibnamefont
  {{Bundy}}}, \bibinfo {author} {\bibfnamefont {N.~G.}\ \bibnamefont
  {{Busca}}}, \bibinfo {author} {\bibfnamefont {W.}~\bibnamefont
  {{Carithers}}}, \bibinfo {author} {\bibfnamefont {A.~R.}\ \bibnamefont
  {{Carnero}}}, \bibinfo {author} {\bibfnamefont {M.~A.}\ \bibnamefont
  {{Carr}}}, \bibinfo {author} {\bibfnamefont {Y.}~\bibnamefont {{Chen}}},
  \bibinfo {author} {\bibfnamefont {J.}~\bibnamefont {{Comparat}}}, \bibinfo
  {author} {\bibfnamefont {N.}~\bibnamefont {{Connolly}}}, \bibinfo {author}
  {\bibfnamefont {F.}~\bibnamefont {{Cope}}}, \bibinfo {author} {\bibfnamefont
  {R.~A.~C.}\ \bibnamefont {{Croft}}}, \bibinfo {author} {\bibfnamefont
  {A.~J.}\ \bibnamefont {{Cuesta}}}, \bibinfo {author} {\bibfnamefont {L.~N.}\
  \bibnamefont {{da Costa}}}, \bibinfo {author} {\bibfnamefont {J.~R.~A.}\
  \bibnamefont {{Davenport}}}, \bibinfo {author} {\bibfnamefont
  {T.}~\bibnamefont {{Delubac}}}, \bibinfo {author} {\bibfnamefont
  {R.}~\bibnamefont {{de Putter}}}, \bibinfo {author} {\bibfnamefont
  {S.}~\bibnamefont {{Dhital}}}, \bibinfo {author} {\bibfnamefont
  {A.}~\bibnamefont {{Ealet}}}, \bibinfo {author} {\bibfnamefont {G.~L.}\
  \bibnamefont {{Ebelke}}}, \bibinfo {author} {\bibfnamefont {D.~J.}\
  \bibnamefont {{Eisenstein}}}, \bibinfo {author} {\bibfnamefont
  {S.}~\bibnamefont {{Escoffier}}}, \bibinfo {author} {\bibfnamefont
  {X.}~\bibnamefont {{Fan}}}, \bibinfo {author} {\bibfnamefont
  {N.}~\bibnamefont {{Filiz Ak}}}, \bibinfo {author} {\bibfnamefont
  {H.}~\bibnamefont {{Finley}}}, \bibinfo {author} {\bibfnamefont
  {A.}~\bibnamefont {{Font-Ribera}}}, \bibinfo {author} {\bibfnamefont
  {R.}~\bibnamefont {{G{\'e}nova-Santos}}}, \bibinfo {author} {\bibfnamefont
  {J.~E.}\ \bibnamefont {{Gunn}}}, \bibinfo {author} {\bibfnamefont
  {H.}~\bibnamefont {{Guo}}}, \bibinfo {author} {\bibfnamefont
  {D.}~\bibnamefont {{Haggard}}}, \bibinfo {author} {\bibfnamefont {P.~B.}\
  \bibnamefont {{Hall}}}, \bibinfo {author} {\bibfnamefont {J.-C.}\
  \bibnamefont {{Hamilton}}}, \bibinfo {author} {\bibfnamefont
  {B.}~\bibnamefont {{Harris}}}, \bibinfo {author} {\bibfnamefont {D.~W.}\
  \bibnamefont {{Harris}}}, \bibinfo {author} {\bibfnamefont {S.}~\bibnamefont
  {{Ho}}}, \bibinfo {author} {\bibfnamefont {D.~W.}\ \bibnamefont {{Hogg}}},
  \bibinfo {author} {\bibfnamefont {D.}~\bibnamefont {{Holder}}}, \bibinfo
  {author} {\bibfnamefont {K.}~\bibnamefont {{Honscheid}}}, \bibinfo {author}
  {\bibfnamefont {J.}~\bibnamefont {{Huehnerhoff}}}, \bibinfo {author}
  {\bibfnamefont {B.}~\bibnamefont {{Jordan}}}, \bibinfo {author}
  {\bibfnamefont {W.~P.}\ \bibnamefont {{Jordan}}}, \bibinfo {author}
  {\bibfnamefont {G.}~\bibnamefont {{Kauffmann}}}, \bibinfo {author}
  {\bibfnamefont {E.~A.}\ \bibnamefont {{Kazin}}}, \bibinfo {author}
  {\bibfnamefont {D.}~\bibnamefont {{Kirkby}}}, \bibinfo {author}
  {\bibfnamefont {M.~A.}\ \bibnamefont {{Klaene}}}, \bibinfo {author}
  {\bibfnamefont {J.-P.}\ \bibnamefont {{Kneib}}}, \bibinfo {author}
  {\bibfnamefont {J.-M.}\ \bibnamefont {{Le Goff}}}, \bibinfo {author}
  {\bibfnamefont {K.-G.}\ \bibnamefont {{Lee}}}, \bibinfo {author}
  {\bibfnamefont {D.~C.}\ \bibnamefont {{Long}}}, \bibinfo {author}
  {\bibfnamefont {C.~P.}\ \bibnamefont {{Loomis}}}, \bibinfo {author}
  {\bibfnamefont {B.}~\bibnamefont {{Lundgren}}}, \bibinfo {author}
  {\bibfnamefont {R.~H.}\ \bibnamefont {{Lupton}}}, \bibinfo {author}
  {\bibfnamefont {M.~A.~G.}\ \bibnamefont {{Maia}}}, \bibinfo {author}
  {\bibfnamefont {M.}~\bibnamefont {{Makler}}}, \bibinfo {author}
  {\bibfnamefont {E.}~\bibnamefont {{Malanushenko}}}, \bibinfo {author}
  {\bibfnamefont {V.}~\bibnamefont {{Malanushenko}}}, \bibinfo {author}
  {\bibfnamefont {R.}~\bibnamefont {{Mandelbaum}}}, \bibinfo {author}
  {\bibfnamefont {M.}~\bibnamefont {{Manera}}}, \bibinfo {author}
  {\bibfnamefont {C.}~\bibnamefont {{Maraston}}}, \bibinfo {author}
  {\bibfnamefont {D.}~\bibnamefont {{Margala}}}, \bibinfo {author}
  {\bibfnamefont {K.~L.}\ \bibnamefont {{Masters}}}, \bibinfo {author}
  {\bibfnamefont {C.~K.}\ \bibnamefont {{McBride}}}, \bibinfo {author}
  {\bibfnamefont {P.}~\bibnamefont {{McDonald}}}, \bibinfo {author}
  {\bibfnamefont {I.~D.}\ \bibnamefont {{McGreer}}}, \bibinfo {author}
  {\bibfnamefont {R.~G.}\ \bibnamefont {{McMahon}}}, \bibinfo {author}
  {\bibfnamefont {O.}~\bibnamefont {{Mena}}}, \bibinfo {author} {\bibfnamefont
  {J.}~\bibnamefont {{Miralda-Escud{\'e}}}}, \bibinfo {author} {\bibfnamefont
  {A.~D.}\ \bibnamefont {{Montero-Dorta}}}, \bibinfo {author} {\bibfnamefont
  {F.}~\bibnamefont {{Montesano}}}, \bibinfo {author} {\bibfnamefont
  {D.}~\bibnamefont {{Muna}}}, \bibinfo {author} {\bibfnamefont {A.~D.}\
  \bibnamefont {{Myers}}}, \bibinfo {author} {\bibfnamefont {T.}~\bibnamefont
  {{Naugle}}}, \bibinfo {author} {\bibfnamefont {R.~C.}\ \bibnamefont
  {{Nichol}}}, \bibinfo {author} {\bibfnamefont {P.}~\bibnamefont
  {{Noterdaeme}}}, \bibinfo {author} {\bibfnamefont {S.~E.}\ \bibnamefont
  {{Nuza}}}, \bibinfo {author} {\bibfnamefont {M.~D.}\ \bibnamefont
  {{Olmstead}}}, \bibinfo {author} {\bibfnamefont {A.}~\bibnamefont
  {{Oravetz}}}, \bibinfo {author} {\bibfnamefont {D.~J.}\ \bibnamefont
  {{Oravetz}}}, \bibinfo {author} {\bibfnamefont {R.}~\bibnamefont {{Owen}}},
  \bibinfo {author} {\bibfnamefont {N.}~\bibnamefont {{Padmanabhan}}}, \bibinfo
  {author} {\bibfnamefont {N.}~\bibnamefont {{Palanque-Delabrouille}}},
  \bibinfo {author} {\bibfnamefont {K.}~\bibnamefont {{Pan}}}, \bibinfo
  {author} {\bibfnamefont {J.~K.}\ \bibnamefont {{Parejko}}}, \bibinfo {author}
  {\bibfnamefont {I.}~\bibnamefont {{P{\^a}ris}}}, \bibinfo {author}
  {\bibfnamefont {W.~J.}\ \bibnamefont {{Percival}}}, \bibinfo {author}
  {\bibfnamefont {I.}~\bibnamefont {{P{\'e}rez-Fournon}}}, \bibinfo {author}
  {\bibfnamefont {I.}~\bibnamefont {{P{\'e}rez-R{\`a}fols}}}, \bibinfo {author}
  {\bibfnamefont {P.}~\bibnamefont {{Petitjean}}}, \bibinfo {author}
  {\bibfnamefont {R.}~\bibnamefont {{Pfaffenberger}}}, \bibinfo {author}
  {\bibfnamefont {J.}~\bibnamefont {{Pforr}}}, \bibinfo {author} {\bibfnamefont
  {M.~M.}\ \bibnamefont {{Pieri}}}, \bibinfo {author} {\bibfnamefont
  {F.}~\bibnamefont {{Prada}}}, \bibinfo {author} {\bibfnamefont {A.~M.}\
  \bibnamefont {{Price-Whelan}}}, \bibinfo {author} {\bibfnamefont {M.~J.}\
  \bibnamefont {{Raddick}}}, \bibinfo {author} {\bibfnamefont {R.}~\bibnamefont
  {{Rebolo}}}, \bibinfo {author} {\bibfnamefont {J.}~\bibnamefont {{Rich}}},
  \bibinfo {author} {\bibfnamefont {G.~T.}\ \bibnamefont {{Richards}}},
  \bibinfo {author} {\bibfnamefont {C.~M.}\ \bibnamefont {{Rockosi}}}, \bibinfo
  {author} {\bibfnamefont {N.~A.}\ \bibnamefont {{Roe}}}, \bibinfo {author}
  {\bibfnamefont {A.~J.}\ \bibnamefont {{Ross}}}, \bibinfo {author}
  {\bibfnamefont {N.~P.}\ \bibnamefont {{Ross}}}, \bibinfo {author}
  {\bibfnamefont {G.}~\bibnamefont {{Rossi}}}, \bibinfo {author} {\bibfnamefont
  {J.~A.}\ \bibnamefont {{Rubi{\~n}o-Martin}}}, \bibinfo {author}
  {\bibfnamefont {L.}~\bibnamefont {{Samushia}}}, \bibinfo {author}
  {\bibfnamefont {A.~G.}\ \bibnamefont {{S{\'a}nchez}}}, \bibinfo {author}
  {\bibfnamefont {C.}~\bibnamefont {{Sayres}}}, \bibinfo {author}
  {\bibfnamefont {S.~J.}\ \bibnamefont {{Schmidt}}}, \bibinfo {author}
  {\bibfnamefont {D.~P.}\ \bibnamefont {{Schneider}}}, \bibinfo {author}
  {\bibfnamefont {C.~G.}\ \bibnamefont {{Sc{\'o}ccola}}}, \bibinfo {author}
  {\bibfnamefont {H.-J.}\ \bibnamefont {{Seo}}}, \bibinfo {author}
  {\bibfnamefont {A.}~\bibnamefont {{Shelden}}}, \bibinfo {author}
  {\bibfnamefont {E.}~\bibnamefont {{Sheldon}}}, \bibinfo {author}
  {\bibfnamefont {Y.}~\bibnamefont {{Shen}}}, \bibinfo {author} {\bibfnamefont
  {Y.}~\bibnamefont {{Shu}}}, \bibinfo {author} {\bibfnamefont
  {A.}~\bibnamefont {{Slosar}}}, \bibinfo {author} {\bibfnamefont {S.~A.}\
  \bibnamefont {{Smee}}}, \bibinfo {author} {\bibfnamefont {S.~A.}\
  \bibnamefont {{Snedden}}}, \bibinfo {author} {\bibfnamefont {F.}~\bibnamefont
  {{Stauffer}}}, \bibinfo {author} {\bibfnamefont {O.}~\bibnamefont
  {{Steele}}}, \bibinfo {author} {\bibfnamefont {M.~A.}\ \bibnamefont
  {{Strauss}}}, \bibinfo {author} {\bibfnamefont {A.}~\bibnamefont
  {{Streblyanska}}}, \bibinfo {author} {\bibfnamefont {N.}~\bibnamefont
  {{Suzuki}}}, \bibinfo {author} {\bibfnamefont {M.~E.~C.}\ \bibnamefont
  {{Swanson}}}, \bibinfo {author} {\bibfnamefont {T.}~\bibnamefont {{Tal}}},
  \bibinfo {author} {\bibfnamefont {M.}~\bibnamefont {{Tanaka}}}, \bibinfo
  {author} {\bibfnamefont {D.}~\bibnamefont {{Thomas}}}, \bibinfo {author}
  {\bibfnamefont {J.~L.}\ \bibnamefont {{Tinker}}}, \bibinfo {author}
  {\bibfnamefont {R.}~\bibnamefont {{Tojeiro}}}, \bibinfo {author}
  {\bibfnamefont {C.~A.}\ \bibnamefont {{Tremonti}}}, \bibinfo {author}
  {\bibfnamefont {M.}~\bibnamefont {{Vargas Maga{\~n}a}}}, \bibinfo {author}
  {\bibfnamefont {L.}~\bibnamefont {{Verde}}}, \bibinfo {author} {\bibfnamefont
  {M.}~\bibnamefont {{Viel}}}, \bibinfo {author} {\bibfnamefont {D.~A.}\
  \bibnamefont {{Wake}}}, \bibinfo {author} {\bibfnamefont {M.}~\bibnamefont
  {{Watson}}}, \bibinfo {author} {\bibfnamefont {B.~A.}\ \bibnamefont
  {{Weaver}}}, \bibinfo {author} {\bibfnamefont {D.~H.}\ \bibnamefont
  {{Weinberg}}}, \bibinfo {author} {\bibfnamefont {B.~J.}\ \bibnamefont
  {{Weiner}}}, \bibinfo {author} {\bibfnamefont {A.~A.}\ \bibnamefont
  {{West}}}, \bibinfo {author} {\bibfnamefont {M.}~\bibnamefont {{White}}},
  \bibinfo {author} {\bibfnamefont {W.~M.}\ \bibnamefont {{Wood-Vasey}}},
  \bibinfo {author} {\bibfnamefont {C.}~\bibnamefont {{Yeche}}}, \bibinfo
  {author} {\bibfnamefont {I.}~\bibnamefont {{Zehavi}}}, \bibinfo {author}
  {\bibfnamefont {G.-B.}\ \bibnamefont {{Zhao}}}, \ and\ \bibinfo {author}
  {\bibfnamefont {Z.}~\bibnamefont {{Zheng}}},\ }\href {\doibase
  10.1088/0004-6256/145/1/10} {\bibfield  {journal} {\bibinfo  {journal} {\aj}\
  }\textbf {\bibinfo {volume} {145}},\ \bibinfo {eid} {10} (\bibinfo {year}
  {2013})},\ \Eprint {http://arxiv.org/abs/1208.0022} {arXiv:1208.0022
  [astro-ph.CO]} \BibitemShut {NoStop}%
\bibitem [{\citenamefont {{Reid}}\ \emph {et~al.}(2012)\citenamefont {{Reid}},
  \citenamefont {{Samushia}}, \citenamefont {{White}}, \citenamefont
  {{Percival}}, \citenamefont {{Manera}}, \citenamefont {{Padmanabhan}},
  \citenamefont {{Ross}}, \citenamefont {{S{\'a}nchez}}, \citenamefont
  {{Bailey}}, \citenamefont {{Bizyaev}}, \citenamefont {{Bolton}},
  \citenamefont {{Brewington}}, \citenamefont {{Brinkmann}}, \citenamefont
  {{Brownstein}}, \citenamefont {{Cuesta}}, \citenamefont {{Eisenstein}},
  \citenamefont {{Gunn}}, \citenamefont {{Honscheid}}, \citenamefont
  {{Malanushenko}}, \citenamefont {{Malanushenko}}, \citenamefont {{Maraston}},
  \citenamefont {{McBride}}, \citenamefont {{Muna}}, \citenamefont {{Nichol}},
  \citenamefont {{Oravetz}}, \citenamefont {{Pan}}, \citenamefont {{de
  Putter}}, \citenamefont {{Roe}}, \citenamefont {{Ross}}, \citenamefont
  {{Schlegel}}, \citenamefont {{Schneider}}, \citenamefont {{Seo}},
  \citenamefont {{Shelden}}, \citenamefont {{Sheldon}}, \citenamefont
  {{Simmons}}, \citenamefont {{Skibba}}, \citenamefont {{Snedden}},
  \citenamefont {{Swanson}}, \citenamefont {{Thomas}}, \citenamefont
  {{Tinker}}, \citenamefont {{Tojeiro}}, \citenamefont {{Verde}}, \citenamefont
  {{Wake}}, \citenamefont {{Weaver}}, \citenamefont {{Weinberg}}, \citenamefont
  {{Zehavi}},\ and\ \citenamefont {{Zhao}}}]{2012MNRAS.426.2719R}%
  \BibitemOpen
  \bibfield  {author} {\bibinfo {author} {\bibfnamefont {B.~A.}\ \bibnamefont
  {{Reid}}}, \bibinfo {author} {\bibfnamefont {L.}~\bibnamefont {{Samushia}}},
  \bibinfo {author} {\bibfnamefont {M.}~\bibnamefont {{White}}}, \bibinfo
  {author} {\bibfnamefont {W.~J.}\ \bibnamefont {{Percival}}}, \bibinfo
  {author} {\bibfnamefont {M.}~\bibnamefont {{Manera}}}, \bibinfo {author}
  {\bibfnamefont {N.}~\bibnamefont {{Padmanabhan}}}, \bibinfo {author}
  {\bibfnamefont {A.~J.}\ \bibnamefont {{Ross}}}, \bibinfo {author}
  {\bibfnamefont {A.~G.}\ \bibnamefont {{S{\'a}nchez}}}, \bibinfo {author}
  {\bibfnamefont {S.}~\bibnamefont {{Bailey}}}, \bibinfo {author}
  {\bibfnamefont {D.}~\bibnamefont {{Bizyaev}}}, \bibinfo {author}
  {\bibfnamefont {A.~S.}\ \bibnamefont {{Bolton}}}, \bibinfo {author}
  {\bibfnamefont {H.}~\bibnamefont {{Brewington}}}, \bibinfo {author}
  {\bibfnamefont {J.}~\bibnamefont {{Brinkmann}}}, \bibinfo {author}
  {\bibfnamefont {J.~R.}\ \bibnamefont {{Brownstein}}}, \bibinfo {author}
  {\bibfnamefont {A.~J.}\ \bibnamefont {{Cuesta}}}, \bibinfo {author}
  {\bibfnamefont {D.~J.}\ \bibnamefont {{Eisenstein}}}, \bibinfo {author}
  {\bibfnamefont {J.~E.}\ \bibnamefont {{Gunn}}}, \bibinfo {author}
  {\bibfnamefont {K.}~\bibnamefont {{Honscheid}}}, \bibinfo {author}
  {\bibfnamefont {E.}~\bibnamefont {{Malanushenko}}}, \bibinfo {author}
  {\bibfnamefont {V.}~\bibnamefont {{Malanushenko}}}, \bibinfo {author}
  {\bibfnamefont {C.}~\bibnamefont {{Maraston}}}, \bibinfo {author}
  {\bibfnamefont {C.~K.}\ \bibnamefont {{McBride}}}, \bibinfo {author}
  {\bibfnamefont {D.}~\bibnamefont {{Muna}}}, \bibinfo {author} {\bibfnamefont
  {R.~C.}\ \bibnamefont {{Nichol}}}, \bibinfo {author} {\bibfnamefont
  {D.}~\bibnamefont {{Oravetz}}}, \bibinfo {author} {\bibfnamefont
  {K.}~\bibnamefont {{Pan}}}, \bibinfo {author} {\bibfnamefont
  {R.}~\bibnamefont {{de Putter}}}, \bibinfo {author} {\bibfnamefont {N.~A.}\
  \bibnamefont {{Roe}}}, \bibinfo {author} {\bibfnamefont {N.~P.}\ \bibnamefont
  {{Ross}}}, \bibinfo {author} {\bibfnamefont {D.~J.}\ \bibnamefont
  {{Schlegel}}}, \bibinfo {author} {\bibfnamefont {D.~P.}\ \bibnamefont
  {{Schneider}}}, \bibinfo {author} {\bibfnamefont {H.-J.}\ \bibnamefont
  {{Seo}}}, \bibinfo {author} {\bibfnamefont {A.}~\bibnamefont {{Shelden}}},
  \bibinfo {author} {\bibfnamefont {E.~S.}\ \bibnamefont {{Sheldon}}}, \bibinfo
  {author} {\bibfnamefont {A.}~\bibnamefont {{Simmons}}}, \bibinfo {author}
  {\bibfnamefont {R.~A.}\ \bibnamefont {{Skibba}}}, \bibinfo {author}
  {\bibfnamefont {S.}~\bibnamefont {{Snedden}}}, \bibinfo {author}
  {\bibfnamefont {M.~E.~C.}\ \bibnamefont {{Swanson}}}, \bibinfo {author}
  {\bibfnamefont {D.}~\bibnamefont {{Thomas}}}, \bibinfo {author}
  {\bibfnamefont {J.}~\bibnamefont {{Tinker}}}, \bibinfo {author}
  {\bibfnamefont {R.}~\bibnamefont {{Tojeiro}}}, \bibinfo {author}
  {\bibfnamefont {L.}~\bibnamefont {{Verde}}}, \bibinfo {author} {\bibfnamefont
  {D.~A.}\ \bibnamefont {{Wake}}}, \bibinfo {author} {\bibfnamefont {B.~A.}\
  \bibnamefont {{Weaver}}}, \bibinfo {author} {\bibfnamefont {D.~H.}\
  \bibnamefont {{Weinberg}}}, \bibinfo {author} {\bibfnamefont
  {I.}~\bibnamefont {{Zehavi}}}, \ and\ \bibinfo {author} {\bibfnamefont
  {G.-B.}\ \bibnamefont {{Zhao}}},\ }\href {\doibase
  10.1111/j.1365-2966.2012.21779.x} {\bibfield  {journal} {\bibinfo  {journal}
  {\mnras}\ }\textbf {\bibinfo {volume} {426}},\ \bibinfo {pages} {2719}
  (\bibinfo {year} {2012})},\ \Eprint {http://arxiv.org/abs/1203.6641}
  {arXiv:1203.6641 [astro-ph.CO]} \BibitemShut {NoStop}%
\bibitem [{\citenamefont {{Jiang}}\ \emph {et~al.}(2006)\citenamefont
  {{Jiang}}, \citenamefont {{Fan}}, \citenamefont {{Cool}}, \citenamefont
  {{Eisenstein}}, \citenamefont {{Zehavi}}, \citenamefont {{Richards}},
  \citenamefont {{Scranton}}, \citenamefont {{Johnston}}, \citenamefont
  {{Strauss}}, \citenamefont {{Schneider}},\ and\ \citenamefont
  {{Brinkmann}}}]{2006AJ....131.2788J}%
  \BibitemOpen
  \bibfield  {author} {\bibinfo {author} {\bibfnamefont {L.}~\bibnamefont
  {{Jiang}}}, \bibinfo {author} {\bibfnamefont {X.}~\bibnamefont {{Fan}}},
  \bibinfo {author} {\bibfnamefont {R.~J.}\ \bibnamefont {{Cool}}}, \bibinfo
  {author} {\bibfnamefont {D.~J.}\ \bibnamefont {{Eisenstein}}}, \bibinfo
  {author} {\bibfnamefont {I.}~\bibnamefont {{Zehavi}}}, \bibinfo {author}
  {\bibfnamefont {G.~T.}\ \bibnamefont {{Richards}}}, \bibinfo {author}
  {\bibfnamefont {R.}~\bibnamefont {{Scranton}}}, \bibinfo {author}
  {\bibfnamefont {D.}~\bibnamefont {{Johnston}}}, \bibinfo {author}
  {\bibfnamefont {M.~A.}\ \bibnamefont {{Strauss}}}, \bibinfo {author}
  {\bibfnamefont {D.~P.}\ \bibnamefont {{Schneider}}}, \ and\ \bibinfo {author}
  {\bibfnamefont {J.}~\bibnamefont {{Brinkmann}}},\ }\href {\doibase
  10.1086/503745} {\bibfield  {journal} {\bibinfo  {journal} {\aj}\ }\textbf
  {\bibinfo {volume} {131}},\ \bibinfo {pages} {2788} (\bibinfo {year}
  {2006})},\ \Eprint {http://arxiv.org/abs/arXiv:astro-ph/0602569}
  {arXiv:astro-ph/0602569} \BibitemShut {NoStop}%
\bibitem [{\citenamefont {{Anderson}}\ \emph {et~al.}(2012)\citenamefont
  {{Anderson}}, \citenamefont {{Aubourg}}, \citenamefont {{Bailey}},
  \citenamefont {{Bizyaev}}, \citenamefont {{Blanton}}, \citenamefont
  {{Bolton}}, \citenamefont {{Brinkmann}}, \citenamefont {{Brownstein}},
  \citenamefont {{Burden}}, \citenamefont {{Cuesta}}, \citenamefont {{da
  Costa}}, \citenamefont {{Dawson}}, \citenamefont {{de Putter}}, \citenamefont
  {{Eisenstein}}, \citenamefont {{Gunn}}, \citenamefont {{Guo}}, \citenamefont
  {{Hamilton}}, \citenamefont {{Harding}}, \citenamefont {{Ho}}, \citenamefont
  {{Honscheid}}, \citenamefont {{Kazin}}, \citenamefont {{Kirkby}},
  \citenamefont {{Kneib}}, \citenamefont {{Labatie}}, \citenamefont {{Loomis}},
  \citenamefont {{Lupton}}, \citenamefont {{Malanushenko}}, \citenamefont
  {{Malanushenko}}, \citenamefont {{Mandelbaum}}, \citenamefont {{Manera}},
  \citenamefont {{Maraston}}, \citenamefont {{McBride}}, \citenamefont
  {{Mehta}}, \citenamefont {{Mena}}, \citenamefont {{Montesano}}, \citenamefont
  {{Muna}}, \citenamefont {{Nichol}}, \citenamefont {{Nuza}}, \citenamefont
  {{Olmstead}}, \citenamefont {{Oravetz}}, \citenamefont {{Padmanabhan}},
  \citenamefont {{Palanque-Delabrouille}}, \citenamefont {{Pan}}, \citenamefont
  {{Parejko}}, \citenamefont {{P{\^a}ris}}, \citenamefont {{Percival}},
  \citenamefont {{Petitjean}}, \citenamefont {{Prada}}, \citenamefont {{Reid}},
  \citenamefont {{Roe}}, \citenamefont {{Ross}}, \citenamefont {{Ross}},
  \citenamefont {{Samushia}}, \citenamefont {{S{\'a}nchez}}, \citenamefont
  {{Schlegel}}, \citenamefont {{Schneider}}, \citenamefont {{Sc{\'o}ccola}},
  \citenamefont {{Seo}}, \citenamefont {{Sheldon}}, \citenamefont {{Simmons}},
  \citenamefont {{Skibba}}, \citenamefont {{Strauss}}, \citenamefont
  {{Swanson}}, \citenamefont {{Thomas}}, \citenamefont {{Tinker}},
  \citenamefont {{Tojeiro}}, \citenamefont {{Maga{\~n}a}}, \citenamefont
  {{Verde}}, \citenamefont {{Wagner}}, \citenamefont {{Wake}}, \citenamefont
  {{Weaver}}, \citenamefont {{Weinberg}}, \citenamefont {{White}},
  \citenamefont {{Xu}}, \citenamefont {{Y{\`e}che}}, \citenamefont {{Zehavi}},\
  and\ \citenamefont {{Zhao}}}]{2012MNRAS.427.3435A}%
  \BibitemOpen
  \bibfield  {author} {\bibinfo {author} {\bibfnamefont {L.}~\bibnamefont
  {{Anderson}}}, \bibinfo {author} {\bibfnamefont {E.}~\bibnamefont
  {{Aubourg}}}, \bibinfo {author} {\bibfnamefont {S.}~\bibnamefont {{Bailey}}},
  \bibinfo {author} {\bibfnamefont {D.}~\bibnamefont {{Bizyaev}}}, \bibinfo
  {author} {\bibfnamefont {M.}~\bibnamefont {{Blanton}}}, \bibinfo {author}
  {\bibfnamefont {A.~S.}\ \bibnamefont {{Bolton}}}, \bibinfo {author}
  {\bibfnamefont {J.}~\bibnamefont {{Brinkmann}}}, \bibinfo {author}
  {\bibfnamefont {J.~R.}\ \bibnamefont {{Brownstein}}}, \bibinfo {author}
  {\bibfnamefont {A.}~\bibnamefont {{Burden}}}, \bibinfo {author}
  {\bibfnamefont {A.~J.}\ \bibnamefont {{Cuesta}}}, \bibinfo {author}
  {\bibfnamefont {L.~A.~N.}\ \bibnamefont {{da Costa}}}, \bibinfo {author}
  {\bibfnamefont {K.~S.}\ \bibnamefont {{Dawson}}}, \bibinfo {author}
  {\bibfnamefont {R.}~\bibnamefont {{de Putter}}}, \bibinfo {author}
  {\bibfnamefont {D.~J.}\ \bibnamefont {{Eisenstein}}}, \bibinfo {author}
  {\bibfnamefont {J.~E.}\ \bibnamefont {{Gunn}}}, \bibinfo {author}
  {\bibfnamefont {H.}~\bibnamefont {{Guo}}}, \bibinfo {author} {\bibfnamefont
  {J.-C.}\ \bibnamefont {{Hamilton}}}, \bibinfo {author} {\bibfnamefont
  {P.}~\bibnamefont {{Harding}}}, \bibinfo {author} {\bibfnamefont
  {S.}~\bibnamefont {{Ho}}}, \bibinfo {author} {\bibfnamefont {K.}~\bibnamefont
  {{Honscheid}}}, \bibinfo {author} {\bibfnamefont {E.}~\bibnamefont
  {{Kazin}}}, \bibinfo {author} {\bibfnamefont {D.}~\bibnamefont {{Kirkby}}},
  \bibinfo {author} {\bibfnamefont {J.-P.}\ \bibnamefont {{Kneib}}}, \bibinfo
  {author} {\bibfnamefont {A.}~\bibnamefont {{Labatie}}}, \bibinfo {author}
  {\bibfnamefont {C.}~\bibnamefont {{Loomis}}}, \bibinfo {author}
  {\bibfnamefont {R.~H.}\ \bibnamefont {{Lupton}}}, \bibinfo {author}
  {\bibfnamefont {E.}~\bibnamefont {{Malanushenko}}}, \bibinfo {author}
  {\bibfnamefont {V.}~\bibnamefont {{Malanushenko}}}, \bibinfo {author}
  {\bibfnamefont {R.}~\bibnamefont {{Mandelbaum}}}, \bibinfo {author}
  {\bibfnamefont {M.}~\bibnamefont {{Manera}}}, \bibinfo {author}
  {\bibfnamefont {C.}~\bibnamefont {{Maraston}}}, \bibinfo {author}
  {\bibfnamefont {C.~K.}\ \bibnamefont {{McBride}}}, \bibinfo {author}
  {\bibfnamefont {K.~T.}\ \bibnamefont {{Mehta}}}, \bibinfo {author}
  {\bibfnamefont {O.}~\bibnamefont {{Mena}}}, \bibinfo {author} {\bibfnamefont
  {F.}~\bibnamefont {{Montesano}}}, \bibinfo {author} {\bibfnamefont
  {D.}~\bibnamefont {{Muna}}}, \bibinfo {author} {\bibfnamefont {R.~C.}\
  \bibnamefont {{Nichol}}}, \bibinfo {author} {\bibfnamefont {S.~E.}\
  \bibnamefont {{Nuza}}}, \bibinfo {author} {\bibfnamefont {M.~D.}\
  \bibnamefont {{Olmstead}}}, \bibinfo {author} {\bibfnamefont
  {D.}~\bibnamefont {{Oravetz}}}, \bibinfo {author} {\bibfnamefont
  {N.}~\bibnamefont {{Padmanabhan}}}, \bibinfo {author} {\bibfnamefont
  {N.}~\bibnamefont {{Palanque-Delabrouille}}}, \bibinfo {author}
  {\bibfnamefont {K.}~\bibnamefont {{Pan}}}, \bibinfo {author} {\bibfnamefont
  {J.}~\bibnamefont {{Parejko}}}, \bibinfo {author} {\bibfnamefont
  {I.}~\bibnamefont {{P{\^a}ris}}}, \bibinfo {author} {\bibfnamefont {W.~J.}\
  \bibnamefont {{Percival}}}, \bibinfo {author} {\bibfnamefont
  {P.}~\bibnamefont {{Petitjean}}}, \bibinfo {author} {\bibfnamefont
  {F.}~\bibnamefont {{Prada}}}, \bibinfo {author} {\bibfnamefont
  {B.}~\bibnamefont {{Reid}}}, \bibinfo {author} {\bibfnamefont {N.~A.}\
  \bibnamefont {{Roe}}}, \bibinfo {author} {\bibfnamefont {A.~J.}\ \bibnamefont
  {{Ross}}}, \bibinfo {author} {\bibfnamefont {N.~P.}\ \bibnamefont {{Ross}}},
  \bibinfo {author} {\bibfnamefont {L.}~\bibnamefont {{Samushia}}}, \bibinfo
  {author} {\bibfnamefont {A.~G.}\ \bibnamefont {{S{\'a}nchez}}}, \bibinfo
  {author} {\bibfnamefont {D.~J.}\ \bibnamefont {{Schlegel}}}, \bibinfo
  {author} {\bibfnamefont {D.~P.}\ \bibnamefont {{Schneider}}}, \bibinfo
  {author} {\bibfnamefont {C.~G.}\ \bibnamefont {{Sc{\'o}ccola}}}, \bibinfo
  {author} {\bibfnamefont {H.-J.}\ \bibnamefont {{Seo}}}, \bibinfo {author}
  {\bibfnamefont {E.~S.}\ \bibnamefont {{Sheldon}}}, \bibinfo {author}
  {\bibfnamefont {A.}~\bibnamefont {{Simmons}}}, \bibinfo {author}
  {\bibfnamefont {R.~A.}\ \bibnamefont {{Skibba}}}, \bibinfo {author}
  {\bibfnamefont {M.~A.}\ \bibnamefont {{Strauss}}}, \bibinfo {author}
  {\bibfnamefont {M.~E.~C.}\ \bibnamefont {{Swanson}}}, \bibinfo {author}
  {\bibfnamefont {D.}~\bibnamefont {{Thomas}}}, \bibinfo {author}
  {\bibfnamefont {J.~L.}\ \bibnamefont {{Tinker}}}, \bibinfo {author}
  {\bibfnamefont {R.}~\bibnamefont {{Tojeiro}}}, \bibinfo {author}
  {\bibfnamefont {M.~V.}\ \bibnamefont {{Maga{\~n}a}}}, \bibinfo {author}
  {\bibfnamefont {L.}~\bibnamefont {{Verde}}}, \bibinfo {author} {\bibfnamefont
  {C.}~\bibnamefont {{Wagner}}}, \bibinfo {author} {\bibfnamefont {D.~A.}\
  \bibnamefont {{Wake}}}, \bibinfo {author} {\bibfnamefont {B.~A.}\
  \bibnamefont {{Weaver}}}, \bibinfo {author} {\bibfnamefont {D.~H.}\
  \bibnamefont {{Weinberg}}}, \bibinfo {author} {\bibfnamefont
  {M.}~\bibnamefont {{White}}}, \bibinfo {author} {\bibfnamefont
  {X.}~\bibnamefont {{Xu}}}, \bibinfo {author} {\bibfnamefont {C.}~\bibnamefont
  {{Y{\`e}che}}}, \bibinfo {author} {\bibfnamefont {I.}~\bibnamefont
  {{Zehavi}}}, \ and\ \bibinfo {author} {\bibfnamefont {G.-B.}\ \bibnamefont
  {{Zhao}}},\ }\href {\doibase 10.1111/j.1365-2966.2012.22066.x} {\bibfield
  {journal} {\bibinfo  {journal} {\mnras}\ }\textbf {\bibinfo {volume} {427}},\
  \bibinfo {pages} {3435} (\bibinfo {year} {2012})},\ \Eprint
  {http://arxiv.org/abs/1203.6594} {arXiv:1203.6594 [astro-ph.CO]} \BibitemShut
  {NoStop}%
\bibitem [{\citenamefont {{Takahashi}}\ \emph {et~al.}(2009)\citenamefont
  {{Takahashi}}, \citenamefont {{Yoshida}}, \citenamefont {{Takada}},
  \citenamefont {{Matsubara}}, \citenamefont {{Sugiyama}}, \citenamefont
  {{Kayo}}, \citenamefont {{Nishizawa}}, \citenamefont {{Nishimichi}},
  \citenamefont {{Saito}},\ and\ \citenamefont
  {{Taruya}}}]{2009ApJ...700..479T}%
  \BibitemOpen
  \bibfield  {author} {\bibinfo {author} {\bibfnamefont {R.}~\bibnamefont
  {{Takahashi}}}, \bibinfo {author} {\bibfnamefont {N.}~\bibnamefont
  {{Yoshida}}}, \bibinfo {author} {\bibfnamefont {M.}~\bibnamefont {{Takada}}},
  \bibinfo {author} {\bibfnamefont {T.}~\bibnamefont {{Matsubara}}}, \bibinfo
  {author} {\bibfnamefont {N.}~\bibnamefont {{Sugiyama}}}, \bibinfo {author}
  {\bibfnamefont {I.}~\bibnamefont {{Kayo}}}, \bibinfo {author} {\bibfnamefont
  {A.~J.}\ \bibnamefont {{Nishizawa}}}, \bibinfo {author} {\bibfnamefont
  {T.}~\bibnamefont {{Nishimichi}}}, \bibinfo {author} {\bibfnamefont
  {S.}~\bibnamefont {{Saito}}}, \ and\ \bibinfo {author} {\bibfnamefont
  {A.}~\bibnamefont {{Taruya}}},\ }\href {\doibase 10.1088/0004-637X/700/1/479}
  {\bibfield  {journal} {\bibinfo  {journal} {\apj}\ }\textbf {\bibinfo
  {volume} {700}},\ \bibinfo {pages} {479} (\bibinfo {year} {2009})},\ \Eprint
  {http://arxiv.org/abs/0902.0371} {arXiv:0902.0371 [astro-ph.CO]} \BibitemShut
  {NoStop}%
\bibitem [{\citenamefont {{Anderson}}\ \emph {et~al.}(2013)\citenamefont
  {{Anderson}}, \citenamefont {{Aubourg}}, \citenamefont {{Bailey}},
  \citenamefont {{Beutler}}, \citenamefont {{Bhardwaj}}, \citenamefont
  {{Blanton}}, \citenamefont {{Bolton}}, \citenamefont {{Brinkmann}},
  \citenamefont {{Brownstein}}, \citenamefont {{Burden}}, \citenamefont
  {{Chuang}}, \citenamefont {{Cuesta}}, \citenamefont {{Dawson}}, \citenamefont
  {{Eisenstein}}, \citenamefont {{Escoffier}}, \citenamefont {{Gunn}},
  \citenamefont {{Guo}}, \citenamefont {{Ho}}, \citenamefont {{Honscheid}},
  \citenamefont {{Howlett}}, \citenamefont {{Kirkby}}, \citenamefont
  {{Lupton}}, \citenamefont {{Manera}}, \citenamefont {{Maraston}},
  \citenamefont {{McBride}}, \citenamefont {{Mena}}, \citenamefont
  {{Montesano}}, \citenamefont {{Nichol}}, \citenamefont {{Nuza}},
  \citenamefont {{Olmstead}}, \citenamefont {{Padmanabhan}}, \citenamefont
  {{Palanque-Delabrouille}}, \citenamefont {{Parejko}}, \citenamefont
  {{Percival}}, \citenamefont {{Petitjean}}, \citenamefont {{Prada}},
  \citenamefont {{Price-Whelan}}, \citenamefont {{Reid}}, \citenamefont
  {{Roe}}, \citenamefont {{Ross}}, \citenamefont {{Ross}}, \citenamefont
  {{Sabiu}}, \citenamefont {{Saito}}, \citenamefont {{Samushia}}, \citenamefont
  {{Sanchez}}, \citenamefont {{Schlegel}}, \citenamefont {{Schneider}},
  \citenamefont {{Scoccola}}, \citenamefont {{Seo}}, \citenamefont {{Skibba}},
  \citenamefont {{Strauss}}, \citenamefont {{Swanson}}, \citenamefont
  {{Thomas}}, \citenamefont {{Tinker}}, \citenamefont {{Tojeiro}},
  \citenamefont {{Vargas Magana}}, \citenamefont {{Verde}}, \citenamefont
  {{Wake}}, \citenamefont {{Weaver}}, \citenamefont {{Weinberg}}, \citenamefont
  {{White}}, \citenamefont {{Xu}}, \citenamefont {{Yeche}}, \citenamefont
  {{Zehavi}},\ and\ \citenamefont {{Zhao}}}]{2013arXiv1312.4877A}%
  \BibitemOpen
  \bibfield  {author} {\bibinfo {author} {\bibfnamefont {L.}~\bibnamefont
  {{Anderson}}}, \bibinfo {author} {\bibfnamefont {E.}~\bibnamefont
  {{Aubourg}}}, \bibinfo {author} {\bibfnamefont {S.}~\bibnamefont {{Bailey}}},
  \bibinfo {author} {\bibfnamefont {F.}~\bibnamefont {{Beutler}}}, \bibinfo
  {author} {\bibfnamefont {V.}~\bibnamefont {{Bhardwaj}}}, \bibinfo {author}
  {\bibfnamefont {M.}~\bibnamefont {{Blanton}}}, \bibinfo {author}
  {\bibfnamefont {A.~S.}\ \bibnamefont {{Bolton}}}, \bibinfo {author}
  {\bibfnamefont {J.}~\bibnamefont {{Brinkmann}}}, \bibinfo {author}
  {\bibfnamefont {J.~R.}\ \bibnamefont {{Brownstein}}}, \bibinfo {author}
  {\bibfnamefont {A.}~\bibnamefont {{Burden}}}, \bibinfo {author}
  {\bibfnamefont {C.-H.}\ \bibnamefont {{Chuang}}}, \bibinfo {author}
  {\bibfnamefont {A.~J.}\ \bibnamefont {{Cuesta}}}, \bibinfo {author}
  {\bibfnamefont {K.~S.}\ \bibnamefont {{Dawson}}}, \bibinfo {author}
  {\bibfnamefont {D.~J.}\ \bibnamefont {{Eisenstein}}}, \bibinfo {author}
  {\bibfnamefont {S.}~\bibnamefont {{Escoffier}}}, \bibinfo {author}
  {\bibfnamefont {J.~E.}\ \bibnamefont {{Gunn}}}, \bibinfo {author}
  {\bibfnamefont {H.}~\bibnamefont {{Guo}}}, \bibinfo {author} {\bibfnamefont
  {S.}~\bibnamefont {{Ho}}}, \bibinfo {author} {\bibfnamefont {K.}~\bibnamefont
  {{Honscheid}}}, \bibinfo {author} {\bibfnamefont {C.}~\bibnamefont
  {{Howlett}}}, \bibinfo {author} {\bibfnamefont {D.}~\bibnamefont {{Kirkby}}},
  \bibinfo {author} {\bibfnamefont {R.~H.}\ \bibnamefont {{Lupton}}}, \bibinfo
  {author} {\bibfnamefont {M.}~\bibnamefont {{Manera}}}, \bibinfo {author}
  {\bibfnamefont {C.}~\bibnamefont {{Maraston}}}, \bibinfo {author}
  {\bibfnamefont {C.~K.}\ \bibnamefont {{McBride}}}, \bibinfo {author}
  {\bibfnamefont {O.}~\bibnamefont {{Mena}}}, \bibinfo {author} {\bibfnamefont
  {F.}~\bibnamefont {{Montesano}}}, \bibinfo {author} {\bibfnamefont {R.~C.}\
  \bibnamefont {{Nichol}}}, \bibinfo {author} {\bibfnamefont {S.~E.}\
  \bibnamefont {{Nuza}}}, \bibinfo {author} {\bibfnamefont {M.~D.}\
  \bibnamefont {{Olmstead}}}, \bibinfo {author} {\bibfnamefont
  {N.}~\bibnamefont {{Padmanabhan}}}, \bibinfo {author} {\bibfnamefont
  {N.}~\bibnamefont {{Palanque-Delabrouille}}}, \bibinfo {author}
  {\bibfnamefont {J.}~\bibnamefont {{Parejko}}}, \bibinfo {author}
  {\bibfnamefont {W.~J.}\ \bibnamefont {{Percival}}}, \bibinfo {author}
  {\bibfnamefont {P.}~\bibnamefont {{Petitjean}}}, \bibinfo {author}
  {\bibfnamefont {F.}~\bibnamefont {{Prada}}}, \bibinfo {author} {\bibfnamefont
  {A.~M.}\ \bibnamefont {{Price-Whelan}}}, \bibinfo {author} {\bibfnamefont
  {B.}~\bibnamefont {{Reid}}}, \bibinfo {author} {\bibfnamefont {N.~A.}\
  \bibnamefont {{Roe}}}, \bibinfo {author} {\bibfnamefont {A.~J.}\ \bibnamefont
  {{Ross}}}, \bibinfo {author} {\bibfnamefont {N.~P.}\ \bibnamefont {{Ross}}},
  \bibinfo {author} {\bibfnamefont {C.~G.}\ \bibnamefont {{Sabiu}}}, \bibinfo
  {author} {\bibfnamefont {S.}~\bibnamefont {{Saito}}}, \bibinfo {author}
  {\bibfnamefont {L.}~\bibnamefont {{Samushia}}}, \bibinfo {author}
  {\bibfnamefont {A.~G.}\ \bibnamefont {{Sanchez}}}, \bibinfo {author}
  {\bibfnamefont {D.~J.}\ \bibnamefont {{Schlegel}}}, \bibinfo {author}
  {\bibfnamefont {D.~P.}\ \bibnamefont {{Schneider}}}, \bibinfo {author}
  {\bibfnamefont {C.~G.}\ \bibnamefont {{Scoccola}}}, \bibinfo {author}
  {\bibfnamefont {H.-J.}\ \bibnamefont {{Seo}}}, \bibinfo {author}
  {\bibfnamefont {R.~A.}\ \bibnamefont {{Skibba}}}, \bibinfo {author}
  {\bibfnamefont {M.~A.}\ \bibnamefont {{Strauss}}}, \bibinfo {author}
  {\bibfnamefont {M.~E.~C.}\ \bibnamefont {{Swanson}}}, \bibinfo {author}
  {\bibfnamefont {D.}~\bibnamefont {{Thomas}}}, \bibinfo {author}
  {\bibfnamefont {J.~L.}\ \bibnamefont {{Tinker}}}, \bibinfo {author}
  {\bibfnamefont {R.}~\bibnamefont {{Tojeiro}}}, \bibinfo {author}
  {\bibfnamefont {M.}~\bibnamefont {{Vargas Magana}}}, \bibinfo {author}
  {\bibfnamefont {L.}~\bibnamefont {{Verde}}}, \bibinfo {author} {\bibfnamefont
  {D.~A.}\ \bibnamefont {{Wake}}}, \bibinfo {author} {\bibfnamefont {B.~A.}\
  \bibnamefont {{Weaver}}}, \bibinfo {author} {\bibfnamefont {D.~H.}\
  \bibnamefont {{Weinberg}}}, \bibinfo {author} {\bibfnamefont
  {M.}~\bibnamefont {{White}}}, \bibinfo {author} {\bibfnamefont
  {X.}~\bibnamefont {{Xu}}}, \bibinfo {author} {\bibfnamefont {C.}~\bibnamefont
  {{Yeche}}}, \bibinfo {author} {\bibfnamefont {I.}~\bibnamefont {{Zehavi}}}, \
  and\ \bibinfo {author} {\bibfnamefont {G.-B.}\ \bibnamefont {{Zhao}}},\
  }\href@noop {} {\bibfield  {journal} {\bibinfo  {journal} {ArXiv e-prints}\ }
  (\bibinfo {year} {2013})},\ \Eprint {http://arxiv.org/abs/1312.4877}
  {arXiv:1312.4877 [astro-ph.CO]} \BibitemShut {NoStop}%
\bibitem [{\citenamefont {{Reid}}\ and\ \citenamefont
  {{White}}(2011)}]{2011MNRAS.417.1913R}%
  \BibitemOpen
  \bibfield  {author} {\bibinfo {author} {\bibfnamefont {B.~A.}\ \bibnamefont
  {{Reid}}}\ and\ \bibinfo {author} {\bibfnamefont {M.}~\bibnamefont
  {{White}}},\ }\href {\doibase 10.1111/j.1365-2966.2011.19379.x} {\bibfield
  {journal} {\bibinfo  {journal} {\mnras}\ }\textbf {\bibinfo {volume} {417}},\
  \bibinfo {pages} {1913} (\bibinfo {year} {2011})},\ \Eprint
  {http://arxiv.org/abs/1105.4165} {arXiv:1105.4165 [astro-ph.CO]} \BibitemShut
  {NoStop}%
\bibitem [{\citenamefont {{Zhao}}\ \emph {et~al.}(2014)\citenamefont {{Zhao}}
  \emph {et~al.}}]{eBOSS}%
  \BibitemOpen
  \bibfield  {author} {\bibinfo {author} {\bibfnamefont {G.-B.}\ \bibnamefont
  {{Zhao}}} \emph {et~al.},\ }\href@noop {} {\enquote {\bibinfo {title} {{The
  Extended BOSS Survey (eBOSS)}},}\ }\bibinfo {howpublished} {, in preparation}
  (\bibinfo {year} {2014})\BibitemShut {NoStop}%
\bibitem [{\citenamefont {{Chiang}}\ \emph {et~al.}(2013)\citenamefont
  {{Chiang}}, \citenamefont {{Wullstein}}, \citenamefont {{Jeong}},
  \citenamefont {{Komatsu}}, \citenamefont {{Blanc}}, \citenamefont
  {{Ciardullo}}, \citenamefont {{Drory}}, \citenamefont {{Fabricius}},
  \citenamefont {{Finkelstein}}, \citenamefont {{Gebhardt}}, \citenamefont
  {{Gronwall}}, \citenamefont {{Hagen}}, \citenamefont {{Hill}}, \citenamefont
  {{Jogee}}, \citenamefont {{Landriau}}, \citenamefont {{Mentuch Cooper}},
  \citenamefont {{Schneider}},\ and\ \citenamefont
  {{Tuttle}}}]{2013arXiv1306.4157C}%
  \BibitemOpen
  \bibfield  {author} {\bibinfo {author} {\bibfnamefont {C.-T.}\ \bibnamefont
  {{Chiang}}}, \bibinfo {author} {\bibfnamefont {P.}~\bibnamefont
  {{Wullstein}}}, \bibinfo {author} {\bibfnamefont {D.}~\bibnamefont
  {{Jeong}}}, \bibinfo {author} {\bibfnamefont {E.}~\bibnamefont {{Komatsu}}},
  \bibinfo {author} {\bibfnamefont {G.~A.}\ \bibnamefont {{Blanc}}}, \bibinfo
  {author} {\bibfnamefont {R.}~\bibnamefont {{Ciardullo}}}, \bibinfo {author}
  {\bibfnamefont {N.}~\bibnamefont {{Drory}}}, \bibinfo {author} {\bibfnamefont
  {M.}~\bibnamefont {{Fabricius}}}, \bibinfo {author} {\bibfnamefont
  {S.}~\bibnamefont {{Finkelstein}}}, \bibinfo {author} {\bibfnamefont
  {K.}~\bibnamefont {{Gebhardt}}}, \bibinfo {author} {\bibfnamefont
  {C.}~\bibnamefont {{Gronwall}}}, \bibinfo {author} {\bibfnamefont
  {A.}~\bibnamefont {{Hagen}}}, \bibinfo {author} {\bibfnamefont {G.~J.}\
  \bibnamefont {{Hill}}}, \bibinfo {author} {\bibfnamefont {S.}~\bibnamefont
  {{Jogee}}}, \bibinfo {author} {\bibfnamefont {M.}~\bibnamefont {{Landriau}}},
  \bibinfo {author} {\bibfnamefont {E.}~\bibnamefont {{Mentuch Cooper}}},
  \bibinfo {author} {\bibfnamefont {D.~P.}\ \bibnamefont {{Schneider}}}, \ and\
  \bibinfo {author} {\bibfnamefont {S.}~\bibnamefont {{Tuttle}}},\ }\href@noop
  {} {\bibfield  {journal} {\bibinfo  {journal} {ArXiv e-prints}\ } (\bibinfo
  {year} {2013})},\ \Eprint {http://arxiv.org/abs/1306.4157} {arXiv:1306.4157
  [astro-ph.CO]} \BibitemShut {NoStop}%
\bibitem [{\citenamefont {{Levi}}\ \emph {et~al.}(2013)\citenamefont {{Levi}},
  \citenamefont {{Bebek}}, \citenamefont {{Beers}}, \citenamefont {{Blum}},
  \citenamefont {{Cahn}}, \citenamefont {{Eisenstein}}, \citenamefont
  {{Flaugher}}, \citenamefont {{Honscheid}}, \citenamefont {{Kron}},
  \citenamefont {{Lahav}}, \citenamefont {{McDonald}}, \citenamefont {{Roe}},
  \citenamefont {{Schlegel}},\ and\ \citenamefont {{representing the DESI
  collaboration}}}]{2013arXiv1308.0847L}%
  \BibitemOpen
  \bibfield  {author} {\bibinfo {author} {\bibfnamefont {M.}~\bibnamefont
  {{Levi}}}, \bibinfo {author} {\bibfnamefont {C.}~\bibnamefont {{Bebek}}},
  \bibinfo {author} {\bibfnamefont {T.}~\bibnamefont {{Beers}}}, \bibinfo
  {author} {\bibfnamefont {R.}~\bibnamefont {{Blum}}}, \bibinfo {author}
  {\bibfnamefont {R.}~\bibnamefont {{Cahn}}}, \bibinfo {author} {\bibfnamefont
  {D.}~\bibnamefont {{Eisenstein}}}, \bibinfo {author} {\bibfnamefont
  {B.}~\bibnamefont {{Flaugher}}}, \bibinfo {author} {\bibfnamefont
  {K.}~\bibnamefont {{Honscheid}}}, \bibinfo {author} {\bibfnamefont
  {R.}~\bibnamefont {{Kron}}}, \bibinfo {author} {\bibfnamefont
  {O.}~\bibnamefont {{Lahav}}}, \bibinfo {author} {\bibfnamefont
  {P.}~\bibnamefont {{McDonald}}}, \bibinfo {author} {\bibfnamefont
  {N.}~\bibnamefont {{Roe}}}, \bibinfo {author} {\bibfnamefont
  {D.}~\bibnamefont {{Schlegel}}}, \ and\ \bibinfo {author} {\bibnamefont
  {{representing the DESI collaboration}}},\ }\href@noop {} {\bibfield
  {journal} {\bibinfo  {journal} {ArXiv e-prints}\ } (\bibinfo {year}
  {2013})},\ \Eprint {http://arxiv.org/abs/1308.0847} {arXiv:1308.0847
  [astro-ph.CO]} \BibitemShut {NoStop}%
\bibitem [{\citenamefont {{Schlegel}}\ \emph {et~al.}(2011)\citenamefont
  {{Schlegel}}, \citenamefont {{Abdalla}}, \citenamefont {{Abraham}},
  \citenamefont {{Ahn}}, \citenamefont {{Allende Prieto}}, \citenamefont
  {{Annis}}, \citenamefont {{Aubourg}}, \citenamefont {{Azzaro}}, \citenamefont
  {{Baltay}}, \citenamefont {{Baugh}}, \citenamefont {{Bebek}}, \citenamefont
  {{Becerril}}, \citenamefont {{Blanton}}, \citenamefont {{Bolton}},
  \citenamefont {{Bromley}}, \citenamefont {{Cahn}}, \citenamefont {{Carton}},
  \citenamefont {{Cervantes-Cota}}, \citenamefont {{Chu}}, \citenamefont
  {{Cortes}}, \citenamefont {{Dawson}}, \citenamefont {{Dey}}, \citenamefont
  {{Dickinson}}, \citenamefont {{Diehl}}, \citenamefont {{Doel}}, \citenamefont
  {{Ealet}}, \citenamefont {{Edelstein}}, \citenamefont {{Eppelle}},
  \citenamefont {{Escoffier}}, \citenamefont {{Evrard}}, \citenamefont
  {{Faccioli}}, \citenamefont {{Frenk}}, \citenamefont {{Geha}}, \citenamefont
  {{Gerdes}}, \citenamefont {{Gondolo}}, \citenamefont {{Gonzalez-Arroyo}},
  \citenamefont {{Grossan}}, \citenamefont {{Heckman}}, \citenamefont
  {{Heetderks}}, \citenamefont {{Ho}}, \citenamefont {{Honscheid}},
  \citenamefont {{Huterer}}, \citenamefont {{Ilbert}}, \citenamefont {{Ivans}},
  \citenamefont {{Jelinsky}}, \citenamefont {{Jing}}, \citenamefont {{Joyce}},
  \citenamefont {{Kennedy}}, \citenamefont {{Kent}}, \citenamefont {{Kieda}},
  \citenamefont {{Kim}}, \citenamefont {{Kim}}, \citenamefont {{Kneib}},
  \citenamefont {{Kong}}, \citenamefont {{Kosowsky}}, \citenamefont
  {{Krishnan}}, \citenamefont {{Lahav}}, \citenamefont {{Lampton}},
  \citenamefont {{LeBohec}}, \citenamefont {{Le Brun}}, \citenamefont {{Levi}},
  \citenamefont {{Li}}, \citenamefont {{Liang}}, \citenamefont {{Lim}},
  \citenamefont {{Lin}}, \citenamefont {{Linder}}, \citenamefont {{Lorenzon}},
  \citenamefont {{de la Macorra}}, \citenamefont {{Magneville}}, \citenamefont
  {{Malina}}, \citenamefont {{Marinoni}}, \citenamefont {{Martinez}},
  \citenamefont {{Majewski}}, \citenamefont {{Matheson}}, \citenamefont
  {{McCloskey}}, \citenamefont {{McDonald}}, \citenamefont {{McKay}},
  \citenamefont {{McMahon}}, \citenamefont {{Menard}}, \citenamefont
  {{Miralda-Escude}}, \citenamefont {{Modjaz}}, \citenamefont
  {{Montero-Dorta}}, \citenamefont {{Morales}}, \citenamefont {{Mostek}},
  \citenamefont {{Newman}}, \citenamefont {{Nichol}}, \citenamefont {{Nugent}},
  \citenamefont {{Olsen}}, \citenamefont {{Padmanabhan}}, \citenamefont
  {{Palanque-Delabrouille}}, \citenamefont {{Park}}, \citenamefont {{Peacock}},
  \citenamefont {{Percival}}, \citenamefont {{Perlmutter}}, \citenamefont
  {{Peroux}}, \citenamefont {{Petitjean}}, \citenamefont {{Prada}},
  \citenamefont {{Prieto}}, \citenamefont {{Prochaska}}, \citenamefont
  {{Reil}}, \citenamefont {{Rockosi}}, \citenamefont {{Roe}}, \citenamefont
  {{Rollinde}}, \citenamefont {{Roodman}}, \citenamefont {{Ross}},
  \citenamefont {{Rudnick}}, \citenamefont {{Ruhlmann-Kleider}}, \citenamefont
  {{Sanchez}}, \citenamefont {{Sawyer}}, \citenamefont {{Schimd}},
  \citenamefont {{Schubnell}}, \citenamefont {{Scoccimaro}}, \citenamefont
  {{Seljak}}, \citenamefont {{Seo}}, \citenamefont {{Sheldon}}, \citenamefont
  {{Sholl}}, \citenamefont {{Shulte-Ladbeck}}, \citenamefont {{Slosar}},
  \citenamefont {{Smith}}, \citenamefont {{Smoot}}, \citenamefont {{Springer}},
  \citenamefont {{Stril}}, \citenamefont {{Szalay}}, \citenamefont {{Tao}},
  \citenamefont {{Tarle}}, \citenamefont {{Taylor}}, \citenamefont {{Tilquin}},
  \citenamefont {{Tinker}}, \citenamefont {{Valdes}}, \citenamefont {{Wang}},
  \citenamefont {{Wang}}, \citenamefont {{Weaver}}, \citenamefont {{Weinberg}},
  \citenamefont {{White}}, \citenamefont {{Wood-Vasey}}, \citenamefont
  {{Yang}}, \citenamefont {{Yeche}}, \citenamefont {{Zakamska}}, \citenamefont
  {{Zentner}}, \citenamefont {{Zhai}},\ and\ \citenamefont
  {{Zhang}}}]{2011arXiv1106.1706S}%
  \BibitemOpen
  \bibfield  {author} {\bibinfo {author} {\bibfnamefont {D.}~\bibnamefont
  {{Schlegel}}}, \bibinfo {author} {\bibfnamefont {F.}~\bibnamefont
  {{Abdalla}}}, \bibinfo {author} {\bibfnamefont {T.}~\bibnamefont
  {{Abraham}}}, \bibinfo {author} {\bibfnamefont {C.}~\bibnamefont {{Ahn}}},
  \bibinfo {author} {\bibfnamefont {C.}~\bibnamefont {{Allende Prieto}}},
  \bibinfo {author} {\bibfnamefont {J.}~\bibnamefont {{Annis}}}, \bibinfo
  {author} {\bibfnamefont {E.}~\bibnamefont {{Aubourg}}}, \bibinfo {author}
  {\bibfnamefont {M.}~\bibnamefont {{Azzaro}}}, \bibinfo {author}
  {\bibfnamefont {S.~B.~C.}\ \bibnamefont {{Baltay}}}, \bibinfo {author}
  {\bibfnamefont {C.}~\bibnamefont {{Baugh}}}, \bibinfo {author} {\bibfnamefont
  {C.}~\bibnamefont {{Bebek}}}, \bibinfo {author} {\bibfnamefont
  {S.}~\bibnamefont {{Becerril}}}, \bibinfo {author} {\bibfnamefont
  {M.}~\bibnamefont {{Blanton}}}, \bibinfo {author} {\bibfnamefont
  {A.}~\bibnamefont {{Bolton}}}, \bibinfo {author} {\bibfnamefont
  {B.}~\bibnamefont {{Bromley}}}, \bibinfo {author} {\bibfnamefont
  {R.}~\bibnamefont {{Cahn}}}, \bibinfo {author} {\bibfnamefont {P.~.}\
  \bibnamefont {{Carton}}}, \bibinfo {author} {\bibfnamefont {J.~L.}\
  \bibnamefont {{Cervantes-Cota}}}, \bibinfo {author} {\bibfnamefont
  {Y.}~\bibnamefont {{Chu}}}, \bibinfo {author} {\bibfnamefont
  {M.}~\bibnamefont {{Cortes}}}, \bibinfo {author} {\bibfnamefont
  {K.}~\bibnamefont {{Dawson}}}, \bibinfo {author} {\bibfnamefont
  {A.}~\bibnamefont {{Dey}}}, \bibinfo {author} {\bibfnamefont
  {M.}~\bibnamefont {{Dickinson}}}, \bibinfo {author} {\bibfnamefont {H.~T.}\
  \bibnamefont {{Diehl}}}, \bibinfo {author} {\bibfnamefont {P.}~\bibnamefont
  {{Doel}}}, \bibinfo {author} {\bibfnamefont {A.}~\bibnamefont {{Ealet}}},
  \bibinfo {author} {\bibfnamefont {J.}~\bibnamefont {{Edelstein}}}, \bibinfo
  {author} {\bibfnamefont {D.}~\bibnamefont {{Eppelle}}}, \bibinfo {author}
  {\bibfnamefont {S.}~\bibnamefont {{Escoffier}}}, \bibinfo {author}
  {\bibfnamefont {A.}~\bibnamefont {{Evrard}}}, \bibinfo {author}
  {\bibfnamefont {L.}~\bibnamefont {{Faccioli}}}, \bibinfo {author}
  {\bibfnamefont {C.}~\bibnamefont {{Frenk}}}, \bibinfo {author} {\bibfnamefont
  {M.}~\bibnamefont {{Geha}}}, \bibinfo {author} {\bibfnamefont
  {D.}~\bibnamefont {{Gerdes}}}, \bibinfo {author} {\bibfnamefont
  {P.}~\bibnamefont {{Gondolo}}}, \bibinfo {author} {\bibfnamefont
  {A.}~\bibnamefont {{Gonzalez-Arroyo}}}, \bibinfo {author} {\bibfnamefont
  {B.}~\bibnamefont {{Grossan}}}, \bibinfo {author} {\bibfnamefont
  {T.}~\bibnamefont {{Heckman}}}, \bibinfo {author} {\bibfnamefont
  {H.}~\bibnamefont {{Heetderks}}}, \bibinfo {author} {\bibfnamefont
  {S.}~\bibnamefont {{Ho}}}, \bibinfo {author} {\bibfnamefont {K.}~\bibnamefont
  {{Honscheid}}}, \bibinfo {author} {\bibfnamefont {D.}~\bibnamefont
  {{Huterer}}}, \bibinfo {author} {\bibfnamefont {O.}~\bibnamefont {{Ilbert}}},
  \bibinfo {author} {\bibfnamefont {I.}~\bibnamefont {{Ivans}}}, \bibinfo
  {author} {\bibfnamefont {P.}~\bibnamefont {{Jelinsky}}}, \bibinfo {author}
  {\bibfnamefont {Y.}~\bibnamefont {{Jing}}}, \bibinfo {author} {\bibfnamefont
  {D.}~\bibnamefont {{Joyce}}}, \bibinfo {author} {\bibfnamefont
  {R.}~\bibnamefont {{Kennedy}}}, \bibinfo {author} {\bibfnamefont
  {S.}~\bibnamefont {{Kent}}}, \bibinfo {author} {\bibfnamefont
  {D.}~\bibnamefont {{Kieda}}}, \bibinfo {author} {\bibfnamefont
  {A.}~\bibnamefont {{Kim}}}, \bibinfo {author} {\bibfnamefont
  {C.}~\bibnamefont {{Kim}}}, \bibinfo {author} {\bibfnamefont {J.~.}\
  \bibnamefont {{Kneib}}}, \bibinfo {author} {\bibfnamefont {X.}~\bibnamefont
  {{Kong}}}, \bibinfo {author} {\bibfnamefont {A.}~\bibnamefont {{Kosowsky}}},
  \bibinfo {author} {\bibfnamefont {K.}~\bibnamefont {{Krishnan}}}, \bibinfo
  {author} {\bibfnamefont {O.}~\bibnamefont {{Lahav}}}, \bibinfo {author}
  {\bibfnamefont {M.}~\bibnamefont {{Lampton}}}, \bibinfo {author}
  {\bibfnamefont {S.}~\bibnamefont {{LeBohec}}}, \bibinfo {author}
  {\bibfnamefont {V.}~\bibnamefont {{Le Brun}}}, \bibinfo {author}
  {\bibfnamefont {M.}~\bibnamefont {{Levi}}}, \bibinfo {author} {\bibfnamefont
  {C.}~\bibnamefont {{Li}}}, \bibinfo {author} {\bibfnamefont {M.}~\bibnamefont
  {{Liang}}}, \bibinfo {author} {\bibfnamefont {H.}~\bibnamefont {{Lim}}},
  \bibinfo {author} {\bibfnamefont {W.}~\bibnamefont {{Lin}}}, \bibinfo
  {author} {\bibfnamefont {E.}~\bibnamefont {{Linder}}}, \bibinfo {author}
  {\bibfnamefont {W.}~\bibnamefont {{Lorenzon}}}, \bibinfo {author}
  {\bibfnamefont {A.}~\bibnamefont {{de la Macorra}}}, \bibinfo {author}
  {\bibfnamefont {C.}~\bibnamefont {{Magneville}}}, \bibinfo {author}
  {\bibfnamefont {R.}~\bibnamefont {{Malina}}}, \bibinfo {author}
  {\bibfnamefont {C.}~\bibnamefont {{Marinoni}}}, \bibinfo {author}
  {\bibfnamefont {V.}~\bibnamefont {{Martinez}}}, \bibinfo {author}
  {\bibfnamefont {S.}~\bibnamefont {{Majewski}}}, \bibinfo {author}
  {\bibfnamefont {T.}~\bibnamefont {{Matheson}}}, \bibinfo {author}
  {\bibfnamefont {R.}~\bibnamefont {{McCloskey}}}, \bibinfo {author}
  {\bibfnamefont {P.}~\bibnamefont {{McDonald}}}, \bibinfo {author}
  {\bibfnamefont {T.}~\bibnamefont {{McKay}}}, \bibinfo {author} {\bibfnamefont
  {J.}~\bibnamefont {{McMahon}}}, \bibinfo {author} {\bibfnamefont
  {B.}~\bibnamefont {{Menard}}}, \bibinfo {author} {\bibfnamefont
  {J.}~\bibnamefont {{Miralda-Escude}}}, \bibinfo {author} {\bibfnamefont
  {M.}~\bibnamefont {{Modjaz}}}, \bibinfo {author} {\bibfnamefont
  {A.}~\bibnamefont {{Montero-Dorta}}}, \bibinfo {author} {\bibfnamefont
  {I.}~\bibnamefont {{Morales}}}, \bibinfo {author} {\bibfnamefont
  {N.}~\bibnamefont {{Mostek}}}, \bibinfo {author} {\bibfnamefont
  {J.}~\bibnamefont {{Newman}}}, \bibinfo {author} {\bibfnamefont
  {R.}~\bibnamefont {{Nichol}}}, \bibinfo {author} {\bibfnamefont
  {P.}~\bibnamefont {{Nugent}}}, \bibinfo {author} {\bibfnamefont
  {K.}~\bibnamefont {{Olsen}}}, \bibinfo {author} {\bibfnamefont
  {N.}~\bibnamefont {{Padmanabhan}}}, \bibinfo {author} {\bibfnamefont
  {N.}~\bibnamefont {{Palanque-Delabrouille}}}, \bibinfo {author}
  {\bibfnamefont {I.}~\bibnamefont {{Park}}}, \bibinfo {author} {\bibfnamefont
  {J.}~\bibnamefont {{Peacock}}}, \bibinfo {author} {\bibfnamefont
  {W.}~\bibnamefont {{Percival}}}, \bibinfo {author} {\bibfnamefont
  {S.}~\bibnamefont {{Perlmutter}}}, \bibinfo {author} {\bibfnamefont
  {C.}~\bibnamefont {{Peroux}}}, \bibinfo {author} {\bibfnamefont
  {P.}~\bibnamefont {{Petitjean}}}, \bibinfo {author} {\bibfnamefont
  {F.}~\bibnamefont {{Prada}}}, \bibinfo {author} {\bibfnamefont
  {E.}~\bibnamefont {{Prieto}}}, \bibinfo {author} {\bibfnamefont
  {J.}~\bibnamefont {{Prochaska}}}, \bibinfo {author} {\bibfnamefont
  {K.}~\bibnamefont {{Reil}}}, \bibinfo {author} {\bibfnamefont
  {C.}~\bibnamefont {{Rockosi}}}, \bibinfo {author} {\bibfnamefont
  {N.}~\bibnamefont {{Roe}}}, \bibinfo {author} {\bibfnamefont
  {E.}~\bibnamefont {{Rollinde}}}, \bibinfo {author} {\bibfnamefont
  {A.}~\bibnamefont {{Roodman}}}, \bibinfo {author} {\bibfnamefont
  {N.}~\bibnamefont {{Ross}}}, \bibinfo {author} {\bibfnamefont
  {G.}~\bibnamefont {{Rudnick}}}, \bibinfo {author} {\bibfnamefont
  {V.}~\bibnamefont {{Ruhlmann-Kleider}}}, \bibinfo {author} {\bibfnamefont
  {J.}~\bibnamefont {{Sanchez}}}, \bibinfo {author} {\bibfnamefont
  {D.}~\bibnamefont {{Sawyer}}}, \bibinfo {author} {\bibfnamefont
  {C.}~\bibnamefont {{Schimd}}}, \bibinfo {author} {\bibfnamefont
  {M.}~\bibnamefont {{Schubnell}}}, \bibinfo {author} {\bibfnamefont
  {R.}~\bibnamefont {{Scoccimaro}}}, \bibinfo {author} {\bibfnamefont
  {U.}~\bibnamefont {{Seljak}}}, \bibinfo {author} {\bibfnamefont
  {H.}~\bibnamefont {{Seo}}}, \bibinfo {author} {\bibfnamefont
  {E.}~\bibnamefont {{Sheldon}}}, \bibinfo {author} {\bibfnamefont
  {M.}~\bibnamefont {{Sholl}}}, \bibinfo {author} {\bibfnamefont
  {R.}~\bibnamefont {{Shulte-Ladbeck}}}, \bibinfo {author} {\bibfnamefont
  {A.}~\bibnamefont {{Slosar}}}, \bibinfo {author} {\bibfnamefont {D.~S.}\
  \bibnamefont {{Smith}}}, \bibinfo {author} {\bibfnamefont {G.}~\bibnamefont
  {{Smoot}}}, \bibinfo {author} {\bibfnamefont {W.}~\bibnamefont {{Springer}}},
  \bibinfo {author} {\bibfnamefont {A.}~\bibnamefont {{Stril}}}, \bibinfo
  {author} {\bibfnamefont {A.~S.}\ \bibnamefont {{Szalay}}}, \bibinfo {author}
  {\bibfnamefont {C.}~\bibnamefont {{Tao}}}, \bibinfo {author} {\bibfnamefont
  {G.}~\bibnamefont {{Tarle}}}, \bibinfo {author} {\bibfnamefont
  {E.}~\bibnamefont {{Taylor}}}, \bibinfo {author} {\bibfnamefont
  {A.}~\bibnamefont {{Tilquin}}}, \bibinfo {author} {\bibfnamefont
  {J.}~\bibnamefont {{Tinker}}}, \bibinfo {author} {\bibfnamefont
  {F.}~\bibnamefont {{Valdes}}}, \bibinfo {author} {\bibfnamefont
  {J.}~\bibnamefont {{Wang}}}, \bibinfo {author} {\bibfnamefont
  {T.}~\bibnamefont {{Wang}}}, \bibinfo {author} {\bibfnamefont {B.~A.}\
  \bibnamefont {{Weaver}}}, \bibinfo {author} {\bibfnamefont {D.}~\bibnamefont
  {{Weinberg}}}, \bibinfo {author} {\bibfnamefont {M.}~\bibnamefont {{White}}},
  \bibinfo {author} {\bibfnamefont {M.}~\bibnamefont {{Wood-Vasey}}}, \bibinfo
  {author} {\bibfnamefont {J.}~\bibnamefont {{Yang}}}, \bibinfo {author}
  {\bibfnamefont {X.~Y.~C.}\ \bibnamefont {{Yeche}}}, \bibinfo {author}
  {\bibfnamefont {N.}~\bibnamefont {{Zakamska}}}, \bibinfo {author}
  {\bibfnamefont {A.}~\bibnamefont {{Zentner}}}, \bibinfo {author}
  {\bibfnamefont {C.}~\bibnamefont {{Zhai}}}, \ and\ \bibinfo {author}
  {\bibfnamefont {P.}~\bibnamefont {{Zhang}}},\ }\href@noop {} {\bibfield
  {journal} {\bibinfo  {journal} {ArXiv e-prints}\ } (\bibinfo {year}
  {2011})},\ \Eprint {http://arxiv.org/abs/1106.1706} {arXiv:1106.1706
  [astro-ph.IM]} \BibitemShut {NoStop}%
\bibitem [{\citenamefont {{Mostek}}\ \emph
  {et~al.}(2012{\natexlab{a}})\citenamefont {{Mostek}}, \citenamefont
  {{Barbary}}, \citenamefont {{Bebek}}, \citenamefont {{Dey}}, \citenamefont
  {{Edelstein}}, \citenamefont {{Jelinsky}}, \citenamefont {{Kim}},
  \citenamefont {{Lampton}}, \citenamefont {{Levi}}, \citenamefont
  {{McDonald}}, \citenamefont {{Poppett}}, \citenamefont {{Roe}}, \citenamefont
  {{Schlegel}},\ and\ \citenamefont {{Sholl}}}]{2012SPIE.8446E..0QM}%
  \BibitemOpen
  \bibfield  {author} {\bibinfo {author} {\bibfnamefont {N.}~\bibnamefont
  {{Mostek}}}, \bibinfo {author} {\bibfnamefont {K.}~\bibnamefont {{Barbary}}},
  \bibinfo {author} {\bibfnamefont {C.~J.}\ \bibnamefont {{Bebek}}}, \bibinfo
  {author} {\bibfnamefont {A.~T.}\ \bibnamefont {{Dey}}}, \bibinfo {author}
  {\bibfnamefont {J.}~\bibnamefont {{Edelstein}}}, \bibinfo {author}
  {\bibfnamefont {P.}~\bibnamefont {{Jelinsky}}}, \bibinfo {author}
  {\bibfnamefont {A.~G.}\ \bibnamefont {{Kim}}}, \bibinfo {author}
  {\bibfnamefont {M.~L.}\ \bibnamefont {{Lampton}}}, \bibinfo {author}
  {\bibfnamefont {M.~E.}\ \bibnamefont {{Levi}}}, \bibinfo {author}
  {\bibfnamefont {P.}~\bibnamefont {{McDonald}}}, \bibinfo {author}
  {\bibfnamefont {C.}~\bibnamefont {{Poppett}}}, \bibinfo {author}
  {\bibfnamefont {N.~A.}\ \bibnamefont {{Roe}}}, \bibinfo {author}
  {\bibfnamefont {D.~J.}\ \bibnamefont {{Schlegel}}}, \ and\ \bibinfo {author}
  {\bibfnamefont {M.~J.}\ \bibnamefont {{Sholl}}},\ }in\ \href {\doibase
  10.1117/12.924915} {\emph {\bibinfo {booktitle} {Society of Photo-Optical
  Instrumentation Engineers (SPIE) Conference Series}}},\ \bibinfo {series}
  {Society of Photo-Optical Instrumentation Engineers (SPIE) Conference
  Series}, Vol.\ \bibinfo {volume} {8446}\ (\bibinfo {year} {2012})\BibitemShut
  {NoStop}%
\bibitem [{\citenamefont {{Mostek}}\ \emph
  {et~al.}(2012{\natexlab{b}})\citenamefont {{Mostek}}, \citenamefont
  {{Barbary}}, \citenamefont {{Dey}}, \citenamefont {{Kennedy}}, \citenamefont
  {{Kim}}, \citenamefont {{Kneib}}, \citenamefont {{Newman}}, \citenamefont
  {{Nugent}}, \citenamefont {{Padmanabhan}}, \citenamefont {{Schlegel}},\ and\
  \citenamefont {{BigBOSS Collaboration}}}]{2012AAS...21933513M}%
  \BibitemOpen
  \bibfield  {author} {\bibinfo {author} {\bibfnamefont {N.~J.}\ \bibnamefont
  {{Mostek}}}, \bibinfo {author} {\bibfnamefont {K.}~\bibnamefont {{Barbary}}},
  \bibinfo {author} {\bibfnamefont {A.}~\bibnamefont {{Dey}}}, \bibinfo
  {author} {\bibfnamefont {R.}~\bibnamefont {{Kennedy}}}, \bibinfo {author}
  {\bibfnamefont {A.}~\bibnamefont {{Kim}}}, \bibinfo {author} {\bibfnamefont
  {J.}~\bibnamefont {{Kneib}}}, \bibinfo {author} {\bibfnamefont
  {J.}~\bibnamefont {{Newman}}}, \bibinfo {author} {\bibfnamefont
  {P.}~\bibnamefont {{Nugent}}}, \bibinfo {author} {\bibfnamefont
  {N.}~\bibnamefont {{Padmanabhan}}}, \bibinfo {author} {\bibfnamefont
  {D.}~\bibnamefont {{Schlegel}}}, \ and\ \bibinfo {author} {\bibnamefont
  {{BigBOSS Collaboration}}},\ }in\ \href@noop {} {\emph {\bibinfo {booktitle}
  {American Astronomical Society Meeting Abstracts \#219}}},\ \bibinfo {series}
  {American Astronomical Society Meeting Abstracts}, Vol.\ \bibinfo {volume}
  {219}\ (\bibinfo {year} {2012})\ p.\ \bibinfo {pages} {\#335.13}\BibitemShut
  {NoStop}%
\bibitem [{\citenamefont {{Mostek}}\ \emph {et~al.}(2013)\citenamefont
  {{Mostek}}, \citenamefont {{Coil}}, \citenamefont {{Cooper}}, \citenamefont
  {{Davis}}, \citenamefont {{Newman}},\ and\ \citenamefont
  {{Weiner}}}]{2013ApJ...767...89M}%
  \BibitemOpen
  \bibfield  {author} {\bibinfo {author} {\bibfnamefont {N.}~\bibnamefont
  {{Mostek}}}, \bibinfo {author} {\bibfnamefont {A.~L.}\ \bibnamefont
  {{Coil}}}, \bibinfo {author} {\bibfnamefont {M.}~\bibnamefont {{Cooper}}},
  \bibinfo {author} {\bibfnamefont {M.}~\bibnamefont {{Davis}}}, \bibinfo
  {author} {\bibfnamefont {J.~A.}\ \bibnamefont {{Newman}}}, \ and\ \bibinfo
  {author} {\bibfnamefont {B.~J.}\ \bibnamefont {{Weiner}}},\ }\href {\doibase
  10.1088/0004-637X/767/1/89} {\bibfield  {journal} {\bibinfo  {journal}
  {\apj}\ }\textbf {\bibinfo {volume} {767}},\ \bibinfo {eid} {89} (\bibinfo
  {year} {2013})},\ \Eprint {http://arxiv.org/abs/1210.6694} {arXiv:1210.6694
  [astro-ph.CO]} \BibitemShut {NoStop}%
\bibitem [{\citenamefont {{Ross}}\ \emph {et~al.}(2009)\citenamefont {{Ross}},
  \citenamefont {{Shen}}, \citenamefont {{Strauss}}, \citenamefont {{Vanden
  Berk}}, \citenamefont {{Connolly}}, \citenamefont {{Richards}}, \citenamefont
  {{Schneider}}, \citenamefont {{Weinberg}}, \citenamefont {{Hall}},
  \citenamefont {{Bahcall}},\ and\ \citenamefont
  {{Brunner}}}]{2009ApJ...697.1634R}%
  \BibitemOpen
  \bibfield  {author} {\bibinfo {author} {\bibfnamefont {N.~P.}\ \bibnamefont
  {{Ross}}}, \bibinfo {author} {\bibfnamefont {Y.}~\bibnamefont {{Shen}}},
  \bibinfo {author} {\bibfnamefont {M.~A.}\ \bibnamefont {{Strauss}}}, \bibinfo
  {author} {\bibfnamefont {D.~E.}\ \bibnamefont {{Vanden Berk}}}, \bibinfo
  {author} {\bibfnamefont {A.~J.}\ \bibnamefont {{Connolly}}}, \bibinfo
  {author} {\bibfnamefont {G.~T.}\ \bibnamefont {{Richards}}}, \bibinfo
  {author} {\bibfnamefont {D.~P.}\ \bibnamefont {{Schneider}}}, \bibinfo
  {author} {\bibfnamefont {D.~H.}\ \bibnamefont {{Weinberg}}}, \bibinfo
  {author} {\bibfnamefont {P.~B.}\ \bibnamefont {{Hall}}}, \bibinfo {author}
  {\bibfnamefont {N.~A.}\ \bibnamefont {{Bahcall}}}, \ and\ \bibinfo {author}
  {\bibfnamefont {R.~J.}\ \bibnamefont {{Brunner}}},\ }\href {\doibase
  10.1088/0004-637X/697/2/1634} {\bibfield  {journal} {\bibinfo  {journal}
  {\apj}\ }\textbf {\bibinfo {volume} {697}},\ \bibinfo {pages} {1634}
  (\bibinfo {year} {2009})},\ \Eprint {http://arxiv.org/abs/0903.3230}
  {arXiv:0903.3230 [astro-ph.CO]} \BibitemShut {NoStop}%
\bibitem [{\citenamefont {{Palanque-Delabrouille}}\ \emph
  {et~al.}(2013{\natexlab{b}})\citenamefont {{Palanque-Delabrouille}},
  \citenamefont {{Magneville}}, \citenamefont {{Y{\`e}che}}, \citenamefont
  {{Eftekharzadeh}}, \citenamefont {{Myers}}, \citenamefont {{Petitjean}},
  \citenamefont {{P{\^a}ris}}, \citenamefont {{Aubourg}}, \citenamefont
  {{McGreer}}, \citenamefont {{Fan}}, \citenamefont {{Dey}}, \citenamefont
  {{Schlegel}}, \citenamefont {{Bailey}}, \citenamefont {{Bizayev}},
  \citenamefont {{Bolton}}, \citenamefont {{Dawson}}, \citenamefont {{Ebelke}},
  \citenamefont {{Ge}}, \citenamefont {{Malanushenko}}, \citenamefont
  {{Malanushenko}}, \citenamefont {{Oravetz}}, \citenamefont {{Pan}},
  \citenamefont {{Ross}}, \citenamefont {{Schneider}}, \citenamefont
  {{Sheldon}}, \citenamefont {{Simmons}}, \citenamefont {{Tinker}},
  \citenamefont {{White}},\ and\ \citenamefont
  {{Willmer}}}]{2013A&A...551A..29P}%
  \BibitemOpen
  \bibfield  {author} {\bibinfo {author} {\bibfnamefont {N.}~\bibnamefont
  {{Palanque-Delabrouille}}}, \bibinfo {author} {\bibfnamefont
  {C.}~\bibnamefont {{Magneville}}}, \bibinfo {author} {\bibfnamefont
  {C.}~\bibnamefont {{Y{\`e}che}}}, \bibinfo {author} {\bibfnamefont
  {S.}~\bibnamefont {{Eftekharzadeh}}}, \bibinfo {author} {\bibfnamefont
  {A.~D.}\ \bibnamefont {{Myers}}}, \bibinfo {author} {\bibfnamefont
  {P.}~\bibnamefont {{Petitjean}}}, \bibinfo {author} {\bibfnamefont
  {I.}~\bibnamefont {{P{\^a}ris}}}, \bibinfo {author} {\bibfnamefont
  {E.}~\bibnamefont {{Aubourg}}}, \bibinfo {author} {\bibfnamefont
  {I.}~\bibnamefont {{McGreer}}}, \bibinfo {author} {\bibfnamefont
  {X.}~\bibnamefont {{Fan}}}, \bibinfo {author} {\bibfnamefont
  {A.}~\bibnamefont {{Dey}}}, \bibinfo {author} {\bibfnamefont
  {D.}~\bibnamefont {{Schlegel}}}, \bibinfo {author} {\bibfnamefont
  {S.}~\bibnamefont {{Bailey}}}, \bibinfo {author} {\bibfnamefont
  {D.}~\bibnamefont {{Bizayev}}}, \bibinfo {author} {\bibfnamefont
  {A.}~\bibnamefont {{Bolton}}}, \bibinfo {author} {\bibfnamefont
  {K.}~\bibnamefont {{Dawson}}}, \bibinfo {author} {\bibfnamefont
  {G.}~\bibnamefont {{Ebelke}}}, \bibinfo {author} {\bibfnamefont
  {J.}~\bibnamefont {{Ge}}}, \bibinfo {author} {\bibfnamefont {E.}~\bibnamefont
  {{Malanushenko}}}, \bibinfo {author} {\bibfnamefont {V.}~\bibnamefont
  {{Malanushenko}}}, \bibinfo {author} {\bibfnamefont {D.}~\bibnamefont
  {{Oravetz}}}, \bibinfo {author} {\bibfnamefont {K.}~\bibnamefont {{Pan}}},
  \bibinfo {author} {\bibfnamefont {N.~P.}\ \bibnamefont {{Ross}}}, \bibinfo
  {author} {\bibfnamefont {D.~P.}\ \bibnamefont {{Schneider}}}, \bibinfo
  {author} {\bibfnamefont {E.}~\bibnamefont {{Sheldon}}}, \bibinfo {author}
  {\bibfnamefont {A.}~\bibnamefont {{Simmons}}}, \bibinfo {author}
  {\bibfnamefont {J.}~\bibnamefont {{Tinker}}}, \bibinfo {author}
  {\bibfnamefont {M.}~\bibnamefont {{White}}}, \ and\ \bibinfo {author}
  {\bibfnamefont {C.}~\bibnamefont {{Willmer}}},\ }\href {\doibase
  10.1051/0004-6361/201220379} {\bibfield  {journal} {\bibinfo  {journal}
  {\aap}\ }\textbf {\bibinfo {volume} {551}},\ \bibinfo {eid} {A29} (\bibinfo
  {year} {2013}{\natexlab{b}})},\ \Eprint {http://arxiv.org/abs/1209.3968}
  {arXiv:1209.3968 [astro-ph.CO]} \BibitemShut {NoStop}%
\bibitem [{\citenamefont {{Laureijs}}\ \emph {et~al.}(2011)\citenamefont
  {{Laureijs}}, \citenamefont {{Amiaux}}, \citenamefont {{Arduini}},
  \citenamefont {{Augu{\`e}res}}, \citenamefont {{Brinchmann}}, \citenamefont
  {{Cole}}, \citenamefont {{Cropper}}, \citenamefont {{Dabin}}, \citenamefont
  {{Duvet}}, \citenamefont {{Ealet}},\ and\ \citenamefont
  {et~al.}}]{2011arXiv1110.3193L}%
  \BibitemOpen
  \bibfield  {author} {\bibinfo {author} {\bibfnamefont {R.}~\bibnamefont
  {{Laureijs}}}, \bibinfo {author} {\bibfnamefont {J.}~\bibnamefont
  {{Amiaux}}}, \bibinfo {author} {\bibfnamefont {S.}~\bibnamefont {{Arduini}}},
  \bibinfo {author} {\bibfnamefont {J.~.}\ \bibnamefont {{Augu{\`e}res}}},
  \bibinfo {author} {\bibfnamefont {J.}~\bibnamefont {{Brinchmann}}}, \bibinfo
  {author} {\bibfnamefont {R.}~\bibnamefont {{Cole}}}, \bibinfo {author}
  {\bibfnamefont {M.}~\bibnamefont {{Cropper}}}, \bibinfo {author}
  {\bibfnamefont {C.}~\bibnamefont {{Dabin}}}, \bibinfo {author} {\bibfnamefont
  {L.}~\bibnamefont {{Duvet}}}, \bibinfo {author} {\bibfnamefont
  {A.}~\bibnamefont {{Ealet}}}, \ and\ \bibinfo {author} {\bibnamefont
  {et~al.}},\ }\href@noop {} {\bibfield  {journal} {\bibinfo  {journal} {ArXiv
  e-prints}\ } (\bibinfo {year} {2011})},\ \Eprint
  {http://arxiv.org/abs/1110.3193} {arXiv:1110.3193 [astro-ph.CO]} \BibitemShut
  {NoStop}%
\bibitem [{\citenamefont {{Spergel}}\ \emph {et~al.}(2013)\citenamefont
  {{Spergel}}, \citenamefont {{Gehrels}}, \citenamefont {{Breckinridge}},
  \citenamefont {{Donahue}}, \citenamefont {{Dressler}}, \citenamefont
  {{Gaudi}}, \citenamefont {{Greene}}, \citenamefont {{Guyon}}, \citenamefont
  {{Hirata}}, \citenamefont {{Kalirai}}, \citenamefont {{Kasdin}},
  \citenamefont {{Moos}}, \citenamefont {{Perlmutter}}, \citenamefont
  {{Postman}}, \citenamefont {{Rauscher}}, \citenamefont {{Rhodes}},
  \citenamefont {{Wang}}, \citenamefont {{Weinberg}}, \citenamefont
  {{Centrella}}, \citenamefont {{Traub}}, \citenamefont {{Baltay}},
  \citenamefont {{Colbert}}, \citenamefont {{Bennett}}, \citenamefont
  {{Kiessling}}, \citenamefont {{Macintosh}}, \citenamefont {{Merten}},
  \citenamefont {{Mortonson}}, \citenamefont {{Penny}}, \citenamefont {{Rozo}},
  \citenamefont {{Savransky}}, \citenamefont {{Stapelfeldt}}, \citenamefont
  {{Zu}}, \citenamefont {{Baker}}, \citenamefont {{Cheng}}, \citenamefont
  {{Content}}, \citenamefont {{Dooley}}, \citenamefont {{Foote}}, \citenamefont
  {{Goullioud}}, \citenamefont {{Grady}}, \citenamefont {{Jackson}},
  \citenamefont {{Kruk}}, \citenamefont {{Levine}}, \citenamefont {{Melton}},
  \citenamefont {{Peddie}}, \citenamefont {{Ruffa}},\ and\ \citenamefont
  {{Shaklan}}}]{2013arXiv1305.5422S}%
  \BibitemOpen
  \bibfield  {author} {\bibinfo {author} {\bibfnamefont {D.}~\bibnamefont
  {{Spergel}}}, \bibinfo {author} {\bibfnamefont {N.}~\bibnamefont
  {{Gehrels}}}, \bibinfo {author} {\bibfnamefont {J.}~\bibnamefont
  {{Breckinridge}}}, \bibinfo {author} {\bibfnamefont {M.}~\bibnamefont
  {{Donahue}}}, \bibinfo {author} {\bibfnamefont {A.}~\bibnamefont
  {{Dressler}}}, \bibinfo {author} {\bibfnamefont {B.~S.}\ \bibnamefont
  {{Gaudi}}}, \bibinfo {author} {\bibfnamefont {T.}~\bibnamefont {{Greene}}},
  \bibinfo {author} {\bibfnamefont {O.}~\bibnamefont {{Guyon}}}, \bibinfo
  {author} {\bibfnamefont {C.}~\bibnamefont {{Hirata}}}, \bibinfo {author}
  {\bibfnamefont {J.}~\bibnamefont {{Kalirai}}}, \bibinfo {author}
  {\bibfnamefont {N.~J.}\ \bibnamefont {{Kasdin}}}, \bibinfo {author}
  {\bibfnamefont {W.}~\bibnamefont {{Moos}}}, \bibinfo {author} {\bibfnamefont
  {S.}~\bibnamefont {{Perlmutter}}}, \bibinfo {author} {\bibfnamefont
  {M.}~\bibnamefont {{Postman}}}, \bibinfo {author} {\bibfnamefont
  {B.}~\bibnamefont {{Rauscher}}}, \bibinfo {author} {\bibfnamefont
  {J.}~\bibnamefont {{Rhodes}}}, \bibinfo {author} {\bibfnamefont
  {Y.}~\bibnamefont {{Wang}}}, \bibinfo {author} {\bibfnamefont
  {D.}~\bibnamefont {{Weinberg}}}, \bibinfo {author} {\bibfnamefont
  {J.}~\bibnamefont {{Centrella}}}, \bibinfo {author} {\bibfnamefont
  {W.}~\bibnamefont {{Traub}}}, \bibinfo {author} {\bibfnamefont
  {C.}~\bibnamefont {{Baltay}}}, \bibinfo {author} {\bibfnamefont
  {J.}~\bibnamefont {{Colbert}}}, \bibinfo {author} {\bibfnamefont
  {D.}~\bibnamefont {{Bennett}}}, \bibinfo {author} {\bibfnamefont
  {A.}~\bibnamefont {{Kiessling}}}, \bibinfo {author} {\bibfnamefont
  {B.}~\bibnamefont {{Macintosh}}}, \bibinfo {author} {\bibfnamefont
  {J.}~\bibnamefont {{Merten}}}, \bibinfo {author} {\bibfnamefont
  {M.}~\bibnamefont {{Mortonson}}}, \bibinfo {author} {\bibfnamefont
  {M.}~\bibnamefont {{Penny}}}, \bibinfo {author} {\bibfnamefont
  {E.}~\bibnamefont {{Rozo}}}, \bibinfo {author} {\bibfnamefont
  {D.}~\bibnamefont {{Savransky}}}, \bibinfo {author} {\bibfnamefont
  {K.}~\bibnamefont {{Stapelfeldt}}}, \bibinfo {author} {\bibfnamefont
  {Y.}~\bibnamefont {{Zu}}}, \bibinfo {author} {\bibfnamefont {C.}~\bibnamefont
  {{Baker}}}, \bibinfo {author} {\bibfnamefont {E.}~\bibnamefont {{Cheng}}},
  \bibinfo {author} {\bibfnamefont {D.}~\bibnamefont {{Content}}}, \bibinfo
  {author} {\bibfnamefont {J.}~\bibnamefont {{Dooley}}}, \bibinfo {author}
  {\bibfnamefont {M.}~\bibnamefont {{Foote}}}, \bibinfo {author} {\bibfnamefont
  {R.}~\bibnamefont {{Goullioud}}}, \bibinfo {author} {\bibfnamefont
  {K.}~\bibnamefont {{Grady}}}, \bibinfo {author} {\bibfnamefont
  {C.}~\bibnamefont {{Jackson}}}, \bibinfo {author} {\bibfnamefont
  {J.}~\bibnamefont {{Kruk}}}, \bibinfo {author} {\bibfnamefont
  {M.}~\bibnamefont {{Levine}}}, \bibinfo {author} {\bibfnamefont
  {M.}~\bibnamefont {{Melton}}}, \bibinfo {author} {\bibfnamefont
  {C.}~\bibnamefont {{Peddie}}}, \bibinfo {author} {\bibfnamefont
  {J.}~\bibnamefont {{Ruffa}}}, \ and\ \bibinfo {author} {\bibfnamefont
  {S.}~\bibnamefont {{Shaklan}}},\ }\href@noop {} {\bibfield  {journal}
  {\bibinfo  {journal} {ArXiv e-prints}\ } (\bibinfo {year} {2013})},\ \Eprint
  {http://arxiv.org/abs/1305.5422} {arXiv:1305.5422 [astro-ph.IM]} \BibitemShut
  {NoStop}%
\bibitem [{\citenamefont {{Nakajima}}\ \emph {et~al.}(2012)\citenamefont
  {{Nakajima}}, \citenamefont {{Mandelbaum}}, \citenamefont {{Seljak}},
  \citenamefont {{Cohn}}, \citenamefont {{Reyes}},\ and\ \citenamefont
  {{Cool}}}]{2012MNRAS.420.3240N}%
  \BibitemOpen
  \bibfield  {author} {\bibinfo {author} {\bibfnamefont {R.}~\bibnamefont
  {{Nakajima}}}, \bibinfo {author} {\bibfnamefont {R.}~\bibnamefont
  {{Mandelbaum}}}, \bibinfo {author} {\bibfnamefont {U.}~\bibnamefont
  {{Seljak}}}, \bibinfo {author} {\bibfnamefont {J.~D.}\ \bibnamefont
  {{Cohn}}}, \bibinfo {author} {\bibfnamefont {R.}~\bibnamefont {{Reyes}}}, \
  and\ \bibinfo {author} {\bibfnamefont {R.}~\bibnamefont {{Cool}}},\ }\href
  {\doibase 10.1111/j.1365-2966.2011.20249.x} {\bibfield  {journal} {\bibinfo
  {journal} {\mnras}\ }\textbf {\bibinfo {volume} {420}},\ \bibinfo {pages}
  {3240} (\bibinfo {year} {2012})},\ \Eprint {http://arxiv.org/abs/1107.1395}
  {arXiv:1107.1395 [astro-ph.CO]} \BibitemShut {NoStop}%
\bibitem [{\citenamefont {{Yoo}}\ and\ \citenamefont
  {{Seljak}}(2012)}]{2012PhRvD..86h3504Y}%
  \BibitemOpen
  \bibfield  {author} {\bibinfo {author} {\bibfnamefont {J.}~\bibnamefont
  {{Yoo}}}\ and\ \bibinfo {author} {\bibfnamefont {U.}~\bibnamefont
  {{Seljak}}},\ }\href {\doibase 10.1103/PhysRevD.86.083504} {\bibfield
  {journal} {\bibinfo  {journal} {\prd}\ }\textbf {\bibinfo {volume} {86}},\
  \bibinfo {eid} {083504} (\bibinfo {year} {2012})},\ \Eprint
  {http://arxiv.org/abs/1207.2471} {arXiv:1207.2471 [astro-ph.CO]} \BibitemShut
  {NoStop}%
\bibitem [{\citenamefont {{Sato}}\ \emph {et~al.}(2009)\citenamefont {{Sato}},
  \citenamefont {{Hamana}}, \citenamefont {{Takahashi}}, \citenamefont
  {{Takada}}, \citenamefont {{Yoshida}}, \citenamefont {{Matsubara}},\ and\
  \citenamefont {{Sugiyama}}}]{2009ApJ...701..945S}%
  \BibitemOpen
  \bibfield  {author} {\bibinfo {author} {\bibfnamefont {M.}~\bibnamefont
  {{Sato}}}, \bibinfo {author} {\bibfnamefont {T.}~\bibnamefont {{Hamana}}},
  \bibinfo {author} {\bibfnamefont {R.}~\bibnamefont {{Takahashi}}}, \bibinfo
  {author} {\bibfnamefont {M.}~\bibnamefont {{Takada}}}, \bibinfo {author}
  {\bibfnamefont {N.}~\bibnamefont {{Yoshida}}}, \bibinfo {author}
  {\bibfnamefont {T.}~\bibnamefont {{Matsubara}}}, \ and\ \bibinfo {author}
  {\bibfnamefont {N.}~\bibnamefont {{Sugiyama}}},\ }\href {\doibase
  10.1088/0004-637X/701/2/945} {\bibfield  {journal} {\bibinfo  {journal}
  {\apj}\ }\textbf {\bibinfo {volume} {701}},\ \bibinfo {pages} {945} (\bibinfo
  {year} {2009})},\ \Eprint {http://arxiv.org/abs/0906.2237} {arXiv:0906.2237
  [astro-ph.CO]} \BibitemShut {NoStop}%
\bibitem [{\citenamefont {{Cooray}}\ and\ \citenamefont
  {{Hu}}(2001)}]{2001ApJ...554...56C}%
  \BibitemOpen
  \bibfield  {author} {\bibinfo {author} {\bibfnamefont {A.}~\bibnamefont
  {{Cooray}}}\ and\ \bibinfo {author} {\bibfnamefont {W.}~\bibnamefont
  {{Hu}}},\ }\href {\doibase 10.1086/321376} {\bibfield  {journal} {\bibinfo
  {journal} {\apj}\ }\textbf {\bibinfo {volume} {554}},\ \bibinfo {pages} {56}
  (\bibinfo {year} {2001})},\ \Eprint
  {http://arxiv.org/abs/arXiv:astro-ph/0012087} {arXiv:astro-ph/0012087}
  \BibitemShut {NoStop}%
\bibitem [{\citenamefont {{Guzik}}\ \emph {et~al.}(2010)\citenamefont
  {{Guzik}}, \citenamefont {{Jain}},\ and\ \citenamefont
  {{Takada}}}]{2010PhRvD..81b3503G}%
  \BibitemOpen
  \bibfield  {author} {\bibinfo {author} {\bibfnamefont {J.}~\bibnamefont
  {{Guzik}}}, \bibinfo {author} {\bibfnamefont {B.}~\bibnamefont {{Jain}}}, \
  and\ \bibinfo {author} {\bibfnamefont {M.}~\bibnamefont {{Takada}}},\ }\href
  {\doibase 10.1103/PhysRevD.81.023503} {\bibfield  {journal} {\bibinfo
  {journal} {\prd}\ }\textbf {\bibinfo {volume} {81}},\ \bibinfo {eid} {023503}
  (\bibinfo {year} {2010})},\ \Eprint {http://arxiv.org/abs/0906.2221}
  {arXiv:0906.2221 [astro-ph.CO]} \BibitemShut {NoStop}%
\bibitem [{\citenamefont {{Hu}}\ and\ \citenamefont
  {{Jain}}(2004)}]{2004PhRvD..70d3009H}%
  \BibitemOpen
  \bibfield  {author} {\bibinfo {author} {\bibfnamefont {W.}~\bibnamefont
  {{Hu}}}\ and\ \bibinfo {author} {\bibfnamefont {B.}~\bibnamefont {{Jain}}},\
  }\href {\doibase 10.1103/PhysRevD.70.043009} {\bibfield  {journal} {\bibinfo
  {journal} {\prd}\ }\textbf {\bibinfo {volume} {70}},\ \bibinfo {eid} {043009}
  (\bibinfo {year} {2004})},\ \Eprint
  {http://arxiv.org/abs/arXiv:astro-ph/0312395} {arXiv:astro-ph/0312395}
  \BibitemShut {NoStop}%
\bibitem [{\citenamefont {{McDonald}}\ \emph
  {et~al.}(2006{\natexlab{b}})\citenamefont {{McDonald}}, \citenamefont
  {{Trac}},\ and\ \citenamefont {{Contaldi}}}]{2006MNRAS.366..547M}%
  \BibitemOpen
  \bibfield  {author} {\bibinfo {author} {\bibfnamefont {P.}~\bibnamefont
  {{McDonald}}}, \bibinfo {author} {\bibfnamefont {H.}~\bibnamefont {{Trac}}},
  \ and\ \bibinfo {author} {\bibfnamefont {C.}~\bibnamefont {{Contaldi}}},\
  }\href {\doibase 10.1111/j.1365-2966.2005.09881.x} {\bibfield  {journal}
  {\bibinfo  {journal} {\mnras}\ }\textbf {\bibinfo {volume} {366}},\ \bibinfo
  {pages} {547} (\bibinfo {year} {2006}{\natexlab{b}})}\BibitemShut {NoStop}%
\bibitem [{\citenamefont {{Smith}}\ \emph {et~al.}(2003)\citenamefont
  {{Smith}}, \citenamefont {{Peacock}}, \citenamefont {{Jenkins}},
  \citenamefont {{White}}, \citenamefont {{Frenk}}, \citenamefont {{Pearce}},
  \citenamefont {{Thomas}}, \citenamefont {{Efstathiou}},\ and\ \citenamefont
  {{Couchman}}}]{2003MNRAS.341.1311S}%
  \BibitemOpen
  \bibfield  {author} {\bibinfo {author} {\bibfnamefont {R.~E.}\ \bibnamefont
  {{Smith}}}, \bibinfo {author} {\bibfnamefont {J.~A.}\ \bibnamefont
  {{Peacock}}}, \bibinfo {author} {\bibfnamefont {A.}~\bibnamefont
  {{Jenkins}}}, \bibinfo {author} {\bibfnamefont {S.~D.~M.}\ \bibnamefont
  {{White}}}, \bibinfo {author} {\bibfnamefont {C.~S.}\ \bibnamefont
  {{Frenk}}}, \bibinfo {author} {\bibfnamefont {F.~R.}\ \bibnamefont
  {{Pearce}}}, \bibinfo {author} {\bibfnamefont {P.~A.}\ \bibnamefont
  {{Thomas}}}, \bibinfo {author} {\bibfnamefont {G.}~\bibnamefont
  {{Efstathiou}}}, \ and\ \bibinfo {author} {\bibfnamefont {H.~M.~P.}\
  \bibnamefont {{Couchman}}},\ }\href@noop {} {\bibfield  {journal} {\bibinfo
  {journal} {\mnras}\ }\textbf {\bibinfo {volume} {341}},\ \bibinfo {pages}
  {1311} (\bibinfo {year} {2003})}\BibitemShut {NoStop}%
\bibitem [{\citenamefont {{Ma}}(1998)}]{1998ApJ...508L...5M}%
  \BibitemOpen
  \bibfield  {author} {\bibinfo {author} {\bibfnamefont {C.}~\bibnamefont
  {{Ma}}},\ }\href@noop {} {\bibfield  {journal} {\bibinfo  {journal} {\apjl}\
  }\textbf {\bibinfo {volume} {508}},\ \bibinfo {pages} {L5} (\bibinfo {year}
  {1998})}\BibitemShut {NoStop}%
\bibitem [{\citenamefont {{Hearin}}\ \emph {et~al.}(2012)\citenamefont
  {{Hearin}}, \citenamefont {{Zentner}},\ and\ \citenamefont
  {{Ma}}}]{2012JCAP...04..034H}%
  \BibitemOpen
  \bibfield  {author} {\bibinfo {author} {\bibfnamefont {A.~P.}\ \bibnamefont
  {{Hearin}}}, \bibinfo {author} {\bibfnamefont {A.~R.}\ \bibnamefont
  {{Zentner}}}, \ and\ \bibinfo {author} {\bibfnamefont {Z.}~\bibnamefont
  {{Ma}}},\ }\href {\doibase 10.1088/1475-7516/2012/04/034} {\bibfield
  {journal} {\bibinfo  {journal} {\jcap}\ }\textbf {\bibinfo {volume} {4}},\
  \bibinfo {eid} {034} (\bibinfo {year} {2012})},\ \Eprint
  {http://arxiv.org/abs/1111.0052} {arXiv:1111.0052 [astro-ph.CO]} \BibitemShut
  {NoStop}%
\bibitem [{\citenamefont {{Bernstein}}\ and\ \citenamefont
  {{Huterer}}(2010)}]{2010MNRAS.401.1399B}%
  \BibitemOpen
  \bibfield  {author} {\bibinfo {author} {\bibfnamefont {G.}~\bibnamefont
  {{Bernstein}}}\ and\ \bibinfo {author} {\bibfnamefont {D.}~\bibnamefont
  {{Huterer}}},\ }\href {\doibase 10.1111/j.1365-2966.2009.15748.x} {\bibfield
  {journal} {\bibinfo  {journal} {\mnras}\ }\textbf {\bibinfo {volume} {401}},\
  \bibinfo {pages} {1399} (\bibinfo {year} {2010})},\ \Eprint
  {http://arxiv.org/abs/0902.2782} {arXiv:0902.2782 [astro-ph.CO]} \BibitemShut
  {NoStop}%
\bibitem [{\citenamefont {{Zhang}}\ \emph {et~al.}(2010)\citenamefont
  {{Zhang}}, \citenamefont {{Pen}},\ and\ \citenamefont
  {{Bernstein}}}]{2010MNRAS.405..359Z}%
  \BibitemOpen
  \bibfield  {author} {\bibinfo {author} {\bibfnamefont {P.}~\bibnamefont
  {{Zhang}}}, \bibinfo {author} {\bibfnamefont {U.-L.}\ \bibnamefont {{Pen}}},
  \ and\ \bibinfo {author} {\bibfnamefont {G.}~\bibnamefont {{Bernstein}}},\
  }\href {\doibase 10.1111/j.1365-2966.2010.16445.x} {\bibfield  {journal}
  {\bibinfo  {journal} {\mnras}\ }\textbf {\bibinfo {volume} {405}},\ \bibinfo
  {pages} {359} (\bibinfo {year} {2010})},\ \Eprint
  {http://arxiv.org/abs/0910.4181} {arXiv:0910.4181 [astro-ph.CO]} \BibitemShut
  {NoStop}%
\bibitem [{\citenamefont {{Newman}}(2008)}]{2008ApJ...684...88N}%
  \BibitemOpen
  \bibfield  {author} {\bibinfo {author} {\bibfnamefont {J.~A.}\ \bibnamefont
  {{Newman}}},\ }\href {\doibase 10.1086/589982} {\bibfield  {journal}
  {\bibinfo  {journal} {\apj}\ }\textbf {\bibinfo {volume} {684}},\ \bibinfo
  {pages} {88} (\bibinfo {year} {2008})},\ \Eprint
  {http://arxiv.org/abs/0805.1409} {arXiv:0805.1409} \BibitemShut {NoStop}%
\bibitem [{\citenamefont {{Ma}}\ and\ \citenamefont
  {{Bernstein}}(2008)}]{2008ApJ...682...39M}%
  \BibitemOpen
  \bibfield  {author} {\bibinfo {author} {\bibfnamefont {Z.}~\bibnamefont
  {{Ma}}}\ and\ \bibinfo {author} {\bibfnamefont {G.}~\bibnamefont
  {{Bernstein}}},\ }\href {\doibase 10.1086/588214} {\bibfield  {journal}
  {\bibinfo  {journal} {\apj}\ }\textbf {\bibinfo {volume} {682}},\ \bibinfo
  {pages} {39} (\bibinfo {year} {2008})},\ \Eprint
  {http://arxiv.org/abs/0712.1562} {arXiv:0712.1562} \BibitemShut {NoStop}%
\bibitem [{\citenamefont {{Ma}}\ \emph {et~al.}(2006)\citenamefont {{Ma}},
  \citenamefont {{Hu}},\ and\ \citenamefont {{Huterer}}}]{2006ApJ...636...21M}%
  \BibitemOpen
  \bibfield  {author} {\bibinfo {author} {\bibfnamefont {Z.}~\bibnamefont
  {{Ma}}}, \bibinfo {author} {\bibfnamefont {W.}~\bibnamefont {{Hu}}}, \ and\
  \bibinfo {author} {\bibfnamefont {D.}~\bibnamefont {{Huterer}}},\ }\href
  {\doibase 10.1086/497068} {\bibfield  {journal} {\bibinfo  {journal} {\apj}\
  }\textbf {\bibinfo {volume} {636}},\ \bibinfo {pages} {21} (\bibinfo {year}
  {2006})},\ \Eprint {http://arxiv.org/abs/arXiv:astro-ph/0506614}
  {arXiv:astro-ph/0506614} \BibitemShut {NoStop}%
\bibitem [{\citenamefont {{LSST Science Collaboration}}\ \emph
  {et~al.}(2009)\citenamefont {{LSST Science Collaboration}}, \citenamefont
  {{Abell}}, \citenamefont {{Allison}}, \citenamefont {{Anderson}},
  \citenamefont {{Andrew}}, \citenamefont {{Angel}}, \citenamefont {{Armus}},
  \citenamefont {{Arnett}}, \citenamefont {{Asztalos}}, \citenamefont
  {{Axelrod}},\ and\ \citenamefont {et~al.}}]{2009arXiv0912.0201L}%
  \BibitemOpen
  \bibfield  {author} {\bibinfo {author} {\bibnamefont {{LSST Science
  Collaboration}}}, \bibinfo {author} {\bibfnamefont {P.~A.}\ \bibnamefont
  {{Abell}}}, \bibinfo {author} {\bibfnamefont {J.}~\bibnamefont {{Allison}}},
  \bibinfo {author} {\bibfnamefont {S.~F.}\ \bibnamefont {{Anderson}}},
  \bibinfo {author} {\bibfnamefont {J.~R.}\ \bibnamefont {{Andrew}}}, \bibinfo
  {author} {\bibfnamefont {J.~R.~P.}\ \bibnamefont {{Angel}}}, \bibinfo
  {author} {\bibfnamefont {L.}~\bibnamefont {{Armus}}}, \bibinfo {author}
  {\bibfnamefont {D.}~\bibnamefont {{Arnett}}}, \bibinfo {author}
  {\bibfnamefont {S.~J.}\ \bibnamefont {{Asztalos}}}, \bibinfo {author}
  {\bibfnamefont {T.~S.}\ \bibnamefont {{Axelrod}}}, \ and\ \bibinfo {author}
  {\bibnamefont {et~al.}},\ }\href@noop {} {\bibfield  {journal} {\bibinfo
  {journal} {ArXiv e-prints}\ } (\bibinfo {year} {2009})},\ \Eprint
  {http://arxiv.org/abs/0912.0201} {arXiv:0912.0201 [astro-ph.IM]} \BibitemShut
  {NoStop}%
\bibitem [{\citenamefont {{Lesgourgues}}\ and\ \citenamefont
  {{Pastor}}(2012)}]{2012arXiv1212.6154L}%
  \BibitemOpen
  \bibfield  {author} {\bibinfo {author} {\bibfnamefont {J.}~\bibnamefont
  {{Lesgourgues}}}\ and\ \bibinfo {author} {\bibfnamefont {S.}~\bibnamefont
  {{Pastor}}},\ }\href@noop {} {\bibfield  {journal} {\bibinfo  {journal}
  {ArXiv e-prints}\ } (\bibinfo {year} {2012})},\ \Eprint
  {http://arxiv.org/abs/1212.6154} {arXiv:1212.6154 [hep-ph]} \BibitemShut
  {NoStop}%
\bibitem [{\citenamefont {{Carbone}}(2013)}]{2013NuPhS.237...50C}%
  \BibitemOpen
  \bibfield  {author} {\bibinfo {author} {\bibfnamefont {C.}~\bibnamefont
  {{Carbone}}},\ }\href {\doibase 10.1016/j.nuclphysbps.2013.04.056} {\bibfield
   {journal} {\bibinfo  {journal} {Nuclear Physics B Proceedings Supplements}\
  }\textbf {\bibinfo {volume} {237}},\ \bibinfo {pages} {50} (\bibinfo {year}
  {2013})}\BibitemShut {NoStop}%
\bibitem [{\citenamefont {{Santos}}\ \emph {et~al.}(2013)\citenamefont
  {{Santos}}, \citenamefont {{Cabella}}, \citenamefont {{Balbi}},\ and\
  \citenamefont {{Vittorio}}}]{2013arXiv1307.2919S}%
  \BibitemOpen
  \bibfield  {author} {\bibinfo {author} {\bibfnamefont {L.}~\bibnamefont
  {{Santos}}}, \bibinfo {author} {\bibfnamefont {P.}~\bibnamefont {{Cabella}}},
  \bibinfo {author} {\bibfnamefont {A.}~\bibnamefont {{Balbi}}}, \ and\
  \bibinfo {author} {\bibfnamefont {N.}~\bibnamefont {{Vittorio}}},\
  }\href@noop {} {\bibfield  {journal} {\bibinfo  {journal} {ArXiv e-prints}\ }
  (\bibinfo {year} {2013})},\ \Eprint {http://arxiv.org/abs/1307.2919}
  {arXiv:1307.2919 [astro-ph.CO]} \BibitemShut {NoStop}%
\bibitem [{\citenamefont {{Giusarma}}\ \emph {et~al.}(2011)\citenamefont
  {{Giusarma}}, \citenamefont {{Corsi}}, \citenamefont {{Archidiacono}},
  \citenamefont {{de Putter}}, \citenamefont {{Melchiorri}}, \citenamefont
  {{Mena}},\ and\ \citenamefont {{Pandolfi}}}]{2011PhRvD..83k5023G}%
  \BibitemOpen
  \bibfield  {author} {\bibinfo {author} {\bibfnamefont {E.}~\bibnamefont
  {{Giusarma}}}, \bibinfo {author} {\bibfnamefont {M.}~\bibnamefont {{Corsi}}},
  \bibinfo {author} {\bibfnamefont {M.}~\bibnamefont {{Archidiacono}}},
  \bibinfo {author} {\bibfnamefont {R.}~\bibnamefont {{de Putter}}}, \bibinfo
  {author} {\bibfnamefont {A.}~\bibnamefont {{Melchiorri}}}, \bibinfo {author}
  {\bibfnamefont {O.}~\bibnamefont {{Mena}}}, \ and\ \bibinfo {author}
  {\bibfnamefont {S.}~\bibnamefont {{Pandolfi}}},\ }\href {\doibase
  10.1103/PhysRevD.83.115023} {\bibfield  {journal} {\bibinfo  {journal}
  {\prd}\ }\textbf {\bibinfo {volume} {83}},\ \bibinfo {eid} {115023} (\bibinfo
  {year} {2011})},\ \Eprint {http://arxiv.org/abs/1102.4774} {arXiv:1102.4774
  [astro-ph.CO]} \BibitemShut {NoStop}%
\bibitem [{\citenamefont {{Wu}}\ \emph {et~al.}(2014)\citenamefont {{Wu}},
  \citenamefont {{Errard}}, \citenamefont {{Dvorkin}}, \citenamefont {{Kuo}},
  \citenamefont {{Lee}}, \citenamefont {{McDonald}}, \citenamefont {{Slosar}},\
  and\ \citenamefont {{Zahn}}}]{2014arXiv1402.4108W}%
  \BibitemOpen
  \bibfield  {author} {\bibinfo {author} {\bibfnamefont {W.~L.~K.}\
  \bibnamefont {{Wu}}}, \bibinfo {author} {\bibfnamefont {J.}~\bibnamefont
  {{Errard}}}, \bibinfo {author} {\bibfnamefont {C.}~\bibnamefont {{Dvorkin}}},
  \bibinfo {author} {\bibfnamefont {C.~L.}\ \bibnamefont {{Kuo}}}, \bibinfo
  {author} {\bibfnamefont {A.~T.}\ \bibnamefont {{Lee}}}, \bibinfo {author}
  {\bibfnamefont {P.}~\bibnamefont {{McDonald}}}, \bibinfo {author}
  {\bibfnamefont {A.}~\bibnamefont {{Slosar}}}, \ and\ \bibinfo {author}
  {\bibfnamefont {O.}~\bibnamefont {{Zahn}}},\ }\href@noop {} {\bibfield
  {journal} {\bibinfo  {journal} {ArXiv e-prints}\ } (\bibinfo {year}
  {2014})},\ \Eprint {http://arxiv.org/abs/1402.4108} {arXiv:1402.4108
  [astro-ph.CO]} \BibitemShut {NoStop}%
\bibitem [{\citenamefont {{Cahn}}\ \emph {et~al.}(2013)\citenamefont {{Cahn}},
  \citenamefont {{Dwyer}}, \citenamefont {{Freedman}}, \citenamefont
  {{Haxton}}, \citenamefont {{Kadel}}, \citenamefont {{Kolomensky}},
  \citenamefont {{Luk}}, \citenamefont {{McDonald}}, \citenamefont {{Orebi
  Gann}},\ and\ \citenamefont {{Poon}}}]{2013arXiv1307.5487C}%
  \BibitemOpen
  \bibfield  {author} {\bibinfo {author} {\bibfnamefont {R.~N.}\ \bibnamefont
  {{Cahn}}}, \bibinfo {author} {\bibfnamefont {D.~A.}\ \bibnamefont {{Dwyer}}},
  \bibinfo {author} {\bibfnamefont {S.~J.}\ \bibnamefont {{Freedman}}},
  \bibinfo {author} {\bibfnamefont {W.~C.}\ \bibnamefont {{Haxton}}}, \bibinfo
  {author} {\bibfnamefont {R.~W.}\ \bibnamefont {{Kadel}}}, \bibinfo {author}
  {\bibfnamefont {Y.~G.}\ \bibnamefont {{Kolomensky}}}, \bibinfo {author}
  {\bibfnamefont {K.~B.}\ \bibnamefont {{Luk}}}, \bibinfo {author}
  {\bibfnamefont {P.}~\bibnamefont {{McDonald}}}, \bibinfo {author}
  {\bibfnamefont {G.~D.}\ \bibnamefont {{Orebi Gann}}}, \ and\ \bibinfo
  {author} {\bibfnamefont {A.~W.~P.}\ \bibnamefont {{Poon}}},\ }\href@noop {}
  {\bibfield  {journal} {\bibinfo  {journal} {ArXiv e-prints}\ } (\bibinfo
  {year} {2013})},\ \Eprint {http://arxiv.org/abs/1307.5487} {arXiv:1307.5487
  [hep-ex]} \BibitemShut {NoStop}%
\bibitem [{\citenamefont {{Slosar}}\ \emph {et~al.}(2012)\citenamefont
  {{Slosar}}, \citenamefont {{McDonald}},\ and\ \citenamefont {{BigBOSS
  Team}}}]{2012AAS...21933501S}%
  \BibitemOpen
  \bibfield  {author} {\bibinfo {author} {\bibfnamefont {A.}~\bibnamefont
  {{Slosar}}}, \bibinfo {author} {\bibfnamefont {P.}~\bibnamefont
  {{McDonald}}}, \ and\ \bibinfo {author} {\bibnamefont {{BigBOSS Team}}},\
  }in\ \href@noop {} {\emph {\bibinfo {booktitle} {American Astronomical
  Society Meeting Abstracts \#219}}},\ \bibinfo {series} {American Astronomical
  Society Meeting Abstracts}, Vol.\ \bibinfo {volume} {219}\ (\bibinfo {year}
  {2012})\ p.\ \bibinfo {pages} {\#335.01}\BibitemShut {NoStop}%
\bibitem [{\citenamefont {{Slosar}}(2006)}]{2006PhRvD..73l3501S}%
  \BibitemOpen
  \bibfield  {author} {\bibinfo {author} {\bibfnamefont {A.}~\bibnamefont
  {{Slosar}}},\ }\href {\doibase 10.1103/PhysRevD.73.123501} {\bibfield
  {journal} {\bibinfo  {journal} {\prd}\ }\textbf {\bibinfo {volume} {73}},\
  \bibinfo {pages} {123501} (\bibinfo {year} {2006})},\ \Eprint
  {http://arxiv.org/abs/arXiv:astro-ph/0602133} {arXiv:astro-ph/0602133}
  \BibitemShut {NoStop}%
\bibitem [{\citenamefont {{Jimenez}}\ \emph {et~al.}(2010)\citenamefont
  {{Jimenez}}, \citenamefont {{Kitching}}, \citenamefont {{Pe{\~n}a-Garay}},\
  and\ \citenamefont {{Verde}}}]{2010JCAP...05..035J}%
  \BibitemOpen
  \bibfield  {author} {\bibinfo {author} {\bibfnamefont {R.}~\bibnamefont
  {{Jimenez}}}, \bibinfo {author} {\bibfnamefont {T.}~\bibnamefont
  {{Kitching}}}, \bibinfo {author} {\bibfnamefont {C.}~\bibnamefont
  {{Pe{\~n}a-Garay}}}, \ and\ \bibinfo {author} {\bibfnamefont
  {L.}~\bibnamefont {{Verde}}},\ }\href {\doibase
  10.1088/1475-7516/2010/05/035} {\bibfield  {journal} {\bibinfo  {journal}
  {\jcap}\ }\textbf {\bibinfo {volume} {5}},\ \bibinfo {eid} {035} (\bibinfo
  {year} {2010})},\ \Eprint {http://arxiv.org/abs/1003.5918} {arXiv:1003.5918
  [astro-ph.CO]} \BibitemShut {NoStop}%
\bibitem [{\citenamefont {{Wagner}}\ \emph {et~al.}(2012)\citenamefont
  {{Wagner}}, \citenamefont {{Verde}},\ and\ \citenamefont
  {{Jimenez}}}]{2012ApJ...752L..31W}%
  \BibitemOpen
  \bibfield  {author} {\bibinfo {author} {\bibfnamefont {C.}~\bibnamefont
  {{Wagner}}}, \bibinfo {author} {\bibfnamefont {L.}~\bibnamefont {{Verde}}}, \
  and\ \bibinfo {author} {\bibfnamefont {R.}~\bibnamefont {{Jimenez}}},\ }\href
  {\doibase 10.1088/2041-8205/752/2/L31} {\bibfield  {journal} {\bibinfo
  {journal} {\apjl}\ }\textbf {\bibinfo {volume} {752}},\ \bibinfo {eid} {L31}
  (\bibinfo {year} {2012})},\ \Eprint {http://arxiv.org/abs/1203.5342}
  {arXiv:1203.5342 [astro-ph.CO]} \BibitemShut {NoStop}%
\bibitem [{\citenamefont {{de Bernardis}}\ \emph {et~al.}(2009)\citenamefont
  {{de Bernardis}}, \citenamefont {{Kitching}}, \citenamefont {{Heavens}},\
  and\ \citenamefont {{Melchiorri}}}]{2009PhRvD..80l3509D}%
  \BibitemOpen
  \bibfield  {author} {\bibinfo {author} {\bibfnamefont {F.}~\bibnamefont {{de
  Bernardis}}}, \bibinfo {author} {\bibfnamefont {T.~D.}\ \bibnamefont
  {{Kitching}}}, \bibinfo {author} {\bibfnamefont {A.}~\bibnamefont
  {{Heavens}}}, \ and\ \bibinfo {author} {\bibfnamefont {A.}~\bibnamefont
  {{Melchiorri}}},\ }\href {\doibase 10.1103/PhysRevD.80.123509} {\bibfield
  {journal} {\bibinfo  {journal} {\prd}\ }\textbf {\bibinfo {volume} {80}},\
  \bibinfo {eid} {123509} (\bibinfo {year} {2009})},\ \Eprint
  {http://arxiv.org/abs/0907.1917} {arXiv:0907.1917 [astro-ph.CO]} \BibitemShut
  {NoStop}%
\bibitem [{\citenamefont {{Beringer}}\ and\ \citenamefont {et~al. (Particle
  Data~Group)}(2012)}]{2012PhRvD..86a0001B}%
  \BibitemOpen
  \bibfield  {author} {\bibinfo {author} {\bibfnamefont {J.}~\bibnamefont
  {{Beringer}}}\ and\ \bibinfo {author} {\bibnamefont {et~al. (Particle
  Data~Group)}},\ }\href {\doibase 10.1103/PhysRevD.86.010001} {\bibfield
  {journal} {\bibinfo  {journal} {\prd}\ }\textbf {\bibinfo {volume} {86}},\
  \bibinfo {eid} {010001} (\bibinfo {year} {2012})},\ \bibinfo {note} {{and
  2013 partial update for the 2014 edition.}}\BibitemShut {Stop}%
\bibitem [{\citenamefont {{McDonald}}(2008)}]{2008PhRvD..78l3519M}%
  \BibitemOpen
  \bibfield  {author} {\bibinfo {author} {\bibfnamefont {P.}~\bibnamefont
  {{McDonald}}},\ }\href {\doibase 10.1103/PhysRevD.78.123519} {\bibfield
  {journal} {\bibinfo  {journal} {\prd}\ }\textbf {\bibinfo {volume} {78}},\
  \bibinfo {pages} {123519} (\bibinfo {year} {2008})},\ \Eprint
  {http://arxiv.org/abs/0806.1061} {arXiv:0806.1061} \BibitemShut {NoStop}%
\bibitem [{\citenamefont {{Dalal}}\ \emph {et~al.}(2008)\citenamefont
  {{Dalal}}, \citenamefont {{White}}, \citenamefont {{Bond}},\ and\
  \citenamefont {{Shirokov}}}]{2008ApJ...687...12D}%
  \BibitemOpen
  \bibfield  {author} {\bibinfo {author} {\bibfnamefont {N.}~\bibnamefont
  {{Dalal}}}, \bibinfo {author} {\bibfnamefont {M.}~\bibnamefont {{White}}},
  \bibinfo {author} {\bibfnamefont {J.~R.}\ \bibnamefont {{Bond}}}, \ and\
  \bibinfo {author} {\bibfnamefont {A.}~\bibnamefont {{Shirokov}}},\ }\href
  {\doibase 10.1086/591512} {\bibfield  {journal} {\bibinfo  {journal} {\apj}\
  }\textbf {\bibinfo {volume} {687}},\ \bibinfo {pages} {12} (\bibinfo {year}
  {2008})},\ \Eprint {http://arxiv.org/abs/0803.3453} {arXiv:0803.3453}
  \BibitemShut {NoStop}%
\bibitem [{\citenamefont {{Matarrese}}\ and\ \citenamefont
  {{Verde}}(2008)}]{2008ApJ...677L..77M}%
  \BibitemOpen
  \bibfield  {author} {\bibinfo {author} {\bibfnamefont {S.}~\bibnamefont
  {{Matarrese}}}\ and\ \bibinfo {author} {\bibfnamefont {L.}~\bibnamefont
  {{Verde}}},\ }\href {\doibase 10.1086/587840} {\bibfield  {journal} {\bibinfo
   {journal} {\apjl}\ }\textbf {\bibinfo {volume} {677}},\ \bibinfo {pages}
  {L77} (\bibinfo {year} {2008})},\ \Eprint
  {http://arxiv.org/abs/arXiv:0801.4826} {arXiv:0801.4826} \BibitemShut
  {NoStop}%
\bibitem [{\citenamefont {{Afshordi}}\ and\ \citenamefont
  {{Tolley}}(2008)}]{2008PhRvD..78l3507A}%
  \BibitemOpen
  \bibfield  {author} {\bibinfo {author} {\bibfnamefont {N.}~\bibnamefont
  {{Afshordi}}}\ and\ \bibinfo {author} {\bibfnamefont {A.~J.}\ \bibnamefont
  {{Tolley}}},\ }\href {\doibase 10.1103/PhysRevD.78.123507} {\bibfield
  {journal} {\bibinfo  {journal} {\prd}\ }\textbf {\bibinfo {volume} {78}},\
  \bibinfo {pages} {123507} (\bibinfo {year} {2008})},\ \Eprint
  {http://arxiv.org/abs/0806.1046} {arXiv:0806.1046} \BibitemShut {NoStop}%
\bibitem [{\citenamefont {{Slosar}}\ \emph {et~al.}(2008)\citenamefont
  {{Slosar}}, \citenamefont {{Hirata}}, \citenamefont {{Seljak}}, \citenamefont
  {{Ho}},\ and\ \citenamefont {{Padmanabhan}}}]{2008JCAP...08..031S}%
  \BibitemOpen
  \bibfield  {author} {\bibinfo {author} {\bibfnamefont {A.}~\bibnamefont
  {{Slosar}}}, \bibinfo {author} {\bibfnamefont {C.}~\bibnamefont {{Hirata}}},
  \bibinfo {author} {\bibfnamefont {U.}~\bibnamefont {{Seljak}}}, \bibinfo
  {author} {\bibfnamefont {S.}~\bibnamefont {{Ho}}}, \ and\ \bibinfo {author}
  {\bibfnamefont {N.}~\bibnamefont {{Padmanabhan}}},\ }\href {\doibase
  10.1088/1475-7516/2008/08/031} {\bibfield  {journal} {\bibinfo  {journal}
  {Journal of Cosmology and Astro-Particle Physics}\ }\textbf {\bibinfo
  {volume} {8}},\ \bibinfo {pages} {31} (\bibinfo {year} {2008})},\ \Eprint
  {http://arxiv.org/abs/0805.3580} {arXiv:0805.3580} \BibitemShut {NoStop}%
\bibitem [{\citenamefont {{Planck Collaboration}}\ \emph
  {et~al.}(2013)\citenamefont {{Planck Collaboration}}, \citenamefont {{Ade}},
  \citenamefont {{Aghanim}}, \citenamefont {{Armitage-Caplan}}, \citenamefont
  {{Arnaud}}, \citenamefont {{Ashdown}}, \citenamefont {{Atrio-Barandela}},
  \citenamefont {{Aumont}}, \citenamefont {{Baccigalupi}}, \citenamefont
  {{Banday}},\ and\ \citenamefont {et~al.}}]{2013arXiv1303.5084P}%
  \BibitemOpen
  \bibfield  {author} {\bibinfo {author} {\bibnamefont {{Planck
  Collaboration}}}, \bibinfo {author} {\bibfnamefont {P.~A.~R.}\ \bibnamefont
  {{Ade}}}, \bibinfo {author} {\bibfnamefont {N.}~\bibnamefont {{Aghanim}}},
  \bibinfo {author} {\bibfnamefont {C.}~\bibnamefont {{Armitage-Caplan}}},
  \bibinfo {author} {\bibfnamefont {M.}~\bibnamefont {{Arnaud}}}, \bibinfo
  {author} {\bibfnamefont {M.}~\bibnamefont {{Ashdown}}}, \bibinfo {author}
  {\bibfnamefont {F.}~\bibnamefont {{Atrio-Barandela}}}, \bibinfo {author}
  {\bibfnamefont {J.}~\bibnamefont {{Aumont}}}, \bibinfo {author}
  {\bibfnamefont {C.}~\bibnamefont {{Baccigalupi}}}, \bibinfo {author}
  {\bibfnamefont {A.~J.}\ \bibnamefont {{Banday}}}, \ and\ \bibinfo {author}
  {\bibnamefont {et~al.}},\ }\href@noop {} {\bibfield  {journal} {\bibinfo
  {journal} {ArXiv e-prints}\ } (\bibinfo {year} {2013})},\ \Eprint
  {http://arxiv.org/abs/1303.5084} {arXiv:1303.5084 [astro-ph.CO]} \BibitemShut
  {NoStop}%
\bibitem [{\citenamefont {{Sefusatti}}\ \emph
  {et~al.}(2012{\natexlab{a}})\citenamefont {{Sefusatti}}, \citenamefont
  {{Crocce}},\ and\ \citenamefont {{Desjacques}}}]{2012MNRAS.425.2903S}%
  \BibitemOpen
  \bibfield  {author} {\bibinfo {author} {\bibfnamefont {E.}~\bibnamefont
  {{Sefusatti}}}, \bibinfo {author} {\bibfnamefont {M.}~\bibnamefont
  {{Crocce}}}, \ and\ \bibinfo {author} {\bibfnamefont {V.}~\bibnamefont
  {{Desjacques}}},\ }\href {\doibase 10.1111/j.1365-2966.2012.21271.x}
  {\bibfield  {journal} {\bibinfo  {journal} {\mnras}\ }\textbf {\bibinfo
  {volume} {425}},\ \bibinfo {pages} {2903} (\bibinfo {year}
  {2012}{\natexlab{a}})},\ \Eprint {http://arxiv.org/abs/1111.6966}
  {arXiv:1111.6966 [astro-ph.CO]} \BibitemShut {NoStop}%
\bibitem [{\citenamefont {{Baldauf}}\ \emph {et~al.}(2011)\citenamefont
  {{Baldauf}}, \citenamefont {{Seljak}},\ and\ \citenamefont
  {{Senatore}}}]{2011JCAP...04..006B}%
  \BibitemOpen
  \bibfield  {author} {\bibinfo {author} {\bibfnamefont {T.}~\bibnamefont
  {{Baldauf}}}, \bibinfo {author} {\bibfnamefont {U.}~\bibnamefont {{Seljak}}},
  \ and\ \bibinfo {author} {\bibfnamefont {L.}~\bibnamefont {{Senatore}}},\
  }\href {\doibase 10.1088/1475-7516/2011/04/006} {\bibfield  {journal}
  {\bibinfo  {journal} {\jcap}\ }\textbf {\bibinfo {volume} {4}},\ \bibinfo
  {eid} {006} (\bibinfo {year} {2011})},\ \Eprint
  {http://arxiv.org/abs/1011.1513} {arXiv:1011.1513 [astro-ph.CO]} \BibitemShut
  {NoStop}%
\bibitem [{\citenamefont {{Nishimichi}}\ \emph {et~al.}(2010)\citenamefont
  {{Nishimichi}}, \citenamefont {{Taruya}}, \citenamefont {{Koyama}},\ and\
  \citenamefont {{Sabiu}}}]{2010JCAP...07..002N}%
  \BibitemOpen
  \bibfield  {author} {\bibinfo {author} {\bibfnamefont {T.}~\bibnamefont
  {{Nishimichi}}}, \bibinfo {author} {\bibfnamefont {A.}~\bibnamefont
  {{Taruya}}}, \bibinfo {author} {\bibfnamefont {K.}~\bibnamefont {{Koyama}}},
  \ and\ \bibinfo {author} {\bibfnamefont {C.}~\bibnamefont {{Sabiu}}},\ }\href
  {\doibase 10.1088/1475-7516/2010/07/002} {\bibfield  {journal} {\bibinfo
  {journal} {\jcap}\ }\textbf {\bibinfo {volume} {7}},\ \bibinfo {eid} {002}
  (\bibinfo {year} {2010})},\ \Eprint {http://arxiv.org/abs/0911.4768}
  {arXiv:0911.4768 [astro-ph.CO]} \BibitemShut {NoStop}%
\bibitem [{\citenamefont {{Liguori}}\ \emph {et~al.}(2010)\citenamefont
  {{Liguori}}, \citenamefont {{Sefusatti}}, \citenamefont {{Fergusson}},\ and\
  \citenamefont {{Shellard}}}]{2010AdAst2010E..73L}%
  \BibitemOpen
  \bibfield  {author} {\bibinfo {author} {\bibfnamefont {M.}~\bibnamefont
  {{Liguori}}}, \bibinfo {author} {\bibfnamefont {E.}~\bibnamefont
  {{Sefusatti}}}, \bibinfo {author} {\bibfnamefont {J.~R.}\ \bibnamefont
  {{Fergusson}}}, \ and\ \bibinfo {author} {\bibfnamefont {E.~P.~S.}\
  \bibnamefont {{Shellard}}},\ }\href {\doibase 10.1155/2010/980523} {\bibfield
   {journal} {\bibinfo  {journal} {Advances in Astronomy}\ }\textbf {\bibinfo
  {volume} {2010}},\ \bibinfo {eid} {980523} (\bibinfo {year} {2010}),\
  10.1155/2010/980523},\ \Eprint {http://arxiv.org/abs/1001.4707}
  {arXiv:1001.4707 [astro-ph.CO]} \BibitemShut {NoStop}%
\bibitem [{\citenamefont {{Sefusatti}}(2009)}]{2009PhRvD..80l3002S}%
  \BibitemOpen
  \bibfield  {author} {\bibinfo {author} {\bibfnamefont {E.}~\bibnamefont
  {{Sefusatti}}},\ }\href {\doibase 10.1103/PhysRevD.80.123002} {\bibfield
  {journal} {\bibinfo  {journal} {\prd}\ }\textbf {\bibinfo {volume} {80}},\
  \bibinfo {eid} {123002} (\bibinfo {year} {2009})},\ \Eprint
  {http://arxiv.org/abs/0905.0717} {arXiv:0905.0717 [astro-ph.CO]} \BibitemShut
  {NoStop}%
\bibitem [{\citenamefont {{Jeong}}\ and\ \citenamefont
  {{Komatsu}}(2009{\natexlab{b}})}]{2009ApJ...703.1230J}%
  \BibitemOpen
  \bibfield  {author} {\bibinfo {author} {\bibfnamefont {D.}~\bibnamefont
  {{Jeong}}}\ and\ \bibinfo {author} {\bibfnamefont {E.}~\bibnamefont
  {{Komatsu}}},\ }\href {\doibase 10.1088/0004-637X/703/2/1230} {\bibfield
  {journal} {\bibinfo  {journal} {\apj}\ }\textbf {\bibinfo {volume} {703}},\
  \bibinfo {pages} {1230} (\bibinfo {year} {2009}{\natexlab{b}})},\ \Eprint
  {http://arxiv.org/abs/0904.0497} {arXiv:0904.0497 [astro-ph.CO]} \BibitemShut
  {NoStop}%
\bibitem [{\citenamefont {{Smith}}\ \emph {et~al.}(2012)\citenamefont
  {{Smith}}, \citenamefont {{Das}},\ and\ \citenamefont
  {{Zahn}}}]{2012PhRvD..85b3001S}%
  \BibitemOpen
  \bibfield  {author} {\bibinfo {author} {\bibfnamefont {T.~L.}\ \bibnamefont
  {{Smith}}}, \bibinfo {author} {\bibfnamefont {S.}~\bibnamefont {{Das}}}, \
  and\ \bibinfo {author} {\bibfnamefont {O.}~\bibnamefont {{Zahn}}},\ }\href
  {\doibase 10.1103/PhysRevD.85.023001} {\bibfield  {journal} {\bibinfo
  {journal} {\prd}\ }\textbf {\bibinfo {volume} {85}},\ \bibinfo {eid} {023001}
  (\bibinfo {year} {2012})},\ \Eprint {http://arxiv.org/abs/1105.3246}
  {arXiv:1105.3246 [astro-ph.CO]} \BibitemShut {NoStop}%
\bibitem [{\citenamefont {{Hou}}\ \emph {et~al.}(2013)\citenamefont {{Hou}},
  \citenamefont {{Keisler}}, \citenamefont {{Knox}}, \citenamefont {{Millea}},\
  and\ \citenamefont {{Reichardt}}}]{2013PhRvD..87h3008H}%
  \BibitemOpen
  \bibfield  {author} {\bibinfo {author} {\bibfnamefont {Z.}~\bibnamefont
  {{Hou}}}, \bibinfo {author} {\bibfnamefont {R.}~\bibnamefont {{Keisler}}},
  \bibinfo {author} {\bibfnamefont {L.}~\bibnamefont {{Knox}}}, \bibinfo
  {author} {\bibfnamefont {M.}~\bibnamefont {{Millea}}}, \ and\ \bibinfo
  {author} {\bibfnamefont {C.}~\bibnamefont {{Reichardt}}},\ }\href {\doibase
  10.1103/PhysRevD.87.083008} {\bibfield  {journal} {\bibinfo  {journal}
  {\prd}\ }\textbf {\bibinfo {volume} {87}},\ \bibinfo {eid} {083008} (\bibinfo
  {year} {2013})},\ \Eprint {http://arxiv.org/abs/1104.2333} {arXiv:1104.2333
  [astro-ph.CO]} \BibitemShut {NoStop}%
\bibitem [{\citenamefont {{Brust}}\ \emph {et~al.}(2013)\citenamefont
  {{Brust}}, \citenamefont {{Kaplan}},\ and\ \citenamefont
  {{Walters}}}]{2013arXiv1303.5379B}%
  \BibitemOpen
  \bibfield  {author} {\bibinfo {author} {\bibfnamefont {C.}~\bibnamefont
  {{Brust}}}, \bibinfo {author} {\bibfnamefont {D.~E.}\ \bibnamefont
  {{Kaplan}}}, \ and\ \bibinfo {author} {\bibfnamefont {M.~T.}\ \bibnamefont
  {{Walters}}},\ }\href@noop {} {\bibfield  {journal} {\bibinfo  {journal}
  {ArXiv e-prints}\ } (\bibinfo {year} {2013})},\ \Eprint
  {http://arxiv.org/abs/1303.5379} {arXiv:1303.5379 [hep-ph]} \BibitemShut
  {NoStop}%
\bibitem [{\citenamefont {{Boehm}}\ \emph {et~al.}(2012)\citenamefont
  {{Boehm}}, \citenamefont {{Dolan}},\ and\ \citenamefont
  {{McCabe}}}]{2012JCAP...12..027H}%
  \BibitemOpen
  \bibfield  {author} {\bibinfo {author} {\bibfnamefont {C.}~\bibnamefont
  {{Boehm}}}, \bibinfo {author} {\bibfnamefont {M.~J.}\ \bibnamefont
  {{Dolan}}}, \ and\ \bibinfo {author} {\bibfnamefont {C.}~\bibnamefont
  {{McCabe}}},\ }\href {\doibase 10.1088/1475-7516/2012/12/027} {\bibfield
  {journal} {\bibinfo  {journal} {\jcap}\ }\textbf {\bibinfo {volume} {12}},\
  \bibinfo {eid} {027} (\bibinfo {year} {2012})},\ \Eprint
  {http://arxiv.org/abs/1207.0497} {arXiv:1207.0497 [astro-ph.CO]} \BibitemShut
  {NoStop}%
\bibitem [{\citenamefont {{Wyman}}\ \emph {et~al.}(2013)\citenamefont
  {{Wyman}}, \citenamefont {{Rudd}}, \citenamefont {{Vanderveld}},\ and\
  \citenamefont {{Hu}}}]{2013arXiv1307.7715W}%
  \BibitemOpen
  \bibfield  {author} {\bibinfo {author} {\bibfnamefont {M.}~\bibnamefont
  {{Wyman}}}, \bibinfo {author} {\bibfnamefont {D.~H.}\ \bibnamefont {{Rudd}}},
  \bibinfo {author} {\bibfnamefont {R.~A.}\ \bibnamefont {{Vanderveld}}}, \
  and\ \bibinfo {author} {\bibfnamefont {W.}~\bibnamefont {{Hu}}},\ }\href@noop
  {} {\bibfield  {journal} {\bibinfo  {journal} {ArXiv e-prints}\ } (\bibinfo
  {year} {2013})},\ \Eprint {http://arxiv.org/abs/1307.7715} {arXiv:1307.7715
  [astro-ph.CO]} \BibitemShut {NoStop}%
\bibitem [{\citenamefont {{Hlozek}}\ \emph {et~al.}(2012)\citenamefont
  {{Hlozek}}, \citenamefont {{Dunkley}}, \citenamefont {{Addison}},
  \citenamefont {{Appel}}, \citenamefont {{Bond}}, \citenamefont {{Sofia
  Carvalho}}, \citenamefont {{Das}}, \citenamefont {{Devlin}}, \citenamefont
  {{D{\"u}nner}}, \citenamefont {{Essinger-Hileman}}, \citenamefont {{Fowler}},
  \citenamefont {{Gallardo}}, \citenamefont {{Hajian}}, \citenamefont
  {{Halpern}}, \citenamefont {{Hasselfield}}, \citenamefont {{Hilton}},
  \citenamefont {{Hincks}}, \citenamefont {{Hughes}}, \citenamefont {{Irwin}},
  \citenamefont {{Klein}}, \citenamefont {{Kosowsky}}, \citenamefont
  {{Marriage}}, \citenamefont {{Marsden}}, \citenamefont {{Menanteau}},
  \citenamefont {{Moodley}}, \citenamefont {{Niemack}}, \citenamefont
  {{Nolta}}, \citenamefont {{Page}}, \citenamefont {{Parker}}, \citenamefont
  {{Partridge}}, \citenamefont {{Rojas}}, \citenamefont {{Sehgal}},
  \citenamefont {{Sherwin}}, \citenamefont {{Sievers}}, \citenamefont
  {{Spergel}}, \citenamefont {{Staggs}}, \citenamefont {{Swetz}}, \citenamefont
  {{Switzer}}, \citenamefont {{Thornton}},\ and\ \citenamefont
  {{Wollack}}}]{2012ApJ...749...90H}%
  \BibitemOpen
  \bibfield  {author} {\bibinfo {author} {\bibfnamefont {R.}~\bibnamefont
  {{Hlozek}}}, \bibinfo {author} {\bibfnamefont {J.}~\bibnamefont {{Dunkley}}},
  \bibinfo {author} {\bibfnamefont {G.}~\bibnamefont {{Addison}}}, \bibinfo
  {author} {\bibfnamefont {J.~W.}\ \bibnamefont {{Appel}}}, \bibinfo {author}
  {\bibfnamefont {J.~R.}\ \bibnamefont {{Bond}}}, \bibinfo {author}
  {\bibfnamefont {C.}~\bibnamefont {{Sofia Carvalho}}}, \bibinfo {author}
  {\bibfnamefont {S.}~\bibnamefont {{Das}}}, \bibinfo {author} {\bibfnamefont
  {M.~J.}\ \bibnamefont {{Devlin}}}, \bibinfo {author} {\bibfnamefont
  {R.}~\bibnamefont {{D{\"u}nner}}}, \bibinfo {author} {\bibfnamefont
  {T.}~\bibnamefont {{Essinger-Hileman}}}, \bibinfo {author} {\bibfnamefont
  {J.~W.}\ \bibnamefont {{Fowler}}}, \bibinfo {author} {\bibfnamefont
  {P.}~\bibnamefont {{Gallardo}}}, \bibinfo {author} {\bibfnamefont
  {A.}~\bibnamefont {{Hajian}}}, \bibinfo {author} {\bibfnamefont
  {M.}~\bibnamefont {{Halpern}}}, \bibinfo {author} {\bibfnamefont
  {M.}~\bibnamefont {{Hasselfield}}}, \bibinfo {author} {\bibfnamefont
  {M.}~\bibnamefont {{Hilton}}}, \bibinfo {author} {\bibfnamefont {A.~D.}\
  \bibnamefont {{Hincks}}}, \bibinfo {author} {\bibfnamefont {J.~P.}\
  \bibnamefont {{Hughes}}}, \bibinfo {author} {\bibfnamefont {K.~D.}\
  \bibnamefont {{Irwin}}}, \bibinfo {author} {\bibfnamefont {J.}~\bibnamefont
  {{Klein}}}, \bibinfo {author} {\bibfnamefont {A.}~\bibnamefont {{Kosowsky}}},
  \bibinfo {author} {\bibfnamefont {T.~A.}\ \bibnamefont {{Marriage}}},
  \bibinfo {author} {\bibfnamefont {D.}~\bibnamefont {{Marsden}}}, \bibinfo
  {author} {\bibfnamefont {F.}~\bibnamefont {{Menanteau}}}, \bibinfo {author}
  {\bibfnamefont {K.}~\bibnamefont {{Moodley}}}, \bibinfo {author}
  {\bibfnamefont {M.~D.}\ \bibnamefont {{Niemack}}}, \bibinfo {author}
  {\bibfnamefont {M.~R.}\ \bibnamefont {{Nolta}}}, \bibinfo {author}
  {\bibfnamefont {L.~A.}\ \bibnamefont {{Page}}}, \bibinfo {author}
  {\bibfnamefont {L.}~\bibnamefont {{Parker}}}, \bibinfo {author}
  {\bibfnamefont {B.}~\bibnamefont {{Partridge}}}, \bibinfo {author}
  {\bibfnamefont {F.}~\bibnamefont {{Rojas}}}, \bibinfo {author} {\bibfnamefont
  {N.}~\bibnamefont {{Sehgal}}}, \bibinfo {author} {\bibfnamefont
  {B.}~\bibnamefont {{Sherwin}}}, \bibinfo {author} {\bibfnamefont
  {J.}~\bibnamefont {{Sievers}}}, \bibinfo {author} {\bibfnamefont {D.~N.}\
  \bibnamefont {{Spergel}}}, \bibinfo {author} {\bibfnamefont {S.~T.}\
  \bibnamefont {{Staggs}}}, \bibinfo {author} {\bibfnamefont {D.~S.}\
  \bibnamefont {{Swetz}}}, \bibinfo {author} {\bibfnamefont {E.~R.}\
  \bibnamefont {{Switzer}}}, \bibinfo {author} {\bibfnamefont {R.}~\bibnamefont
  {{Thornton}}}, \ and\ \bibinfo {author} {\bibfnamefont {E.}~\bibnamefont
  {{Wollack}}},\ }\href {\doibase 10.1088/0004-637X/749/1/90} {\bibfield
  {journal} {\bibinfo  {journal} {\apj}\ }\textbf {\bibinfo {volume} {749}},\
  \bibinfo {eid} {90} (\bibinfo {year} {2012})},\ \Eprint
  {http://arxiv.org/abs/1105.4887} {arXiv:1105.4887 [astro-ph.CO]} \BibitemShut
  {NoStop}%
\bibitem [{\citenamefont {{Chang}}\ \emph {et~al.}(2008)\citenamefont
  {{Chang}}, \citenamefont {{Pen}}, \citenamefont {{Peterson}},\ and\
  \citenamefont {{McDonald}}}]{2008PhRvL.100i1303C}%
  \BibitemOpen
  \bibfield  {author} {\bibinfo {author} {\bibfnamefont {T.-C.}\ \bibnamefont
  {{Chang}}}, \bibinfo {author} {\bibfnamefont {U.-L.}\ \bibnamefont {{Pen}}},
  \bibinfo {author} {\bibfnamefont {J.~B.}\ \bibnamefont {{Peterson}}}, \ and\
  \bibinfo {author} {\bibfnamefont {P.}~\bibnamefont {{McDonald}}},\ }\href
  {\doibase 10.1103/PhysRevLett.100.091303} {\bibfield  {journal} {\bibinfo
  {journal} {Physical Review Letters}\ }\textbf {\bibinfo {volume} {100}},\
  \bibinfo {pages} {091303} (\bibinfo {year} {2008})},\ \Eprint
  {http://arxiv.org/abs/0709.3672} {arXiv:0709.3672} \BibitemShut {NoStop}%
\bibitem [{\citenamefont {{Pober}}\ \emph {et~al.}(2013)\citenamefont
  {{Pober}}, \citenamefont {{Parsons}}, \citenamefont {{DeBoer}}, \citenamefont
  {{McDonald}}, \citenamefont {{McQuinn}}, \citenamefont {{Aguirre}},
  \citenamefont {{Ali}}, \citenamefont {{Bradley}}, \citenamefont {{Chang}},\
  and\ \citenamefont {{Morales}}}]{2013AJ....145...65P}%
  \BibitemOpen
  \bibfield  {author} {\bibinfo {author} {\bibfnamefont {J.~C.}\ \bibnamefont
  {{Pober}}}, \bibinfo {author} {\bibfnamefont {A.~R.}\ \bibnamefont
  {{Parsons}}}, \bibinfo {author} {\bibfnamefont {D.~R.}\ \bibnamefont
  {{DeBoer}}}, \bibinfo {author} {\bibfnamefont {P.}~\bibnamefont
  {{McDonald}}}, \bibinfo {author} {\bibfnamefont {M.}~\bibnamefont
  {{McQuinn}}}, \bibinfo {author} {\bibfnamefont {J.~E.}\ \bibnamefont
  {{Aguirre}}}, \bibinfo {author} {\bibfnamefont {Z.}~\bibnamefont {{Ali}}},
  \bibinfo {author} {\bibfnamefont {R.~F.}\ \bibnamefont {{Bradley}}}, \bibinfo
  {author} {\bibfnamefont {T.-C.}\ \bibnamefont {{Chang}}}, \ and\ \bibinfo
  {author} {\bibfnamefont {M.~F.}\ \bibnamefont {{Morales}}},\ }\href {\doibase
  10.1088/0004-6256/145/3/65} {\bibfield  {journal} {\bibinfo  {journal} {\aj}\
  }\textbf {\bibinfo {volume} {145}},\ \bibinfo {eid} {65} (\bibinfo {year}
  {2013})},\ \Eprint {http://arxiv.org/abs/1210.2413} {arXiv:1210.2413
  [astro-ph.CO]} \BibitemShut {NoStop}%
\bibitem [{\citenamefont {{Masui}}\ \emph
  {et~al.}(2010{\natexlab{a}})\citenamefont {{Masui}}, \citenamefont
  {{McDonald}},\ and\ \citenamefont {{Pen}}}]{2010PhRvD..81j3527M}%
  \BibitemOpen
  \bibfield  {author} {\bibinfo {author} {\bibfnamefont {K.~W.}\ \bibnamefont
  {{Masui}}}, \bibinfo {author} {\bibfnamefont {P.}~\bibnamefont {{McDonald}}},
  \ and\ \bibinfo {author} {\bibfnamefont {U.}~\bibnamefont {{Pen}}},\ }\href
  {\doibase 10.1103/PhysRevD.81.103527} {\bibfield  {journal} {\bibinfo
  {journal} {\prd}\ }\textbf {\bibinfo {volume} {81}},\ \bibinfo {pages}
  {103527} (\bibinfo {year} {2010}{\natexlab{a}})},\ \Eprint
  {http://arxiv.org/abs/1001.4811} {arXiv:1001.4811 [astro-ph.CO]} \BibitemShut
  {NoStop}%
\bibitem [{\citenamefont {{Tegmark}}\ and\ \citenamefont
  {{Zaldarriaga}}(2010)}]{2010PhRvD..82j3501T}%
  \BibitemOpen
  \bibfield  {author} {\bibinfo {author} {\bibfnamefont {M.}~\bibnamefont
  {{Tegmark}}}\ and\ \bibinfo {author} {\bibfnamefont {M.}~\bibnamefont
  {{Zaldarriaga}}},\ }\href {\doibase 10.1103/PhysRevD.82.103501} {\bibfield
  {journal} {\bibinfo  {journal} {\prd}\ }\textbf {\bibinfo {volume} {82}},\
  \bibinfo {eid} {103501} (\bibinfo {year} {2010})},\ \Eprint
  {http://arxiv.org/abs/0909.0001} {arXiv:0909.0001 [astro-ph.CO]} \BibitemShut
  {NoStop}%
\bibitem [{\citenamefont {{Masui}}\ \emph
  {et~al.}(2010{\natexlab{b}})\citenamefont {{Masui}}, \citenamefont
  {{Schmidt}}, \citenamefont {{Pen}},\ and\ \citenamefont
  {{McDonald}}}]{2010PhRvD..81f2001M}%
  \BibitemOpen
  \bibfield  {author} {\bibinfo {author} {\bibfnamefont {K.~W.}\ \bibnamefont
  {{Masui}}}, \bibinfo {author} {\bibfnamefont {F.}~\bibnamefont {{Schmidt}}},
  \bibinfo {author} {\bibfnamefont {U.}~\bibnamefont {{Pen}}}, \ and\ \bibinfo
  {author} {\bibfnamefont {P.}~\bibnamefont {{McDonald}}},\ }\href {\doibase
  10.1103/PhysRevD.81.062001} {\bibfield  {journal} {\bibinfo  {journal}
  {\prd}\ }\textbf {\bibinfo {volume} {81}},\ \bibinfo {pages} {062001}
  (\bibinfo {year} {2010}{\natexlab{b}})},\ \Eprint
  {http://arxiv.org/abs/0911.3552} {arXiv:0911.3552} \BibitemShut {NoStop}%
\bibitem [{\citenamefont {{Desjacques}}\ and\ \citenamefont
  {{Seljak}}(2010{\natexlab{a}})}]{2010AdAst2010E..89D}%
  \BibitemOpen
  \bibfield  {author} {\bibinfo {author} {\bibfnamefont {V.}~\bibnamefont
  {{Desjacques}}}\ and\ \bibinfo {author} {\bibfnamefont {U.}~\bibnamefont
  {{Seljak}}},\ }\href {\doibase 10.1155/2010/908640} {\bibfield  {journal}
  {\bibinfo  {journal} {Advances in Astronomy}\ }\textbf {\bibinfo {volume}
  {2010}},\ \bibinfo {eid} {908640} (\bibinfo {year} {2010}{\natexlab{a}}),\
  10.1155/2010/908640},\ \Eprint {http://arxiv.org/abs/1006.4763}
  {arXiv:1006.4763 [astro-ph.CO]} \BibitemShut {NoStop}%
\bibitem [{\citenamefont {{Tseliakhovich}}\ \emph {et~al.}(2010)\citenamefont
  {{Tseliakhovich}}, \citenamefont {{Hirata}},\ and\ \citenamefont
  {{Slosar}}}]{2010PhRvD..82d3531T}%
  \BibitemOpen
  \bibfield  {author} {\bibinfo {author} {\bibfnamefont {D.}~\bibnamefont
  {{Tseliakhovich}}}, \bibinfo {author} {\bibfnamefont {C.}~\bibnamefont
  {{Hirata}}}, \ and\ \bibinfo {author} {\bibfnamefont {A.}~\bibnamefont
  {{Slosar}}},\ }\href {\doibase 10.1103/PhysRevD.82.043531} {\bibfield
  {journal} {\bibinfo  {journal} {\prd}\ }\textbf {\bibinfo {volume} {82}},\
  \bibinfo {eid} {043531} (\bibinfo {year} {2010})},\ \Eprint
  {http://arxiv.org/abs/1004.3302} {arXiv:1004.3302 [astro-ph.CO]} \BibitemShut
  {NoStop}%
\bibitem [{\citenamefont {{Desjacques}}\ and\ \citenamefont
  {{Seljak}}(2010{\natexlab{b}})}]{2010PhRvD..81b3006D}%
  \BibitemOpen
  \bibfield  {author} {\bibinfo {author} {\bibfnamefont {V.}~\bibnamefont
  {{Desjacques}}}\ and\ \bibinfo {author} {\bibfnamefont {U.}~\bibnamefont
  {{Seljak}}},\ }\href {\doibase 10.1103/PhysRevD.81.023006} {\bibfield
  {journal} {\bibinfo  {journal} {\prd}\ }\textbf {\bibinfo {volume} {81}},\
  \bibinfo {eid} {023006} (\bibinfo {year} {2010}{\natexlab{b}})},\ \Eprint
  {http://arxiv.org/abs/0907.2257} {arXiv:0907.2257 [astro-ph.CO]} \BibitemShut
  {NoStop}%
\bibitem [{\citenamefont {{Wagner}}\ \emph {et~al.}(2010)\citenamefont
  {{Wagner}}, \citenamefont {{Verde}},\ and\ \citenamefont
  {{Boubekeur}}}]{2010JCAP...10..022W}%
  \BibitemOpen
  \bibfield  {author} {\bibinfo {author} {\bibfnamefont {C.}~\bibnamefont
  {{Wagner}}}, \bibinfo {author} {\bibfnamefont {L.}~\bibnamefont {{Verde}}}, \
  and\ \bibinfo {author} {\bibfnamefont {L.}~\bibnamefont {{Boubekeur}}},\
  }\href {\doibase 10.1088/1475-7516/2010/10/022} {\bibfield  {journal}
  {\bibinfo  {journal} {\jcap}\ }\textbf {\bibinfo {volume} {10}},\ \bibinfo
  {eid} {022} (\bibinfo {year} {2010})},\ \Eprint
  {http://arxiv.org/abs/1006.5793} {arXiv:1006.5793 [astro-ph.CO]} \BibitemShut
  {NoStop}%
\bibitem [{\citenamefont {{Barnaby}}(2010)}]{2010PhRvD..82j6009B}%
  \BibitemOpen
  \bibfield  {author} {\bibinfo {author} {\bibfnamefont {N.}~\bibnamefont
  {{Barnaby}}},\ }\href {\doibase 10.1103/PhysRevD.82.106009} {\bibfield
  {journal} {\bibinfo  {journal} {\prd}\ }\textbf {\bibinfo {volume} {82}},\
  \bibinfo {eid} {106009} (\bibinfo {year} {2010})},\ \Eprint
  {http://arxiv.org/abs/1006.4615} {arXiv:1006.4615 [astro-ph.CO]} \BibitemShut
  {NoStop}%
\bibitem [{\citenamefont {{Becker}}\ \emph {et~al.}(2011)\citenamefont
  {{Becker}}, \citenamefont {{Huterer}},\ and\ \citenamefont
  {{Kadota}}}]{2011JCAP...01..006B}%
  \BibitemOpen
  \bibfield  {author} {\bibinfo {author} {\bibfnamefont {A.}~\bibnamefont
  {{Becker}}}, \bibinfo {author} {\bibfnamefont {D.}~\bibnamefont {{Huterer}}},
  \ and\ \bibinfo {author} {\bibfnamefont {K.}~\bibnamefont {{Kadota}}},\
  }\href {\doibase 10.1088/1475-7516/2011/01/006} {\bibfield  {journal}
  {\bibinfo  {journal} {\jcap}\ }\textbf {\bibinfo {volume} {1}},\ \bibinfo
  {eid} {006} (\bibinfo {year} {2011})},\ \Eprint
  {http://arxiv.org/abs/1009.4189} {arXiv:1009.4189 [astro-ph.CO]} \BibitemShut
  {NoStop}%
\bibitem [{\citenamefont {{LoVerde}}\ and\ \citenamefont
  {{Smith}}(2011)}]{2011JCAP...08..003L}%
  \BibitemOpen
  \bibfield  {author} {\bibinfo {author} {\bibfnamefont {M.}~\bibnamefont
  {{LoVerde}}}\ and\ \bibinfo {author} {\bibfnamefont {K.~M.}\ \bibnamefont
  {{Smith}}},\ }\href {\doibase 10.1088/1475-7516/2011/08/003} {\bibfield
  {journal} {\bibinfo  {journal} {\jcap}\ }\textbf {\bibinfo {volume} {8}},\
  \bibinfo {eid} {003} (\bibinfo {year} {2011})},\ \Eprint
  {http://arxiv.org/abs/1102.1439} {arXiv:1102.1439 [astro-ph.CO]} \BibitemShut
  {NoStop}%
\bibitem [{\citenamefont {{Smith}}\ and\ \citenamefont
  {{LoVerde}}(2011)}]{2011JCAP...11..009S}%
  \BibitemOpen
  \bibfield  {author} {\bibinfo {author} {\bibfnamefont {K.~M.}\ \bibnamefont
  {{Smith}}}\ and\ \bibinfo {author} {\bibfnamefont {M.}~\bibnamefont
  {{LoVerde}}},\ }\href {\doibase 10.1088/1475-7516/2011/11/009} {\bibfield
  {journal} {\bibinfo  {journal} {\jcap}\ }\textbf {\bibinfo {volume} {11}},\
  \bibinfo {eid} {009} (\bibinfo {year} {2011})},\ \Eprint
  {http://arxiv.org/abs/1010.0055} {arXiv:1010.0055 [astro-ph.CO]} \BibitemShut
  {NoStop}%
\bibitem [{\citenamefont {{Wagner}}\ and\ \citenamefont
  {{Verde}}(2012)}]{2012JCAP...03..002W}%
  \BibitemOpen
  \bibfield  {author} {\bibinfo {author} {\bibfnamefont {C.}~\bibnamefont
  {{Wagner}}}\ and\ \bibinfo {author} {\bibfnamefont {L.}~\bibnamefont
  {{Verde}}},\ }\href {\doibase 10.1088/1475-7516/2012/03/002} {\bibfield
  {journal} {\bibinfo  {journal} {\jcap}\ }\textbf {\bibinfo {volume} {3}},\
  \bibinfo {eid} {002} (\bibinfo {year} {2012})},\ \Eprint
  {http://arxiv.org/abs/1102.3229} {arXiv:1102.3229 [astro-ph.CO]} \BibitemShut
  {NoStop}%
\bibitem [{\citenamefont {{Scoccimarro}}\ \emph {et~al.}(2012)\citenamefont
  {{Scoccimarro}}, \citenamefont {{Hui}}, \citenamefont {{Manera}},\ and\
  \citenamefont {{Chan}}}]{2012PhRvD..85h3002S}%
  \BibitemOpen
  \bibfield  {author} {\bibinfo {author} {\bibfnamefont {R.}~\bibnamefont
  {{Scoccimarro}}}, \bibinfo {author} {\bibfnamefont {L.}~\bibnamefont
  {{Hui}}}, \bibinfo {author} {\bibfnamefont {M.}~\bibnamefont {{Manera}}}, \
  and\ \bibinfo {author} {\bibfnamefont {K.~C.}\ \bibnamefont {{Chan}}},\
  }\href {\doibase 10.1103/PhysRevD.85.083002} {\bibfield  {journal} {\bibinfo
  {journal} {\prd}\ }\textbf {\bibinfo {volume} {85}},\ \bibinfo {eid} {083002}
  (\bibinfo {year} {2012})},\ \Eprint {http://arxiv.org/abs/1108.5512}
  {arXiv:1108.5512 [astro-ph.CO]} \BibitemShut {NoStop}%
\bibitem [{\citenamefont {{Sefusatti}}\ \emph
  {et~al.}(2012{\natexlab{b}})\citenamefont {{Sefusatti}}, \citenamefont
  {{Fergusson}}, \citenamefont {{Chen}},\ and\ \citenamefont
  {{Shellard}}}]{2012JCAP...08..033S}%
  \BibitemOpen
  \bibfield  {author} {\bibinfo {author} {\bibfnamefont {E.}~\bibnamefont
  {{Sefusatti}}}, \bibinfo {author} {\bibfnamefont {J.~R.}\ \bibnamefont
  {{Fergusson}}}, \bibinfo {author} {\bibfnamefont {X.}~\bibnamefont {{Chen}}},
  \ and\ \bibinfo {author} {\bibfnamefont {E.~P.~S.}\ \bibnamefont
  {{Shellard}}},\ }\href {\doibase 10.1088/1475-7516/2012/08/033} {\bibfield
  {journal} {\bibinfo  {journal} {\jcap}\ }\textbf {\bibinfo {volume} {8}},\
  \bibinfo {eid} {033} (\bibinfo {year} {2012}{\natexlab{b}})},\ \Eprint
  {http://arxiv.org/abs/1204.6318} {arXiv:1204.6318 [astro-ph.CO]} \BibitemShut
  {NoStop}%
\bibitem [{\citenamefont {{Biagetti}}\ \emph {et~al.}(2013)\citenamefont
  {{Biagetti}}, \citenamefont {{Desjacques}},\ and\ \citenamefont
  {{Riotto}}}]{2013MNRAS.429.1774B}%
  \BibitemOpen
  \bibfield  {author} {\bibinfo {author} {\bibfnamefont {M.}~\bibnamefont
  {{Biagetti}}}, \bibinfo {author} {\bibfnamefont {V.}~\bibnamefont
  {{Desjacques}}}, \ and\ \bibinfo {author} {\bibfnamefont {A.}~\bibnamefont
  {{Riotto}}},\ }\href {\doibase 10.1093/mnras/sts467} {\bibfield  {journal}
  {\bibinfo  {journal} {\mnras}\ }\textbf {\bibinfo {volume} {429}},\ \bibinfo
  {pages} {1774} (\bibinfo {year} {2013})},\ \Eprint
  {http://arxiv.org/abs/1208.1616} {arXiv:1208.1616 [astro-ph.CO]} \BibitemShut
  {NoStop}%
\bibitem [{\citenamefont {{Nelson}}\ and\ \citenamefont
  {{Shandera}}(2013)}]{2013PhRvL.110m1301N}%
  \BibitemOpen
  \bibfield  {author} {\bibinfo {author} {\bibfnamefont {E.}~\bibnamefont
  {{Nelson}}}\ and\ \bibinfo {author} {\bibfnamefont {S.}~\bibnamefont
  {{Shandera}}},\ }\href {\doibase 10.1103/PhysRevLett.110.131301} {\bibfield
  {journal} {\bibinfo  {journal} {Physical Review Letters}\ }\textbf {\bibinfo
  {volume} {110}},\ \bibinfo {eid} {131301} (\bibinfo {year} {2013})},\ \Eprint
  {http://arxiv.org/abs/1212.4550} {arXiv:1212.4550 [astro-ph.CO]} \BibitemShut
  {NoStop}%
\bibitem [{\citenamefont {{Barnaby}}\ and\ \citenamefont
  {{Huang}}(2009)}]{2009PhRvD..80l6018B}%
  \BibitemOpen
  \bibfield  {author} {\bibinfo {author} {\bibfnamefont {N.}~\bibnamefont
  {{Barnaby}}}\ and\ \bibinfo {author} {\bibfnamefont {Z.}~\bibnamefont
  {{Huang}}},\ }\href {\doibase 10.1103/PhysRevD.80.126018} {\bibfield
  {journal} {\bibinfo  {journal} {\prd}\ }\textbf {\bibinfo {volume} {80}},\
  \bibinfo {eid} {126018} (\bibinfo {year} {2009})},\ \Eprint
  {http://arxiv.org/abs/0909.0751} {arXiv:0909.0751 [astro-ph.CO]} \BibitemShut
  {NoStop}%
\bibitem [{\citenamefont {{Sefusatti}}\ \emph {et~al.}(2009)\citenamefont
  {{Sefusatti}}, \citenamefont {{Liguori}}, \citenamefont {{Yadav}},
  \citenamefont {{Jackson}},\ and\ \citenamefont
  {{Pajer}}}]{2009JCAP...12..022S}%
  \BibitemOpen
  \bibfield  {author} {\bibinfo {author} {\bibfnamefont {E.}~\bibnamefont
  {{Sefusatti}}}, \bibinfo {author} {\bibfnamefont {M.}~\bibnamefont
  {{Liguori}}}, \bibinfo {author} {\bibfnamefont {A.~P.~S.}\ \bibnamefont
  {{Yadav}}}, \bibinfo {author} {\bibfnamefont {M.~G.}\ \bibnamefont
  {{Jackson}}}, \ and\ \bibinfo {author} {\bibfnamefont {E.}~\bibnamefont
  {{Pajer}}},\ }\href {\doibase 10.1088/1475-7516/2009/12/022} {\bibfield
  {journal} {\bibinfo  {journal} {\jcap}\ }\textbf {\bibinfo {volume} {12}},\
  \bibinfo {eid} {022} (\bibinfo {year} {2009})},\ \Eprint
  {http://arxiv.org/abs/0906.0232} {arXiv:0906.0232 [astro-ph.CO]} \BibitemShut
  {NoStop}%
\bibitem [{\citenamefont {{Kawasaki}}\ \emph
  {et~al.}(2009{\natexlab{a}})\citenamefont {{Kawasaki}}, \citenamefont
  {{Nakayama}}, \citenamefont {{Sekiguchi}}, \citenamefont {{Suyama}},\ and\
  \citenamefont {{Takahashi}}}]{2009JCAP...01..042K}%
  \BibitemOpen
  \bibfield  {author} {\bibinfo {author} {\bibfnamefont {M.}~\bibnamefont
  {{Kawasaki}}}, \bibinfo {author} {\bibfnamefont {K.}~\bibnamefont
  {{Nakayama}}}, \bibinfo {author} {\bibfnamefont {T.}~\bibnamefont
  {{Sekiguchi}}}, \bibinfo {author} {\bibfnamefont {T.}~\bibnamefont
  {{Suyama}}}, \ and\ \bibinfo {author} {\bibfnamefont {F.}~\bibnamefont
  {{Takahashi}}},\ }\href {\doibase 10.1088/1475-7516/2009/01/042} {\bibfield
  {journal} {\bibinfo  {journal} {\jcap}\ }\textbf {\bibinfo {volume} {1}},\
  \bibinfo {eid} {042} (\bibinfo {year} {2009}{\natexlab{a}})},\ \Eprint
  {http://arxiv.org/abs/0810.0208} {arXiv:0810.0208} \BibitemShut {NoStop}%
\bibitem [{\citenamefont {{Kawasaki}}\ \emph
  {et~al.}(2009{\natexlab{b}})\citenamefont {{Kawasaki}}, \citenamefont
  {{Nakayama}},\ and\ \citenamefont {{Takahashi}}}]{2009JCAP...01..026K}%
  \BibitemOpen
  \bibfield  {author} {\bibinfo {author} {\bibfnamefont {M.}~\bibnamefont
  {{Kawasaki}}}, \bibinfo {author} {\bibfnamefont {K.}~\bibnamefont
  {{Nakayama}}}, \ and\ \bibinfo {author} {\bibfnamefont {F.}~\bibnamefont
  {{Takahashi}}},\ }\href {\doibase 10.1088/1475-7516/2009/01/026} {\bibfield
  {journal} {\bibinfo  {journal} {\jcap}\ }\textbf {\bibinfo {volume} {1}},\
  \bibinfo {eid} {026} (\bibinfo {year} {2009}{\natexlab{b}})},\ \Eprint
  {http://arxiv.org/abs/0810.1585} {arXiv:0810.1585 [hep-ph]} \BibitemShut
  {NoStop}%
\bibitem [{\citenamefont {{Viel}}\ \emph {et~al.}(2013)\citenamefont {{Viel}},
  \citenamefont {{Becker}}, \citenamefont {{Bolton}},\ and\ \citenamefont
  {{Haehnelt}}}]{2013arXiv1306.2314V}%
  \BibitemOpen
  \bibfield  {author} {\bibinfo {author} {\bibfnamefont {M.}~\bibnamefont
  {{Viel}}}, \bibinfo {author} {\bibfnamefont {G.~D.}\ \bibnamefont
  {{Becker}}}, \bibinfo {author} {\bibfnamefont {J.~S.}\ \bibnamefont
  {{Bolton}}}, \ and\ \bibinfo {author} {\bibfnamefont {M.~G.}\ \bibnamefont
  {{Haehnelt}}},\ }\href@noop {} {\bibfield  {journal} {\bibinfo  {journal}
  {ArXiv e-prints}\ } (\bibinfo {year} {2013})},\ \Eprint
  {http://arxiv.org/abs/1306.2314} {arXiv:1306.2314 [astro-ph.CO]} \BibitemShut
  {NoStop}%
\bibitem [{\citenamefont {{Viel}}\ \emph {et~al.}(2008)\citenamefont {{Viel}},
  \citenamefont {{Becker}}, \citenamefont {{Bolton}}, \citenamefont
  {{Haehnelt}}, \citenamefont {{Rauch}},\ and\ \citenamefont
  {{Sargent}}}]{2008PhRvL.100d1304V}%
  \BibitemOpen
  \bibfield  {author} {\bibinfo {author} {\bibfnamefont {M.}~\bibnamefont
  {{Viel}}}, \bibinfo {author} {\bibfnamefont {G.~D.}\ \bibnamefont
  {{Becker}}}, \bibinfo {author} {\bibfnamefont {J.~S.}\ \bibnamefont
  {{Bolton}}}, \bibinfo {author} {\bibfnamefont {M.~G.}\ \bibnamefont
  {{Haehnelt}}}, \bibinfo {author} {\bibfnamefont {M.}~\bibnamefont {{Rauch}}},
  \ and\ \bibinfo {author} {\bibfnamefont {W.~L.~W.}\ \bibnamefont
  {{Sargent}}},\ }\href {\doibase 10.1103/PhysRevLett.100.041304} {\bibfield
  {journal} {\bibinfo  {journal} {Physical Review Letters}\ }\textbf {\bibinfo
  {volume} {100}},\ \bibinfo {eid} {041304} (\bibinfo {year} {2008})},\ \Eprint
  {http://arxiv.org/abs/0709.0131} {arXiv:0709.0131} \BibitemShut {NoStop}%
\bibitem [{\citenamefont {{Seljak}}\ \emph
  {et~al.}(2006{\natexlab{b}})\citenamefont {{Seljak}}, \citenamefont
  {{Makarov}}, \citenamefont {{McDonald}},\ and\ \citenamefont
  {{Trac}}}]{2006PhRvL..97s1303S}%
  \BibitemOpen
  \bibfield  {author} {\bibinfo {author} {\bibfnamefont {U.}~\bibnamefont
  {{Seljak}}}, \bibinfo {author} {\bibfnamefont {A.}~\bibnamefont {{Makarov}}},
  \bibinfo {author} {\bibfnamefont {P.}~\bibnamefont {{McDonald}}}, \ and\
  \bibinfo {author} {\bibfnamefont {H.}~\bibnamefont {{Trac}}},\ }\href
  {\doibase 10.1103/PhysRevLett.97.191303} {\bibfield  {journal} {\bibinfo
  {journal} {Physical Review Letters}\ }\textbf {\bibinfo {volume} {97}},\
  \bibinfo {pages} {191303} (\bibinfo {year} {2006}{\natexlab{b}})}\BibitemShut
  {NoStop}%
\bibitem [{\citenamefont {{Afshordi}}\ \emph {et~al.}(2003)\citenamefont
  {{Afshordi}}, \citenamefont {{McDonald}},\ and\ \citenamefont
  {{Spergel}}}]{2003ApJ...594L..71A}%
  \BibitemOpen
  \bibfield  {author} {\bibinfo {author} {\bibfnamefont {N.}~\bibnamefont
  {{Afshordi}}}, \bibinfo {author} {\bibfnamefont {P.}~\bibnamefont
  {{McDonald}}}, \ and\ \bibinfo {author} {\bibfnamefont {D.~N.}\ \bibnamefont
  {{Spergel}}},\ }\href@noop {} {\bibfield  {journal} {\bibinfo  {journal}
  {\apjl}\ }\textbf {\bibinfo {volume} {594}},\ \bibinfo {pages} {L71}
  (\bibinfo {year} {2003})}\BibitemShut {NoStop}%
\bibitem [{\citenamefont {{Carr}}\ \emph {et~al.}(2010)\citenamefont {{Carr}},
  \citenamefont {{Kohri}}, \citenamefont {{Sendouda}},\ and\ \citenamefont
  {{Yokoyama}}}]{2010PhRvD..81j4019C}%
  \BibitemOpen
  \bibfield  {author} {\bibinfo {author} {\bibfnamefont {B.~J.}\ \bibnamefont
  {{Carr}}}, \bibinfo {author} {\bibfnamefont {K.}~\bibnamefont {{Kohri}}},
  \bibinfo {author} {\bibfnamefont {Y.}~\bibnamefont {{Sendouda}}}, \ and\
  \bibinfo {author} {\bibfnamefont {J.}~\bibnamefont {{Yokoyama}}},\ }\href
  {\doibase 10.1103/PhysRevD.81.104019} {\bibfield  {journal} {\bibinfo
  {journal} {\prd}\ }\textbf {\bibinfo {volume} {81}},\ \bibinfo {eid} {104019}
  (\bibinfo {year} {2010})},\ \Eprint {http://arxiv.org/abs/0912.5297}
  {arXiv:0912.5297 [astro-ph.CO]} \BibitemShut {NoStop}%
\bibitem [{\citenamefont {{Krause}}\ and\ \citenamefont
  {{Hirata}}(2011)}]{2011MNRAS.410.2730K}%
  \BibitemOpen
  \bibfield  {author} {\bibinfo {author} {\bibfnamefont {E.}~\bibnamefont
  {{Krause}}}\ and\ \bibinfo {author} {\bibfnamefont {C.~M.}\ \bibnamefont
  {{Hirata}}},\ }\href {\doibase 10.1111/j.1365-2966.2010.17638.x} {\bibfield
  {journal} {\bibinfo  {journal} {\mnras}\ }\textbf {\bibinfo {volume} {410}},\
  \bibinfo {pages} {2730} (\bibinfo {year} {2011})},\ \Eprint
  {http://arxiv.org/abs/1004.3611} {arXiv:1004.3611 [astro-ph.CO]} \BibitemShut
  {NoStop}%
\bibitem [{\citenamefont {{Hirata}}(2009)}]{2009MNRAS.399.1074H}%
  \BibitemOpen
  \bibfield  {author} {\bibinfo {author} {\bibfnamefont {C.~M.}\ \bibnamefont
  {{Hirata}}},\ }\href {\doibase 10.1111/j.1365-2966.2009.15353.x} {\bibfield
  {journal} {\bibinfo  {journal} {\mnras}\ }\textbf {\bibinfo {volume} {399}},\
  \bibinfo {pages} {1074} (\bibinfo {year} {2009})},\ \Eprint
  {http://arxiv.org/abs/0903.4929} {arXiv:0903.4929} \BibitemShut {NoStop}%
\bibitem [{\citenamefont {{Gazta{\~n}aga}}\ \emph {et~al.}(2012)\citenamefont
  {{Gazta{\~n}aga}}, \citenamefont {{Eriksen}}, \citenamefont {{Crocce}},
  \citenamefont {{Castander}}, \citenamefont {{Fosalba}}, \citenamefont
  {{Marti}}, \citenamefont {{Miquel}},\ and\ \citenamefont
  {{Cabr{\'e}}}}]{2012MNRAS.422.2904G}%
  \BibitemOpen
  \bibfield  {author} {\bibinfo {author} {\bibfnamefont {E.}~\bibnamefont
  {{Gazta{\~n}aga}}}, \bibinfo {author} {\bibfnamefont {M.}~\bibnamefont
  {{Eriksen}}}, \bibinfo {author} {\bibfnamefont {M.}~\bibnamefont {{Crocce}}},
  \bibinfo {author} {\bibfnamefont {F.~J.}\ \bibnamefont {{Castander}}},
  \bibinfo {author} {\bibfnamefont {P.}~\bibnamefont {{Fosalba}}}, \bibinfo
  {author} {\bibfnamefont {P.}~\bibnamefont {{Marti}}}, \bibinfo {author}
  {\bibfnamefont {R.}~\bibnamefont {{Miquel}}}, \ and\ \bibinfo {author}
  {\bibfnamefont {A.}~\bibnamefont {{Cabr{\'e}}}},\ }\href {\doibase
  10.1111/j.1365-2966.2012.20613.x} {\bibfield  {journal} {\bibinfo  {journal}
  {\mnras}\ }\textbf {\bibinfo {volume} {422}},\ \bibinfo {pages} {2904}
  (\bibinfo {year} {2012})},\ \Eprint {http://arxiv.org/abs/1109.4852}
  {arXiv:1109.4852 [astro-ph.CO]} \BibitemShut {NoStop}%
\bibitem [{\citenamefont {{Kirk}}\ \emph {et~al.}(2013)\citenamefont {{Kirk}},
  \citenamefont {{Lahav}}, \citenamefont {{Bridle}}, \citenamefont {{Jouvel}},
  \citenamefont {{Abdalla}},\ and\ \citenamefont
  {{Frieman}}}]{2013arXiv1307.8062K}%
  \BibitemOpen
  \bibfield  {author} {\bibinfo {author} {\bibfnamefont {D.}~\bibnamefont
  {{Kirk}}}, \bibinfo {author} {\bibfnamefont {O.}~\bibnamefont {{Lahav}}},
  \bibinfo {author} {\bibfnamefont {S.}~\bibnamefont {{Bridle}}}, \bibinfo
  {author} {\bibfnamefont {S.}~\bibnamefont {{Jouvel}}}, \bibinfo {author}
  {\bibfnamefont {F.~B.}\ \bibnamefont {{Abdalla}}}, \ and\ \bibinfo {author}
  {\bibfnamefont {J.~A.}\ \bibnamefont {{Frieman}}},\ }\href@noop {} {\bibfield
   {journal} {\bibinfo  {journal} {ArXiv e-prints}\ } (\bibinfo {year}
  {2013})},\ \Eprint {http://arxiv.org/abs/1307.8062} {arXiv:1307.8062
  [astro-ph.CO]} \BibitemShut {NoStop}%
\bibitem [{\citenamefont {{de Putter}}\ \emph
  {et~al.}(2013{\natexlab{a}})\citenamefont {{de Putter}}, \citenamefont
  {{Dor{\'e}}},\ and\ \citenamefont {{Das}}}]{2013arXiv1306.0534D}%
  \BibitemOpen
  \bibfield  {author} {\bibinfo {author} {\bibfnamefont {R.}~\bibnamefont {{de
  Putter}}}, \bibinfo {author} {\bibfnamefont {O.}~\bibnamefont {{Dor{\'e}}}},
  \ and\ \bibinfo {author} {\bibfnamefont {S.}~\bibnamefont {{Das}}},\
  }\href@noop {} {\bibfield  {journal} {\bibinfo  {journal} {ArXiv e-prints}\ }
  (\bibinfo {year} {2013}{\natexlab{a}})},\ \Eprint
  {http://arxiv.org/abs/1306.0534} {arXiv:1306.0534 [astro-ph.CO]} \BibitemShut
  {NoStop}%
\bibitem [{\citenamefont {{de Putter}}\ \emph
  {et~al.}(2013{\natexlab{b}})\citenamefont {{de Putter}}, \citenamefont
  {{Dor{\'e}}},\ and\ \citenamefont {{Takada}}}]{2013arXiv1308.6070D}%
  \BibitemOpen
  \bibfield  {author} {\bibinfo {author} {\bibfnamefont {R.}~\bibnamefont {{de
  Putter}}}, \bibinfo {author} {\bibfnamefont {O.}~\bibnamefont {{Dor{\'e}}}},
  \ and\ \bibinfo {author} {\bibfnamefont {M.}~\bibnamefont {{Takada}}},\
  }\href@noop {} {\bibfield  {journal} {\bibinfo  {journal} {ArXiv e-prints}\ }
  (\bibinfo {year} {2013}{\natexlab{b}})},\ \Eprint
  {http://arxiv.org/abs/1308.6070} {arXiv:1308.6070 [astro-ph.CO]} \BibitemShut
  {NoStop}%
\bibitem [{\citenamefont {{Cai}}\ and\ \citenamefont
  {{Bernstein}}(2012)}]{2012MNRAS.422.1045C}%
  \BibitemOpen
  \bibfield  {author} {\bibinfo {author} {\bibfnamefont {Y.-C.}\ \bibnamefont
  {{Cai}}}\ and\ \bibinfo {author} {\bibfnamefont {G.}~\bibnamefont
  {{Bernstein}}},\ }\href {\doibase 10.1111/j.1365-2966.2012.20676.x}
  {\bibfield  {journal} {\bibinfo  {journal} {\mnras}\ }\textbf {\bibinfo
  {volume} {422}},\ \bibinfo {pages} {1045} (\bibinfo {year} {2012})},\ \Eprint
  {http://arxiv.org/abs/1112.4478} {arXiv:1112.4478 [astro-ph.CO]} \BibitemShut
  {NoStop}%
\bibitem [{\citenamefont {{Bernstein}}(2009)}]{2009ApJ...695..652B}%
  \BibitemOpen
  \bibfield  {author} {\bibinfo {author} {\bibfnamefont {G.~M.}\ \bibnamefont
  {{Bernstein}}},\ }\href {\doibase 10.1088/0004-637X/695/1/652} {\bibfield
  {journal} {\bibinfo  {journal} {\apj}\ }\textbf {\bibinfo {volume} {695}},\
  \bibinfo {pages} {652} (\bibinfo {year} {2009})},\ \Eprint
  {http://arxiv.org/abs/0808.3400} {arXiv:0808.3400} \BibitemShut {NoStop}%
\end{thebibliography}%

\section{Appendix: Additional calculations}

\subsection{Traditional FoM without neutrino masses}

For comparison to past work,
Table \ref{tableoldFoM} shows traditional FoMs with fixed neutrino mass. 

\begin{table}
\caption{
Original DETF Figures of Merit, i.e., {\it not}  marginalized over $\summnu$.
See Table \ref{tab:experimentabbreviations} for survey codes.}
\label{tableoldFoM} 
\begin{tabular}{lcccc}
\hline
\hline
Surveys & FoM & $a_p$ & $\sigma_{w_p}$ & $\sigma_{\Omega_k}$ \\
\hline
P+BgB+BlB & 37 & 0.65 & 0.055 & 0.0026 \\
P+BgA0.1+BlB & 53 & 0.62 & 0.040 & 0.0025 \\
P+BgA0.1+BlB+ebA0.1 & 80 & 0.65 & 0.032 & 0.0020 \\
P+DES & 45 & 0.73 & 0.032 & 0.0024 \\
P+BgB+BlB+DES & 85 & 0.72 & 0.028 & 0.0014 \\
P+BgA0.1+BlB+DES & 112 & 0.69 & 0.024 & 0.0014 \\
P+BgA0.1+BlB+ebA0.1 & 80 & 0.65 & 0.032 & 0.0020 \\
P+BgA0.1+BlB+ebA0.2 & 106 & 0.66 & 0.026 & 0.0019 \\
P+BgA0.1+BlB+ebA0.1+DES & 137 & 0.70 & 0.021 & 0.0013 \\
P+BBgB & 128 & 0.71 & 0.023 & 0.0013 \\
P+BBgB+BlB & 134 & 0.72 & 0.023 & 0.0012 \\
P+BBlB+BgB & 77 & 0.77 & 0.037 & 0.0013 \\
P+BBgB+BBlB & 166 & 0.73 & 0.023 & 0.0010 \\
P+BBgA0.1 & 291 & 0.73 & 0.016 & 0.0010 \\
P+BBgA0.1+BBlB & 342 & 0.75 & 0.015 & 0.0008 \\
P+BBgA0.1+DES & 360 & 0.74 & 0.013 & 0.0009 \\
P+BBgA0.1+BBlB+DES & 410 & 0.75 & 0.013 & 0.0008 \\
P+BBgA0.2+BBlB & 756 & 0.74 & 0.011 & 0.0007 \\
P+BBgA0.2+BBlB+DES & 948 & 0.74 & 0.009 & 0.0007 \\
P+BB24gB+BB24lB & 252 & 0.72 & 0.019 & 0.0009 \\
P+BB24gA0.1+BB24lB & 557 & 0.75 & 0.012 & 0.0007 \\
P+BB24gA0.1+BB24lB+DES & 647 & 0.75 & 0.010 & 0.0006 \\
P+BB24gA0.2+BB24lB & 1194 & 0.73 & 0.009 & 0.0006 \\
P+BB24gA0.2+BB24lB+DES & 1454 & 0.74 & 0.007 & 0.0005 \\
P+BgB+BlB+euB & 152 & 0.75 & 0.022 & 0.0010 \\
P+BgA0.1+BlB+euA0.1 & 366 & 0.77 & 0.014 & 0.0008 \\
P+BgA0.1+BlB+euA0.1+DES & 448 & 0.77 & 0.012 & 0.0007 \\
P+BgA0.2+BlB+euA0.2 & 844 & 0.76 & 0.010 & 0.0006 \\
P+BgA0.2+BlB+euA0.2+DES & 1080 & 0.76 & 0.008 & 0.0006 \\
P+BB24gA0.1+BB24lB+euA0.1 & 787 & 0.76 & 0.010 & 0.0006 \\
P+BB24gA0.1+BB24lB+euA0.1+DES & 908 & 0.76 & 0.009 & 0.0005 \\
P+BB24gA0.2+BB24lB+euA0.2 & 1695 & 0.74 & 0.008 & 0.0005 \\
P+BB24gA0.2+BB24lB+euA0.2+DES & 2025 & 0.75 & 0.006 & 0.0005 \\
P+LSST & 221 & 0.76 & 0.013 & 0.0015 \\
P+BgB+BlB+LSST & 291 & 0.76 & 0.012 & 0.0010 \\
P+BB24gA0.1+BB24lB+LSST & 934 & 0.75 & 0.008 & 0.0006 \\
P+BB24gA0.2+BB24lB+LSST & 2165 & 0.74 & 0.005 & 0.0005 \\
P+BB24gA0.1+BB24lB+euA0.1+LSST & 1262 & 0.76 & 0.007 & 0.0005 \\
P+BB24gA0.2+BB24lB+euA0.2+LSST & 3124 & 0.76 & 0.004 & 0.0004 \\
\hline
\hline
\end{tabular}
\end{table}

Table \ref{tableFoMSWGFoM} shows traditional FoMs based on FoMSWG parameters
and the FoMSWG Planck Fisher matrix.  

\begin{table}
\caption{
Original DETF Figures of Merit including Planck CMB {\it from FoMSWG} 
(indicated by P3 -- normally we compute our
own Planck Fisher matrix, because FoMSWG did not include neutrinos).
WL3 stands for the FoMSWG Stage III weak lensing Fisher matrix. 
See Table \ref{tab:experimentabbreviations} for other survey codes.}
\label{tableFoMSWGFoM} 
\begin{tabular}{lrccc}
\hline
\hline
Surveys & FoM & $a_p$ & $\sigma_{w_p}$ & $\sigma_{\Omega_k}$ \\
\hline
P3+BgB+BlB & 31 & 0.62 & 0.057 & 0.0030 \\
P3+BBgB+BBlB & 160 & 0.72 & 0.023 & 0.0010 \\
P3+BBgA0.1+BBlB & 265 & 0.73 & 0.018 & 0.0009 \\
P3+DES & 35 & 0.73 & 0.035 & 0.0031 \\
P3+WL3 & 15 & 0.67 & 0.068 & 0.0029 \\
P3+BBgA0.1+BBlB+DES & 382 & 0.74 & 0.013 & 0.0008 \\
P3+BBgA0.1+BBlB+WL3 & 287 & 0.74 & 0.018 & 0.0008 \\
\hline
\hline
\end{tabular}
\end{table}
We see by comparing these two tables that our treatment of Planck
gives very similar results to the FoMSWG treatment (they are not intended to
be identical -- e.g., we use a different parameter set and do not assume the
$\tau$ measurement will be systematics limited). On the other hand, our 
treatment of DES is quite optimistic relative to the FoMSWG Stage III weak
lensing example (we do not include shear calibration uncertainty, 
intrinsic alignments, etc.). 

\subsection{Overlapping redshift and photometric surveys}
\label{sec:overl-redsh-phot}

It has been suggested that dramatic gains in constraining power can 
be achieved when redshift surveys and photometric (lensing) surveys overlap
on the sky \cite{2012MNRAS.422.2904G}, so they can be directly cross-correlated
rather than simply providing complementary parameter constraints. We 
investigated this possibility
but did not find significant gains, so we left it out of our main discussion
for simplicity. We may address this in detail in 
a future paper, but, because it might be expected that this would be a big part
of parameter projections like ours, we give some basic results and discussion 
here. 
(The authors of 
\cite{2012MNRAS.422.2904G} did find some problems with their calculations
(E. Gaztanaga, private communication), although the issue has not been 
entirely resolved.) 

\subsubsection{Overlapping survey Fisher matrices}

We compute Fisher matrices for overlapping surveys by first treating the 
redshift
survey galaxies as more angular clustering tracers, with zero photo-$z$ 
errors, for
the purpose of cross-correlations (i.e., as in 
\S \ref{sec:photometric-surveys}), and then adding a standard 3D power 
spectrum-based redshift survey
Fisher matrix to pick up the small-radial-scale information (as in 
\S \ref{sec:redshift-surveys}).
Ideally, the auto angular clustering at different redshifts 
and cross-correlation between them has the same information as the 3D power
spectrum. We use 3D power Fisher matrices at small scales because the angular 
clustering is evaluated using a wide redshift bin, i.e.,  
$\Delta z=0.2$ is used and the 3D Fisher matrix 
is added to account for radial information internal to the $\Delta z=0.2$ 
slice (it is unnecessarily computationally cumbersome to try to fully resolve
the redshift survey information by making $\Delta z$ very small, e.g., bins
of width $\Delta z\sim 0.001(1+z)$ are required to resolve the BAO feature).
In order to avoid double-counting very low $k_z$ information which is
already included in the angular correlations, we down-weight 3D redshift-survey
modes in overlapping calculations by a factor
$1-\left[\sin\left(f_{k_z} k_z \Delta z\right)/
\left(f_{k_z} k_z \Delta z\right)\right]^2$,
where $\Delta z$ is the width of the angular clustering bins. This damping
factor, which cannot be derived rigorously, was calibrated,
finding $f_{k_z}=1$, by the requirement
that, when using only redshift survey galaxies, Fisher matrix results computed
in purely redshift-survey (3D power spectrum) mode were the same as the
summed Fisher matrix results from redshift-survey mode on small radial scales
and angular clustering mode on large radial scales. The exact value of
$f_{k_z}$ is unimportant, however, because in the end this is
only a tiny fraction of redshift survey modes, as we discuss below.

Results possibly including overlapping LSST and DESI are shown in 
Table \ref{tableoverlap}.
\begin{table}
\caption{
FoM (with neutrinos, i.e., like Table \ref{tableOmkWz}) for
overlapping vs. non-overlapping DESI and LSST (to be clear, DESI is always
14000 sq. deg., and LSST always 20000, but in the overlapping case the 
14000 sq. deg. DESI area is assumed be covered by LSST, while in the 
non-overlapping case it is not).
See Table \ref{tab:experimentabbreviations} for other survey codes.
\label{tableoverlap}}
\begin{center}
\begin{tabular}{lccccc}
\hline
\hline
Surveys & $\Delta z$ & FoM & $a_p$ & $\sigma_{w_p}$ & $\sigma_{\Omega_k}$ \\
\hline
P+BBgA0.1 & 0.2 & 109 & 0.66 & 0.029 & 0.0013 \\
P+LSST & 0.2 & 134 & 0.72 & 0.019 & 0.0019  \\
P+BBgA0.1+LSST-no overlap & 0.2 & 288 & 0.71 & 0.014 & 0.0011 \\
P+BBgA0.1+LSST-overlap & 0.2 & 308 & 0.71 & 0.014 & 0.0010 \\
P+BBgA0.1 & 0.1 &  105 & 0.66 & 0.029 & 0.0014 \\
P+LSST & 0.1 &  162 & 0.73 & 0.018 & 0.0017 \\
P+BBgA0.1+LSST-no overlap & 0.1 & 314 & 0.71 & 0.014 & 0.0010  \\
P+BBgA0.1+LSST-overlap & 0.1 & 329 & 0.71 & 0.013 & 0.0010 \\
\hline
\hline
\end{tabular}
\end{center}
\end{table}
While the redshift and lensing surveys are highly complementary, whether they
overlap or not makes very little difference.
To verify that this conclusion is not sensitive to our redshift binning, we
show results for both $\Delta z=0.2$ and $\Delta z=0.1$. 
The biggest difference is that the 
FoM for Planck+LSST improves by 20\% for the finer binning, presumably because
$\Delta z=0.2$ bins are still wider than our assumed $0.05 (1+z)$ rms photo-$z$
errors, so some information is lost. 
The few percent degradation of the redshift survey with finer 
binning is presumably due to the extra free bias parameters introduced. In any
case, the most relevant combined results change by less than 10\%, and the gain
from overlap is almost the same. 

Note that this calculation is not directly comparable to 
\cite{2013arXiv1307.8062K} because they
compute their spectroscopic survey constraint using exclusively a $C_\ell$ 
method like we use for the photometric constraints, with bins of 
width $\Delta z=0.05$, corresponding to 
$\sim 85 \hMpc$ at $z\sim 1$ -- clearly far from what is necessary to resolve
the BAO-scale structure where we find the bulk of our redshift survey 
information. There are a lot of other differences, e.g., they generally do not 
include Planck constraints, and do not include  
correlations involving photometric galaxy density (i.e., autocorrelations
and what is usually called galaxy-galaxy lensing), which enhance the 
constraining power of the non-overlapping case. 
Our conclusion is consistent with \cite{2013arXiv1306.0534D}, who find gains
in certain scenarios including photo-$z$ systematic uncertainties, but not
in our scenario.
Our results are very similar to \cite{2013arXiv1308.6070D}.

\subsubsection{Qualitative understanding of why cross-correlations are not 
very helpful}

\begin{figure}[tb]
\centering
\includegraphics[width=0.8\textwidth,angle=0]{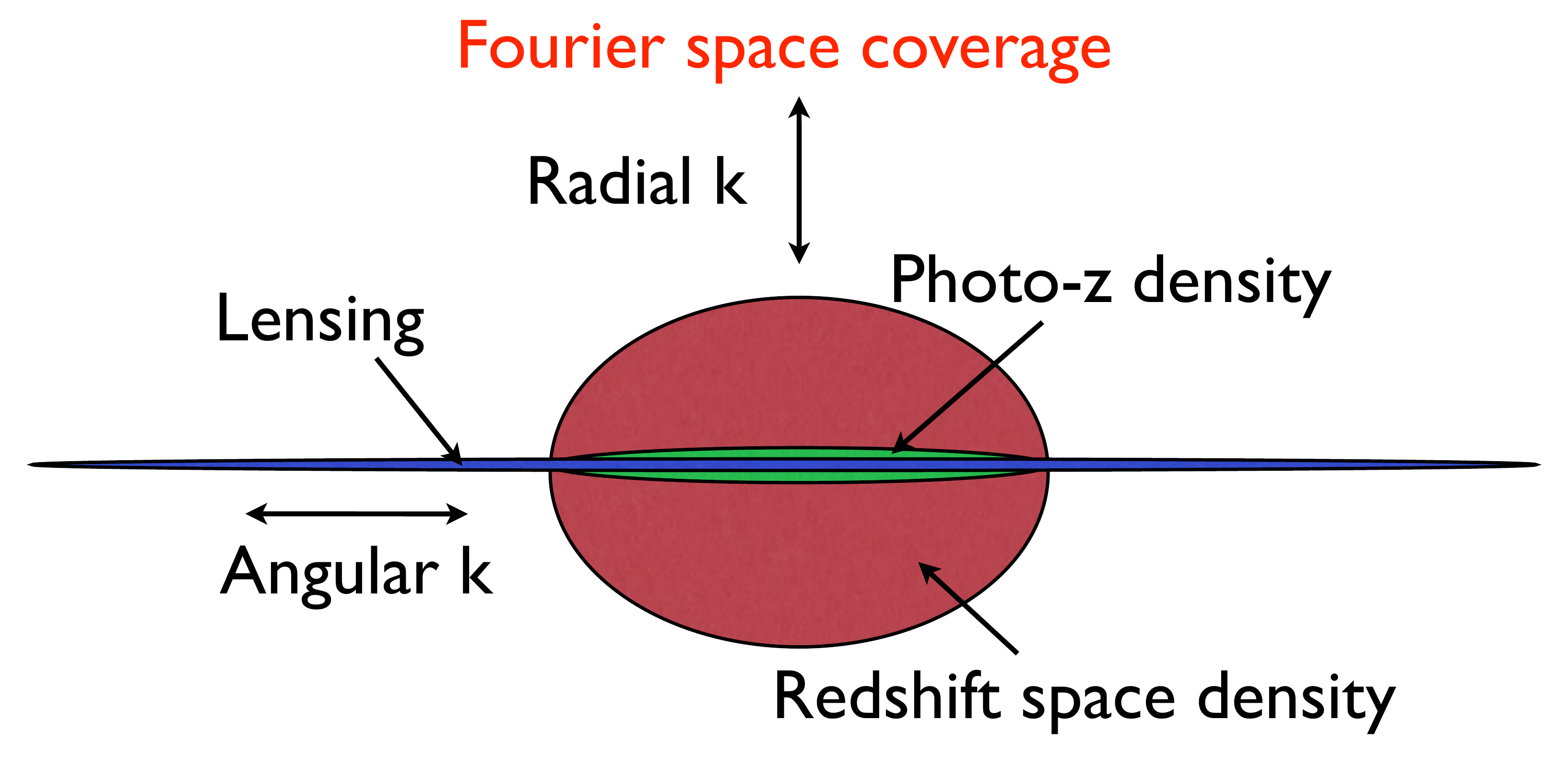}
\caption{
Qualitative diagram of Fourier space coverage of elements of photometric and
spectroscopic surveys.
The red oval shows conceptually the Fourier space coverage of the redshift 
survey 
(the boundary of the red oval would correspond to $k\sim 0.1-0.2 \ihMpc$, 
the scale of
breakdown of perturbation theory),
the green oval shows the coverage of the photo-$z$ density field, and the
blue oval shows (even more qualitatively) the coverage of the lensing 
convergence field.}
\label{fig:overlapslide}
\end{figure}

Given the intuitive appeal of overlapping lensing and redshift surveys, and
the results of \cite{2012MNRAS.422.2904G}, some further effort to qualitatively
understand the results here seems desirable. (\cite{2012MNRAS.422.1045C}
also found that overlap can be valuable, however, they present their results
rather abstractly so it is difficult to project them onto concrete scenarios
-- as far as we can tell our results are probably consistent.)
Figure \ref{fig:overlapslide} attempts to illuminate what is really going on
when surveys overlap.
The Fourier space coverage of the redshift survey is
limited at both high radial and transverse $k$ by non-linearities and the
accompanying scale dependence of bias and stochasticity.
The number density measurements for the photometric survey have similar
coverage in
the transverse direction but have drastically reduced coverage in the radial
direction, set by the photo-$z$ accuracy. Lensing is further limited in the
radial direction because of the breadth of the lensing kernel, however, we
expect that the signal can be predicted to significantly smaller transverse
scales (the radial Fourier space coverage for lensing is not perfectly
well-defined and
has been drawn semi-arbitrarily in Fig. \ref{fig:overlapslide} -- in any case,
it is clearly significantly narrower than the photo-$z$ band for relevant  
surveys). The take away point from Fig. \ref{fig:overlapslide} is that, even
if you locate lensing and redshift surveys on the same patch of sky, the bulk
of their modes, i.e., most of their constraining power, still do not really
overlap. In some sense, the idea that it is even possible for these surveys to
strongly overlap is an illusion. Redshift surveys get their
constraining power
from their fine radial resolution, which opens up a huge number of modes with
substantial radial $k$ component. Lensing surveys get their power from
believing that, being directly sensitive to mass rather than a biased tracer,
their signal is predictable down to much smaller scales, which opens up a large
number of transverse modes (remember that the transverse direction is two
dimensional, so Fig. \ref{fig:overlapslide} does not really do justice to the
large number of modes present in the extended lensing coverage). This
qualitative argument would certainly not be sufficient on its own, but it
suggests that one is facing an uphill battle to find value in overlap. From
this point of view the
full Fisher matrix results should not be too surprising.

There are other ideas for how overlapping photometric and redshift surveys
can be beneficial, mostly for helping to control problems with lensing, e.g., 
calibrating photo-$z$ systematics \cite{2008ApJ...684...88N}, or 
intrinsic alignments, or shear calibration errors 
\cite{2009ApJ...695..652B}.
These are all subject to the limited statistics of overlap, but we 
cannot be sure if they are important or not without detailed calculations 
(first one has to determine whether these are important or not to begin with,
and then, if they are, whether overlap fixes them in practice). 
For photo-$z$ error calibration, however, \cite{2013arXiv1308.6070D} found 
that the value was 
less than hoped for (in agreement with our own preliminary calculations). 

\end{document}